# Advanced Astroinformatics for Variable Star Classification

by

Kyle Burton Johnston

Master of Space Sciences
Department of Physics and Space Sciences
College of Science, Florida Institute of Technology
2006

Bachelor of Astrophysics
Department of Physics and Space Sciences
College of Science, Florida Institute of Technology
2004

A dissertation
submitted to Florida Institute of Technology
in partial fulfillment of the requirements
for the degree of

Doctorate of Philosophy
in
Space Sciences

Melbourne, Florida
April, 2019

We the undersigned committee hereby approve the attached dissertation
Title: Advanced Astroinformatics for Variable Star Classification
Author: Kyle Burton Johnston

---

Saida Caballero-Nieves, Ph.D.
Assistant Professor
Aerospace, Physics and Space Sciences
Committee Chair

---

Adrian M. Peter, Ph.D.
Associate Professor
Computer Engineering and Sciences
Outside Committee Member

---

Véronique Petit, Ph.D.
Assistant Professor
Physics and Astronomy

---

Eric Perlman, Ph.D.
Professor
Aerospace, Physics and Space Sciences

---

Daniel Batcheldor, Ph.D.
Professor and Department Head
Aerospace, Physics and Space Sciences

ABSTRACT

Title: Advanced Astroinformatics for Variable Star Classification

Author: Kyle Burton Johnston

Major Advisor: Saida Caballero-Nieves, Ph.D.

This project outlines the complete development of a variable star classification algorithm methodology. With the advent of Big-Data in astronomy, professional astronomers are left with the problem of how to manage large amounts of data, and how this deluge of information can be studied in order to improve our understanding of the universe. While our focus will be on the development of machine learning methodologies for the identification of variable star type based on light curve data and associated information, one of the goals of this work is the acknowledgment that the development of a true machine learning methodology must include not only study of what goes into the service (features, optimization methods) but a study on how we understand what comes out of the service (performance analysis). The complete development of a beginning-to-end system development strategy is presented as the following individual developments (simulation, training, feature extraction, detection, classification, and performance analysis). We propose that a complete machine learning strategy for use in the upcoming era of big data from the next generation of big telescopes, such as LSST, must consider this type of design integration.

# Table of Contents

# List of Figures

# List of Tables

# List of Symbols, Nomenclature or Abbreviations

| Abbreviations | |
|---|---|
| CCD | Charged Coupled Device |
| SDSS | Sloan Digital Sky Survey |
| 2MASS | Two Micron All-Sky Survey |
| DENIS | Deep Near Infrared Survey of the Southern Sky |
| VISTA | Visible and Infrared Survey Telescope for Astronomy |
| ASAS | All Sky Automated Survey |
| LINEAR | Lincoln Near-Earth Asteroid Research |
| Pan-STARRS | Panoramic Survey Telescope and Rapid Response System |
| LSST | Large Synoptic Survey Telescope |
| AAVSO | American Association of Variable Star Observers |
| IR | Infrared |
| RF | Radio Frequency |
| L-S Periodogram | Lomb-Scargle |
| LSC | LINEAR Supervised Classification |
| FFT | Fast Fourier Transform |
| SSMM | Slotted Symbolic Markov Modeling |
| OCD | O'Connell Effect Detector using Push-Pull Learning |
| DF | Distribution Fields |
| ADASS | Astronomical Data Analysis Software and Systems |
| OGLE | Optical Gravitational Lensing Experiment |
| CoRoT | Convection, Rotation and planetary Transits |
| SN | Super Novae |
| DSP | Digital Signal Processing |
| DFT | Discrete Fourier Transformation |

| Abbreviations | |
|---|---|
| k-NN | k Nearest Neighbor |
| PWC | Parzen Window Classifier |
| CART | Classification and Regression Tree |
| RBF-NN | Radial Basis Function - Neural Network |
| MLP | Multi-Layer Perceptron |
| SLP | Single-Layer Perceptron |
| RF | Random Forest |
| SVD | Singular Value Decomposition |
| LMNN | Large Margin Nearest Neighbors |
| PCA | Principle Component Analysis |
| MMC | Mahalanobis Metric Learning with Application for Clustering |
| NCA | Neighborhood Components Analysis |
| ITML | Information Theoretic Machine Learning |
| S&J | Schultz & Joachims Learning |
| EWMA | Exponentially Weighted Moving Average |
| CCA | Canonical Correlation Analysis |
| EDA | Exploratory Data Analysis |
| SVM | Support Vector Machines |
| KSVM | Kernel Support Vector Machines |
| ROC | Receiver Operating Characteristic Curve |
| PR | Precision - Recall Curve |
| AUC | Area Under the Curve |
| fp | False Positive |
| tp | True Positive |
| OC-SVM | One Class Support Vector Machines |
| OC-PWC | One Class Parzen Window Classifier |
| ANOVA | Analysis of Variance |
| PAA | Piecewise Aggregation Approximation |
| PLS2 | Partial Least Squares |
| UCR | University of California Riverside |
| CDF | Cumulative Distribution Function |
| QDA | Quadratic Discriminate Analysis |

| Abbreviations | |
|---|---|
| ECVA | Extended Canonical Variate Analysis |
| CVA | Canonical Variate Analysis |
| DWT | Discrete Wavelet Transformation |
| KELT | Kilodegree Extremely Little Telescope |
| OEEBs | O'Connell Effect Eclipsing Binaries |
| OER | O'Connell Effect Ratio |
| LCA | Light Curve Asymmetry |
| SNR | Signal to Noise Ratio |
| slf4j | Simple Logging Facade for Java |
| GCVS | general catalog of variable stars |

# Acknowledgements

The author is grateful for the valuable machine learning discussion with S. Wiechecki Vergara and R. Haber. Initial astroinformatics interest was provided by H. Oluseyi. Editing and review provided by G. Langhenry and H. Monteith. Custom scientific graphics were provided by S. Herndon. Inspiration provided by C. Role.

Research was partially supported by Perspecta, Inc. This material is based upon work supported by the NASA Florida Space Grant under 2018 Dissertation And Thesis Improvement Fellowship (No. 202379). The LINEAR program is sponsored by the National Aeronautics and Space Administration (NRA Nos. NNH09ZDA001N, 09-NEOO09-0010) and the United States Air Force under Air Force Contract FA8721-05-C-0002. This material is based upon work supported by the National Science Foundation under Grant No. CNS 09-23050. This research has made use of NASA's Astrophysics Data System.

# Dedication

I would like to thank all of those that supported me in my pursuit of this work and this goal.

First to my loving wife Caroline and son Edison who have been eternally patient, supportive and loving with me over all these years. Second, my parents and family members who have been with me time and time again, helping me along this journey; thank you for your love and support.

Thanks Dr. Saida Caballero-Nieves, Dr. Véronique Petit, and Dr. Adrian Peter for the support and encouragement. Thanks also to Dr. Stephen Wiechecki Vergara and Dr. Kevin Hutchenson who encouraged me to start this journey to begin with.

Thanks also to Dr. Cori Fletcher and Dr. Trisha Doyle, the best graduate school friends/support system I could have asked for. To Dr. Nicole Silvestri, the first astronomer I ever worked for, I dedicate Appendix D of this work.

To those reading, this work was meant to support yours; use it, expand on it, learn from it.

*The Cloths of Heaven*

"Had I the heaven's embroidered cloths, Enwrought with golden and silver light, The blue and the dim and the dark cloths Of night and light and the half-light; I would spread the cloths under your feet: But I, being poor, have only my dreams; I have spread my dreams under your feet; Tread softly because you tread on my dreams."

– W. B. Yeats

# Chapter 1

# Introduction

With the advent of digital astronomy, new benefits and new challenges have been presented to the modern-day astronomer. While data are captured in a more efficient and accurate manner using digital means, the efficiency of data retrieval has led to an overload of scientific data for processing and storage. That means that more stars, in more detail, are captured per night; but increasing data capture begets exponentially increasing data processing. Database management, digital signal processing, automated image reduction, and statistical analysis of data have all made their way to the forefront of tools for modern astronomy. This cross-disciplinary approach of leveraging statistical analysis and data mining methods to analyze astronomical data is often referred to as *astrostatistics* or *astroinformatics*.

The data captured by the modern astronomer can take on many forms but fall into two basic categories: observed, that is, resulting from the physical detection of either particles or waves emanating from an astrophysical source, or simulated, that is, resulting from computation or synthetic representation of a hypothetical astrophysical system. Observed data can be from almost any point on the electro-magnetic spectrum (photons), can be any emanation of particles (e.g., neutrinos), and, more recently, can even be resultant from more exotic emanations, such as gravitational waves. Within photon detection, we can further break down the categories into specific techniques: imaging, photometry, spectroscopy, polarimetry, and so on. These methods further analyze the character of the emanation be it either: the rate (flux) at which the photons are emitted, the energy of the individual photons, the relative distribution of photons emitted, or the phase orientation of the photons. Each emanation type–method pairing allows for the inferred measurement of certain associated properties of the astrophysical source of interest. This study focuses on optical time-domain analysis—the measurement of how radiation flux in the visible (∼390–700 nm) range changes over time. If these data, the amplitude of the measurement, do change over time, and the astrophysical source being observed is a stellar object, then we state that the source is a *variable star*.

The requirements for variability are fairly straightforward: the change in flux over time must exceed in amplitude any other variations from contributions that are not the source and must change at a rate that is large enough to be statistically noticeable. Thus a star might vary, but the amplitude of the variation might be negligible compared to background noise variation observed along with the star, or the resolution of the sensor (sensitivity) might be lower than what is needed to discern a change in amplitude, and therefore no variation would have been observed. Alternatively, a star might vary, but the change as a function of time might be so long as to be unnoticeable in the local time frame or too short compared to the sample rate of the observations, causing the star's amplitude to appear constant. To some degree, all stars have a flux output that is variable; much of what we categorize as "variability" is dependent on the equipment we are using to make observations, the received energy, and the structure of the variation itself.

## 1.1 Utility of Variable Star Automated Identification

It is precisely the dependence on equipment to detect variability that results in the pressing need for more advanced methods for time domain analysis. It is no surprise that one of the first variable stars, Omicron Ceti (Mira), was both bright and varied slowly; this allowed for discovery by eye of the variation [Wilk, 1996]. As more advanced methods of detection have become common, stars that vary faster, are dimmer, or have a smaller amplitude variation are all now discoverable. The improvement to detector efficiency allowed the necessary exposure rate to decrease while leaving the signal-to-noise ratio of the observation of the same star unchanged, thus increasing the sampling rate of the observation. Similarly, the economics of astronomical observations have become favorable to increasing the sampling rate, be it either the decreasing cost of detectors (CCDs) or the ability to create detectors of increasing size. Even telescope automation has had a hand in increasing the sampling rate, allowing for more sky to be observed, more frequently, without the added expense of having a human in the loop. Between being able to increase the sampling rate for all observations, the increase in image size, and the prevalence of larger detector optics, the increase in astronomical wide-field survey projects has resulted in an exponential increase in the number of stellar observations in general.

Surveys such as the Two Micron All-Sky Survey [2MASS, Cutri et al., 2003], the Deep Near Infrared Survey of the Southern Sky [DENIS, Epchtein et al., 1997], the Visible and Infrared Survey Telescope for Astronomy [VISTA, Sutherland et al., 2015], and the Sloan Digital Sky Survey [SDSS, Abazajian et al., 2009] attempt to image (photometry) a wide field in their respective frequency regions. While some surveys are designed for photometric depth, others attempt to observe variability, either in position or in brightness. The All Sky Automated Survey (ASAS) was designed to measure brightness changes specifically, while surveys like the Lincoln Near-Earth Asteroid Research (LINEAR) survey and the Catalina Sky Survey were originally designed to be used for near-Earth object tracking but have since been exploited for stellar variability detection.

This all-sky, time-domain data collection, in its extreme, can be best demonstrated by two surveys: the Panoramic Survey Telescope and Rapid Response System (Pan-STARRS) and the Large Synoptic Survey Telescope (LSST):

- Pan-STARRS is designed as a near-Earth asteroid detection survey. Each image taken requires approximately 2 GB of storage, with exposure times between 30 and 60 s and with an additional minute or so used for computer processing. Since images are taken on a continuous basis, the total data collection is roughly 10 TB the PS4 telescopes every night. The very large field of view of the telescope and the short exposure times enable approximately 6000 square degrees of sky to be imaged every night. Roughly the entire observable sky can be imaged in a period of 40 hours (or approximately 10 hours per night over 4 days). Given the need to avoid times when the Moon is bright, this means that an area equivalent to the entire sky is surveyed four times a month.

- LSST is designed specifically for astronomical observation and has a 3.2-gigapixel (GP) prime focus digital camera that will take a 15-s exposure every 20-s. Repointing such a large telescope (including settling time) within 5 s requires an exceptionally short and stiff structure. This in turn implies a very small f-number, which requires very precise focusing of the camera. The focal plane will be 64 cm in diameter and will include 189 CCD detectors, each of 16 megapixels (MP). Allowing for maintenance, bad weather, and other contingencies, the camera is expected to take more than 200,000 pictures (1.28 PB uncompressed) per year.

It is apparent from the estimates of per year/per night output from two of the most modern all-sky surveys that the rate of data output vastly exceeds the rate at which astronomical analysis can be performed by-hand. Seeing as how there is no indication that this trend of big data will change anytime soon, automated processing is a necessity if meaningful use is to be made of the data. This includes automated reduction (magnitudes, images, colors, etc.), association (matching objects across surveys), and interpretation of observed objects.

Beyond automated reduction being necessary for processing such a large data influx, automated reduction also lends itself to being more consistent and deterministic than having a human in the loop. Likewise, increasing scale requires improved error reduction for analysis methods to be justifiably efficient. **Our project here focuses on the automated categorization of variable stars and the ability to determine what kind of variable we are observing based on its time-domain signal and a prelabeled set of comparison data**. This was once a manual process; given a light curve, an expert in the field of stellar variability would compare by eye the data to known data sources and determine an estimated label. This sort of one-by-one analysis is still done as part of the development of training data for many of the supervised classification efforts currently in development and for all of the data used in this discussion.

## 1.2 Proposal for an Advanced Astroinformatics Design

In general, the combination of signal processing and classification results in the generic linear flow shown in Figure 1.1.

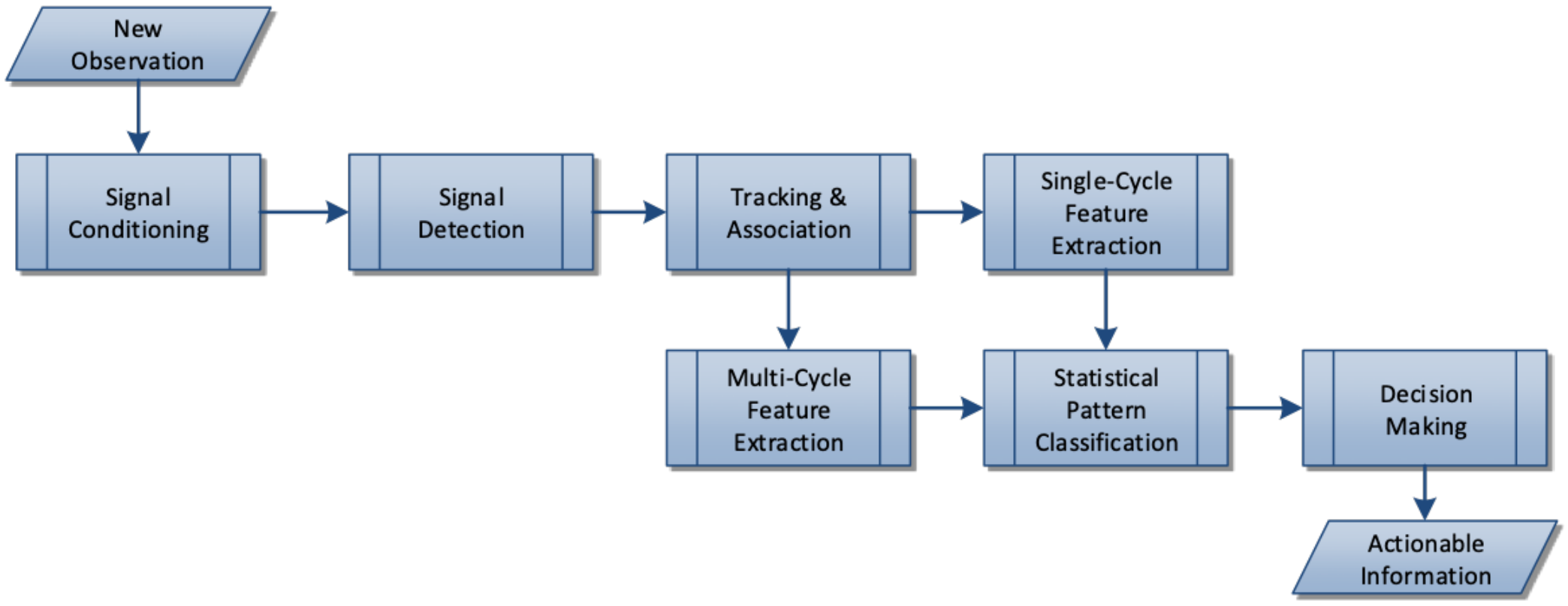

Figure 1.1: A standard system engineering design (flow chart) of an automated remote detection algorithm

The individual measurements of flux over time constitute the signal observed. Some form of signal conditioning can be applied (but need not be) to the signal; this is often an attempt to normalize the waveform or to reduce the noise via filtering. The measurements made can be transformed or mapped to new representations—features—which are optimal or of interest for a given goal. For consistency, we make the following definitions:

- The astrophysical object being measured is the source.

- The tool performing the measuring is the receiver.
- A path was taken by the photons to get from the source to the receiver.
- The signal is the measurement of the flux, and that signal measured, over time, results in the time-domain waveform.
- This collection of measurements operated on is referred to here as the feature space (e.g. Fourier domain, wavelet domain, time domain statistics, photometry, etc.).

Much of the initial effort is the disentanglement of the three functions (source, path, sensor) from one another to understand the source function. These challenges, however, are not unique to astronomy. This effort focuses on addressing four specific issues: class space definition, incomplete measurements, continuous signal and secondary information, and performance evaluation.

While the AAVSO keeps a catalog of variable star types [Samus' et al., 2017], this listing is dynamic. Variable star classifications have been added, removed, and updated since its inception [Percy, 2007]. These changes have included the discovery of new variables stars, new variable types, the determination that two classes are the same class, and the determination that one class is two different classes. This was the case with Delta Scuti stars [Sterken and Jaschek, 2005]: originally classified as RR Lyrae subtype RRs, they were eventually identified as their own class. Furthermore, a set of high-frequency Delta Scuti stars that were found to be metal poor were subdivided into an additional class now called SX Phoenicis stars.

We present Figure 1.2 for a high-level view of variability types: some variability classes are uniquely different, resulting from different physical processes (Beta Cep vs. Eclipsing Binaries, instability vs. occultation), some variability classes result from the same physical process but different periods. For example, Ceph vs. SX Phoenicis, both result from pulsations causes by He ionization, but different periods resulting from temperature and metallicity differs. Furthermore some variable classes result from similar physical processes but different underlying causes. For example both Beta Cep vs. Ceph pulsate resulting from instability but one from Fe ionization and the other He ionization (respectfully).

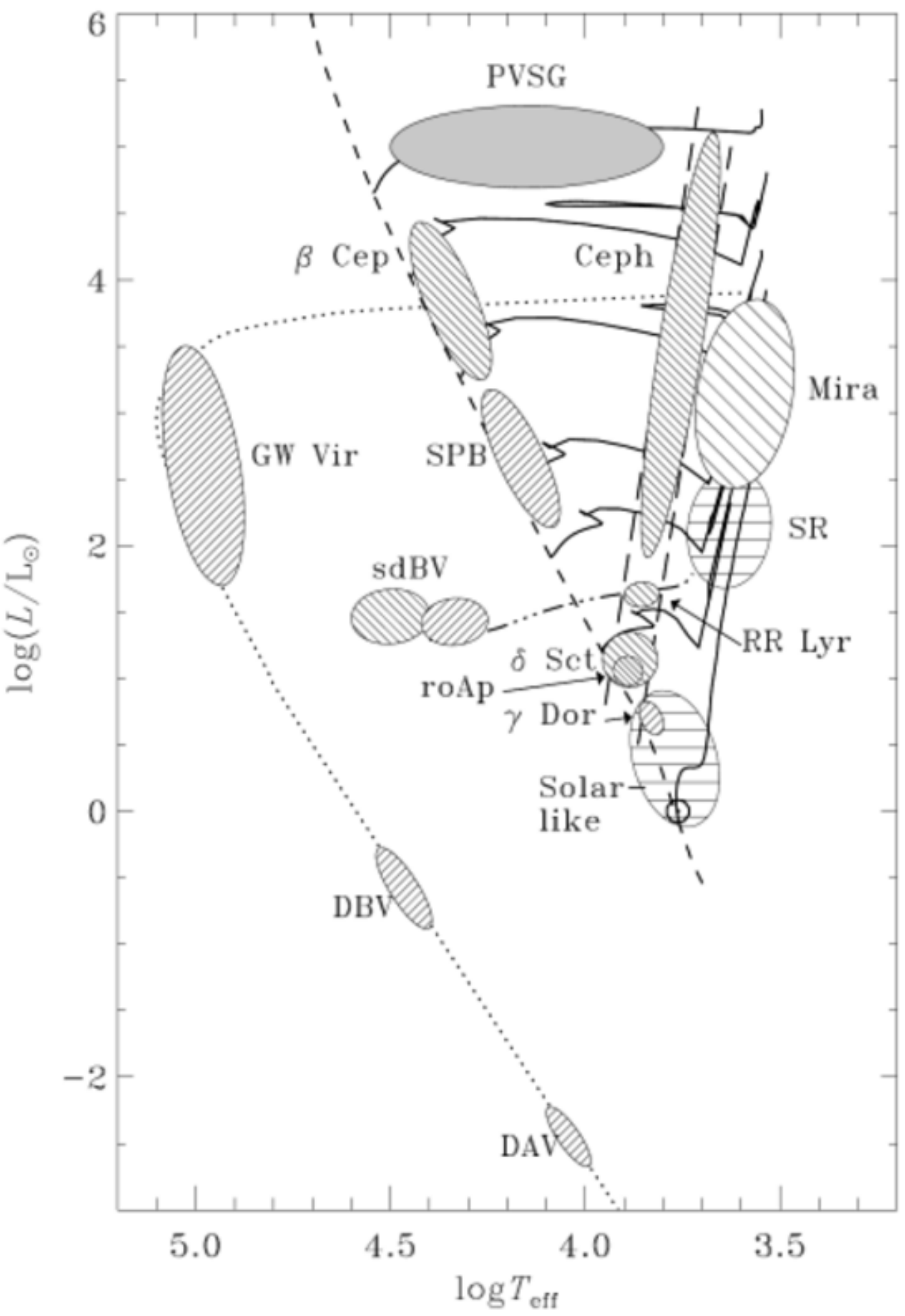

Figure 1.2: Variability in the Hertzsprung-Russell (HR) Diagram , A pulsation HR Diagram showing many classes of pulsating stars for which asteroseismology is possible [Kurtz, 2005].

In short, the definitions of variability are based on a mix of physical underlying parameter boundaries that have been empirically set by experts and on similar time-domain features. This results in ambiguity that can complicate the variable star classification problem. The expert definitions have an additional effect on the construction of the classifier: biases from observations affect the sampling of each individual classes. For example, brighter variables are more likely to be observed and therefore, without additional filtering, will be over-represented compared to other variables. This will result in a class imbalance [Longadge and Dongre, 2013] and will ultimately degrade the performance of the classifier, especially in the case of multiclass classifiers [see Johnston and Oluseyi, 2017]. How we evaluate the performance of classifiers and compare the classifiers knowing there is a class imbalance is of major concern.

Stellar variable time series data can roughly be described as passively observed time series snippets, extracted from what is a contiguous signal over multiple nights or sets of observations. The continuous nature of the time series provides both complications and opportunities for time series analysis. The time series signature has the potential to change over time, and new observations mean increased opportunity for an unstable signature over the long term. If the time signature does not change, then new observations will result in additive information that will be used to further define the signature function associated with the class. Implementing a methodology that will address both issues, potential for change and potential for additional information, would be beneficial. If the sampling were regular (and continuous), short-time Fourier transforms (spectrograms) or periodiograms would be ideal as they would be direct transforms from time to frequency domains. The data analyzed cannot necessarily be represented in Fourier space (perfectly), and while the wavelet version of the spectrogram or scalogram [Rioul and Vetterli, 1991, Szatmary et al., 1994, Bolós and Benítez, 2014] could be used, the data are often irregularly sampled, further complicating the analysis. Methods for obtaining regularly spaced samples from irregular samples are known [Broersen, 2009, Rehfeld and Kurths, 2014]; however, these methods have unforeseen effects on the frequency domain signature being extracted, thereby corrupting the signature pattern.

Astronomical time series data are also frequently irregular; that is, there is no associated fixed $\triangle t$ over the whole of the data that is consistent with the observation. Even when there is a consistent observation rate, this rate is often broken up because of a given observational plan, daylight interference, or weather-related constraints. Whatever classification method is used must be able to handle various irregular sampling rates and observational dropouts, without introducing biases and artifacts into the derived feature space that will be used for classification. Most analysis methods require regularized samples. Those that do not either require some form of transformation from irregular to regular sample rates by a defined methodology or apply some assumption about the time-domain function that generated the variation to begin with (such as a Lomb–Scargle periodogram). Irregular sampling solutions [Bos et al., 2002, Broersen, 2009] to address this problem can be defined in one of three ways:

- slotting methods, which model points along the timeline using fuzzy or hard models [Rehfeld et al., 2011, Rehfeld and Kurths, 2014];
- resampling estimators, which use interpolation to generate the missing points and obtain a consistent sample rate;
- Lomb–Scargle periodogram–like estimators that apply a model or basis function across the time series and maximize the coefficients of the basis function to find an accurate representation of the time series.

Astronomical surveys produce equally astronomical amounts of data. Processing these data is a time-intensive process, and not necessarily a repeatable one; while unsupervised classification will not be discussed as part of this analysis, and though we assume that the initial construction of training data requires a human in the loop, the act of using the training data needs to be automated. Not only does the effort need to leverage automated computational processing but said processing must have an error rate that is meaningful with respect to the scale and cost of the survey itself.

For example, if we have a survey of 100 stars, 10% of which are of a specific class type (e.g., type A), and we have a false alarm rate of 1%, it can be expected that from the original survey, ∼11 stars will be selected to be of interest, and of these 11 stars, 1 star is likely to be falsely identified. If a manual identification process takes 30 min to look at each star, then 5.5 hours later, a small team can confirm that there is one star that was inappropriately labeled, and at $7.25 an hour, the cost of reevaluation becomes ∼$40. If the scale of the survey is 1 million objects, then 110,000 objects will be identified as of interest, and of those, 10,000 objects will be false alarms. Using the cost estimates outlined, then, to review all stars identified as of interest given a standard 2080-hour work year would require 26.4 years and roughly $400,000 at minimum wage.

In this simplified example, only detection is considered; if the problem is total classification of all sources, the resources necessary to process the work manually become unmanageable. Thus, to resolve what is primarily a resource problem, supervised classification is necessary. Servers are faster and less expensive than humans-in-the-loop, and algorithms are standardizable and are manageable and defensible in terms of their decision-making processes, thus the particular task of consistency and repetition is well suited to automation. Likewise, performance estimates are quantifiable, and often, characteristics such as false alarms and missed detection rates are manageable. We present here an answer to the big data variable star astrophysics problem: a well-documented high-performance automated supervised classification algorithm tuned with cost in mind.

## 1.3 Outline of the Effort

This project focuses on the supervised machine learning domain as it applies to astrophysics. This research intends to construct a supervised classification system for variable star observations that is tailored to the unique challenges faced by the astrophysics community. To accomplish this will require a focus on an interdisciplinary approach, split between modern astrophysics and machine learning research (i.e., astrostatistics and astroinformatics).

We will be dealing with cases where the data we are using to train have been hand-reviewed by an expert. We assume that the additional information provided by the expert is correct; likewise, we make the assumption that the data themselves are not defective (e.g., photometry from two different stars listed as one); handling of mislabeled data is possible but is beyond the scope of this specific research effort. Given a new input observation, the algorithm shall generate an estimate of label, that is, of the type of variable star the new observation corresponds to, given training data.

We focus on everything else, following the trend established by the literature. In areas where novel research occurs, the culmination of the efforts results in an associated publication on the topic (and therefore the contribution). This work is

organized as follows:

1. We review stellar variability in chapter 2, signal processing, and machine learning in chapter 3. This establishes a baseline of necessary understanding for introducing the other key components.

2. In chapter 4, we discuss system design and performance of an automated classifier.

    (a) *Description.* We test various industry standard methods that have been published within the astroinformatics community, and provide improved performance analysis estimates geared toward the challenges faced by astronomers (e.g., class imbalance, large class space, low population representation, high variance in class pattern). This includes extending the supervised classification system into an anomaly detection algorithm, to be used in the discovery and identification of new, previously unobserved variable star representations.

    (b) *New developments.* Our novel contributions to the field included the testing of LINEAR data against multiple classifiers, one vs. all/multi-class classification performance comparison, detector performance quantification via ROC-AUC and PR-AUC, application of detectors to unlabeled LINEAR data

    (c) *Article.* The work in this chapter was published in Johnston, K. B., & Oluseyi, H. M. (2017). Generation of a supervised classification algorithm for time-series variable stars with an application to the LINEAR dataset. *New Astronomy, 52*, 35–47.

(d) *LSC*. The code developed in this chapter was published in Johnston [2018]. LSC (LINEAR Supervised Classification) trains a number of classifiers, including random forest and K-nearest neighbor, to classify variable stars and compares the results to determine which classifier is most successful. Written in R, the package includes anomaly detection code for testing the application of the selected classifier to new data, thus enabling the creation of highly reliable data sets of classified variable stars[1].

3. In chapter 5, we discuss novel feature space implementation.

   (a) *Description*. Looking beyond traditional time series feature extraction methodologies (FFT and light curve folding), we instead focus on time-invariant feature spaces and novel digital signal processing methods tailored to variable star observations. Ideally, these are feature spaces that are easily implemented for various survey conditions, rapid to compute given various sizes of observations, and easy to optimize in terms of the linear separability of the identified variable star class space.

   (b) *New developments*. Our novel contributions to the field included the SSMM, as well as the application of SSMM to LINEAR and UCR data.

   (c) *Article*. The work in this chapter was published in Johnston, K. B., & Peter, A. M. (2017). Variable star signature classification using slotted symbolic Markov modeling. *New Astronomy, 50*, 1–11. In addition, the work here was presented as the poster Johnston, K. B., & Peter, A.

[1] https://github.com/kjohnston82/LINEARSupervisedClassification

M. (2016). *Variable star signature classification using slotted symbolic Markov modeling.* Presented at AAS 227, Kissimmee, FL.

(d) *SSMM.* The code developed in this chapter was published in Johnston and Peter [2018]. SSMM (Slotted Symbolic Markov Modeling) reduces time-domain stellar variable observations to classify stellar variables. The method can be applied to both folded and unfolded data and does not require time warping for waveform alignment. Written in MATLAB, the performance of the supervised classification code is quantifiable and consistent, and the rate at which new data are processed is dependent only on the computational processing power available[2].

4. In chapter 6, we discuss a detector for O'Connell-type eclipsing Binaries (using metric learning and DF features).

   (a) *New developments.* Our novel contributions to the field included the development of O'Connell-type Eclipsing Binary detector based on Kepler data and detection of new O'Connell-type eclipsing Binaries to be used in defining variable star category (from LINEAR and Kepler datasets)

   (b) *Article.* The work in this chapter is to be published as Johnston, K.B., et al. (2019). A detection metric designed for O'Connell effect eclipsing binaries. *Computational Astrophysics and Cosmology.* In addition, the work here was presented as the poster Johnston, K. B., et al. (2018). *Learning a novel detection metric for the detection of O'Connell effect eclipsing binaries.* Presented at AAS 231, National Harbor, MD.

---

[2]https://github.com/kjohnston82/SSMM

(c) *OCD*. The code developed in this chapter was published in "O'Connell Effect Detector using Push-Pull Learning" Johnston, Kyle B.; Haber, Rana. OCD (O'Connell Effect Detector using Push-Pull Learning) detects eclipsing binaries that demonstrate the O'Connell Effect. This time-domain signature extraction methodology uses a supporting supervised pattern detection algorithm. The methodology maps stellar variable observations (time-domain data) to a new representation known as distribution fields (DF), the properties of which enable efficient handling of issues such as irregular sampling and multiple values per time instance. Using this representation, the code applies a metric learning technique directly on the DF space capable of specifically identifying the stars of interest; the metric is tuned on a set of labeled eclipsing binary data from the Kepler survey, targeting particular systems exhibiting the O'Connell effect. This code is useful for large-scale data volumes such as that expected from next generation telescopes such as LSST[3].

5. In chapter 7, we discuss multi-view classification of variable stars using metric learning.

    (a) *Description*. The processing and analysis of time series features via advanced classification means; introducing to the astrostatistics community improved methods beyond the current standard.

---

[3]https://github.com/kjohnston82/OCDetector

(b) *New developments.* Our novel contributions to the field included the development of a Large margin multi-view metric learning for matrix variates, Barzilai and Borwein [1988] for matrix data

(c) *Article.* The work in this chapter is to be published as Johnston, K. B., et al. (2019). Variable star classification using multi-view metric learning. *Monthly Notices of the Royal Astronomical Society.* In addition, the work here was presented as the poster Johnston, K. B., et al. (2018). *Variable star classification using multi-view metric learning.* Presented at ADASS Conference XXVIII, College Park, MD.

(d) *JVarStar.*The code developed in this chapter was published in Johnston et al. [2019]. Contains Java translations of code designed specifically for analysis and the supervised classification of variable stars[4].

We should note that much of the inspiration for this research comes from the dissertation of Debosscher [2009], "Automated Classification of Variable Stars: Application to the OGLE and CoRoT Databases"; this PhD dissertation has provided many of the current astroinformatics efforts with a baseline methodology and performance. Standard methods (classifiers) were implemented and compared based on standard feature spaces (specifically, frequency domain and "expert"-identified features). While we note (specifically in Johnston and Oluseyi [2017]) that there are some gaps in the implementation of the supervised classification design, overall, the template Debosscher proposes is sound. We seek here to extend much of the research Debosscher outlined as well as the associated papers resulting from that study [Angeloni et al., 2014, Dubath et al., 2011, Richards et al., 2011, 2012, Masci et al., 2014], not by replicating efforts, but by improving either the feature space in which variable stars are considered or the methodologies in developing supervised classification systems for the impending large surveys.

[4] https://github.com/kjohnston82/VariableStarAnalysis

The final design is to do the following: a user will be able to provide time-domain data (waveforms) to an algorithm; additionally, color/differential photometry may be provided. The algorithm will automatically process the waveform and associated information, with minimal interface and loss of data. The user will be able to train, cross-validate, and test a supervised classification system. Well-defined estimates of performance will be derived for the trained system, and the algorithm will, based on the trained data, automatically process observations and produce estimates of label/category on previously unlabeled data. One of the goals of this work is the acknowledgment that the development of a true machine learning methodology must include not only a study of what goes into the service (features, optimization methods) but a study of how we understand what comes out of the service (decision processing, performance analysis).

# Chapter 2

# Variable Stars

The goal for this effort is the classification of variable stars via analysis of raw photometric light curve data. For an informative breakdown of different types of variable stars, see Eyer and Blake [2005]; they categorize variability into extrinsic (something else is causing the variability) and intrinsic (the object is the source of the variability). Extrinsic sources include both asteroids (reflecting light) and stars (eclipsing, rotation/star spots, and the result of microlensing). Intrinsic sources include both active galactic nuclei (radio quiet and radio loud) and stars (eclipsing, eruptive, cataclysmic, pulsation, and secular).

This outline is helpful when considering what types of features are useful for differentiation between class types. For example, most of the types listed under "cataclysmic" are going to be impulsive, and stars in the pulsating category will have repetitive signatures that may or may not have a consistent frequency or event amplitude modulation. What is required for a comprehensive classification design is the selection of diverse features that can provide utility for a variety of targets of interest. The "tree of variability" attempts to categorize the stellar variable

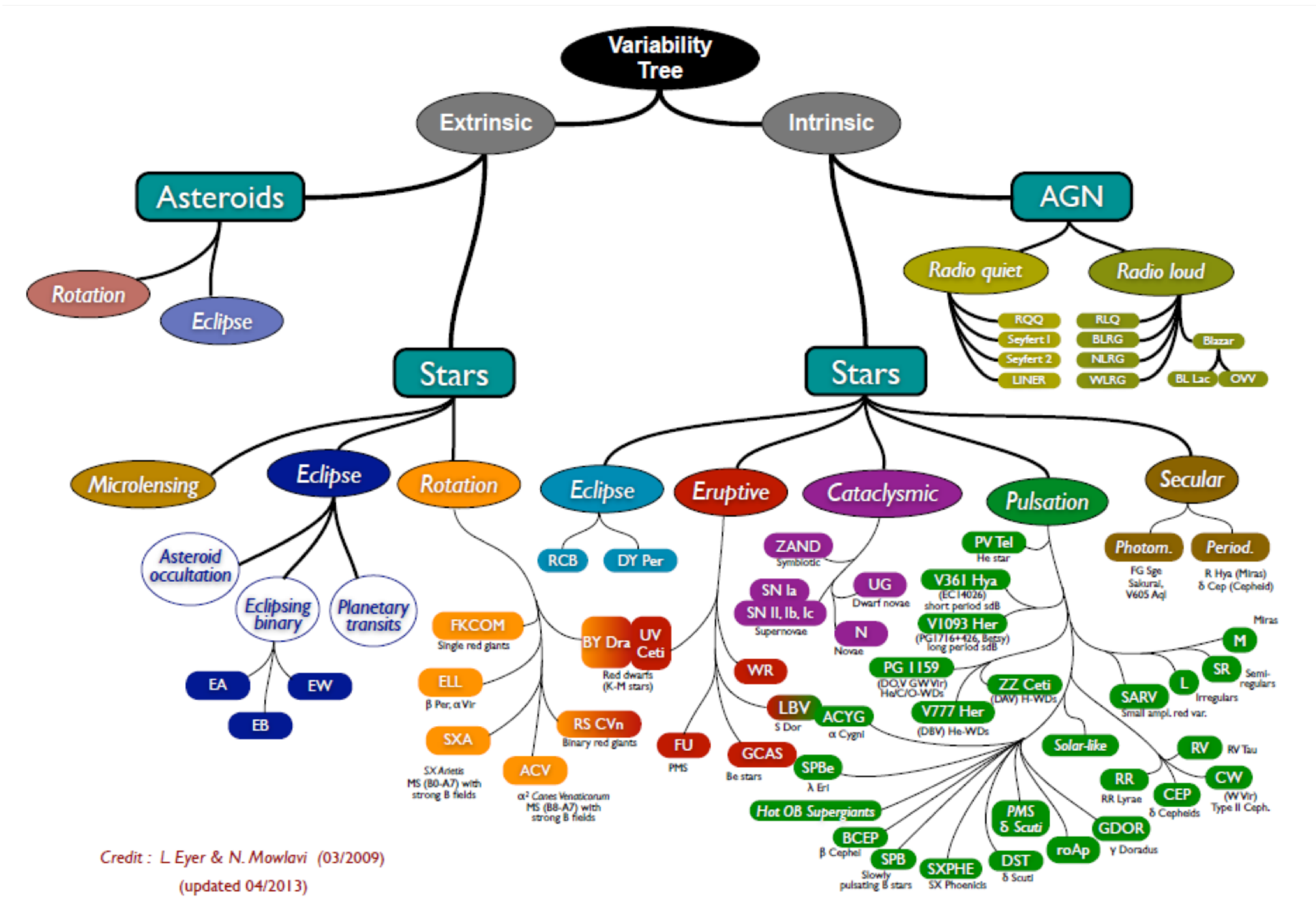

Figure 2.1: A graphical representation of the "Tree of Variability", i.e. the relationship between various astronomical variable types with respect to their physical cause [Eyer and Blake, 2005]

types based on physical cause of variation [Figure 2.1 Eyer and Blake, 2005, Gaia Collaboration et al., 2019]. We can also categorize the stellar variables from a signal point of view as a function of their signal type (see Figure 2.2), including the following:

- stationary processes: variation that is random but "stable" (the statistics of the time-domain signal are consistent at any point in time, e.g., the Sun)
- cyclostationary processes: the signal varies cyclically in time (e.g., RR Lyr, Eclipsing Binary/EB)
- impulsive process: signal variation is a sudden increase (or decrease) in signal, but the change does not cyclically occur (e.g., supernova)

- nonstationary: change of underlying time-domain statistics that are neither impulsive nor cyclic (e.g., Be star)

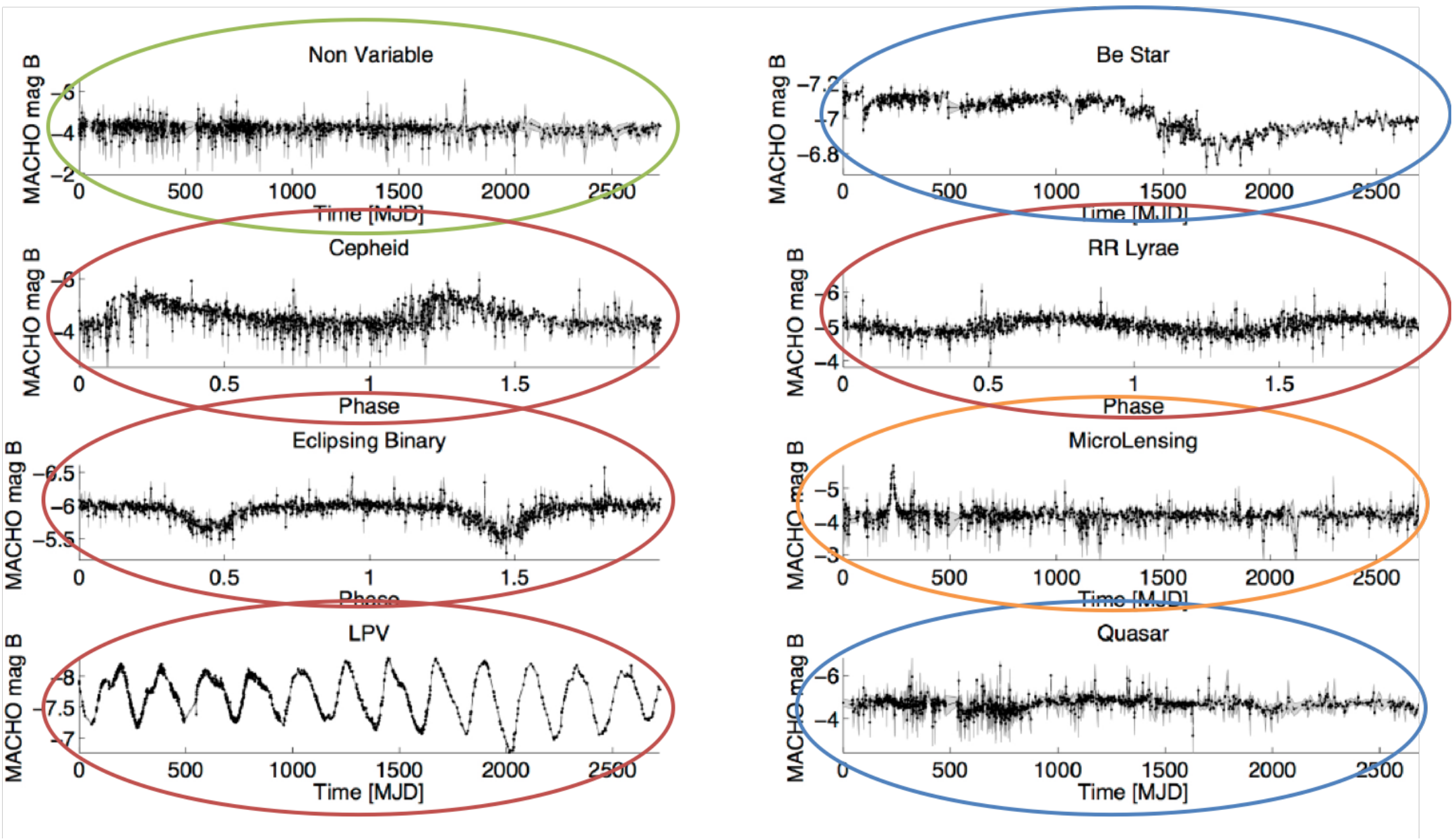

Figure 2.2: Statistical Signal Examples: Nonstationary Signal (blue), Stationary Signal (green), Cyclostationary (red), and Impulsive (orange) [Nun et al., 2014]

While our interest is in those stars that have cyclostationary signal processes, our analysis is applicable for the other processes as well (Super Novae/SN, for example). These definitions are local definitions, that is, based on the data observed. These are real, physical, evolving sources, and the change from one type to a different type is always possible given changes in the physical processes of the source or just given new observations. A valid question, however, with regard to the identification and specifically the grouping of different types of variable stars is, why do we care? One reason is the "general identification of types"—being able to say "these two are similar and these other two are dissimilar" has its own utility in gaining a deeper understanding of the universe. In addition to this, however, the variability of some stars has been found to be directly linked to other physical parameters (luminosity, stellar mass, rotation rate, etc.). This has been shown for cases such as eclipsing binaries, RR Lyrae, SN, and Beta Cephei. In these cases, the variation of luminosity observed can be tied back to some extrinsic or intrinsic value. Many variations result from changes to the star's atmosphere, what we will call "source variation." If we narrow down the tree of variability to just stars and just the most likely causes, we can highlight the following cases:

- *eruptive*: Eruptive variables are stars varying in brightness because of violent processes and flares occurring in their chromospheres and coronae. The light changes are usually accompanied by shell events or mass outflow in the form of stellar winds of variable intensity and/or by interaction with the surrounding interstellar medium.

- *pulsating*: Pulsating variables are stars showing periodic expansion and contraction of their surface layers. The pulsation may be radial or nonradial. A radially pulsating star remains spherical in shape, while in the case of nonradial pulsations, the star's shape periodically deviates from a sphere, and event neighboring zones of its surface may have opposite pulsation phases.

- *rotation*: For variable stars with nonuniform surface brightness and/or ellipsoidal shape whose variability is caused by axial rotation with respect to the observer, the nonuniformity of surface brightness distribution may be caused by the presence of spots. Nonuniformity could also by caused by some thermal or chemical inhomogeneity of the atmosphere caused by a magnetic field whose axis is not coincident with the rotation axis.

- *cataclysmic (explosive and novalike)*: This includes variable stars showing outbursts caused by thermonuclear burst processes in their surface layers or deep in their interiors.

- *eclipsing*: Some variations result from how the light gets from the star to the observer, that is, dimming resulting from the light being absorbed or occulted by another object, what we will call "path variation." Eclipsing binaries and eclipsing planetary systems are two types of variables that fall into this category and that are addressed in this study.

We will not discuss in additional detail all variable star subgroups, but we address here those variable types that are common to our studies, specifically eclipsing binaries and pulsating variables. Much of our decision to select specific variable types has been influenced on representation in surveys. The construction of a classifier when only a few (sometimes one) training data points are available is difficult at best and ill advised or wrong at worst.

## 2.1 Eclipsing Binaries and Contact Binaries

The multiplicity of stars is fairly well studied [Mathieu, 1994, Duchêne and Kraus, 2013]; a high number of stars exists in binary or greater systems. A statistical distribution of binary orbital separation is a current topic of survey research [Poveda et al., 2007], but an approximation of distribution for main sequence visual binaries as log of the semi-major axis based on Öpik [1924] is often cited. This assumed distribution is a gross approximation, and variations exist with respect to a star forming region that generated the binary and the stage of life that the binary is in.

A number of methods can be used to determine if a target star is in a multiple system, radial velocity measurements being one of the main ones. Alternatively, a companion star can be detected if the target star exhibits variability resulting from occultation, that is, the cooler of the pair (the companion) eclipsing the hotter of the pair (the primary). This will result in a cyclostationary periodic time series curve, and at a given primary period, the star will exhibit decreases in flux caused by the stars going in front of one another. These decreases are dependant on a number of factors: the relative size differences of the two stars, their differing effective temperatures, and the inclination of the binary with respect to the viewer.

If we assume no other photometric variability factors (flares, spots, etc.), we can establish some basic points of interest on the phased light curve of a binary star and will take Figure 2.4 as an example. For this discussion, we will assume the binary has a dimmer and brighter star and will refer to the brighter star as the primary and the dimmer as the secondary. At phase ∼0.15 and ∼0.65 are the two minima of the observation. They are unequal, although they need not be, and represent the primary behind the secondary and secondary in front of the primary, respectfully. The times between the minima, where the amplitude is roughly constant, represent those moments when the binary is observed and no visual occultation is occurring [Percy, 2007].

If we assume a circular orbit, we can roughly come up with some basic relations of how star size and brightness (flux times the effective area radiated) affect the binary. Let us consider the sky-projected image surface area of the stars to be $A_p$ and $A_s$ for the primary and secondary; likewise, we define the brightnesses of the stars to be $B_p$ and $B_s$. The total observed power radiated is then $\sim \sum A_i B_i$, and so if $x$ is the maximum observed flux, then $x = A_p B_p + A_s B_s$. Likewise, if we call the total received power at first and second minima $y$ and $z$, then $y = (A_p - A_s k_p)B_p + A_s B_s$ and $z = \max(0, A_s - A_p k_s)B_s + A_p B_p$, where $k_p$ and $k_s$ represent some fraction that allows for an incomplete occultation as a result of inclination. We can relate the differences in flux observed to physical relationships between the two stars in the binary via

$$x - z = A_s B_s - \max(0, A_s - A_p k_s)B_s \tag{2.1}$$

and

$$x - y = A_s k_p B_p; \tag{2.2}$$

thus we can define relationships between the relative depths of the eclipse and the physical properties (surface area and luminosity). This is what makes eclipsing binaries so interesting: they provide insight into the components of the system. Assuming a k of 1, the transition from maxima to minima is a linear relationship.

There are of course a myriad of ways, physically, that these assumptions could go wrong: partial eclipses (as discussed), limb darkening (from the stellar atmosphere), and other atmospheric effects (flares, reflection, spots, etc.) all can effect the relationships defined here and must be taken into consideration prior to analysis of the system properties.

Using limb darkening as an example: given a star with a meaningful stellar atmosphere, it is known that intensity with respect to an observer is not uniform over the observed surface. This can be shown for the simple case of an semi-infinite, radiation emergent, atmosphere. Following Hubeny and Mihalas [2014], the intensity for a given frequency can be shown as Equation 2.3

$$I_\nu(0,\mu) = \int_0^\infty S_\nu(t_\nu) \exp\left(-t_\nu/\mu\right) dt_\nu/\mu, \tag{2.3}$$

where $\mu$ is the angle of incidence ($\mu = cos(\theta)$), $\nu$ is the frequency of the light, S is the source function of the atmosphere, and $t_\nu$ is the optical depth. Assuming a grey atmosphere (opacity that does not vary with frequency) and one that is also in radiative equilibrium, Hubeny and Mihalas [2014] show that (their equation 17.4); $S(\tau) = J(\tau)$, the source function can be shown to be the the mean intensity. Using Hubeny and Mihalas [2014] equation 17.14, i.e. the Eddington approximation:

$$J(\tau) = 3H(\tau + \frac{2}{3}), \tag{2.4}$$

equation 2.3 becomes:

$$I(0,\mu) = 3H(\tau + \frac{2}{3}); \tag{2.5}$$

and the limb-darkening function (the ratio of intensity of the star with respect to intensity of the center of the star), can be shown to be:

$$I(0,\mu)/I(0,1) = \frac{3}{5}(\mu + \frac{2}{3}). \tag{2.6}$$

Given our prior case of $k = 1$, i.e. in-plane occultation, we can see that the decrease in intensity will no longer be a linear function, but instead dependent on $\mu = cos(\theta)$, resulting in curvature in the light curve.

The general catalog of variable stars [GCVS, Samus' et al., 2017] identifies three, traditional, categories of eclipsing binaries: EA, EB, and EW. These represent a gradient of how close the binaries are and more importantly how the Roche lobes have been filled, thus affecting the light curve shape. Roche lobes are defined as the region around a star in which the stellar mass is still gravitationally bound. When stars expand they can reach sizes that exceed this limit resulting in matter, in the binary case, transferring from one star to the other. Additionally, RS Canum is also identified as a subtype but will not be addressed here.

### 2.1.1 $\beta$ Persei (EA, Detached Binaries)

Like many variable star types, these are named after their prototype, Algol ($\beta$ Persei). Algol's variability has been known since the late 1700s; it represents the set of binaries known as semidetached eclipsing variables. Variability is consistent—there is little mass transfer resulting in very little effect on the orbital parameters. Likewise, both components are usually nearly circular in shape. Figure 2.3 provides an overview of the EA binary example, the top line is a graph of the light curve as well as the expected binary orientation with respect to the observer (represented as the eye); the bottom left and right figures are diagrams showing the expected mass transfer relationship, here specifically no-mass transfer is expected. This may not be the case when the secondary is much dimmer/cooler than the primary and therefore contributes little to nothing to the overall light curve, in which case the secondary might be highly distorted.

These systems are not limited to being associated with a given evolutionary stage [Sterken and Jaschek, 2005]; various compositions include two main sequence stars (CM Lac), two evolved components with no Roche lobe overflow (AR Lac), one evolved and one overflowing (RZ Cas), and one evolved and one not evolved (V 1379 Aql). An example is shown in Figure 2.4.

The variable class is identified by the constant maxima as well as by the sharp peak minima. The minima need not be the same depth, however, it is possible. Periodicity can range from fractions of days to multiple years.

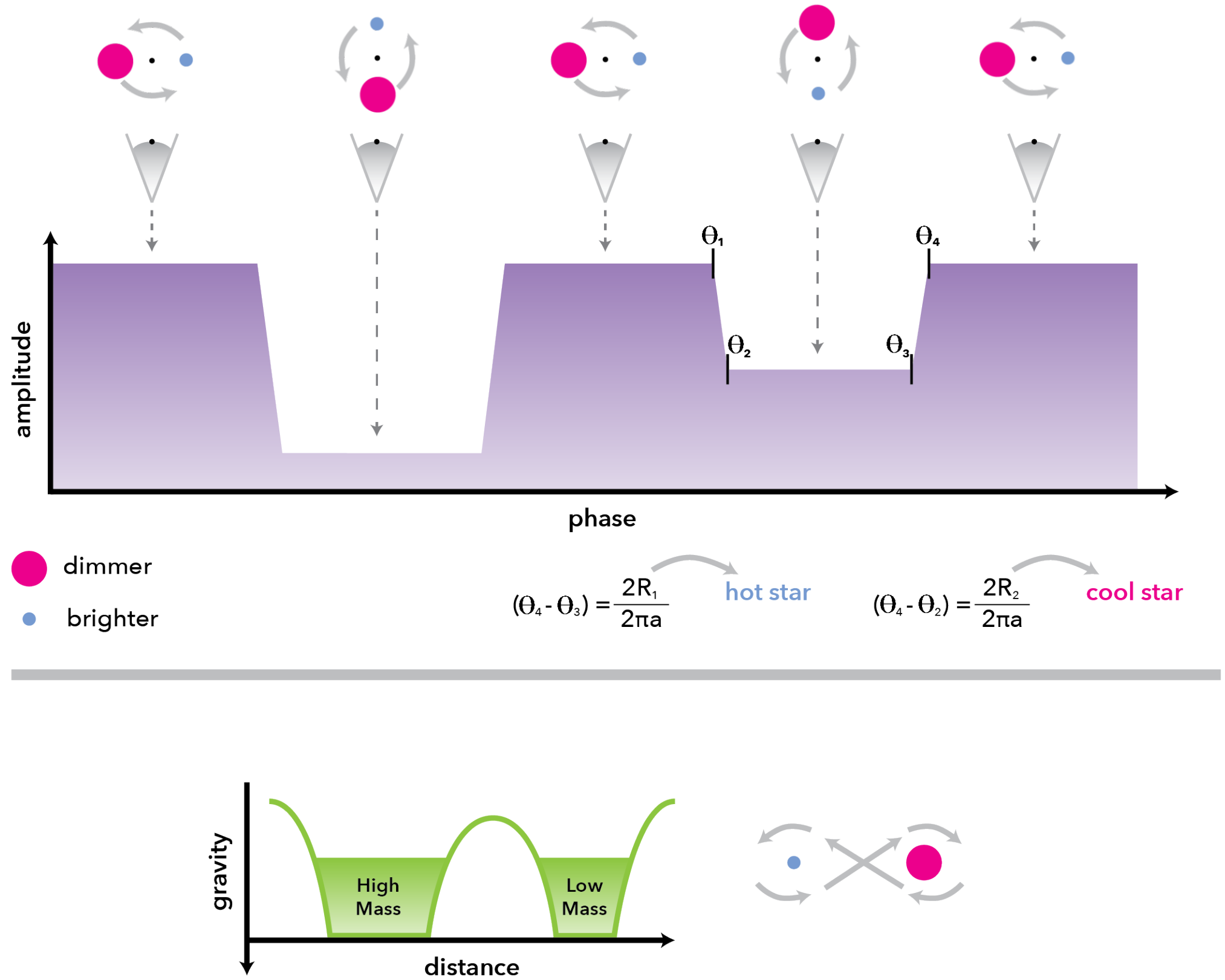

Figure 2.3: Conceptual overview of $\beta$ Persei—EA type—eclipsing binaries (Detached Binaries). Top figure is a representation of the binary orientation with respect to a viewer and the resulting observed light curve. Bottom left: Representation of the the positions of the gravitational potential well for the two components in a binary. Bottom right: depicts the Roche lobe envelope of the system.

### 2.1.2 $\beta$ Lyrae (EB, Semi-Detached Binary)

Interestingly, EB binaries are not a consistent population. Identified by the smooth transition from maximum to minimum, with uneven minima, the causes of these binaries vary. Some of these targets have highly eccentric orbits; others have varying degrees of Roche lobe filling; and others have mismatched evolved pairs. Figure 2.5 provides an overview of the EB binary example, the top line is a graph of the light curve as well as the expected binary orientation with respect to the observer (represented as the eye); the bottom left and right figures are diagrams showing the expected mass transfer relationship, here specifically the larger star has filled its' Roche lobe, and begun to transfer mass to the companion.

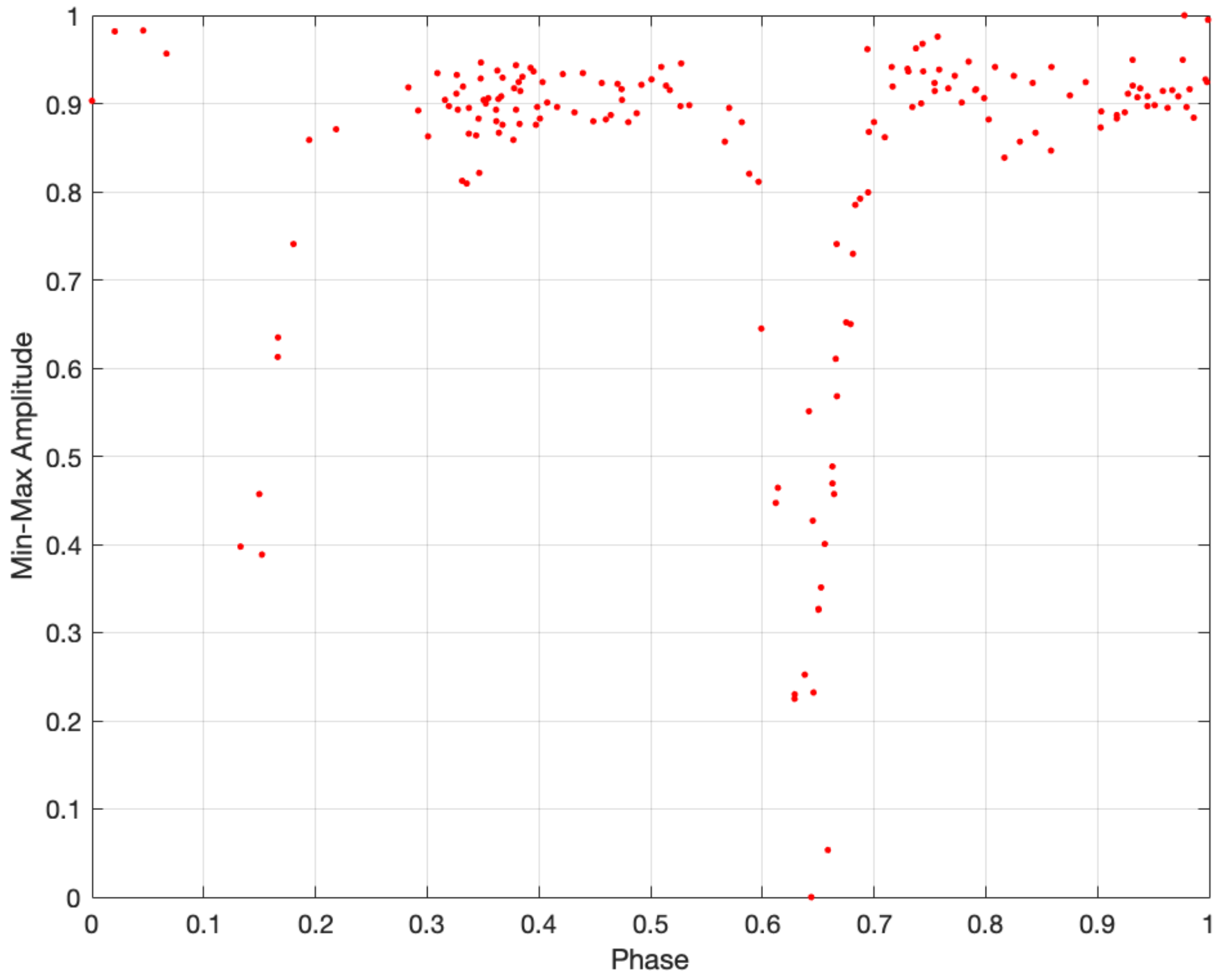

Figure 2.4: Example $\beta$ Persei (EA, Detached Binaries) Phased Time Domain Curve, light curve generated from LINEAR data (LINEAR ID: 10013411).

In short, the archaic nature of the classification system has lumped together a number of stars that look similar but on further inspection have a wide variety of associated physical parameters. We will not enumerate examples here of these various similar cases; however, we do provide an example of a EB type binary in Figure 2.6.

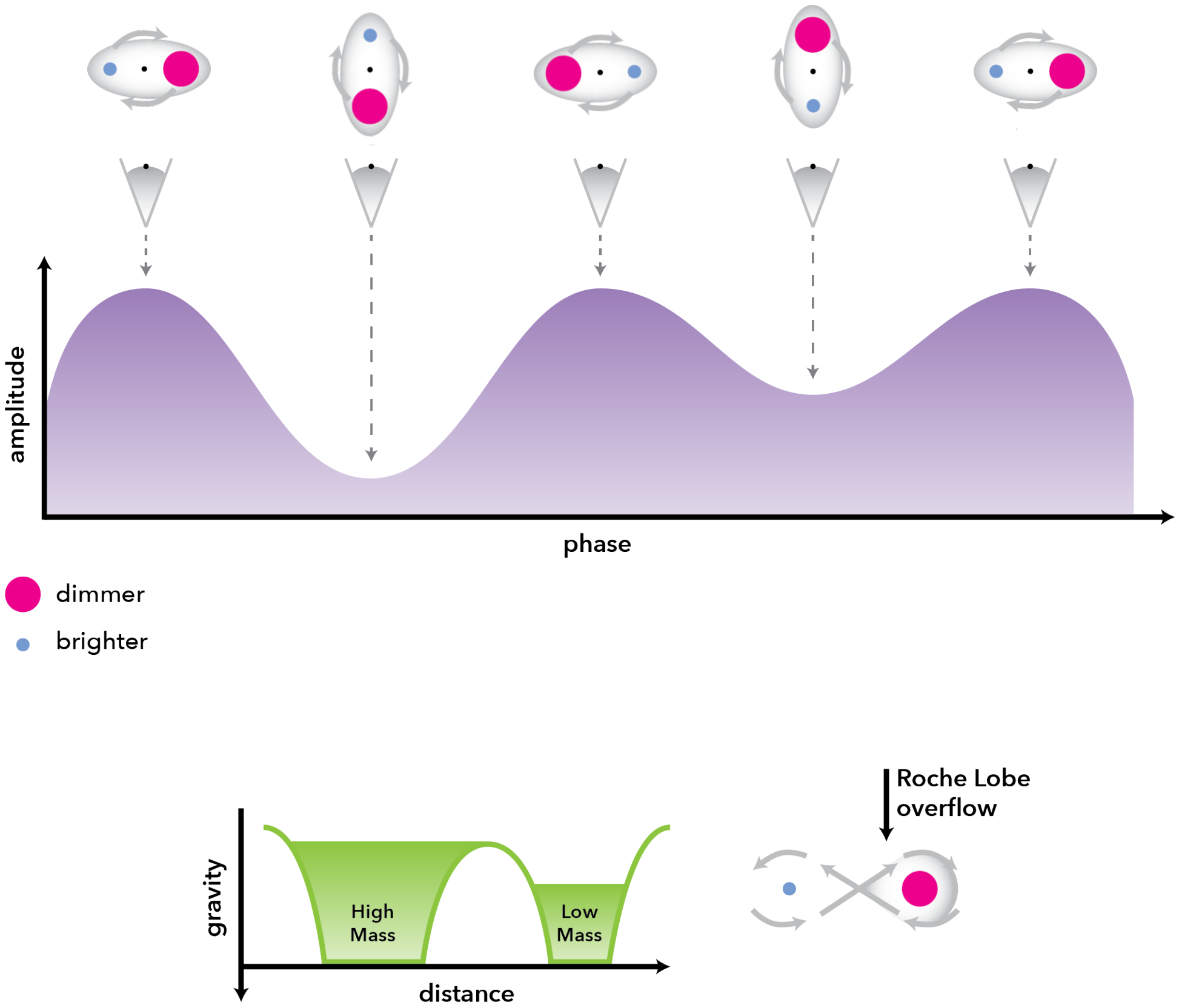

Figure 2.5: Conceptual overview of $\beta$ Lyrae—EB type—eclipsing binaries (Semi-Detached Binaries). Top figure is a representation of the binary orientation with respect to a viewer and the resulting observed light curve. Bottom left: Representation of the the positions of the gravitational potential well for the two components in a binary. Bottom right: depicts the Roche lobe envelope of the system.

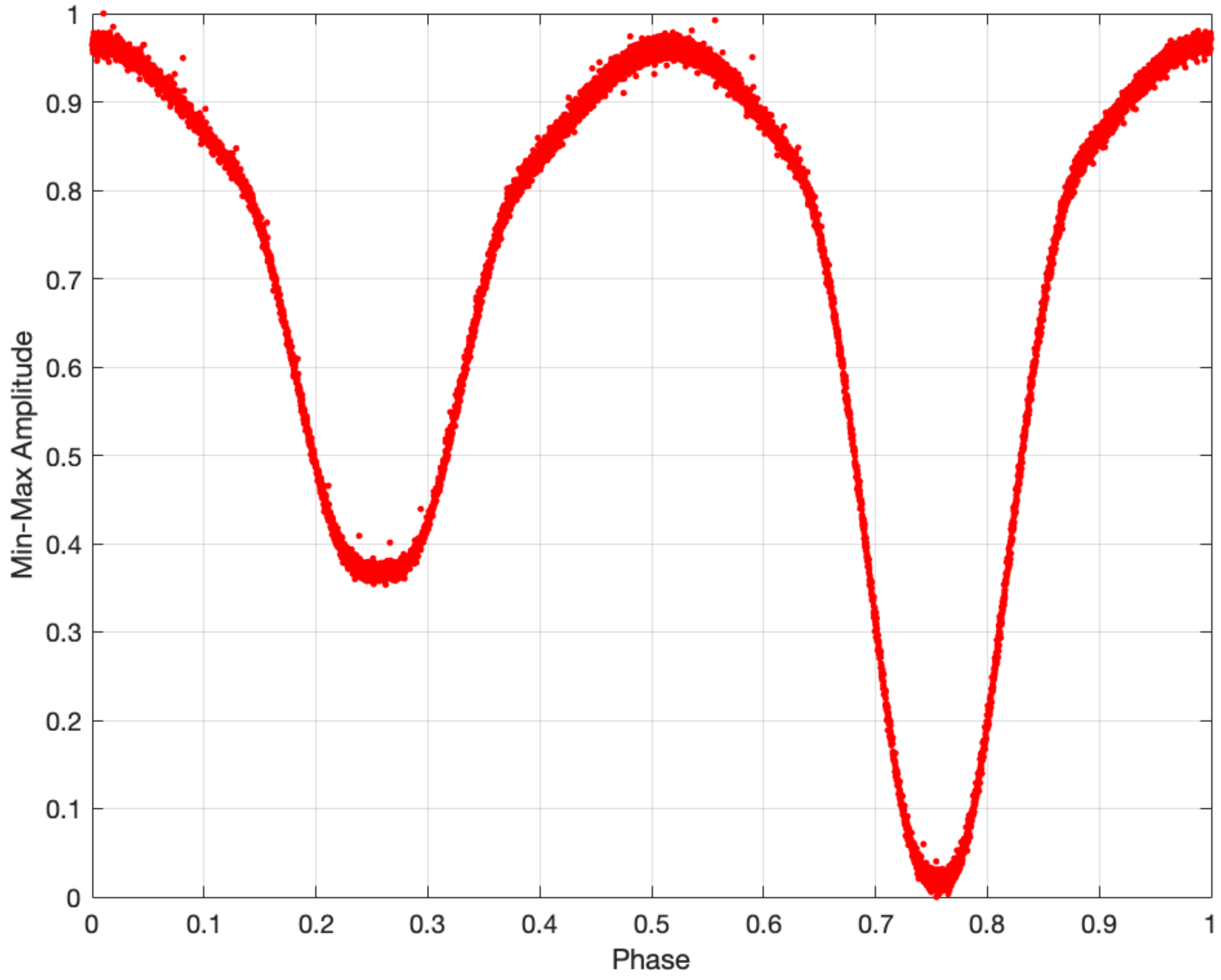

Figure 2.6: Example $\beta$ Lyrae (EB, Semi-Detached Binary) Phased Time Domain Curve, light curve generated from Kepler data (KIC: 02719436)

### 2.1.3 W Ursae Majoris (EW, Contact Binary)

As opposed to the semidetached binaries, contact binaries are those systems where the components fill their Roche lobes. Mass transfer does occur, and because of their closeness in proximity, their periods are relatively short (~0.2–0.8 days). Figure 2.7 provides an overview of the EW binary example, the top line is a graph of the light curve as well as the expected binary orientation with respect to the observer (represented as the eye); the bottom left and right figures are diagrams showing the expected mass transfer relationship, here specifically two stars share a common atmosphere.

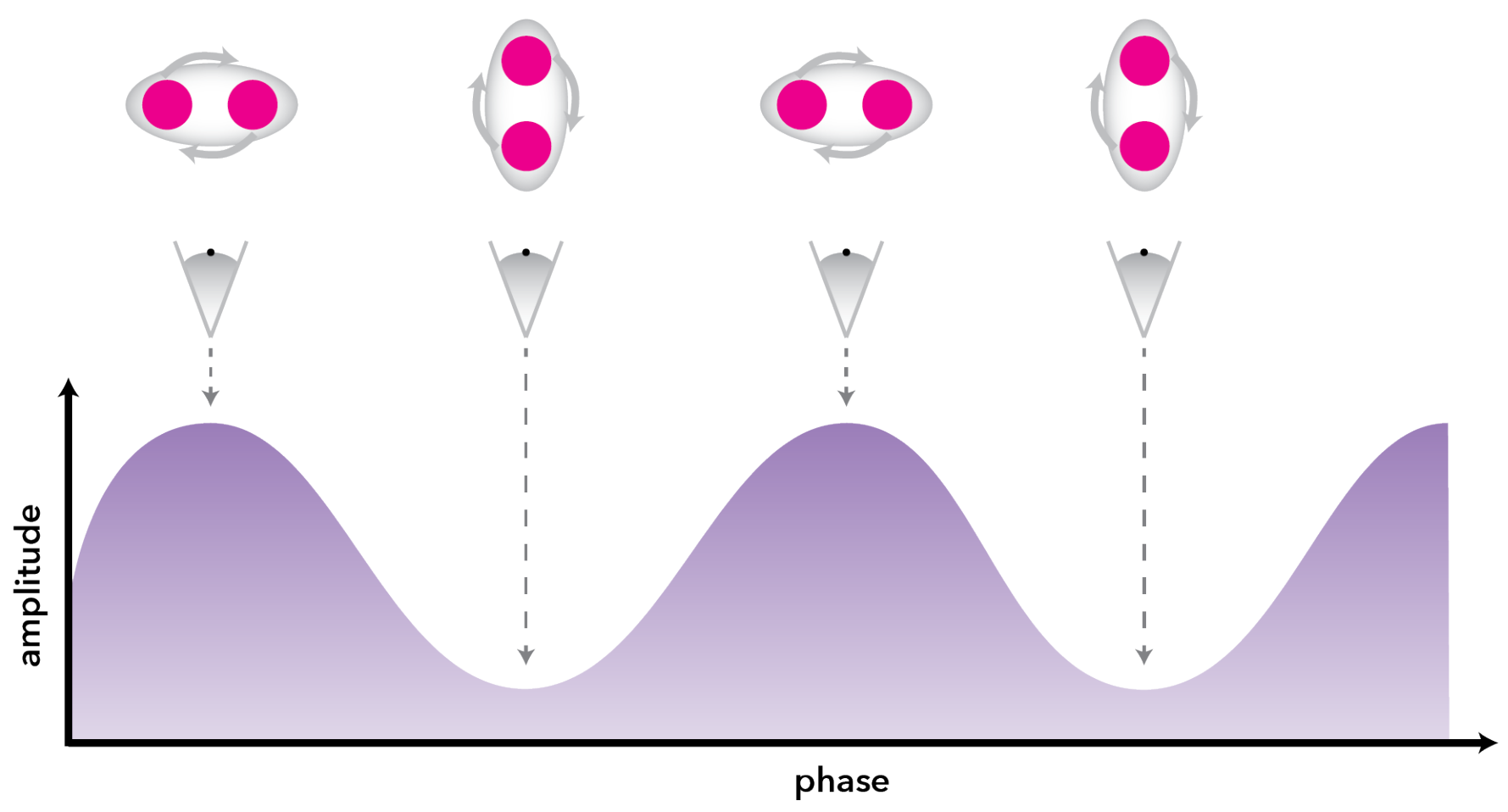

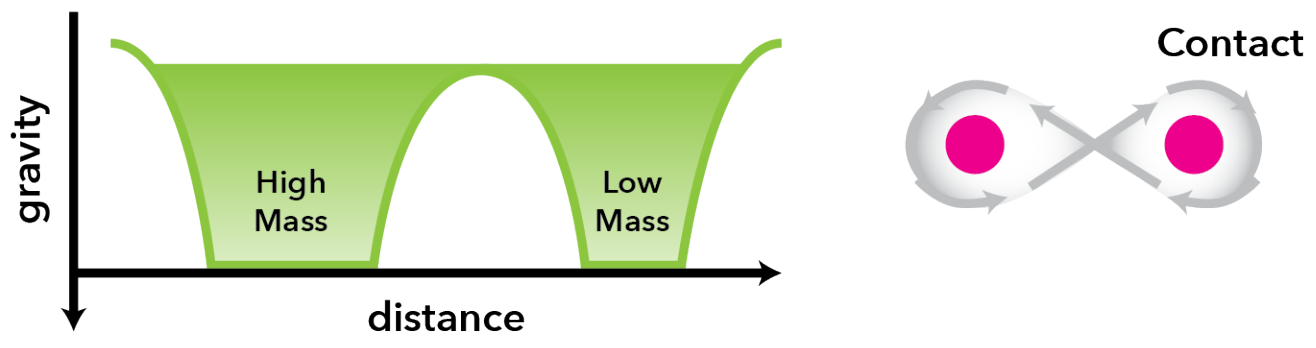

Figure 2.7: Overview of EW, Contact Binary. Top figure is a representation of the binary orientation with respect to a viewer and the resulting observed light curve. Bottom left: Representation of the the positions of the gravitational potential well for the two components in a binary. Bottom right: depicts the Roche lobe envelope of the system.

Because the matter between the stars is effectively shared, their light curves are marked by a smooth entrance and exit out of both minima and maxima [Terrell et al., 2012] and by very similar primary and secondary minima (equalization of effective temperature from mixing) (see Figure 2.8). The components are usually dwarf stars (F-G or later), and have shorter periods compared to EA and EB type binaries.

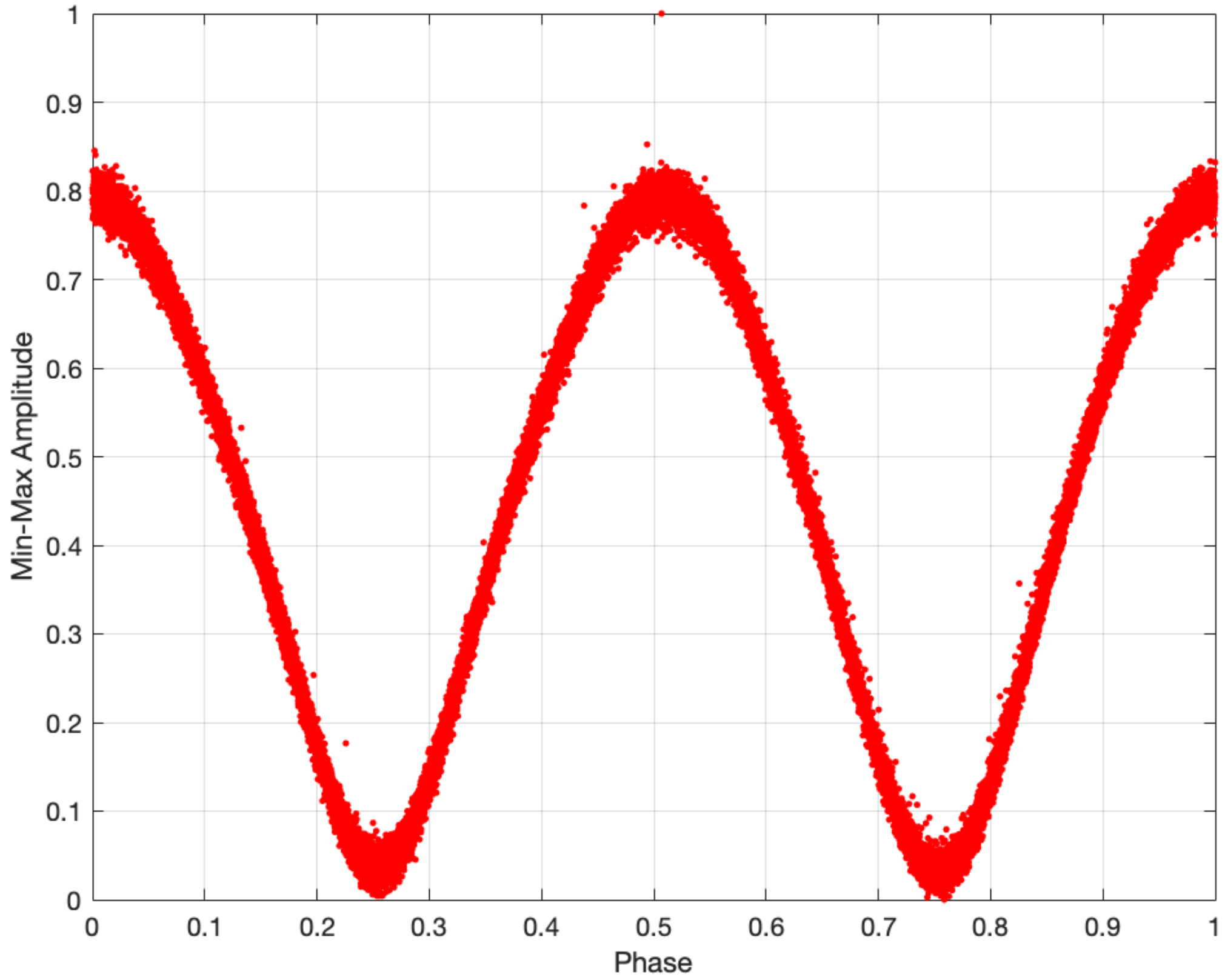

Figure 2.8: Example W Ursae Majoris (EW, Contact Binaries) Phased Time Domain Curve, light curve generated from Kepler data (KIC: 6024572).

## 2.2 Pulsating Variables

The term *pulsating variable* encompasses a large swath of variable subgroups. For our purposes here, we focus on those stars that lie along the instability strip (Cepheid-like). While stars are normally in hydrostatic equilibrium, there are cases where the radiant energy generated by the star's core is converted into stored energy via ionization in a He shell (usually). The mechanism of energy generation, storage, cooling, and heating to generation results in a cycle of radial expansion and contraction. This instability mechanism in stars is a result of an partial ioniza-tion zone; this ionization zone allows for the storage of energy, causing the cyclical process.

We can describe underlying dynamics of the mass shell as follows [Padmanabhan, 2001]:

$$m\ddot{r} = 4\pi r^2 P - \frac{GMm}{r^2}. \tag{2.7}$$

When in equilibrium, $m\ddot{r} = 0$, and $r = r_{\text{eq}}$. We can adiabatically perturb the shell about this equilibrium value to describe small changes:

$$m\frac{d^2\delta r}{dt^2} = 4\pi r_{\text{eq}}^2 P_{\text{eq}} \left(\frac{2\delta r}{r} + \frac{\delta P}{P}\right) + \frac{GMm}{r_{\text{eq}}^2}\left(\frac{2\delta r}{r}\right); \tag{2.8}$$

this defines the radial displacement of the shell. Furthermore, we can substitute the pressure function for density using the adiabatic relationship $\left(\frac{\delta P}{P}\right) = \gamma\left(\frac{\delta\rho}{\rho}\right) = -3\gamma\left(\frac{\delta r}{r}\right)$, where $\rho r^3 = c$ based on mass conservation. Substituting into equation 2.8, we find:

$$\frac{d^2\delta r}{dt^2} = \frac{GM}{r_{\text{eq}}^3}\left(3\gamma - 4\right)\delta r = \omega^2 \delta r. \tag{2.9}$$

Thus the radius of the shell is oscillatory about the equilibrium radius, with angular frequency $\omega^2 = \frac{GM}{r_{\text{eq}}^3}\left(3\gamma - 4\right)$. Revisiting the mass shell dynamics, we can reframe equation 2.7 as

$$\ddot{r} = -\frac{Gm}{r^2} - 4\pi r^2 \frac{\partial P}{\partial m}. \tag{2.10}$$

Similar to how we perturbed the radius, we can perturb the other underlying physical parameters ($P$ and $\rho$), that is,

$$P(m,t) = P_0(m) + P_1(m,t) = P_0(m)\left[1 + p(m)e^{i\omega t}\right], \tag{2.11}$$

$$r(m,t) = r_0(m) + r_1(m,t) = r_0(m)\left[1 + x(m)e^{i\omega t}\right], \tag{2.12}$$

$$\rho(m,t) = \rho_0(m) + \rho_1(m,t) = \rho_0(m)\left[1 + d(m)e^{i\omega t}\right] \tag{2.13}$$

Linearizing the perturbation and substituting into equation 2.10, we get

$$\frac{P_0}{\rho_0}\frac{\partial p}{\partial r_0} = \omega^2 r_o x + g_0\left(p + 4x\right), \tag{2.14}$$

where $g_0 = \left(\frac{Gm}{r_0^2}\right)$; similarly, via linearization of $\left(\frac{\partial r}{\partial m}\right) = \left(\frac{1}{4\pi r^2 \rho}\right)$, we get

$$r_0\frac{\partial x}{\partial r_0} = -3x - d. \tag{2.15}$$

Given the adiabatic relationship $p = \gamma d$, we can write equation 2.15 as

$$p = -3\gamma x - \gamma r_0\frac{\partial x}{\partial r_0}. \tag{2.16}$$

Based on these differential equations, we can establish the governing equation as

$$\frac{\partial}{\partial r_0}\left(\gamma\frac{\partial x}{\partial r_0}\right) + \frac{4}{r_0}\frac{\partial}{\partial r_0}\left(\gamma x\right) - \frac{\rho_0 g_0}{P_0}\gamma\frac{\partial x}{\partial r_0} + \frac{\rho_0}{P_0}\left[\frac{g_0}{r_0}\left(4 - 3\gamma\right) + \omega^2\right]x = 0. \tag{2.17}$$

Rearranging equation 2.17 and rewriting it in a second-order differential equation form, we get equation 2.18:

$$x" + \left(\frac{4}{r_0} - \frac{\rho_0 g_0}{P_0}\right)x' + \frac{\rho_0}{\gamma P_0}\left[\omega^2 + \left(4 - 3\gamma\right)\frac{g_0}{r_0}\right]x = 0. \tag{2.18}$$

We can change the independent variable $r_0$ to $z = Ar_0$, define the polytropic equation of state as $P = K\rho^{(n+1)/n}$, and give the gravitational potential as $\Phi$. Therefore

$$\begin{aligned} g_0 = \tfrac{\partial \Phi_0}{\partial r_0} &= A\Phi \tfrac{dw}{dz}, \qquad & A^2 &= \tfrac{4\pi G}{[(n+1)K]^n} \left(-\Phi\right)^{n-1}, \\ \rho_0 &= \left[\tfrac{-\Phi w}{(n+1)K}\right]^n, \qquad & \tfrac{\rho_0}{P_0} &= \tfrac{1}{K}\rho^{-1/n} = -\tfrac{n+1}{\Phi w}. \end{aligned} \tag{2.19}$$

We can substitute the definitions in equation 2.19 into our second-order equations to get

$$\frac{d^2x}{dz^2} + \left(\frac{4}{z} + \frac{n+1}{w}\frac{dw}{dz}\right)\frac{dx}{dz} + \left[\Omega^2 - \frac{(4-3\gamma)(n+1)}{\gamma}\frac{1}{z}\frac{dw}{dz}\right]\frac{x}{w} = 0, \tag{2.20}$$

where $\Omega^2 = -\frac{n+1}{4\pi G\gamma\rho_c}\omega^2$ is a dimensionless frequency for the polytrop and $\rho_c$ is the central density. The period of oscillation can then be given as

$$\Pi\sqrt{\bar{\rho}} = \left[\frac{(n+1)\pi}{\Omega^2 G\gamma}\left(\frac{\bar{\rho}}{\rho_c}\right)_n\right]^{1/2}. \tag{2.21}$$

Thus, for a fixed mode of oscillation, the period is dependent on density. This allows the estimation of relationships between period, luminosity, color, and mass [Iben, 1971, Freedman et al., 1994, Alcock et al., 1998] for radially pulsating stars. Beyond the He instability strip members, there are pulsating groups, such as those along the iron ionization region [e.g., Beta Cep, SPB; Iglesias et al., 1987], as well as stars along the white dwarf evolutionary curve that experience pulsations (e.g., DOV, DBV, PNNV). For more information, we recommend looking at Percy [2007].

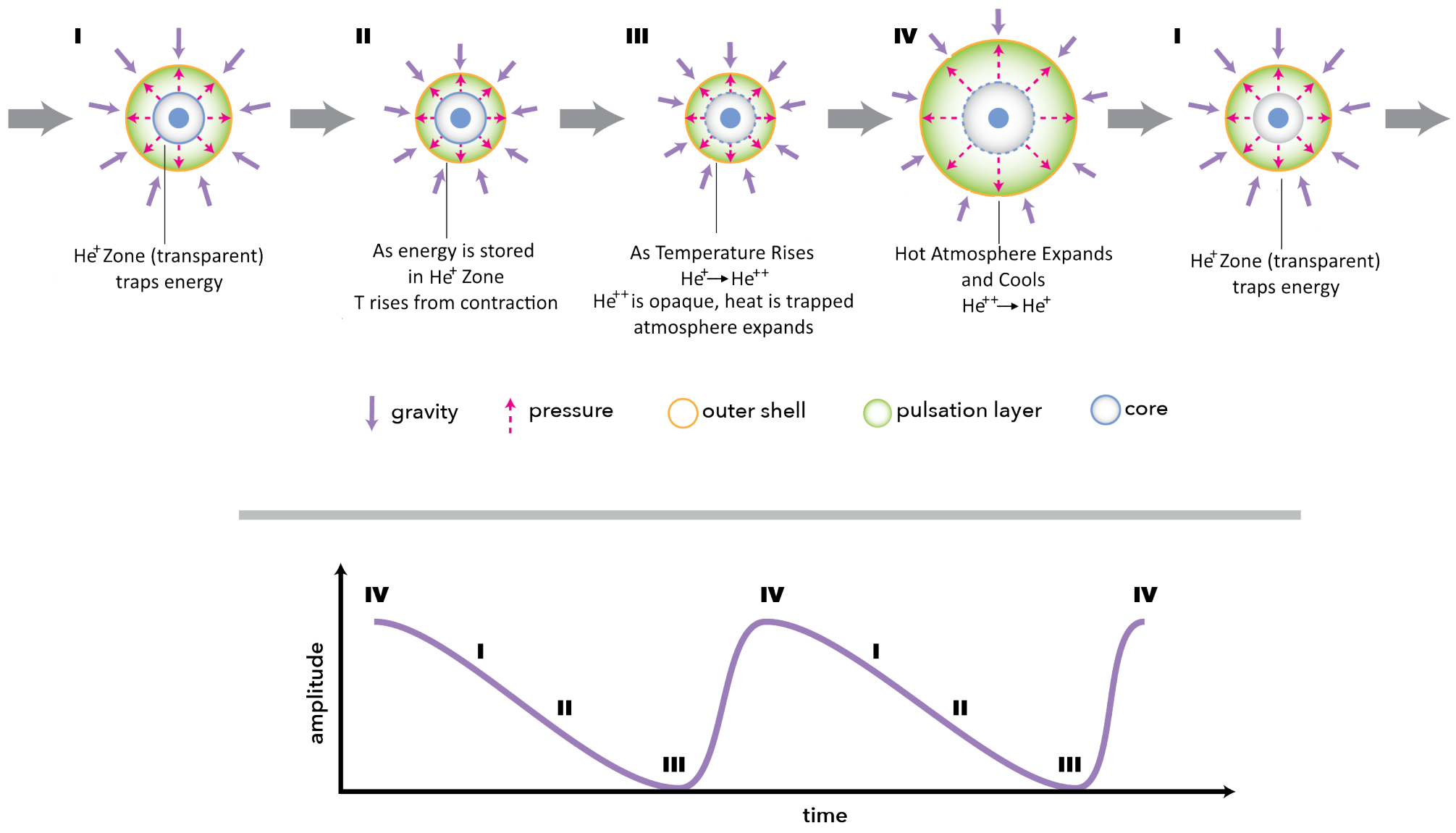

Figure 2.9: Conceptual representation of the cepheid instability strip cycle. Top figure: The cycle of energy absorption, atmospheric heating, atmospheric expansion, and energy release (cyclic process), Bottom figure: the resulting light curve caused by the cyclic process.

The cyclic–sawtooth light curve (Figure 2.9) makes these pulsating variables fairly easy to identify and to discriminate. Additionally, because the variability is rooted in intrinsic physical parameters, that is, the expansion and contraction of the star, in some cases, it is possible to link the periodicity of the star to luminosity and other parameters. Cepheids can be used in this manner and therefore have utility as a standard candle.

### 2.2.1 Cepheid (I & II)

The Cepheid group of pulsating variables lies along the instability strip (see Figure 1.2). They are yellow supergiant pulsating variables with a period between 1 and 100 days. The group is further divided into population I and II; pop. I are younger and more massive than the Sun, whereas pop. II are older and less massive. Our effort here has focused on variables with shorter period oscillations, hence Cepheid (I & II) are not included as part of our analysis.

### 2.2.2 RR Lyr (a, b, & c)

The group of stars referred to as RR Lyr are pulsating variables that lie along the instability strip. They have periods between 0.1 and 1.0 days, with spectral types between A and F. They are further divided into three subtypes: a, b, and c. RR Lyr (a) have mid-range periods, RR Lyr (b) have long periods, and RR Lyr (c) have shorter periods. The RR Lyr (a & b) type stars have very similar light curves; in the phased domain, they are skewed (or asymmetrical) (see Figure 2.10 Left), while RR Lyr (c) have a much more symmetrical shape in the phase domain (nearly sinusoidal) (see Figure 2.10 Right).

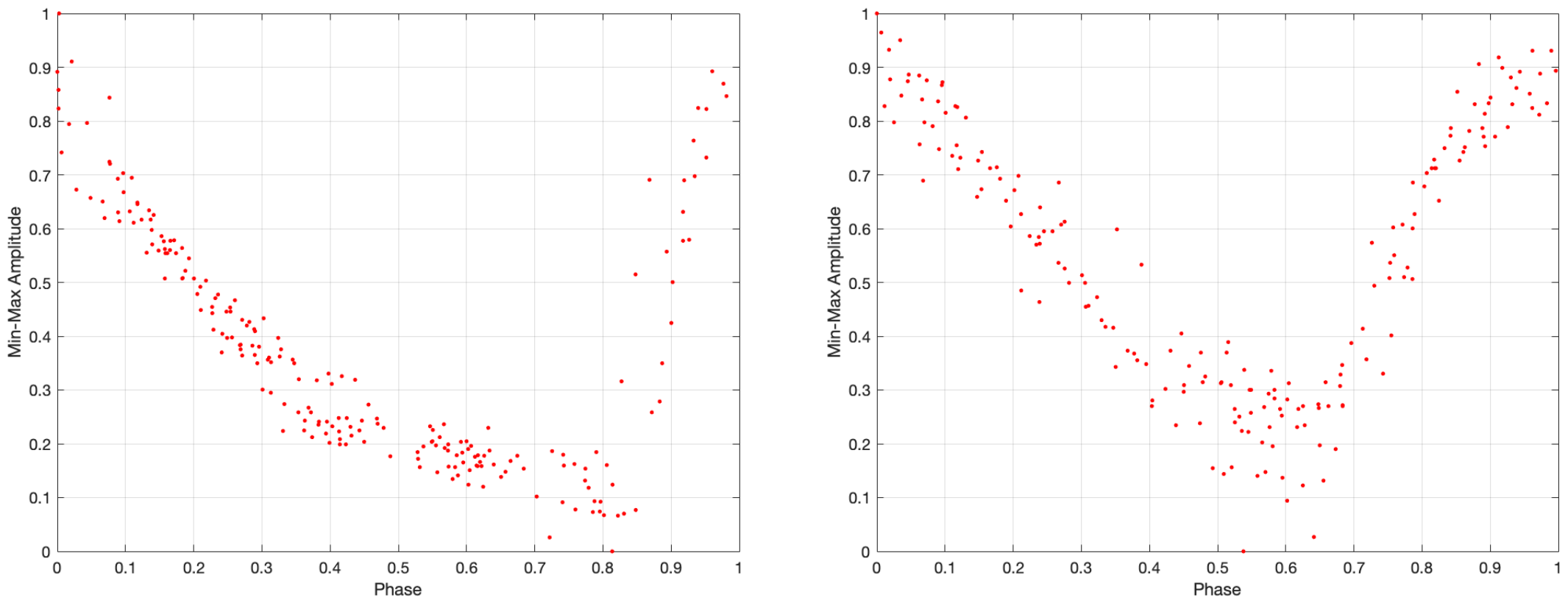

Figure 2.10: Example phased time domain waveform, Left: Example RR Lyr (ab) (LINEAR ID: 10021274), Right: Example RR Lyr (c) (LINEAR ID: 10032668), data collected from LINEAR dataset

### 2.2.3 Delta Scuti SX Phe

Further down the HR diagram (decreasing luminosity) and along the instability strip are Delta Scutis; these are pulsating variables of type A to F, with short periods (on order of 0.02–0.3 days). Variable stars of this type have smaller amplitudes compared to RR Lyr and Cepheids and are often more complex in their light curves. Specifically, they can express multiple periodicities. The expectation that there is an underlying relationship between periodicity and luminosity that we would expect from stars on the Cepheid strip is still true and is based on the fundamental frequency of the radial pulsation [Petersen and Christensen-Dalsgaard, 1999]. An example of a Delta Scuti time curve is provided in Figure 2.11.

In a similar spectra class and location on the HR diagram are older and more metal-poor stars, referred to as SX Phoenicis variables [Cohen and Sarajedini, 2012]. These tend to have larger amplitudes and a shorter range of periods (0.03–0.08 days) but still have a very similar phased light curve shape compared to the Delta Scuti. We have grouped these two classes together for the purposes of this analysis. Similar to Delta Scutis, the SX Phoenicis also have a period–luminosity relationship that can be empirically determined and exploited [Santolamazza et al., 2001].

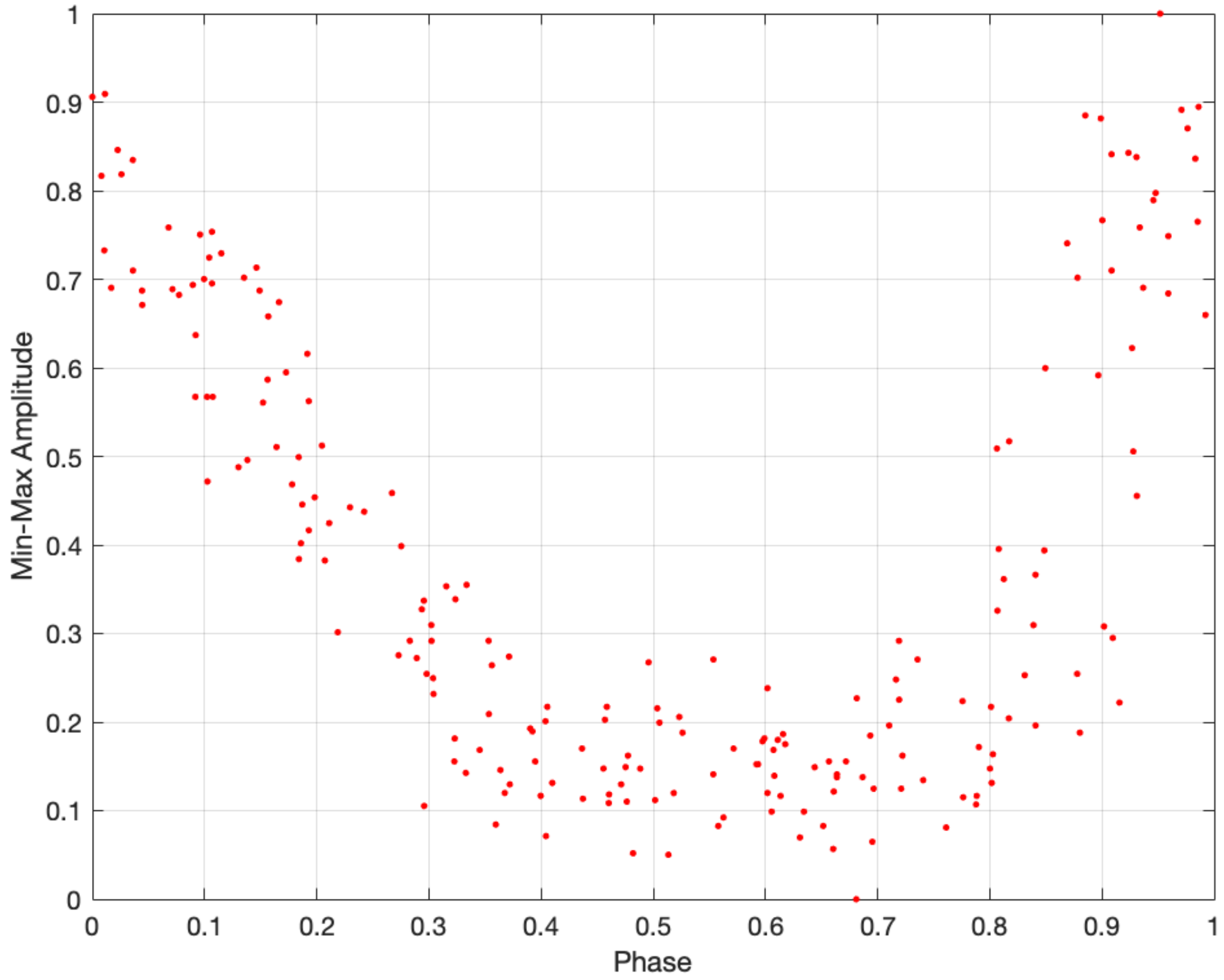

Figure 2.11: Example Delta Scuti phased time domain curve, data collected from LINEAR dataset (LINEAR ID: 1781225)

# Chapter 3

# Tools and Methods

We have discussed variable stars targeted by this effort, and while we have mentioned that they are variable and have even presented their light curves, we have not made a more formal definition of what exactly we are measuring. Time domain data are simply a set of measurements sampled at some interval. The measurement can be any of the standard astronomical observations: magnitude, color, radial velocity, spectrum, polarimetry, and so on, can be sampled over time [Kitchin, 2003]. Within this work, we focus on single-color photometry. Because of the monotonic relationship between flux and magnitude (logarithmic transform), differences between training a classifier in the magnitude domain or flux domain are moot. Thus we will be using both LINEAR magnitudes (in the V-band) and Kepler normalized (and corrected) flux values to train.

While time domain data can be derived from any type of measurement, the analysis performed within the context of this work focuses entirely on photometry in the optical or visual domain. For large, modern surveys, this usually involves a space-based (Kepler) or ground-based (LINEAR) telescope. Depending on the survey, there might be one (LINEAR) or many (SDSS) CCDs, and often the number of detectors will correlate with the number of filters used. Automated slewing, tracking, and targeting all allow for the development of automated surveys. The result is effectively "movies" of the stars, a plurality of digital images of swaths of sky. In addition to the automated mechanical and electric components of modern surveys, an initial digital automation also is common. Kepler, for example [Jenkins et al., 2010, Christiansen et al., 2013], has an automated processing pipeline that operates on the digital images, detecting power distributions of individual sources and measuring the apparent magnitude of the source based on known calibration operations. Quintana et al. [2010] addresses CCD calibration (i.e., dark, bias, flat, smear, etc.), and Twicken et al. [2010] discusses both the initial photometry analysis (flux estimation) and the presearch conditioning (artifact removal).

The result is a relatively systematic noise-free light curve. While the references provided here are specific to Kepler and the space-based survey mission, the generic process of calibration, conditioning, and analysis is common across all surveys. Likewise, there are processes unique to the Kepler mission that do not occur in ground-based missions. For example, Kepler performs a smear correction because it has no shutter, and does not need to perform an atmospheric correction because it is in space. Much of the preprocessing that the surveys perform—the processing pipeline, as it is often referred to—is designed to condition the observed raw data into a format that is common across all surveys (clean, correct, adjust magnitude, time domain data).

## 3.1 DSP with Irregularly Sampled Data

Although they are not an issue with Kepler data, ground-based surveys often suffer from irregular sampling rates; that is, the time between individual samples is not consistent, and downtime occurs because of electronics. A lack of shutter on Kepler means there is no downtime between images, that is, no loss of photons while the CCD processes, and while this then means that part of the CCD calibration included a smear correction operation, it also means that regular sampling rates are possible within a given continuous observation effort. Breaks in time domain curves still occur as targets are revisited. Ground-based surveys, however, still must contend with a rotating Earth (day–night breaks), and most imaging systems have a shutter to contend with. The result is an irregular sampled time domain curve. A comparison of Kepler and LINEAR raw time series data and sample rates is shown in Figure 3.1.

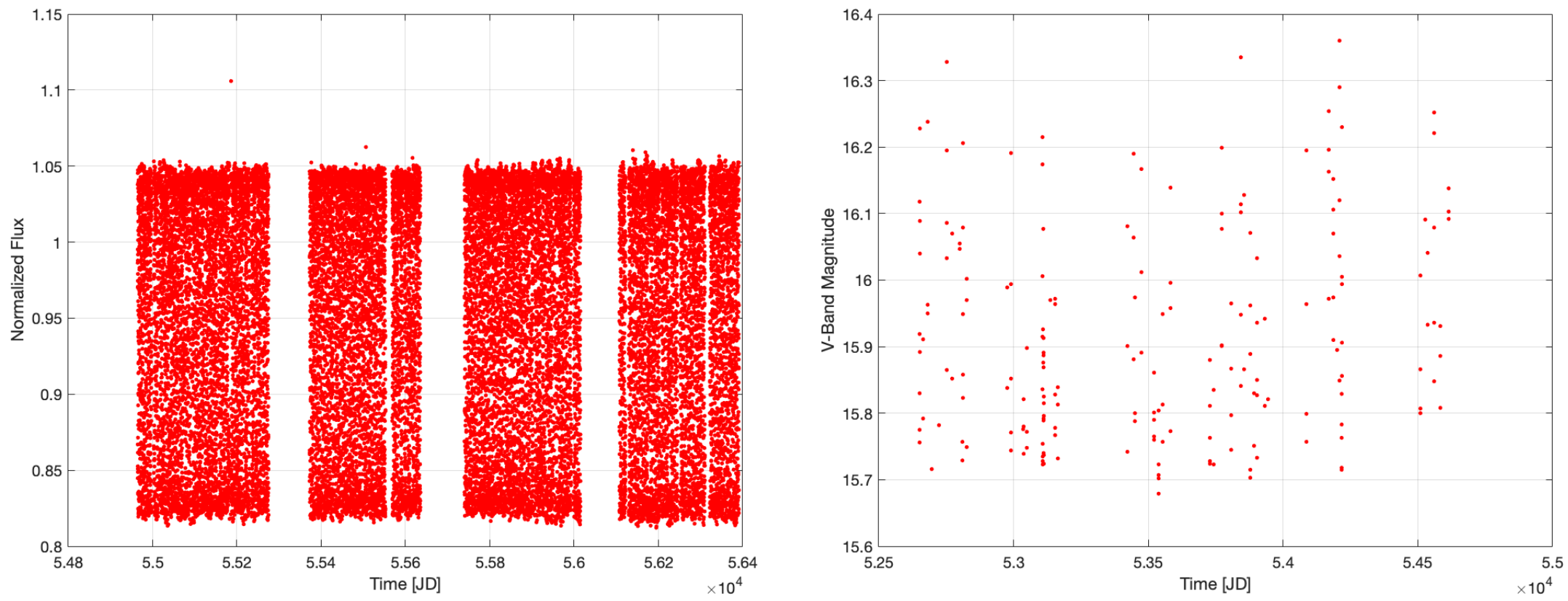

Figure 3.1: Example Raw Time Domain Curves, Left: Kepler Data (KIC: 6024572), Right: LINEAR (LINEAR ID: 10003298)

It is apparent from the figure that the Kepler data have not only a regular sample rate but also fewer breaks and more samples than the LINEAR data. Even within surveys, the number of samples can differ per target, a histogram of number of samples for a given target for all of the Kepler Eclipsing binary data pulled from the Villanova data base is provided in Figure 3.2.

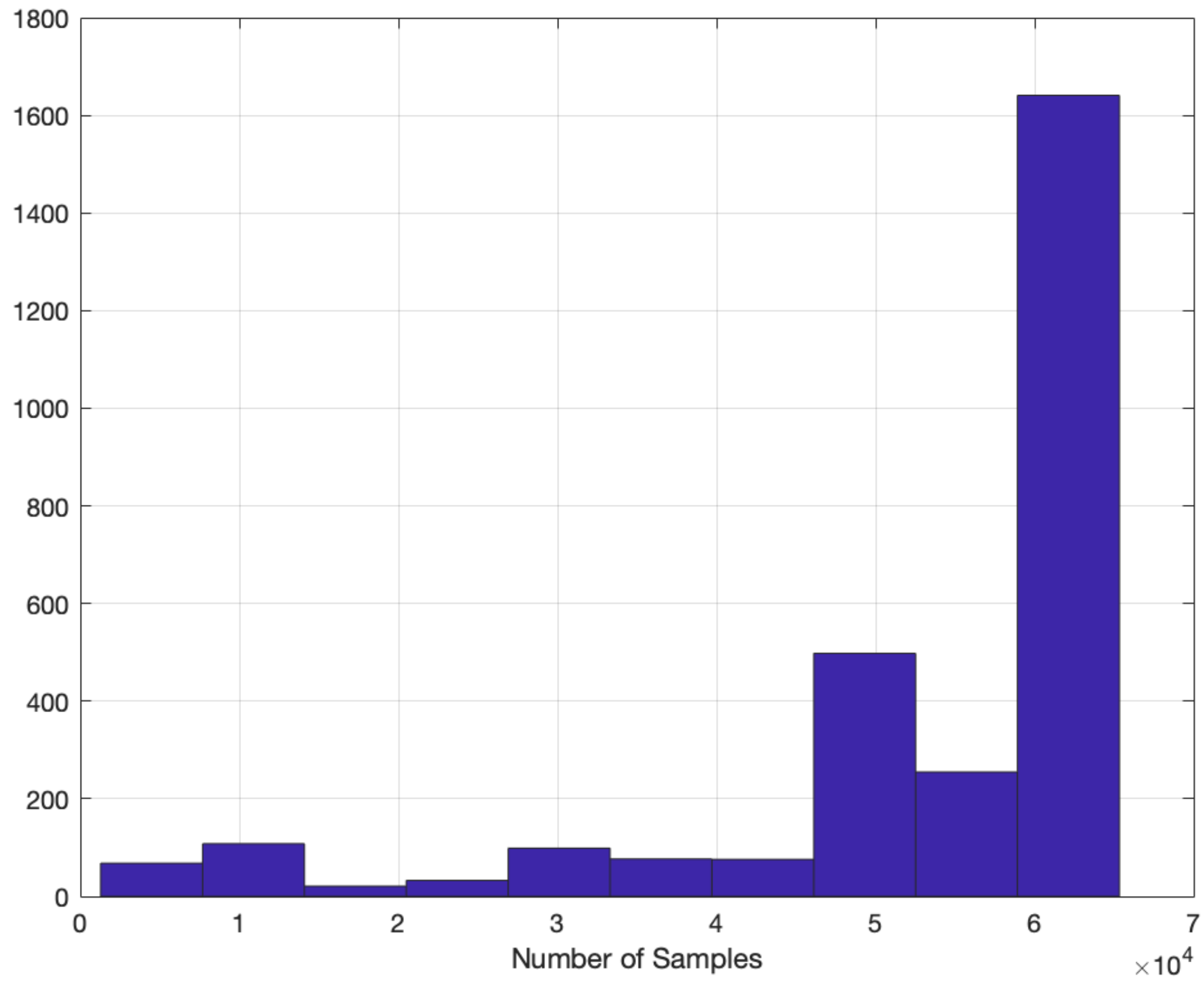

Figure 3.2: Histogram of the number of observations per target star for targets in the Villanova Kepler Eclipsing Binary Catalogue (x-axis: number of observations for given star, y-axis: number of stars with that number of observations). Graph shows expectation of data size per light curve ( 60000 points).

These differences in the number of observations within a given survey, irregularity of sampling, differing time scales, and likewise differing phases all preclude the application of standard classification methodologies directly to the time domain data. If one of our goals is a universal classification utility, we are left with the quandary of how to compare like-to-like both within and across surveys. Time domain transforms allow us to map our raw time domain observations to a feature space in which comparisons can be made, specifically measurements of distance and similarity.

The field of digital signal processing (DSP) contains a number of these transformations: Fourier transformations, wavelet transformations, and autocorrelation, to name a few. As Fulcher et al. [2013] discussed, methods for time series analysis can take on a multitude of forms, with each method designed to quantify some specific value or set of values. As Fulcher et al. [2013] show, not all methods are unique, and many correlate with one another; this is especially true within analysis families (linear correlations, statistical distributions, stationarity evaluations, entropy measures, etc.). The astronomical community, initially focusing on those variable stars that have a consistent, repeating signature, commonly uses Fourier domain statistics to evaluate time domain signals. The transformation into the frequency domain has a twofold effect: (1) mapping to the Fourier (frequency) domain means that within a given range of frequencies, variable stars can be compared one to one, and (2) the data can be phased or folded on themselves about a primary period if one is found, again allowing variable stars to be compared one to one in the phase domain.

This application of Fourier transformation is really an effort to generate a primary period for the variable star, a process referred to here as period finding. Astronomers have developed many methods for period finding since the late 1970s, when time domain astronomy first became feasible. While the Fourier transformation has a much longer history [Heath, 2018], the algorithms designed, such as discrete Fourier transformation (DFT) and fast Fourier transform (FFT), require regular sample rates to operate. The method of Lomb–Scargle [Lomb, 1976, Scargle, 1982] was one of the first methods to address the issues faced by astronomers. Since then, the number of methods has increased, leveraging a variety of different techniques [Graham et al., 2013b].

Most of the period-finding algorithms are roughly methods of spectral transformation with an associated peak/max/min-finding algorithm and include such methods as discrete Fourier transform, wavelet decomposition, least squares approximation, string length, autocorrelation, conditional entropy, and autoregressive methods. Graham et al. [2013b] review these transformation methods (with respect to period finding) and find that the optimal period-finding algorithm is different for different stars. With the history and confidence associated with the Lomb–Scargle method, it was selected as the main method for generating a primary period within this work.

The Lomb–Scargle algorithm computes the Lomb normalized periodogram (spectral power as a function of frequency) of a sequence $H$ of $N$ data points with, sampled at times $T$, which are not necessarily evenly spaced [Scargle, 1982]. $T$ and $H$ must be vectors of equal size. The routine calculates the spectral power for an increasing sequence of frequencies (in reciprocal units of the time array $T$) up to some high frequency threshold input constant times the average Nyquist frequency; the user additionally supplies an an oversampling factor, typically $OFAC \geq 4$. The returned values are arrays of frequencies considered ($f$), the associated spectral power ($P$) and estimated significance of the power values ($\sigma$).

Although the implementation outlined as part of this research is based on that described in Teukolsky et al. [1992], section 13.8, rather than using trigonometric recurrences, this implementation takes advantage of Java's array operators to calculate the exact spectral power as defined in equation 13.8.4 on page 577 of Teukolsky et al. [1992]. For more information, our implementation of the Lomb–Scargle algorithm is provided as part of the Variable Star package.[1]

## 3.2 Phasing, Folding, and Smoothing

One of the key transformations applied to variable star data is a folding of the observations. The procedure is straightforward: a Fourier transform or similar operation generates the distribution of frequencies in the observed time domain data; a maximum power is found [Graham et al., 2013b], and the data are phased about this main period. Maximum power, or dominant period, in the Fourier domain is a proxy for consistency of frequency over the observations. The result is a figure like the one presented in Figure 3.3, a domain bound on $[0, 1]$ and a y-axis range that depends on the original observations or on the transformations applied by the analyst (such as the min-max transformation applied here).[2]

This period found is referred to as the primary period and is the dominant period expressed by the variable. This method allows for a simplification of the cyclic variation of stellar brightness, be it from transiting eclipses, radial pulsation, or other cases that produce a cyclostationary signal. Of course, no signal is perfectly repeated, and variations exist from cycle to cycle caused by either intrinsic or extrinsic factors. The noise of the detector or the atmosphere can cause changes from cycle to cycle. Stars can and often do vary for multiple different reasons [Percy, 2007], often inconsistently, in addition to the dominate period determined from the period-finding algorithms discussed earlier. The result is that for a given bin $(dx)$ in phase, the amplitude $(f(x))$ has a distribution. Ideally, the central tendency of the distribution is the expected amplitude if only the main period existed. If we assume that this main period signal is a defining characteristic to be used in the classification of variable stars, then we require a procedure to isolate or extract the true underlying period of variation. To this end, we briefly review Friedman's variable span smoother, that is, the SUPER-SMOOTHER algorithm [Friedman, 1984].

---

[1] fit.astro.vsa.analysis.feature.LombNormalizedPeriodogram

[2] fit.astro.vsa.analysis.feature.SignalConditioning

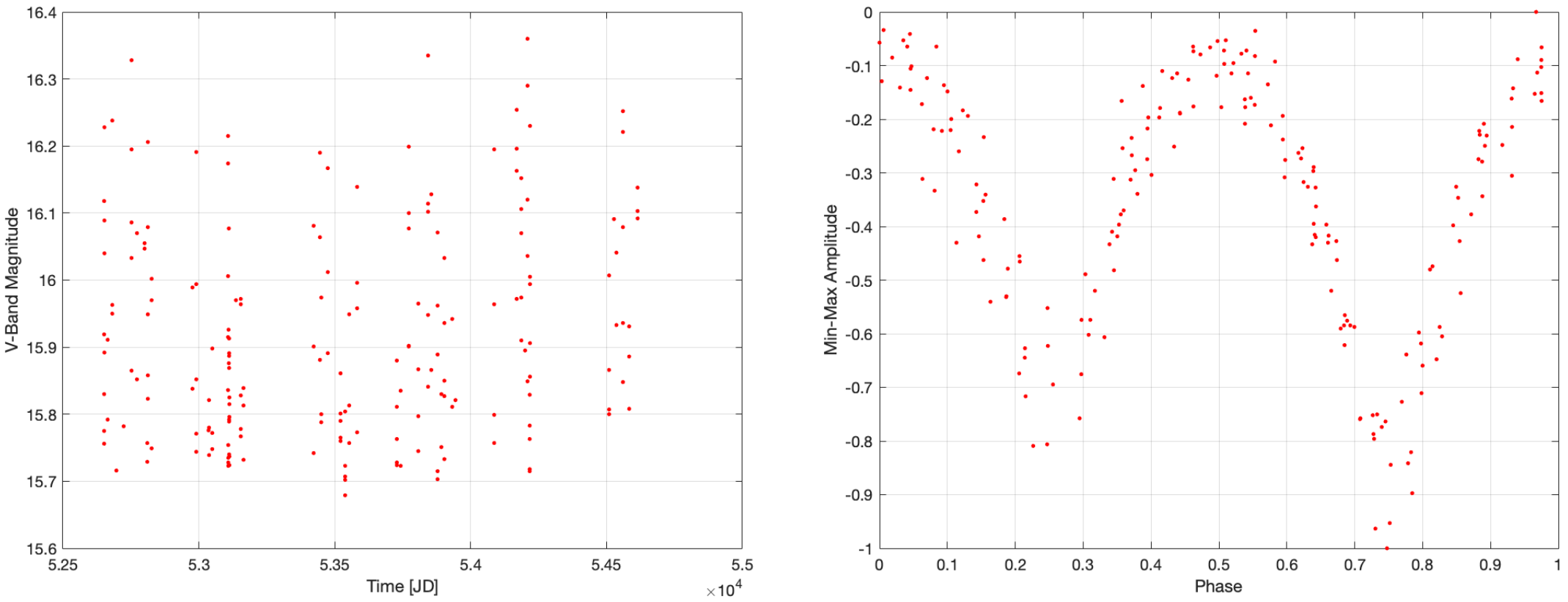

Figure 3.3: Left: Raw Time Domain Data, Right: Corresponding Transformed Phased Time Domain Data (LINEAR ID: 10003298)

Let us assume we have bivariate data $(x_1, y_1), (x_2, y_2) \ldots (x_n, y_n)$; the smoothing function can be given as equation 3.1:

$$y_i = s(x_i) + r_i, \tag{3.1}$$

where $s(x_i)$ is the signal and $r_i$ is the residual error. Likewise, the optimal smooth-ing function is the one that minimizes the sum of the residuals $\sum (r_i)^2$:

$$\min_s \sum (y_i - s(x_i))^2 . \tag{3.2}$$

A simple smoothing function would be the boxcar averaging algorithm [Holcomb and Norberg, 1955]. A fixed window (the span) is slid across the domain; at each iteration, an average is found and used at the smoothed estimate. Friedman [1984] proposed an advancement on this idea, by proposing a variable bandwidth smoother. Given a defined span of $J$, we can define a local linear estimator, our smoothing function, as equation 3.3:

$$\hat{y}_k = \hat{\alpha} + \hat{\beta} x_k, k = 1, ..., n. \tag{3.3}$$

This linear model is applied from local fits of points similar to the boxcar average, $i_{-\frac{J}{2}}, ..., i_{+\frac{J}{2}}$, with $x_i \leq x_{i+1}$ for $i = \frac{J}{2}, ..., n - \frac{J}{2}$. The variable span smoother attempts to optimize with respect to equation 3.4:

$$\min_{s,J} \sum (y_i - s(x_i | J(x_i)))^2 . \tag{3.4}$$

Similar to the constant-span case, the local linear smoother is applied with several discrete values of $J$ in the range $0 < J < n$; optimal solutions for $s$ and $J$ are found using a leave-one-out cross-validation procedure [Duda et al., 2012]. Friedman [1984] originally recommended three values: $J = 0.05n, 0.2n, 0.5n$. These are intended to reproduce the three main parts of the frequency spectrum of $f(x)$ (low-, medium-, and high-frequency components). The algorithm selects the best span based on the error analysis proposed in equation 3.4 for each input $x_i$; these "best" estimates are then smoothed using the medium frequency span. The output is a smooth estimate of the input data, as demonstrated in Figure 3.4.

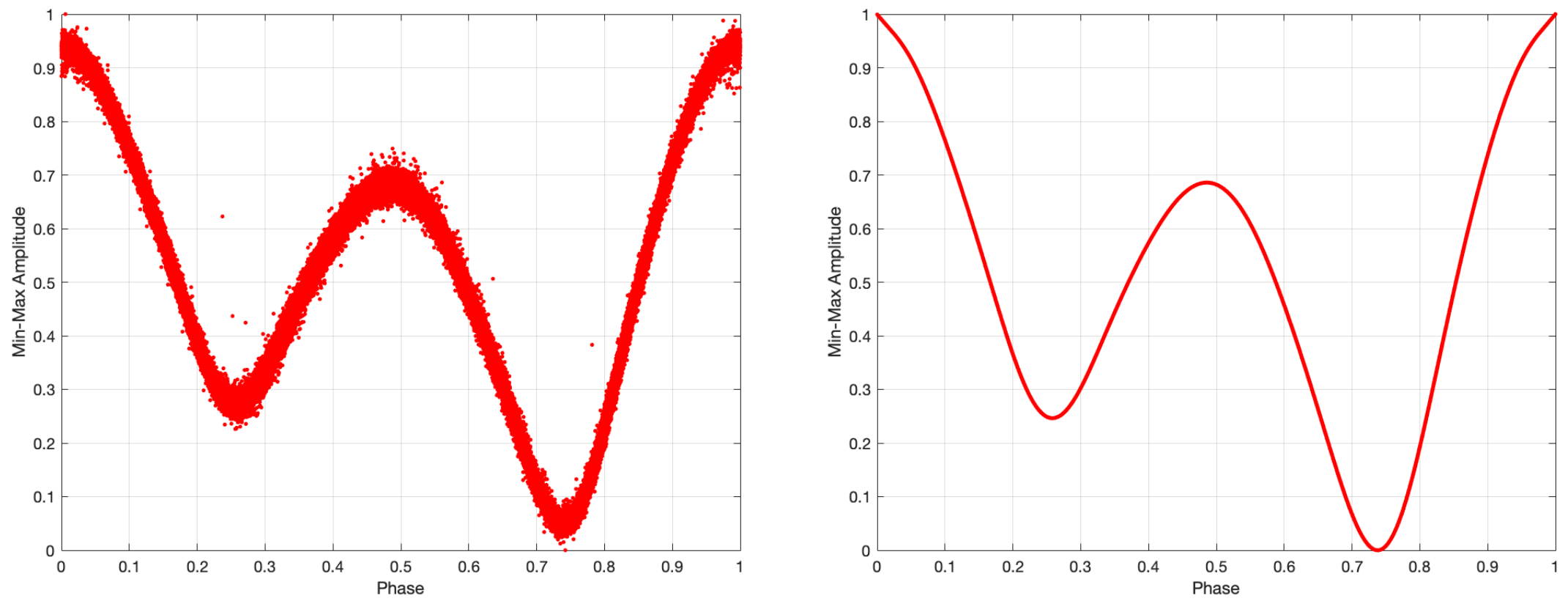

Figure 3.4: Left: Transformed Phased Time Domain Data, Right: Corresponding SUPER-SMOOTHER Phased Time Domain Data (Kepler Data, KIC: 5820209)

## 3.3 Feature Extraction

The information we observe is from raw signals: amplitude in time (time series), spectral amplitudes (color), spectra, and so on. As we have discussed, the raw time domain signal can be difficult to work with, so analysis usually involves a transformation to a different representation (transform, mapping, reduction). Traditional variable star astronomy focuses on "folding" the time domain data (light curve), that is, transforming the data such that subsequent phases of cyclostationary signals are overlapping. This is done by finding the main (primary) frequency of the repeated signal, and the data are then "phased" together, resulting in a phased plot, as shown in Figure 3.3.

We are still left with the challenge of what to do with this information we have generated so far : raw time domain data, a frequency representation of the time domain data (i.e., the periodogram), and a phased data plot, none of which can be used for one-to-one comparison yet. As mentioned, the time domain data can be of unequal sizes (and phases); the frequency domain representation will not be consistent within class (eclipsing binaries, for example, can have a wide range of periods); and the phased data representation, while on a constrained domain space, can have unequal samples, thus precluding one-to-one comparisons. Additional transformations are necessary, then, for a similarity analysis.

Community efforts in the reduction of the time domain data have mostly focused on the development of independent metrics derived from one of the three outputs discussed so far (raw data, frequency data, or phased data). Expert selection of measurable values that are consistent within a given class space (variable star type) have resulted in a multitude of features. These have included the following:

- time domain statistics: This includes the quantification of the statistics associated with raw time domain data. The statistics could be as simple as means, standard deviations, kurtosis, skew, max, min, and so on. They could also be local statistics, either on the phased or unphased data.
- frequency domain statistics: Most common in literature is statistical reduction of the transformation representations, specifically ratios ($f_2/f_1$), differences, levels, and so on.

Most, if not all, standard transformations of this nature stem from an original reference, Debosscher [2009], and have since been added to over time to produce a set of upward of 60 features. A graphical representation of these expertly selected features is provided in Figure 3.5.[3]

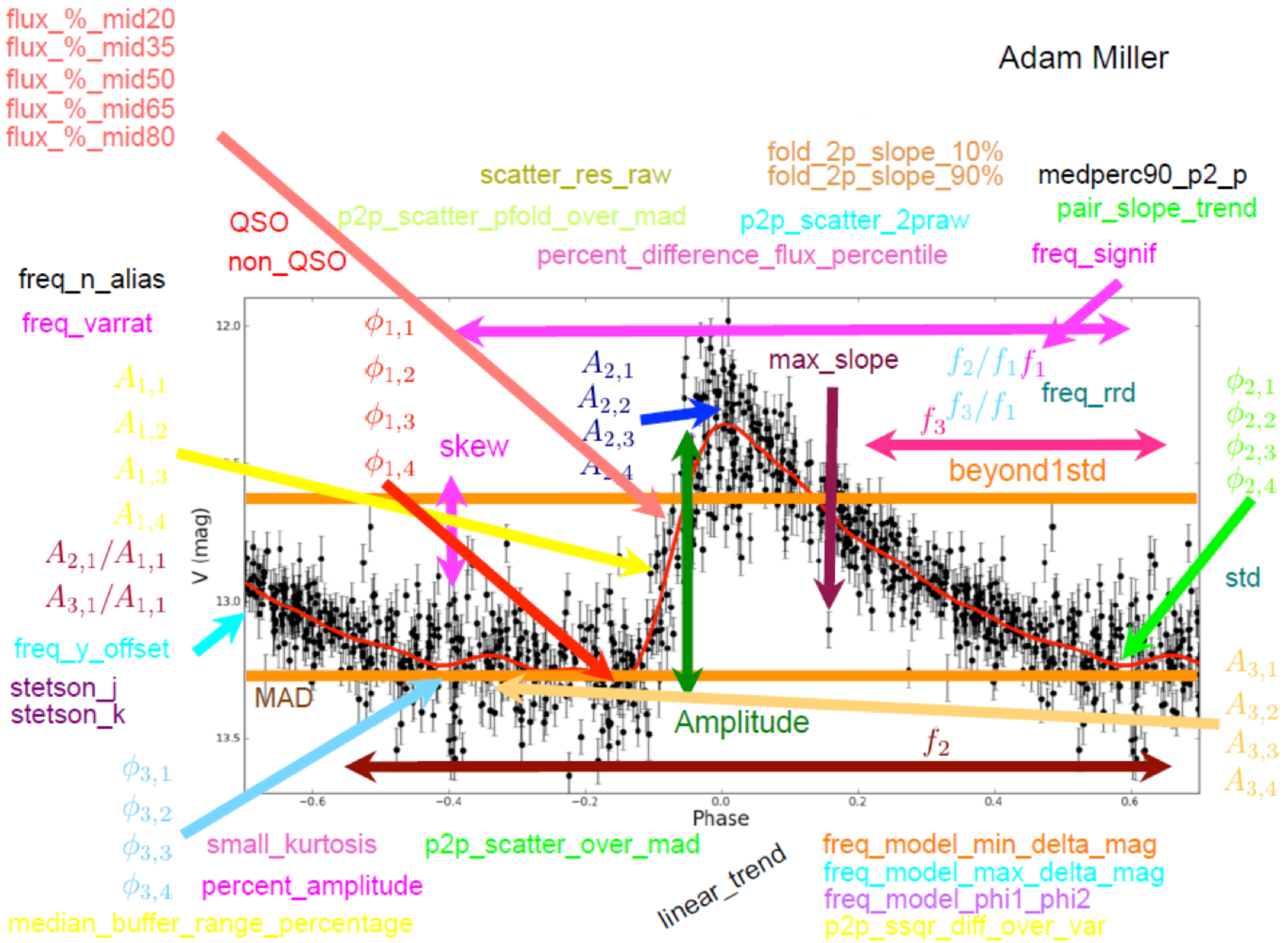

Figure 3.5: Example of community standard features used in classification

These features, however, to the best of our knowledge, are statistical measures that have been implemented to distill the complex time domain data into singular measures and are not necessarily optimized or selected for the purpose of maximizing the ability of a classifier to differentiate between variable stars. The work presented here will propose additional transformations that might be useful in discriminating variable stars.

[3]Mahabal, A. (2016). *Complete classification conundrum.* Presented at Statistical Challenges in Modern Astronomy VI, June 6–10, Carnegie Mellon University.

## 3.4 Machine Learning and Astroinformatics

One of the key tools of this research effort is the leveraging of the past 30+ years of machine learning advancements into the construction of an algorithm for the automated classification or detection of variable stars of interest. The tools reviewed here represent a small fraction of the total possible methods; we focus here on those that have found favor in the astroinformatics community, or those that are predecessors to those methods that we have proposed and implemented. "Astroinformatics: A 21st Century Approach to Astronomy" [Borne et al., 2009] is one of the first community calls for a codification of this field of study. Submitted to the Astronomy and Astrophysics Decadal Survey, it advocates for the formal creation, recognition, and support of a major new discipline called "astroinformatics," defined here as "the formalization of data-intensive astronomy and astrophysics for research and education" [Borne, 2010]. More specifically, this is the field of study that includes

- large database management for astrophysical data
- high-performance computing for astrophysics simulations
- data mining, visualization, or exploratory data analysis specific for astrophysical data
- machine learning for astrophysical problems and observations

We focus here on the fourth bullet point, machine learning. Machine learning, in the context of astroinformatics and specifically our goal (data discovery), can be further broken down into a couple methods of data discovery [see Figure 3.6, Booz Allen Hamilton, 2013].

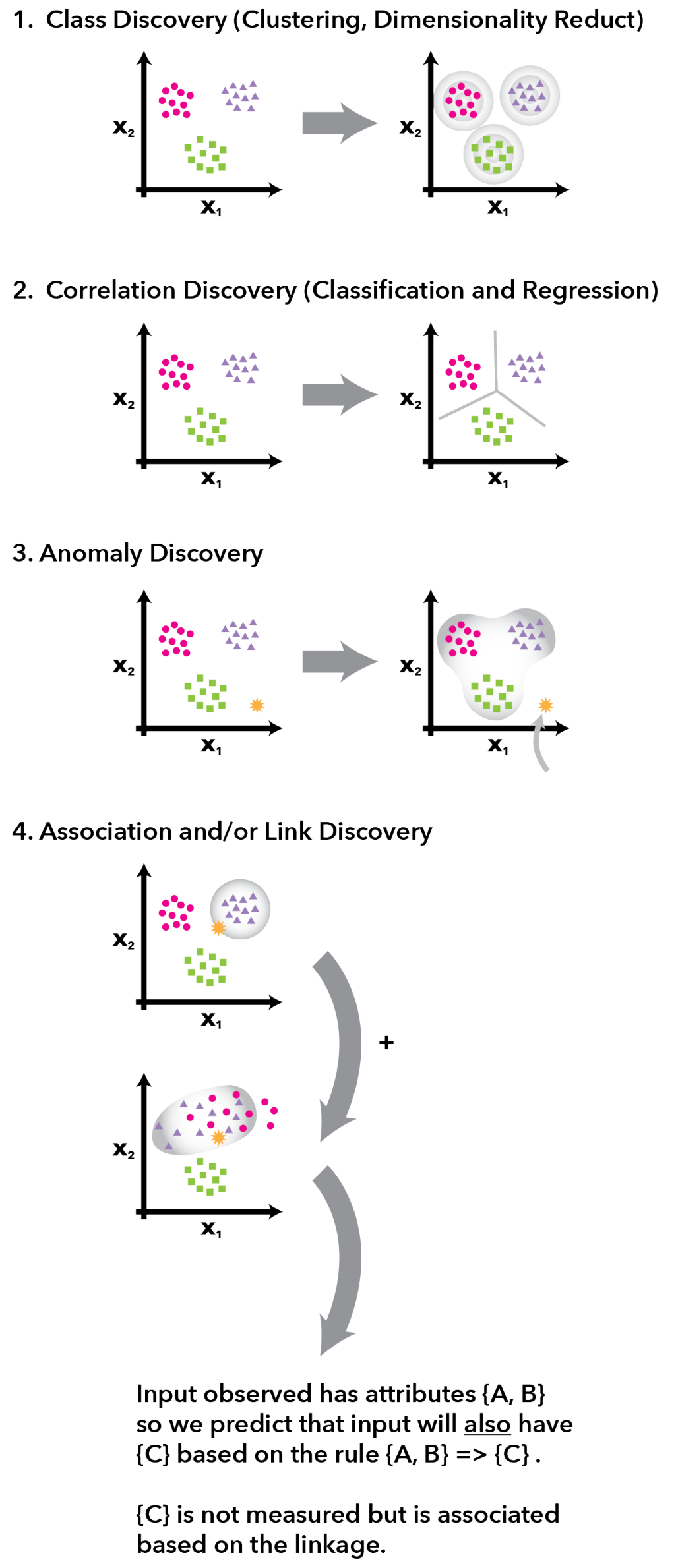

Figure 3.6: High-Level categories and examples of data discovery and machine learning techniques. (1) Class discovery is used to generate groupings from previously analyzed data, (2) Correlation discovery is used to construct models of inference based on prior analysis, (3) Anomaly Discovery is used to determine if new observations follow historical patterns, and (4) Link Discovery is used to make inferences based on collections of data

First, with regard to class discovery, how can we find categories of objects based on observed traits and known prior information? Efforts in this direction focus on unsupervised classification (clustering) methods. Second, with regard to correlation discovery, based on what we know and trends we are able to map and understand, how can we understand new observations? Work here usually falls under the category of supervised classification, which implies training algorithms based on expertly labeled data to infer labels on newly observed data. Third is anomaly discovery, the "unknown unknowns": how do we construct an algorithm to determine when new data fall out of range of our prior observed trends? Last is association or link discovery, or training an algorithm to make connections that are interesting or helpful based on a predesigned set of rules.

### 3.4.1 Example Dataset for Demonstration of Classifiers

For demonstration purposes, we have constructed a test set with which to demonstrate some standard classifiers. Figure 3.7 is a plot of 45 solar neighborhood stars on an HR diagram, which we have expanded by adding 200 artificially generated data points. This expansion was performed via data augmentation, specifically a bivariate jittering of the original data with Gaussian noise; thus the population of each upsampled artificially generated data point has some basis in the original 45 "templates." These data are intended for demonstration purposes only.

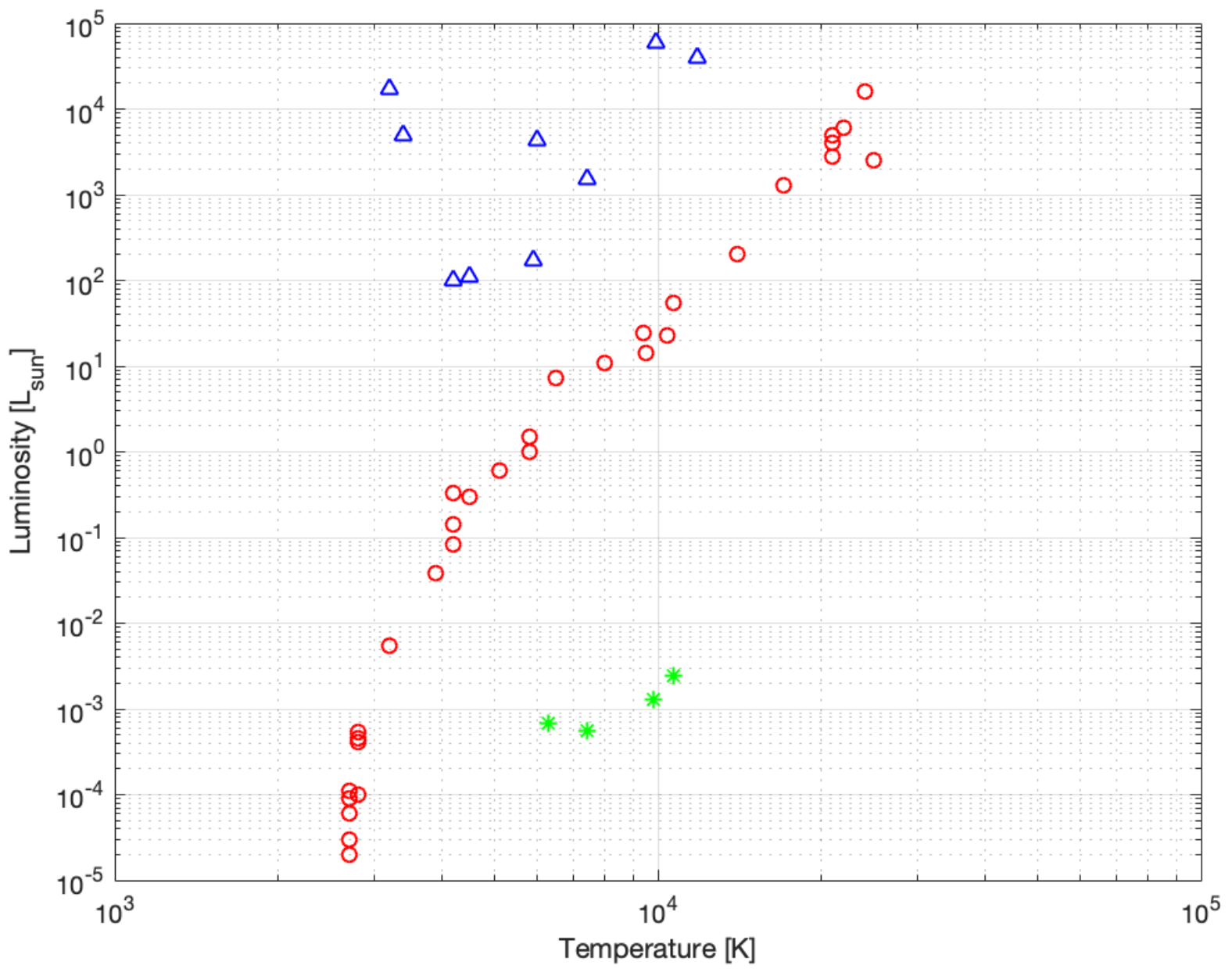

Figure 3.7: The HR Diagram Test Data to be used in the demonstration of standard classifiers. Original dataset shown here contains 45 solar neighborhood stars on an HR diagram (Red: Main Sequence Stars, Green: White Dwarfs, Blue: Supergiants and Giants). Data shown here is for demonstration purposes.

This joint data set of artificial and real observations will be used going forward to demonstrate the concepts presented.

### 3.4.2 Machine Learning Nomenclature

This work will address three of the four machine learning categories addressed earlier, class discovery in the form of clustering and correlation discovery (both discussed here), as well as some initial anomaly detection proposed in section 4.3.3. The correlation discovery can be further broken down into a few basic categories with respect to complexity of implementation; these are presented in Figure 3.8.

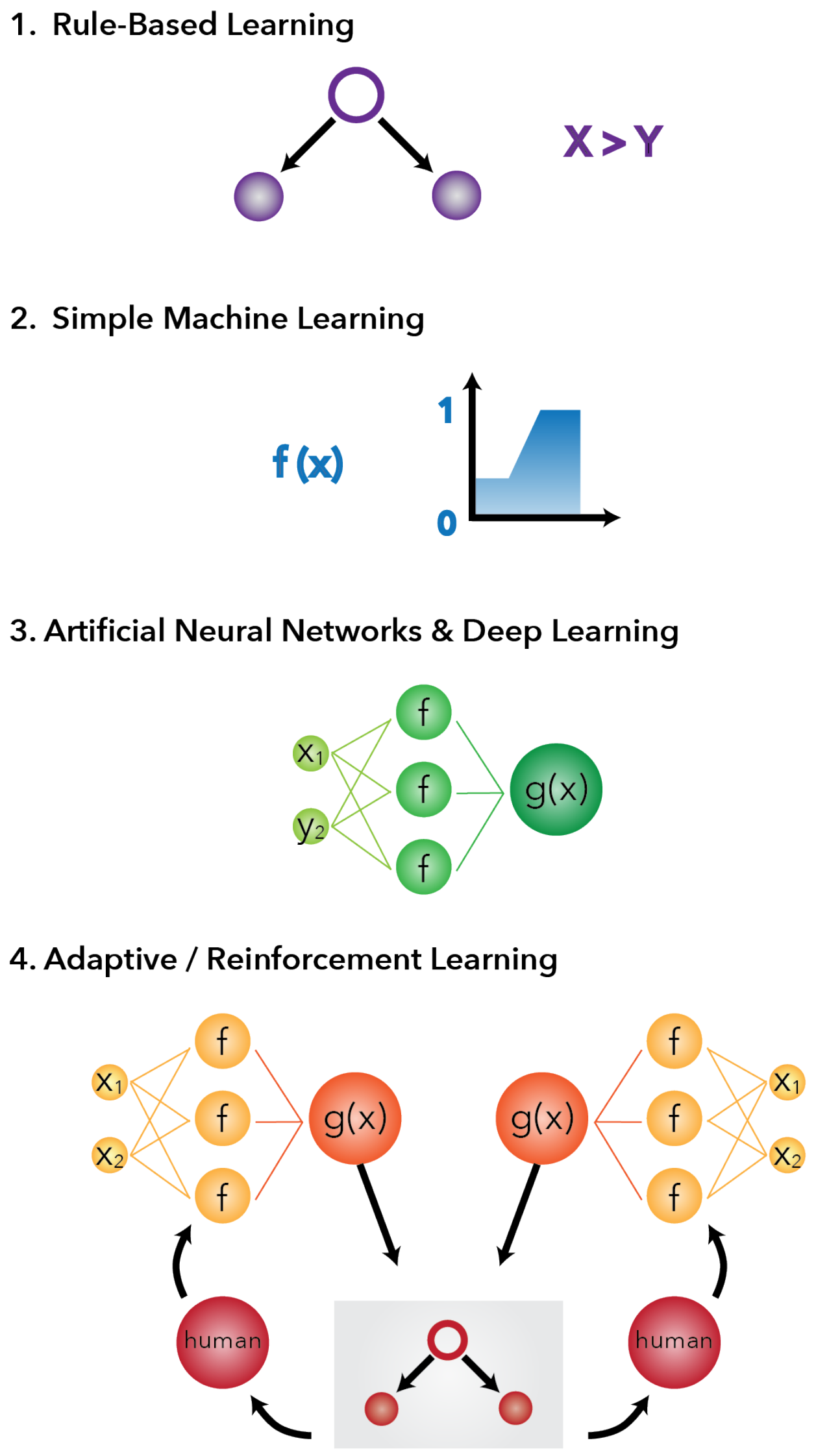

Figure 3.8: High-Level categories and examples of correlation discovery techniques. (1) Rule based learning uses simple if/else decisions to make inferences about data, (2) Simple Machine learning uses statistical models to make inferences about data, (3) Artificial Neural Networks combine multiple simple machine learning models to make inferences about data, (4) Adaptive learning allows for the iterative training/retraining of statistical models at more data becomes available or as definitions change.

We focus on the construction of simple machine learning functions (see Figure

3.8), as they tend to lend themselves better to transparency, which is a necessity for scientific implementation. Simple artificial neural networks are implemented as part of this work.

As a baseline, we adopt the notation and nomenclature of Hastie et al. [2009] when it comes to our overview of the subject of machine learning. As we focus our discussion to metric learning specifically, we will borrow notation from Bellet et al. [2015]. Let us define a simple generic model: $\hat{Y} = \hat{\beta}_0 + \sum_{j=1}^{p} X_j \hat{\beta}_j$. Here we have made some observations as part of an experiment of both an input ($X_j$ also can be referred to as a feature or attribute) and an output or response $\hat{Y}$. Our model is defined both by the functionality and the parameters $\hat{\beta}_j$. The parameters are learned based on our training data, and the training data are the set of pairs ($X_j$, $y_j$) observed as part of the experiment. The feature measured could be a value, a set of values, a continuous series, a matrix, a string, and so on. Note that when the response is qualitative, you might see ($X_j$, $g_i$), denoting that instead of the response being a continuous variable, $g_i$ is from the set $G$ labels or classes, and these are discrete options, such as (1,2,3, . . . ), (a,b,c, . . . ), (red, green, blue, . . . ), and are representative descriptions of what was being measured. Figure 3.9 gives an example of the boundaries resulting from the development of the classifier and the associated regions of class space.

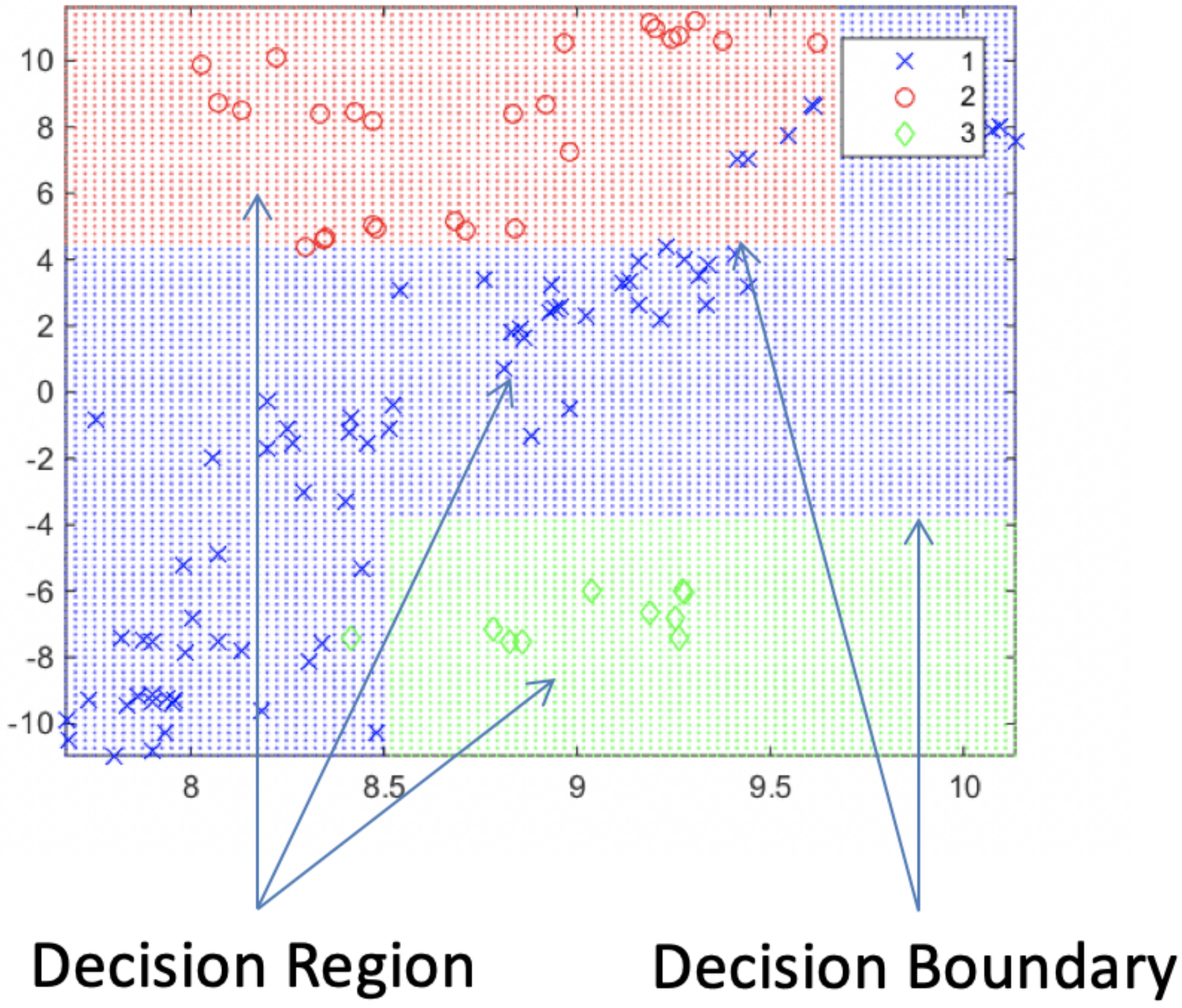

Figure 3.9: Decision space example, for this example the figures axis are merely an example dimension (x1, x2) to represent a generic bivariate space.

### 3.4.3 k-NN Algorithm

The k-Nearest Neighbor algorithm estimates a classification label based on the "closest" samples provided in the training data [Altman, 1992]. If $\{x_n\}$ is a set of training data $n$ big, then we find the distance between a new pattern and each pattern in the training set based on a given measure of distance. Often, the distance between points is estimated via $L^p$-norm, given here as equation 3.5:

$$\|\mathbf{x} - \mathbf{y}\|_p = \left(|x_1 - y_1|^p + |x_2 - y_2|^p + ... + |x_n - y_n|^p\right)^{1/p}. \tag{3.5}$$

The value of $p$ can be adjusted but is frequently given as 2 (Euclidean distance). This is not the only distance measure—a much more complete survey is given in Cha [2007], covering not just the $L^p$-norm family but also lesser known comparison families (intersection, inner product, squared-chord, $\chi^2$, and Shannon's entropy). Based on the distance measurement, the algorithm will classify the new pattern depending on the majority of class labels in the $k$ closest training data sets. The pattern can be rejected ("unknown") when there is not majority rule or if the distance exceeds some threshold value defined by the user.

For example, Figure 3.10 shows the training set (in red and blue) and the test case (in green). If $k = 1$, the test case would be classified as "red," as the nearest training sample is red. If $k = 5$, the test case would be classified as "blue," as the closest five measurements (dashed circle) contain three blue and two red (majority vote wins).[4]

---

[4]https://en.wikipedia.org/wiki/K-nearest_neighbors_algorithm

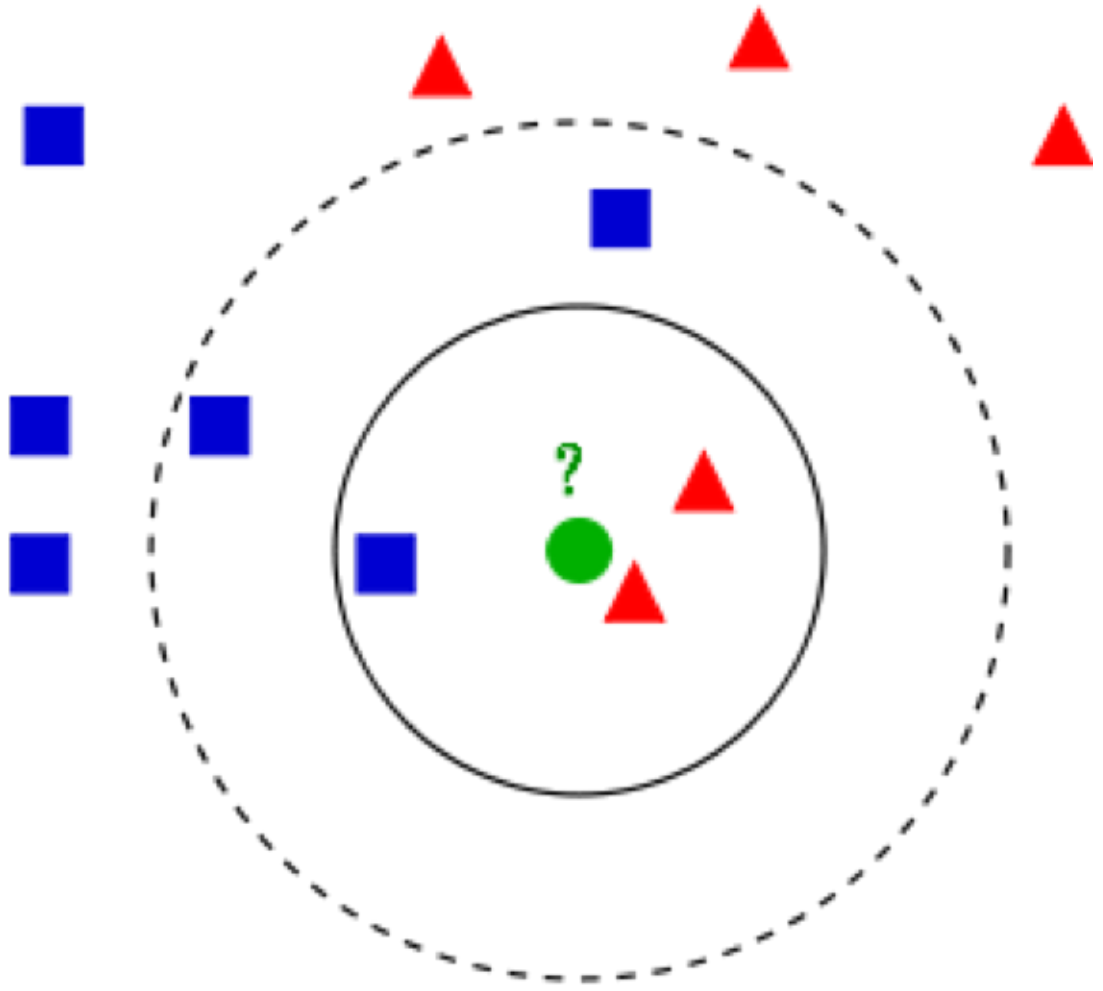

Figure 3.10: An example k-NN decision space, the new data point (in green) is compared to the training data (in red and blue), distances are determined between observation and training data. The circles (solid and dashed) represent the nearest neighbor boundaries, $k = 3$ and $k = 5$ respectfully. Notice how the classification of the new observation changes between these two settings (from red to blue)

While k-NN is nonparametric, posterior probabilities are estimable when $k$ is large and the density of points is large. If we call $v$ the volume of space contained by the $k$ nearest points (the area of the circles in Figure 3.10), $P(\mathbf{x}|\omega_i) \sim \frac{k_i/N}{v}$, where $k_i$ is the number of matching patterns and $N$ is the number of training samples. A variation of the k-Nearest Neighbor algorithm is to weight the contribution of the class votes ($\rho$k-NN). Weighting is often a function of distance; points that are farther away have less effect on the class decision [Hechenbichler and Schliep, 2004]. Using the training data, we have applied the k-NN algorithm[5] and determined that $k = 1$ is the optimal solution. Figure 3.11 is a plot of a decision space and the test data, data which were not used in training.

[5] fit.astro.vsa.utilities.ml.knn

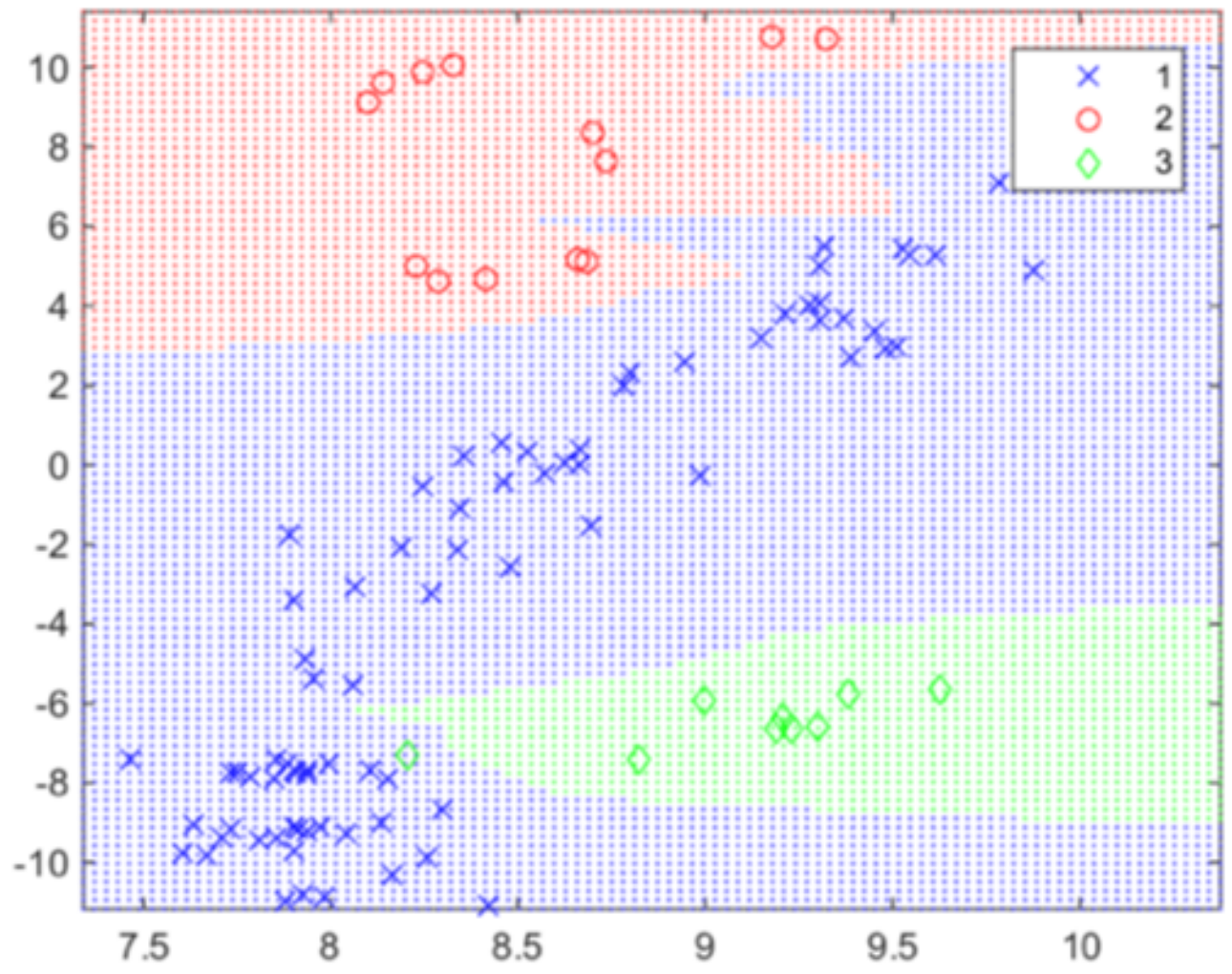

Figure 3.11: k-NN decision space example in our standard generic ($x_1$, $x_2$) coordinate system

### 3.4.4 Parzen Window Classifier

We are interested in $f(x|c = m)$, that is, the distribution of the observable for a given class type. Often we either do not know the distribution or the distribution is known but is not easily expressible (e.g., not Gaussian). We can identify a good approximation of the distribution using kernel density estimation (KDE; aka Parzen window density estimation). Similar to the k-NN classifier, the Parzen window classifier[6] (PWC) is based on defining similarity. The classifier leverages KDE to approximate the probability density function for each class distribution [Rosenblatt, 1956]. The one-dimensional KDE can easily be extended to the mul-

[6] fit.astro.vsa.utilities.ml.pwc

tidimensional case, given here as equation 3.6:

$$\hat{f}\left(\mathbf{x}|\mathbf{x}_n, S\right) = \frac{1}{N}\sum_{n=1}^{N}\frac{1}{S^D}\cdot K\left(\frac{\mathbf{x}-\mathbf{x}_n}{S}\right), \tag{3.6}$$

where $D$ is the dimensionality of the $\mathbf{x}$ vector and $K(u)$ is some function we can define as a window or "kernel." For each new point $\mathbf{x}$, an estimate of $\hat{f}\left(\mathbf{x}|\mathbf{x}_n, S\right)$ can be generated per class. Using the prior approximations, we can estimate the posterior probability of the pattern observed belonging to class $i$ as

$$\hat{p}\left(\omega_i|\mathbf{x}\right) = \frac{\hat{f}\left(\mathbf{x}|\omega_i\right)\hat{p}\left(\omega_i\right)}{\sum_{k=1}^{J}\hat{f}\left(\mathbf{x}|\omega_k\right)\hat{p}\left(\omega_k\right)}. \tag{3.7}$$

The decision space is then based on a comparison of the probability density estimate per class and the prior probabilities [Parzen, 1962, Duda et al., 2012], given here as Equation 3.8,

$$\hat{f}\left(\mathbf{x}|\omega_1\right) > \frac{P\left(\omega_2\right)}{P\left(\omega_1\right)}\hat{f}\left(\mathbf{x}|\omega_2\right). \tag{3.8}$$

see Figure 3.12 for an example [7].

[7] commons.wikimedia.org

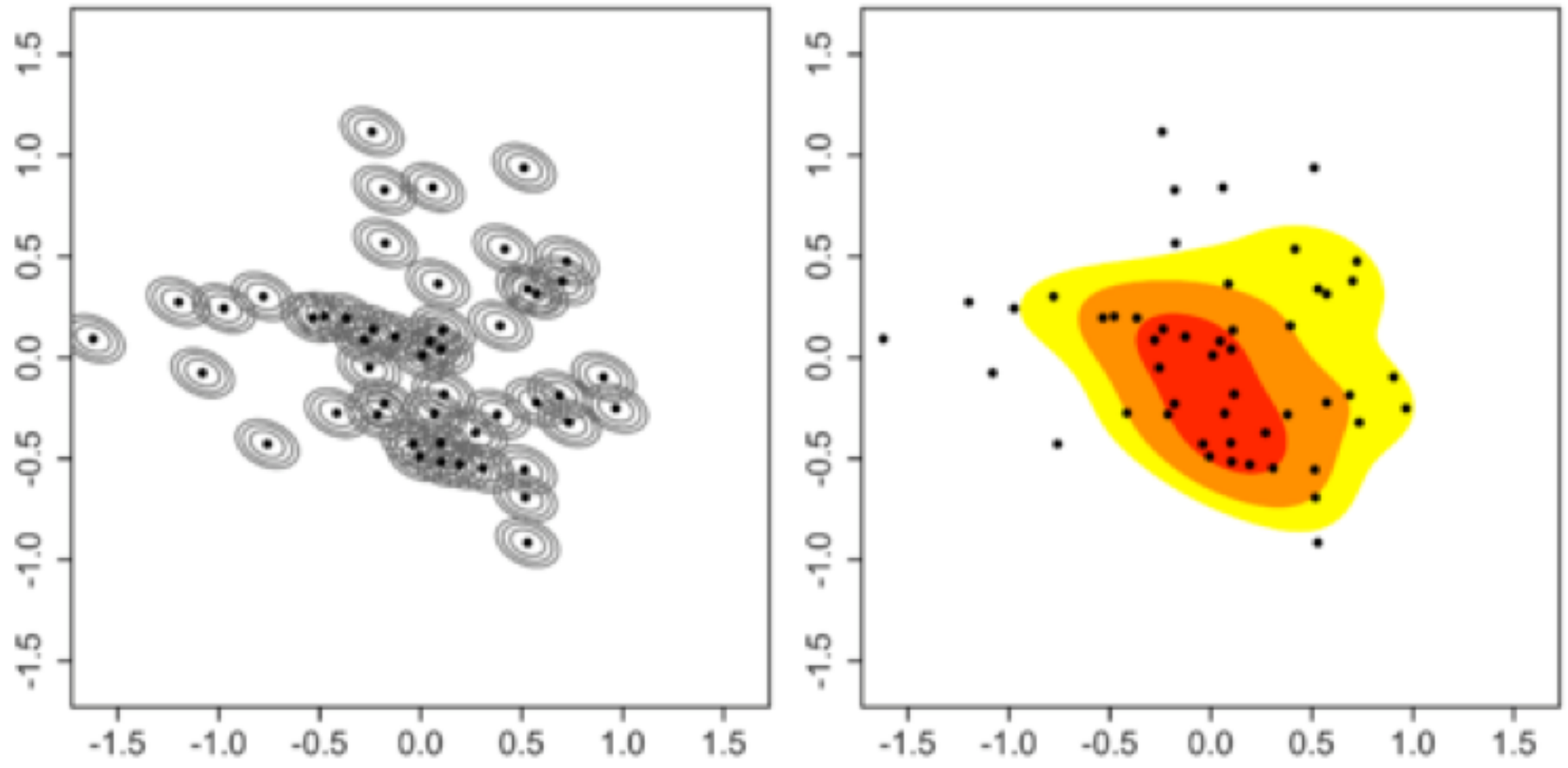

Figure 3.12: PWC Example showing the transition between points with kernels mapped to their spatial (x,y) coordinates (Left), to the approximated probability distribution heat map (Right).

The two-class classification scheme can easily be extended into a multiclass design should it be required. Figure 3.13 is a plot of a decision space and the test data, data which were not used in training..

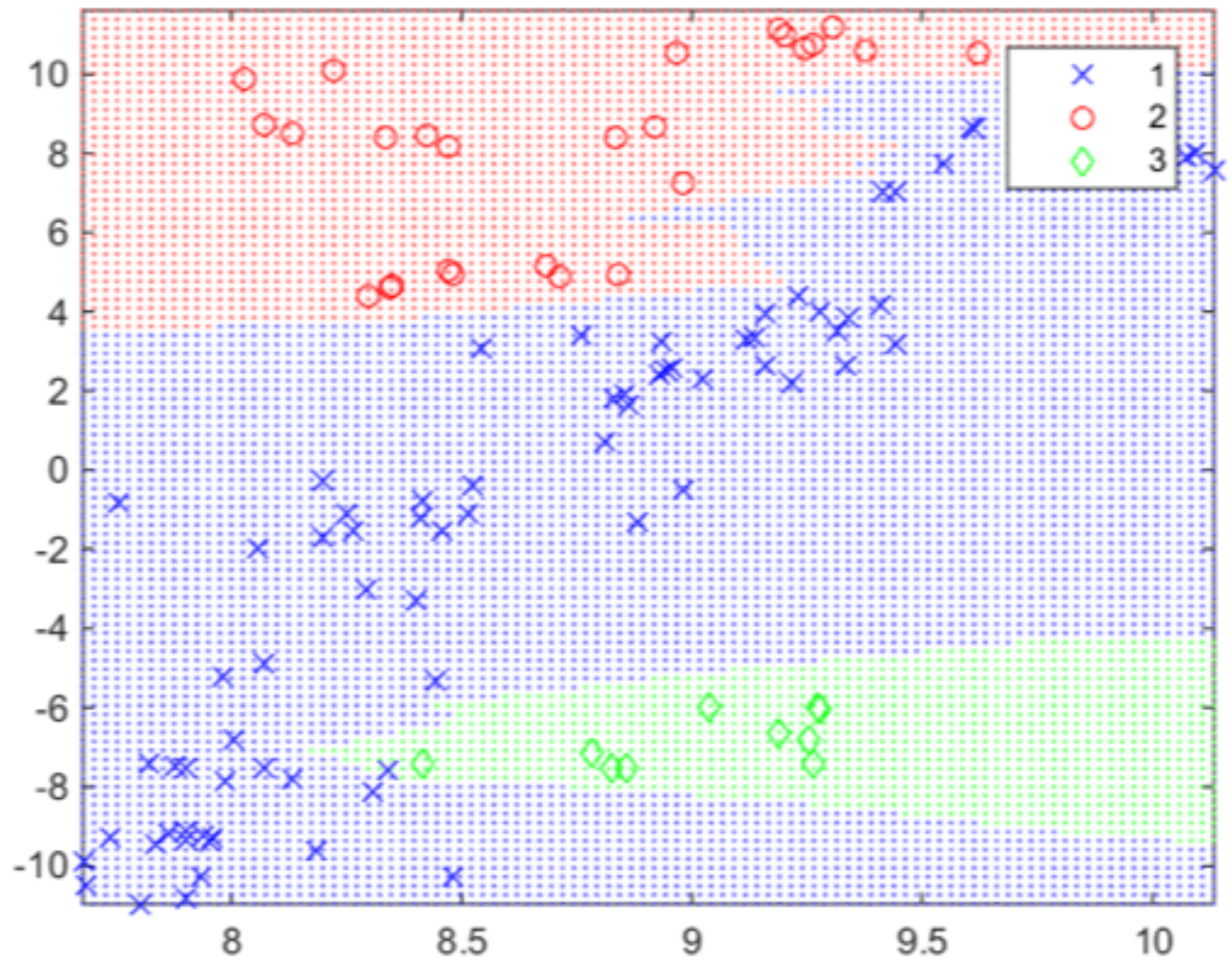

Figure 3.13: PWC decision space example in our standard generic ($x_1$, $x_2$) coordinate system

### 3.4.5 Decision Tree Algorithms

Let us discuss the notion of a classification decision tree algorithm, a set of implemented rules that make discrimination decisions. To set up the idea of a classification decision tree, we provide the following example: I have a basket of fruit (apple, orange, grape), and I wish to describe the process by which I separate the fruit into piles of similar fruits. How do I make my decisions as to which fruit goes into which pile? I could weigh each piece of fruit—grapes are the lightest, but apples and oranges might weigh the same (within the variance of their weights). I could measure the size of the pieces of fruit, but I would run into similar concerns. The most distinguishing feature is the color of the fruit, and the algorithm would go something like "look at the fruit: if red, put into basket A; if orange, put into basket B; if purple or green, put into basket C."

These logical decisions are binary (if/else), and we can graph this algorithm as a series of binary decisions, see Figure 3.14. Each decision node is a binary decision maker that divides the original population (parent) into two new populations (children): true to the left group, false to the right group. The final baskets (e.g., blue circles in Figure 3.14) or populations are referred to as terminal nodes.

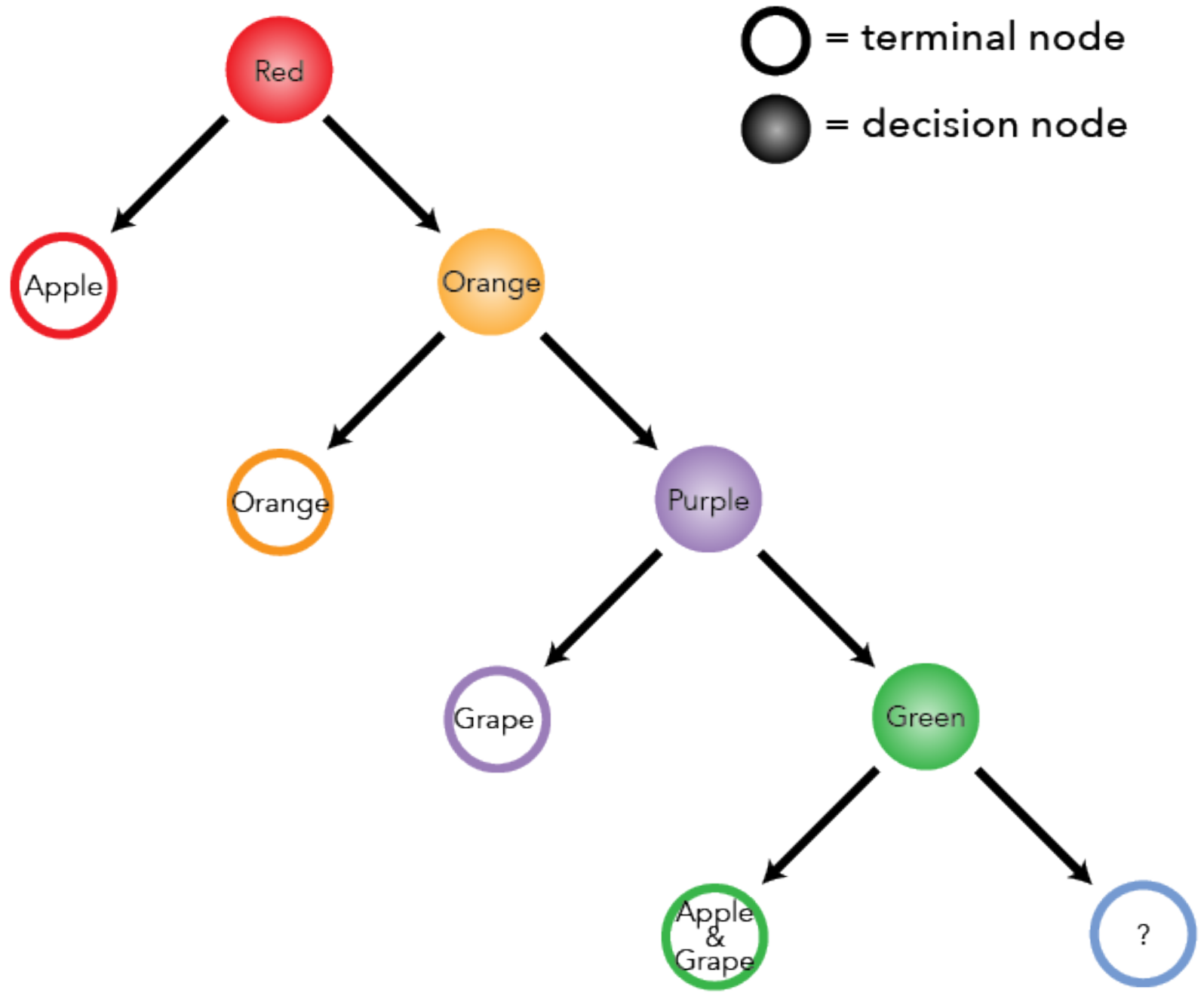

Figure 3.14: A simple example of an decision tree generated based on expert knowledge. The solid circles represent decision nodes (simple if/else statements), and the empty circles represent decisions nodes. Note that ”yes” always goes to the left child node.

#### 3.4.5.1 Classification and Regression Tree

The classification and regression tree (CART) is a method for generating (optimizing) the binary decision tree for the purposes of classification using measures of impurity [Breiman et al., 1984]. When the algorithm makes a decision, the split ($S$) is defined as a function of both the dimensions of the vector ($d$) and the threshold of the decision (some value). If we look at the simplest tree, then the parent node is given as $t$, and the two child nodes are $t_L$ and $t_R$.

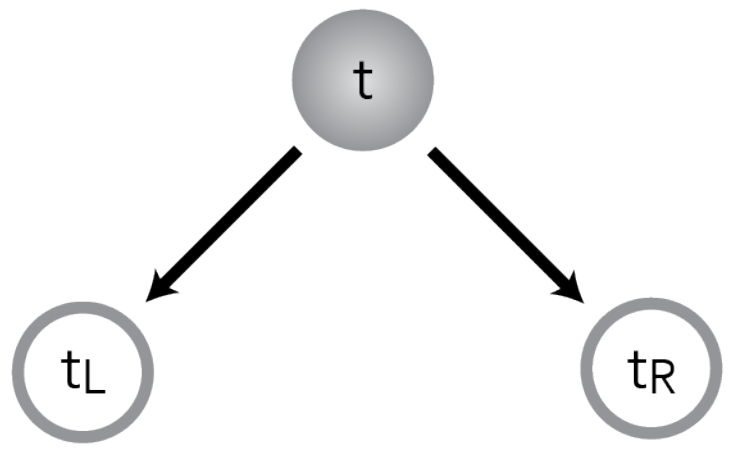

Figure 3.15: An example Simple Decision Tree with parent (t), left and right child ($t_L$ and $t_R$)

The impurity $i$ of any given node is estimated using an impurity metric. Entropy and Gini diversity are two popular methods for determining an impurity metric. Change in impurity is given as equation 3.9:

$$\Delta i(S,t) = i(t) - \left[P_L \cdot i(t_L) + P_R \cdot i(t_R)\right], \tag{3.9}$$

where $P_L = \frac{N_L}{N_t}$ and $p(j|t) = \frac{N_j}{N_t}$. Entropy-based impurity is given as equation 3.10:

$$i(t) = -\sum_j p\left(j|t\right) \cdot \log\left(p\left(j|t\right)\right). \tag{3.10}$$

The CART algorithm attempts to maximize the change in impurity of the parent or maximize the purity of the children. These iterations continue until each terminal node contains a cluster that is pure. The cross-validation process is then implemented to prune the tree (remove nodes from the tree and collapse upward) to minimize the cross-validation error. The resulting classification function is given as $T\left(\mathbf{x};\theta_b\right)$, the tree of binary (if/else) decisions ($\theta_b$) that have been optimized based on the input data set. The output of any given tree is the set of posterior probabilities, that is, likelihoods that the input vector $\mathbf{x}$ is of a given class type

$\hat{C}(x)$, which we show as equation 3.11:

$$\hat{C}(x) = \max\left(T\left(\mathbf{x};\theta_b\right)\right). \tag{3.11}$$

Figure 3.16 is a plot of a decision space and the test data (which were not used in training) based on the CART decisions;[8] note that the if/else decisions result in a purely linear decision boundary (cuts along a given dimension).

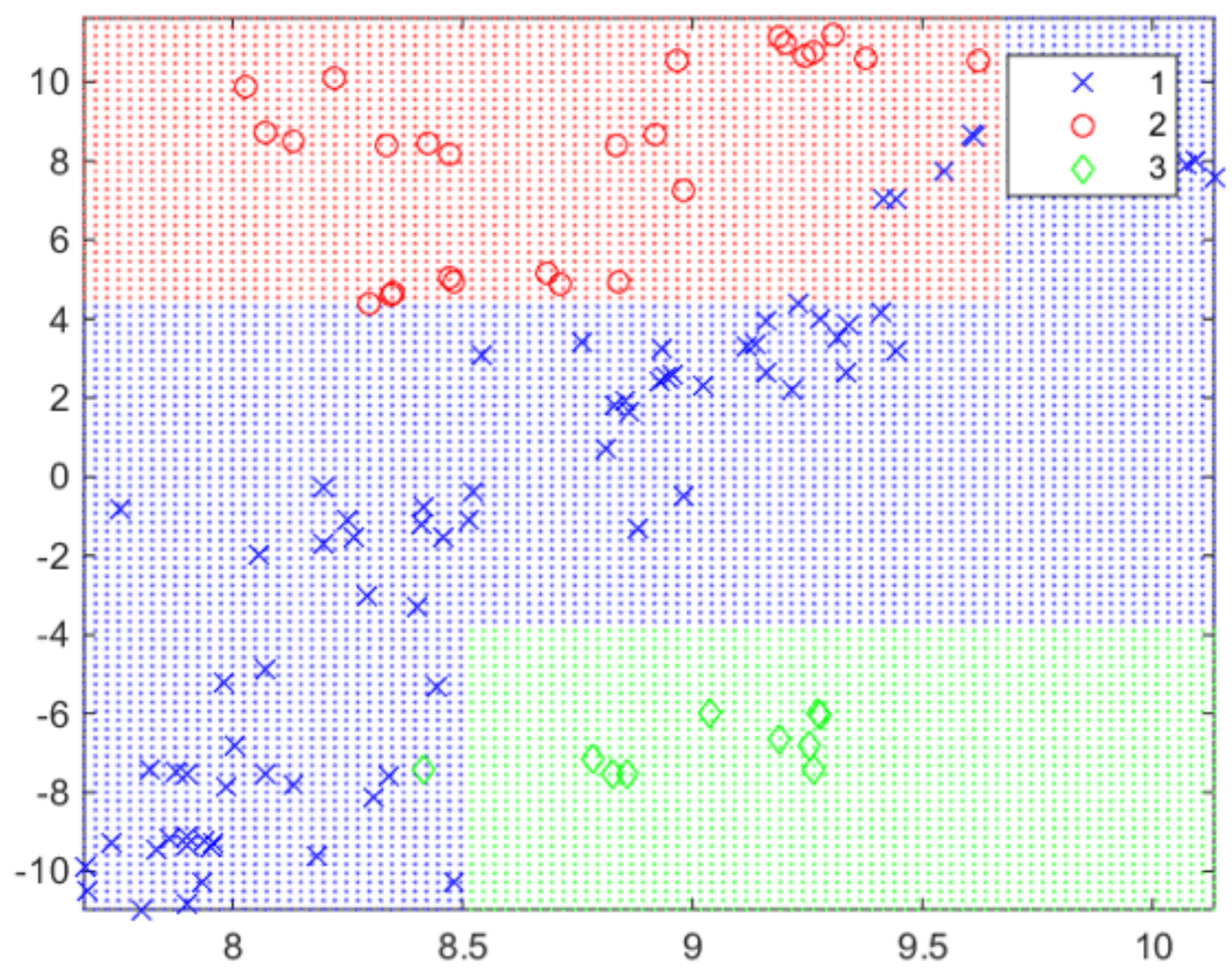

Figure 3.16: CART decision space example in our standard generic ($x_1$, $x_2$) coordinate system

#### 3.4.5.2 Random Forest Classifiers

Bootstrap aggregation can be leveraged to reduce the variance normally associated with the application of the CART algorithm to a given data set. A random sampling of $N$ points is drawn to generate a given set of input training data,; CART is applied to this subset of data to generate the tree $T_b(\mathbf{x})$. This operation is performed $B$ number of times to generate a set of noisy but unbiased models of the class decision space. This set of trees is the random forest classifier. The algorithm classifies by simply taking a consensus decision across all trees (majority rule). Hastie et al. [2009] state that if the variance of the output of an individual tree is $\sigma^2$, then the average output of the random forest is equation 3.12:

$$\rho\sigma^2 + \frac{1-\rho}{B}\sigma^2, \tag{3.12}$$

where $\rho$ is the positive pairwise correlation. As the number of trees increases, the second term goes to zero, and because $\rho \in [-1, 1]$, $\rho\sigma^2 \leq \sigma^2$; the random forest classifier will have a reduced or equal variance compared to the individual CART algorithm. This means that increasing the value of $B$ will have diminishing returns, as $\frac{1-\rho}{B} \to 0$. Optimal size of the value of $B$ can be found via cross-validation.

[8]fit.astro.vsa.utilities.ml.cart

### 3.4.6 Artificial Neural Network

We consider the binary classification case where the labels of the observations, whatever they might be, have been mapped to a 0 or 1. Logistic regression [Hastie et al., 2009] models the posterior probability of the classes, given the input observations, via a linear function in $x$. By specifying the model as a logit transform, the continuous variable input can be mapped to the domain $[0, 1]$, which we can interpret as the posterior probability:

$$Pr\left(G = 1 | X = x\right) = \exp\left(\beta_{10} + \beta_1^T x\right). \tag{3.13}$$

The parameter set $\theta = \left\{\beta_{10}, \beta_1^T\right\}$ can be solved via optimization methods; specifically, a loss function can be established as the sum of the log posterior probabilities (log-likelihood):

$$l(\theta) = \sum_{i=1}^{N} \log p_{g_i}\left(x_i; \theta\right). \tag{3.14}$$

The function can be minimized via Newton–Raphson algorithm [Heath, 2018], where the Hessian matrix and derivatives are derived;[9] the optimization process is then

$$\beta^{new} = \beta^{old} - \left(\frac{\partial^2 l(\beta)}{\partial\beta\partial\beta^T}\right)^{-1} \frac{\partial l(\beta)}{\partial\beta}. \tag{3.15}$$

The result of the optimization and the decision generation process is a linear decision line (or set of linear decision lines in the case of the multiclass problem). Based on the model in equation 3.13, a simple threshold value can be set that determines the mapping to our binary decision (0 or 1). This simple transform model is the single-layer perceptron (SLP) model given graphically in Figure 3.17, also known as a neuron.[10]

---

[9] fit.astro.vsa.utilities.ml.lrc

[10] http://www.saedsayad.com/artificial_neural_network_bkp.htm

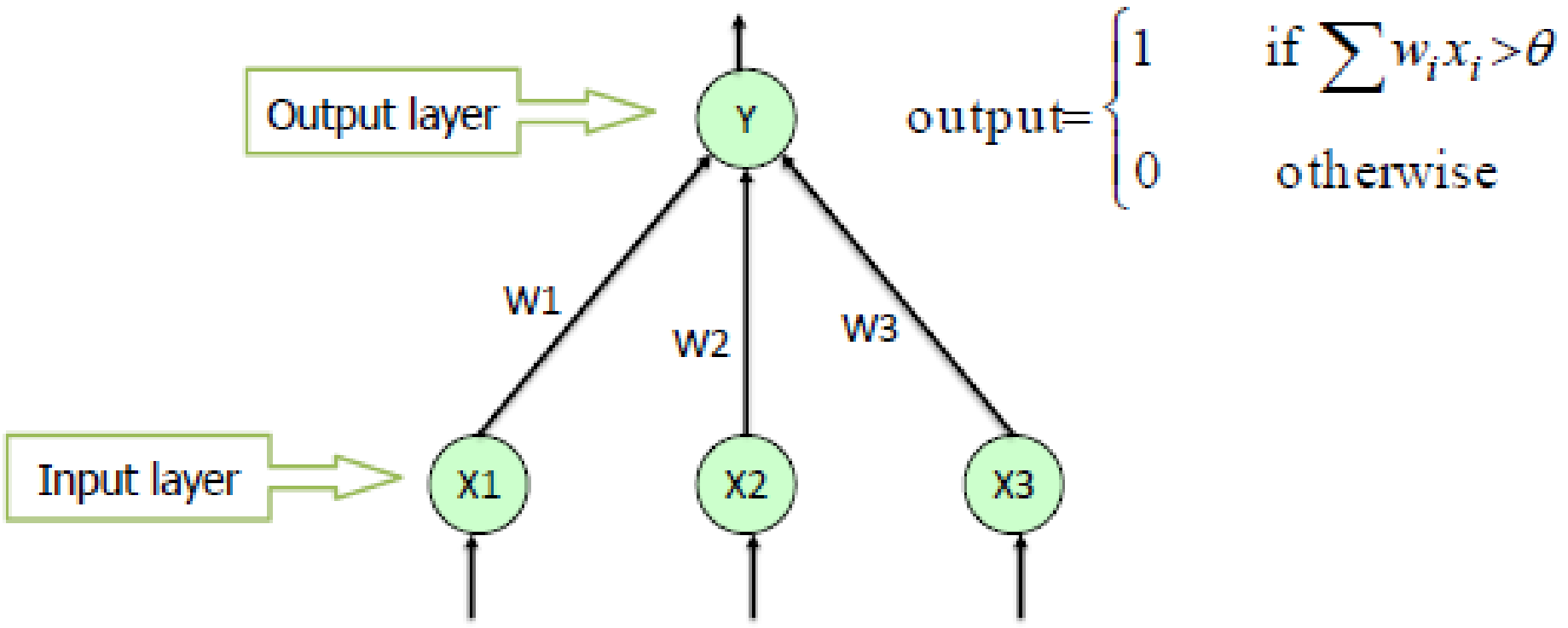

Figure 3.17: An example diagram for a Single-Layer Perceptron with single outer layer and an inner layer with three perceptrons

#### 3.4.6.1 Multilayer Perceptron

Multilayer perceptron (MLP) classification is the extension of the single-layer case in terms of both numbers of perceptrons and layers of perceptrons [Rumelhart et al., 1985]. Specifically, MLP establishes one or more "hidden layers," each one containing a finite set of transformations where each individual transformation is a single-layer perceptron. As demonstrated in Figure 3.18, the input layer (the set of observations) feeds the first hidden layer (the set of SLPs).[11] In most cases, all nodes of one layer feed all nodes of the next. While not shown, the output of the first hidden layer could feed a second hidden layer, the second a third, and so on. The last layer is the output of the algorithm, either an estimated label (classification) or a point estimate (regression).

[11] http://www.saedsayad.com/artificial_neural_network_bkp.htm

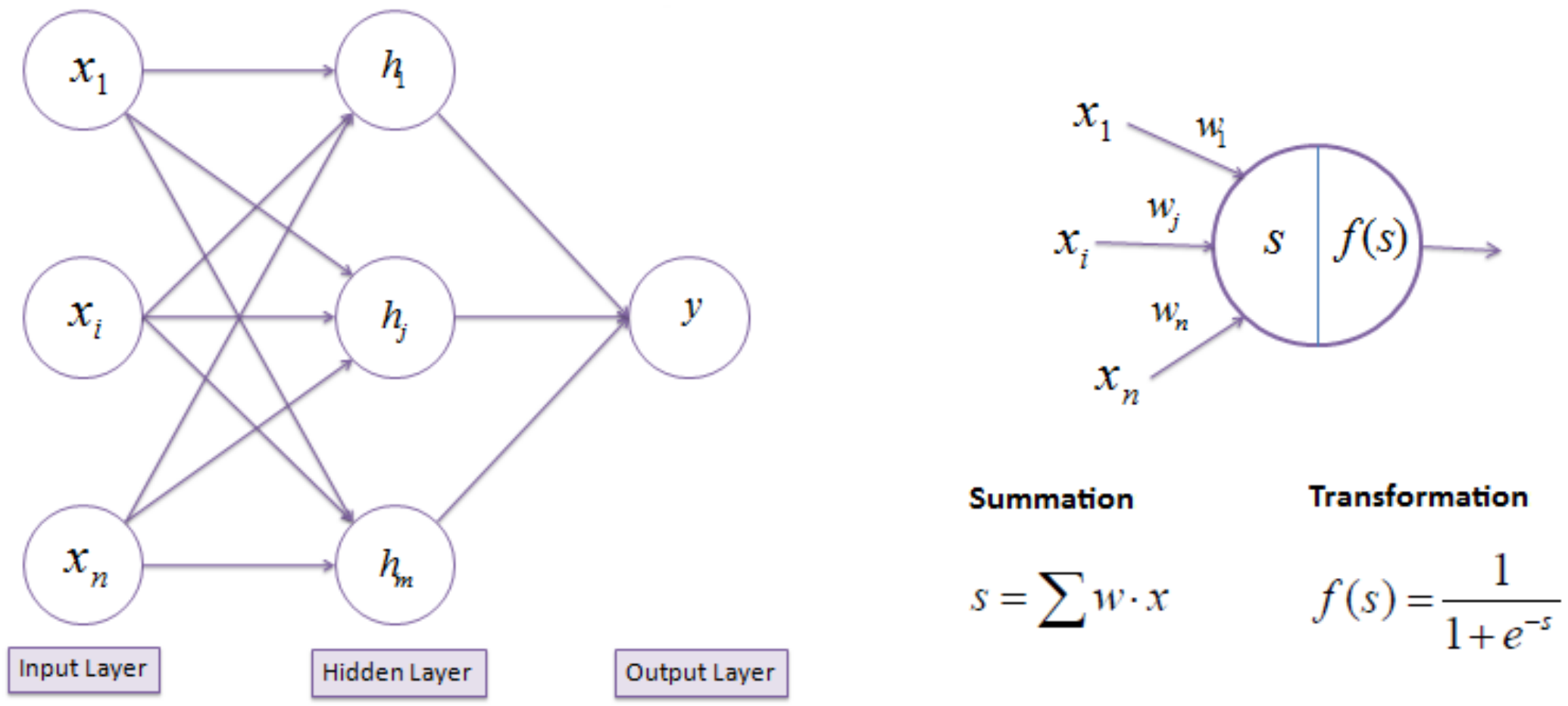

Figure 3.18: An example diagram for a Multilayer Perceptron, the left figure is the multilayer structure, the right is the perceptron/activation node (with summation and transformation)

Training of the individual perceptron weights is performed via a process of forward/backward propagation. Individual observations are passed into the input layer, and an initial set of neurons is constructed with randomly assigned weights. Based on a linear model, the results are combined and set to the output layer. Residuals are generated by comparing the output layer to the observed labels or response values, and these residuals are propagated backward through the layers and applied to each perceptron based on the amount of residual for which the perceptron was responsible. This forward–backward iterative process continues until the residuals computed are minimized to some tolerance. The set of hidden layer and neurons allows the MLP process to expand what was a linear decision space from the SLP or logistic regression algorithm into a nonlinear model. An example MLP has been applied to our test data set, and the resulting decision space is demonstrated in Figure 3.19.

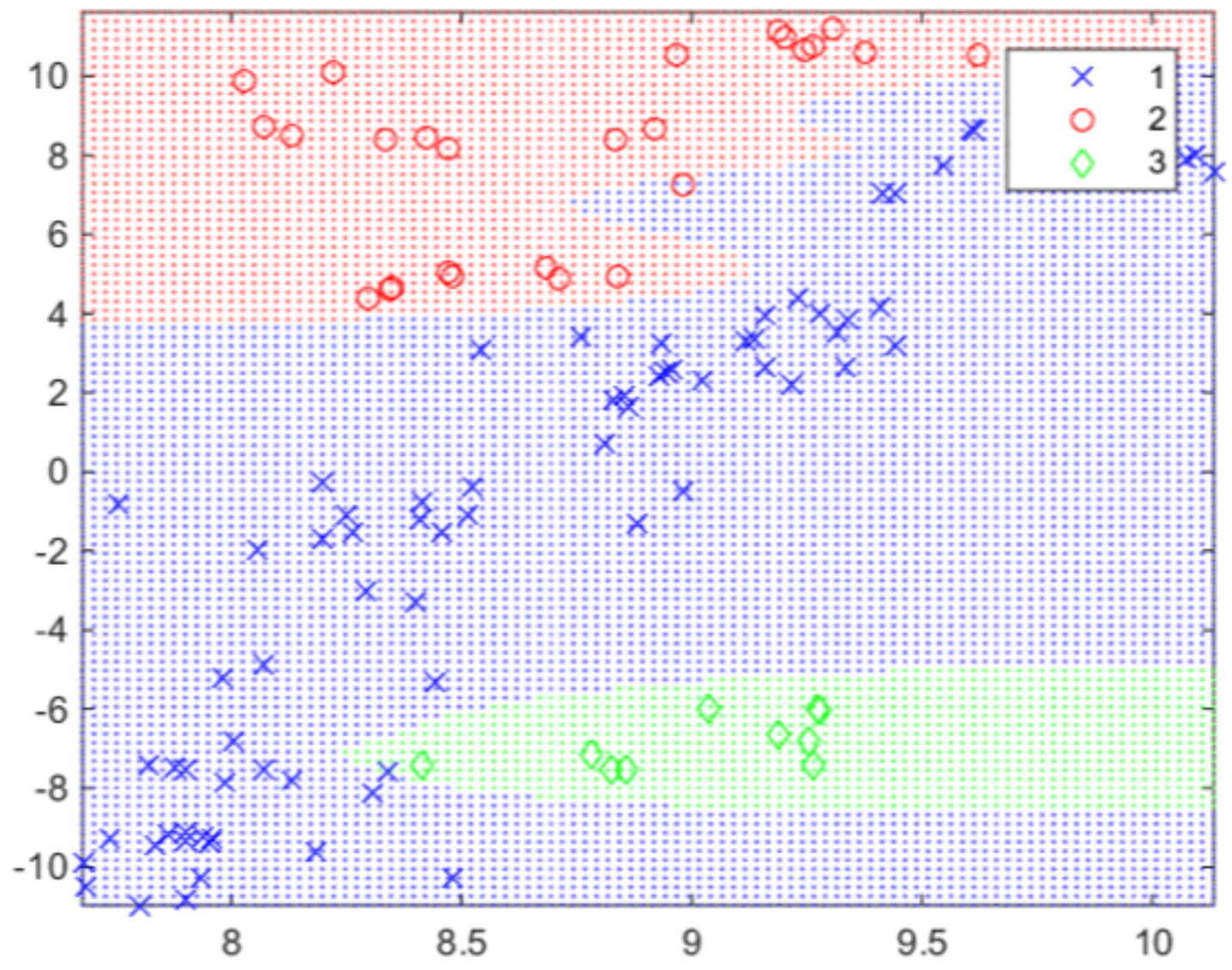

Figure 3.19: MLP decision space in our standard generic ($x_1$, $x_2$) coordinate system

### 3.4.7 k-Means Clustering

As a means of standard clustering, based on vector-variate input (e.g., $\mathbb{R}^{m\times 1}$), k-means is both a straightforward operation and an industry standard [Duda et al., 2012]. The algorithm falls into the category of descent optimization. The algorithm iterates in an attempt to define clusters, that is, groupings of similar inputs where similarity is based on a set of defined rules associated with distance. Although we can define distance however we see fit, the standard k-means clustering algorithm uses the Euclidean distance; this need not be an absolute requirement, however—any estimation of central tendency could be used here [Modak et al., 2018]. Clusters, for the purpose of the algorithm, are defined by a single vector, that is, the centroid of the cluster. Membership $C(i)$ of any given input point to a cluster is based on closest cluster centroid ($m_k$). The algorithm optimizes the within-point scatter given the cluster definitions. This optimization problem is defined as

$$\min_{C,\{m_k\}_1^K} \sum_{k=1}^{K} N_k \sum_{C(i)=k} \|x_i - m_k\|^2 . \tag{3.16}$$

The algorithm operates as follows:

1. Each data point is randomly assigned a cluster membership.
2. For a given cluster assignment (membership of each point to a cluster based on closest centroid), the center of the cluster is estimated (mean of members).
3. Based on the means estimated in the prior step, the points are reassigned membership based on closest cluster centroid: $C(i) = \min_{1\leq k\leq K} \|x_i - m_k\|^2$.
4. Steps 2 and 3 are iterated until membership among the input data set is unchanged.

The resulting output is a list of clusters, their center vectors, and the associated members. Because of the random nature of the initial assignment, the final assignment of cluster number can vary from application to application of the algorithm to the same data set. Likewise, results can vary depending on the number of clusters used in the operation. As part of our analysis, we propose a feature space that is matrix-variate (see section 7.3.4), and likewise our analysis of the resulting observations needs to be able to accept inputs that are on the $\mathbb{R}^{m\times n}$ space. To that end, we propose a k-means algorithm that uses the Forbinus norm instead of the Euclidean. The algorithm can then be directly applied to the observed feature

data.[12]

### 3.4.8 Principal Component Analysis (PCA)

The key to useful feature space reduction is the transformation of an observed raw data set into a new domain with a smaller dimensionality but with roughly the same amount of useful information [Cassisi et al., 2012]. What quantifies "useful" is a point of order that can be disputed, but PCA defines this as a function of correlation [Einasto et al., 2011]. As an example, a feature space in the domain $\mathbb{R}^{n \times 1}$ where all $n$ dimensions are perfectly correlated contains redundant information. So we ask, how much of a "difference" is each feature making? How much am I getting for my effort? How many features do I need? What are the diminishing returns? As an example, we present the Palomar Transient Factory (PTF) and Catalina Real-Time Transient Survey (CRTS) features, two different surveys using two different feature spaces. A correlation matrix of features is presented in Figure 3.20, and it is apparent that there are features that do not correlate (light colors) and features that almost certainly do (dark colors).[13]

---

[12] fit.astro.vsa.common.utilities.clustering.KMeansClustering

[13] Mahabal, A. (2016). *Complete classification conundrum.* Presented at Statistical Challenges in Modern Astronomy VI, June 6–10, Carnegie Mellon University.

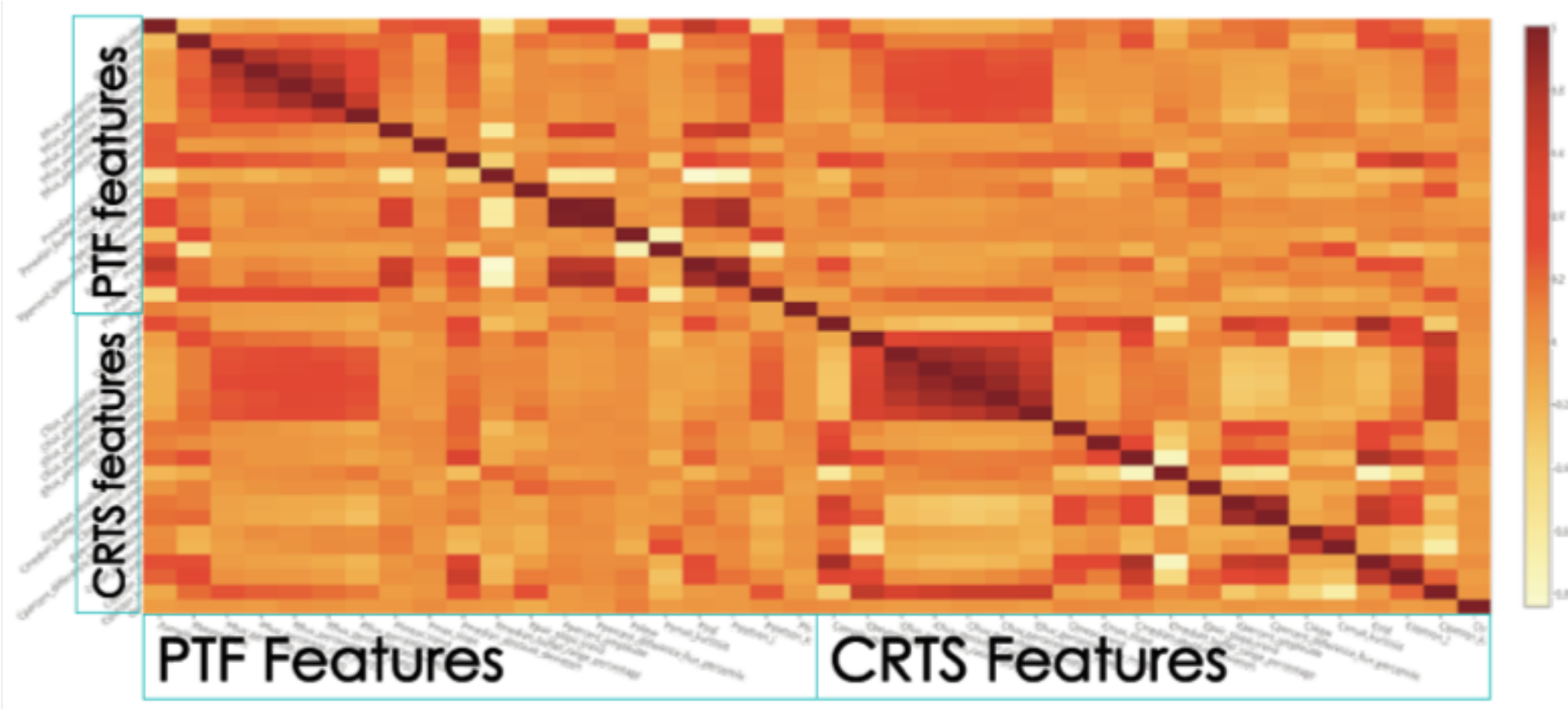

Figure 3.20: Correlation of features between two different surveys, note the features collected from each survey were independent, but on the some of the same targets. Light colors represent no correlation between features, dark colors represent high correlation.

PCA is a rotation of the axes of the original feature space (orthogonal axes) into a domain where the new axes, referred to as the principal components, lie along the directions of maximum variation. This is called the principal axis method and results in orthogonal (uncorrelated) factors. There are a number of means for computing the PCA of a data set [Miranda et al., 2008]; here we briefly review the connection between PCA and the known methodology of singular value decomposition (SVD). The input matrix of observation $X$ is decomposed via $X = U\Sigma W^T$; a number of computationally efficient means of solving for these components exists [e.g., Bosner, 2006]. The set of principal components is then $T = XW$, where each column of $T$ is a principal component, with the first column being the component that has maximum variation. A transformation of $X$ using PCA for the purposes of dimensionality reduction can be performed via $T_L \leftarrow XW_L$, where $T_L$ has been truncated to the first $L$ number of dimensions.[14]

### 3.4.9 k-Means Representation

Much of machine learning, in particular, the preparation of input data prior to training, involves the transformation or mapping of the original observed data and feature space into a domain that is favorable to achieving the machine learning goals. This might mean a dimensionality increase, as would be the case with a radial basis function neural network [Park and Sandberg, 1991, Faloutsos et al., 1994]; the input data set of feature dimension $\mathbb{R}^{m\times 1}$ is compared to a set of neurons that numbers larger than the dimensionality of the original feature space ($n > m$). The components of the neuron can vary depending on implementation, but generally the radial basis function is commonly taken to mean the Gaussian kernel given here as equation 3.17:

$$\rho(x_i; c) = \exp\left(-\beta \left\|x_i - c\right\|^2\right), \tag{3.17}$$

where $\beta$ is the width of the kernel and $c$ is the centroid. The value and number of centroids are provided by the user, though they are often a set or subset of the input training data. The width of the kernel can be optimized based on training procedures. Likewise, it can be naively fixed, as can the weight of each neuron output similar to the multilayer perceptron classifier discussed in section 3.4.6.1. By applying a classifier that might generate linear decision criteria (e.g., logistic regression) to this expanded $\mathbb{R}^{n\times 1}$ feature space, the linear decision mapped back

[14] fit.astro.vsa.analysis.feature.PCA

to the original $\mathbb{R}^{m\times 1}$ can be nonlinear, resulting in a more generic or adaptable classifier.

Conversely, we can decrease the dimensionality of the observational feature space by selecting a set of neurons that numbers smaller than the dimensionality of the original feature space ($n < m$). Let us suppose that the data are already clustered or grouped together prior to implementation of a dimensionality reduction methodology [Park et al., 2003]. This would be the case in the supervised classification problem where the data have an associated label (class); this would also be the case if k-means or some other unsupervised clustering methodology had been applied to unlabeled data (cluster number). Assuming a vector-variate input (e.g., $\mathbb{R}^{m\times 1}$), the centroid of any given grouping can be given as the average of the members:

$$c = \frac{1}{N}\sum_{i=1}^{N} a_i. \tag{3.18}$$

We could just as easily use the median or other measurements of the cluster centroid. The result is the set of centroids that represent our neurons $C = [c_1, c_2, ..., c_k] \in \mathbb{R}^{m\times k}$ and can be used to transform our original observations into a new $k$-dimensional equation

$$Y = \left[\|x_i - c_1\|^2, \|x_i - c_2\|^2, ..., \|x_i - c_k\|^2\right]. \tag{3.19}$$

This is a simplified version of the algorithm presented by Park et al. [2003], who include additional weighting of the individual contributions of each neuron determined by an optimization process. Likewise, we have included here only the Euclidean distance ($\|x_i - c_2\|^2$), but other distance measurements could easily be included or used in its stead [Cha, 2007].

### 3.4.10 Metric Learning

Metric learning has its roots in the understanding of how and why observations are considered similar. The very idea of similarity is based around the numerical measurement of distance, and the computation of a distance is generated via application of a distance function. Bellet et al. [2015] define a distance function as follows: over a set $X$, a distance is a pairwise function $d : X \times X \to \mathbb{R}$ that satisfies the following conditions for all $x, x', x" \in X$:

1. $d(x, x') = d(x', x)$ [symmetry]

2. $d(x, x') \geq 0$ [non-negativity]

3. $d(x, x') = 0$ if and only if $x = x'$ [identity of indiscernibles]

4. $d(x, x') \leq d(x, x') + d(x', x")$ [triangle inequality]

These are standard requirements for a distance and collectively are referred to as the distance axioms. Bellet et al. [2015] further define the specific metric distance as equation 3.20:

$$d(x, x') = \sqrt{(x - x')^T \mathbf{M} (x - x')}, \tag{3.20}$$

where $X \subseteq \mathbb{R}^n$ and the metric is required to be $\mathbf{M} \in \mathbb{S}_+^n$, where $\mathbb{S}_+^n$ is the cone of symmetric positive semi-definite $n \times n$ real-valued matrices. Metric learning seeks to optimize this distance via manipulation of the metric $\mathbf{M}$, based on available side data (see Figure 3.21).

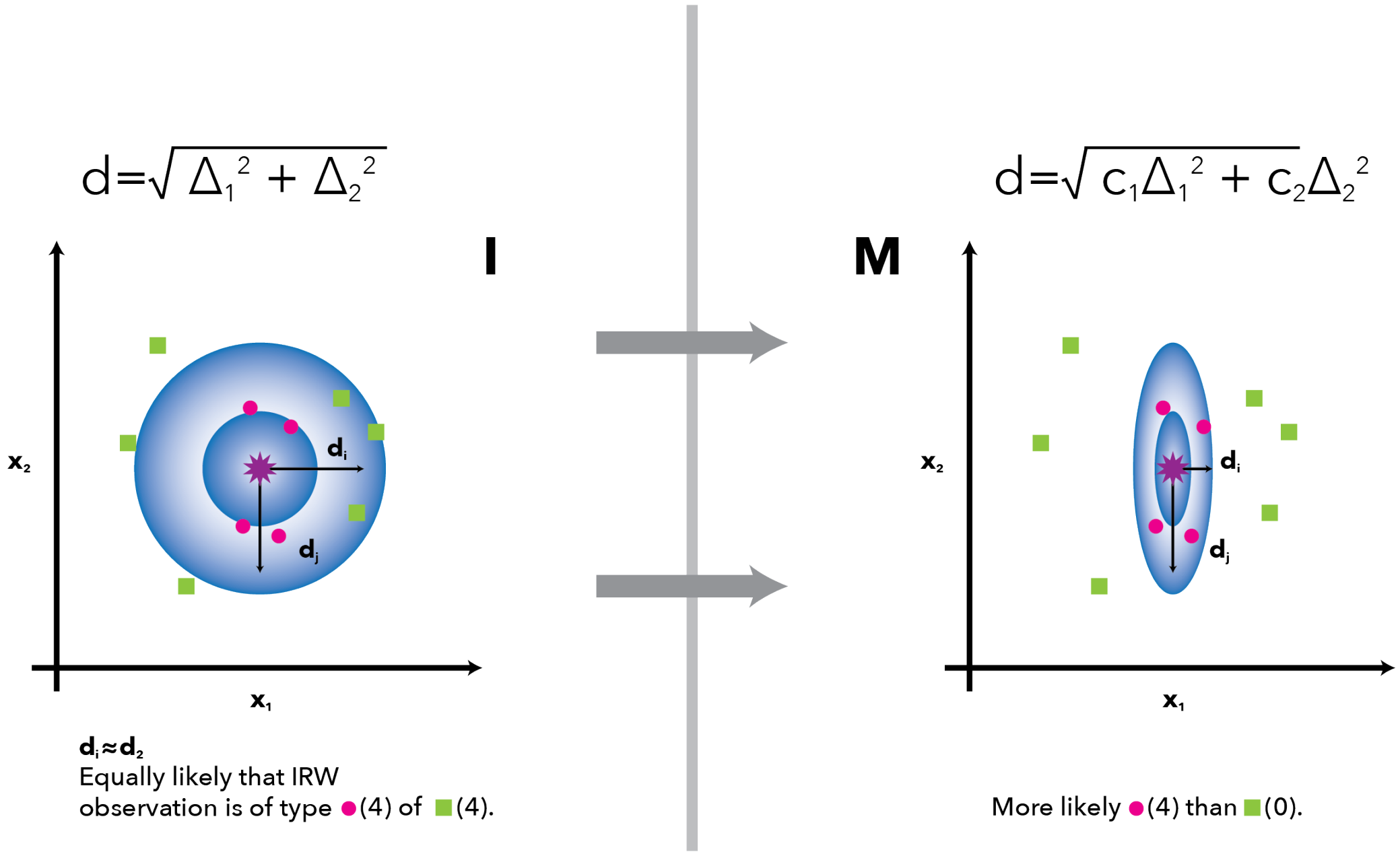

Figure 3.21: Example of the change in distance between points as a results of Metric Learning optimization. Note, left figure is in the Euclidean case, the right figure is with a tailored metric.

How the optimization occurs and what is considered important, that is, the construction of the objective function, together compose the underlying difference between the various metric learning algorithms. The side information leveraged is defined as the set of labeled data $\{x_i, y_i\}_{i=1}^{n}$; furthermore, the triplet is defined as $(x_i, x_j, x_k)$, where $x_i$ and $x_j$ have the same label but $x_i$ and $x_k$ do not. It is expected, then, based on the definition of similarity and distance, that $d(x_i, x_j) < d(x_i, x_k)$, that is, that the distances between similar labels are smaller than the distances between dissimilar labels. Methods like LMNN [Weinberger et al., 2009] leverage this comparison to bring similar things closer together, while pushing dissimilar things further apart.

Given the metric learning optimization process, the result is a tailored distance metric and associated distance function (equation 3.20). This distance function is then leveraged in a standard k-NN classification algorithm. The k-NN algorithm estimates a classification label based on the closest samples provided in training [Altman, 1992]. If $x_n$ is a set of training data $n$ big, then we find the distance between a new pattern $x_i$ and each pattern in the training set. The new pattern is classified depending on the majority of class labels in the closest $k$.[15]

#### 3.4.10.1 Large-Margin Nearest Neighbor Metric Learning (LMNN)

Bellet et al. [2015] review a number of metric learning algorithms, in addition to a suite of algorithms for augmenting and improving the outlined classification techniques. Methods include the fundamental MMC method [Xing et al., 2003] and NCA method [Goldberger et al., 2005], both of which follow the procedure of establishing an objective function and then attempting to optimize the metric with respect to the side information available (input data). That being said, the Large-Margin Nearest Neighbor Metric learning [Weinberger et al., 2009] algorithm is a metric learning design that generalizes well to most situations.

It is also easily extendable, making it a key component in the development of many other, more complex designs. Many of the more advanced methods discussed by Bellet et al. [2015] extend the original LMNN design in some fashion. Simply put, the goal of LMNN is to bring things that are similar closer together, while pushing things that are different further away from one another; for an example, see Figure 3.22. This is done via construction of an objective function with respect to the metric distance.

[15] fit.astro.vsa.utilities.ml.metriclearning

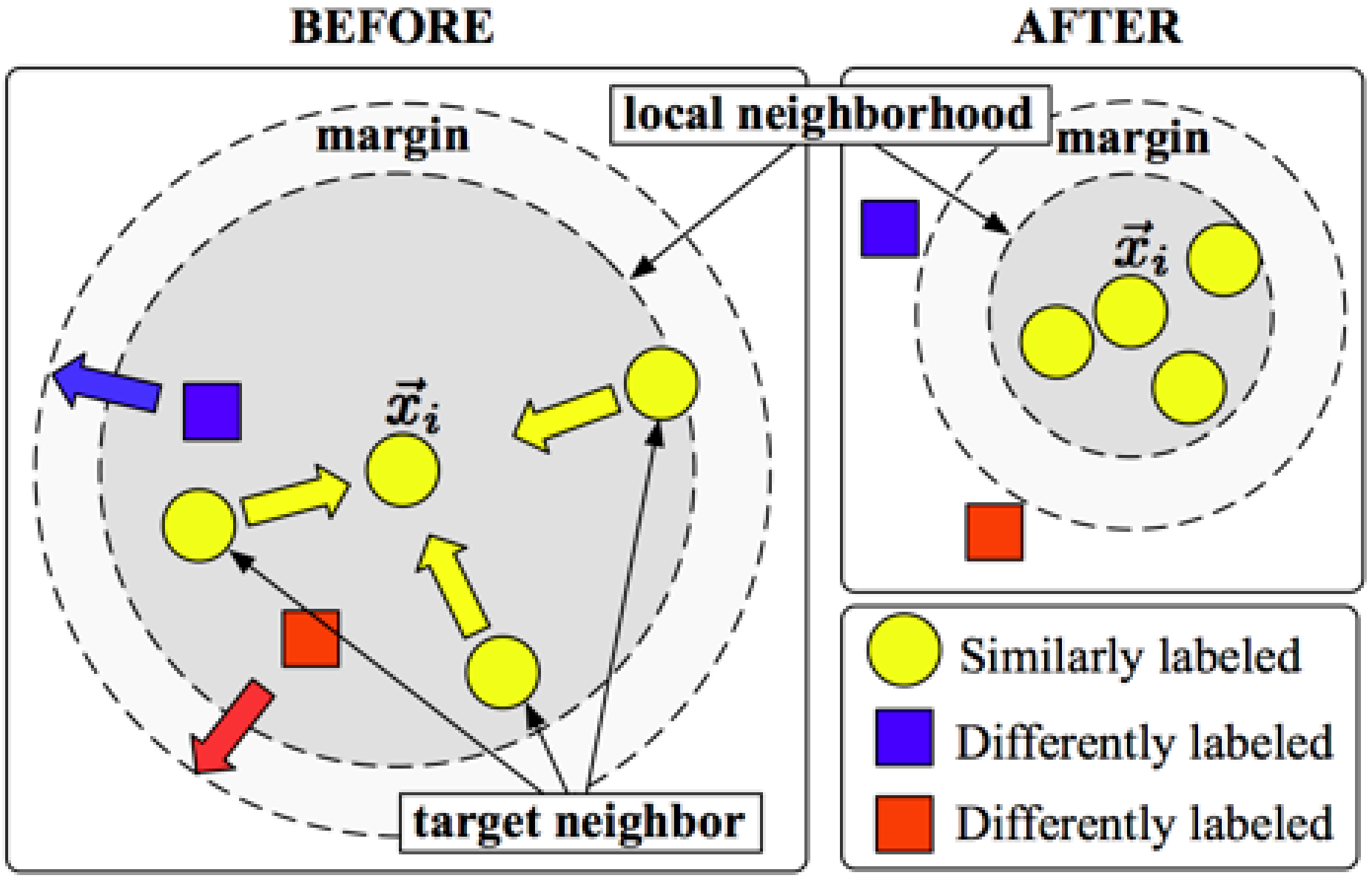

Figure 3.22: Diagram demonstrating the goal of Large-Margin Nearest Neighbor Metric Learning, the objective function is constructed such that observations from the same class are brought closer together, while observations from different classes are pushed further apart. [Weinberger et al., 2009]

We can compose the LMNN problem as the following convex optimization:

$$\begin{aligned} &\min_{\mathbf{M} \in S_+^d, \xi \geq 0} (1-\mu) \textstyle\sum_{(x_i,x_j) \in S_{lmnn}} d_{\mathbf{M}}^2 (x_i, x_j) + \mu \textstyle\sum_{i,j,k} \xi_{ijk} \\ &s.t.\ d_{\mathbf{M}}^2 (x_i, x_j) - d_{\mathbf{M}}^2 (x_i, x_k) \geq 1 - \xi_{i,j,k}\ \forall (x_i, x_j, x_k), \end{aligned} \tag{3.21}$$

where $d_{\mathbf{M}}^2 (x_i, x_j) = \|x_i - x_j\|_{\mathbf{M}}^2$ and $\mu \in [0, 1]$. The convex optimization problem can be reformulated as an objective function that is strictly convex while still maintaining the requirement that the metric be positive semidefinite. This is done via decomposition of the metric, $\mathbf{M} = \mathbf{L}^T\mathbf{L}$, and a replacement of the inequality constraint (i.e., slack function) with the hinge loss function ($[z]_+ = max(z, 0)$). Likewise, we can replace the summation over all similar pairs with a summation of those both similar and "within the neighborhood," which is defined via k-nearest neighbors based on Euclidean distance ($\mathbf{M} = \mathbf{I}$).

This limitation to those points within some limited neighborhood distance allows for a more rapid computation of optimal metric, with little to no loss of performance (the points that are important are the points that are used). The resulting transformations are given as the following objective function:

$$\begin{aligned} \varepsilon(\mathbf{L}) = (1-\mu)\textstyle\sum_{ij}\eta_{ij}\left\|\mathbf{L}(x_i - x_j)\right\|^2 \\ +\mu\textstyle\sum_{ijk}\eta_{ij}(1-y_{ik})\left[1 + \left\|\mathbf{L}(x_i - x_j)\right\|^2 - \left\|\mathbf{L}(x_i - x_k)\right\|^2\right]_+, \end{aligned} \tag{3.22}$$

where $\eta_{ij} = 1$ when $x_i$ and $x_j$ are in the same neighborhood and $y_{ik} = 0$ when $x_i$ and $x_k$ are of the same class-type. Optimization occurs via gradient descent (equation 3.23), where the derivative of the objective function is found and used to iterate the solution:

$$\mathbf{L}^{(t+1)} = \mathbf{L}^{(t)} - \beta \frac{\partial \varepsilon(\mathbf{L})}{\partial \mathbf{L}}. \tag{3.23}$$

A full derivation and algorithm design can be found in Weinberger et al. [2009].[16]

### 3.4.11 Metric Learning Improvements

The LMNN algorithm is the basis on which a large number of proposed metric learning algorithms have been constructed. These additional algorithms introduce additions to the objective function or changes to the training and design of the classifier to provides additional flexibility. For example, the S&J metric learning algorithm [Schultz and Joachims, 2004] introduces a regularization term as part of the loss function for the purpose of favoring a lower-complexity metric. Kernel-LMNN [Chatpatanasiri et al., 2008] and $M^2$-LMNN [Weinberger and Saul, 2008] learn a kernel transformation and multiple metrics (via tree partitioning) to generate a much more complex decision space. Bottou and Bousquet [2008] develop a distributed/stochastic version by estimating an overall metric $M$ by training on smaller segments of a larger training set via averaging or EWMA, depending on whether all the data are available or whether the training process is continuous. Kedem et al. [2012] develop gradient-boosted LMNN, a generalization of LMNN to learn a distance in a nonlinear projection space defined by some transformation based on gradient-boosted regression trees. Likewise, Parameswaran and Weinberger [2010] reformulate LMNN to learn the shared metric based on a given number of $T$ tasks (different data sets). These formulations and extensions allow the underpinning design of LMNN to be extensible to many applications, and while they are mentioned here as examples, the designs we present can of course be modified with these or many of the other designs [Bellet et al., 2015], based on the needs of the task.

[16] fit.astro.vsa.utilities.ml.metriclearning.lmnn

### 3.4.12 Multi-view Learning

We address the following classification problem: given a set of expertly labeled side data containing $C$ different classes, where measurements can be made on the classes in question to extract a set of features for each observation, how do we define a distance metric that optimizes the misclassification rate? As discussed, we have identified a number of features and feature spaces that may provide utility in discriminating between various types of stellar variables. How do we combine this information and generate a decision space, or rather, how do we define the distance $d_{ij} = (x_i - x_j)'\mathbf{M}(x_i - x_j)$ when $x_i$ contains two matrices (SSMM or DF in our case)? Specifically, we attempt to construct a distance metric based on multiple attributes of different dimensions (e.g., $\mathbb{R}^{m \times n}$ and $\mathbb{R}^{m \times 1}$). To respond to this challenge, we investigate the utility of multi-view learning. For our purposes here, we specify each individual measurement as the feature and the individual extractions or representations of the underlying measurement as the view. As an example, if provided the color of a variable star in *ugriz*, the individual measurements of $u - g$ or $r - i$ shall be referred to here as the features, but the collective set of colors is the view.

Xu et al. [2013] review multiview learning and outline some basic definitions. Multiview learning treats the individual views separately but also provides some functionality for joint learning, where the importance of each view is dependent on the others. As an alternative to multi-view learning, the multiple views could be transformed into a single view, usually via concatenation. The costs and benefits of single-view versus multi-view learning are discussed by Xu et al. [2013] and are beyond the scope of this discussion. Likewise, classifier fusion [Tax and Duin, 2001, Tax and Muller, 2003, Kittler et al., 1998] could be viewed as an alternative to multi-view learning. Here each view would be independently learned and result in an independent classification algorithm. The results of the set of these classifiers are combined (mixing of posterior probability) to result in a singular estimate of classification/label. This is similar to the operation of a random forest classifier; that is, results from multiple individual trees combine to form a joint estimate. The single-view learning with concatenation, multi-view learning, and classifier fusion designs *can be differentiated by when the joining of the views is considered: before, during, or after.*

Multi-view learning can be roughly divided into three topic areas: (1) co-training, (2) multiple-kernel learning, and (3) subspace learning. Each method attempts to consider all views *during* the training process. Multiple-kernel learning algorithms attempt to exploit kernels that naturally correspond to different views and combine kernels either linearly or nonlinearly to improve learning performance [Gönen and Alpaydın, 2011]. Subspace learning uses canonical correlation analysis (CCA) or a similar method to generate an optimal latent representation of two views, which can be trained on directly. This CCA method can be performed multiple times for the case in which many views exist; it also frequently results in a dimensionality that is lower than the original space [Hotelling, 1936, Akaho, 2006, Zhu et al., 2012]. This work will focus on the method of co-training, specifically, *metric co-training*. Large-margin multi-view metric learning [Hu et al., 2014, 2018] is an example of metric co-training; the designed objective function minimizes the optimization of the individual view as well as the difference between view distances, simultaneously. LM$^3$L is reviewed, allowing us to establish some basic definitions regarding multi-view learning for the general astronomical community.

#### 3.4.12.1 Large-Margin Multi-view Metric Learning

The multi-view metric distance is defined as:

$$d_M^2(x_i, x_j) = \sum_{k=1}^{K} w_k \left(x_i^k - x_j^k\right)^T \mathbf{M}_k \left(x_i^k - x_j^k\right), \tag{3.24}$$

where $K$ is the number of views and $x_i^k$ is the $i^{\text{th}}$ observation of the $k^{\text{th}}$ view for a given input. A weight of importance $w_k$ is estimated for each view via our co-training, as is the metric for the view $\mathbf{M}_k$. Note that each view can have a different dimensionality, as each $\mathbf{M}_k$ is uniquely defined for the view for which it was trained. It is apparent that the distance is some weighted average of the individual view distances. The objective function for LM$^3$L [Hu et al., 2014, 2018] is defined as:

$$\begin{gathered}\min_{\mathbf{M}_1,\ldots,\mathbf{M}_K} J = \sum_{k=1}^{K} w_k^p I_k + \lambda \sum_{k,l=1,k<l}^{K} \sum_{i,j} \left( d^2_{\mathbf{M}_k}(x_i^k, x_j^k) - d^2_{\mathbf{M}_l}(x_i^l, x_j^l) \right)^2 \\ s.t. \quad \sum_{k=1}^{K} w_k = 1, w_k \geq 0, \lambda > 0,\end{gathered} \tag{3.25}$$

where $I_k$ is the objective function for a given $k^{\text{th}}$ individual view:

$$\min_{\mathbf{M}_k} I_k = \sum_{i,j,} h \left[ \tau_k - y_{ij} \left( \mu_k - d^2_{\mathbf{M}_k}(x_i^k, x_j^k) \right) \right], \tag{3.26}$$

where $h[z]$ is the hinge loss function; $y_{ij} = 1$ when data are from the same class, and $y_{ij} = -1$ otherwise; and $\tau_k$ and $\mu_k$ are threshold parameters that enforce the constraint $y_{ij} \left( \mu_k - d^2_{\mathbf{M}_k}(x_i^k, x_j^k) \right) > \tau_k$. This inequality constraint is the usual large-margin constraint seen in metric learning: that the distance between two similarly labeled data, $d^2_{\mathbf{M}_k}(x_i^k, x_j^k)$, should be smaller than $\mu_k - \tau_k$ and larger than $\tau_k + \mu_k$ if the data are from different classes. The parameter $\lambda$ is a control for the importance of the co-training (i.e., the relationship between distance metrics between different views).

This problem can be solved via a gradient descent scheme. In practice, optimizing $\mathbf{M}_k$ requires enforcing the requirement $\mathbf{M}_k \succ 0$, which can be slow, depending on the methodology used. Hu et al. [2014] transform the metric $\mathbf{M_k}$, following Weinberger et al. [2009], via the decomposition $\mathbf{M} = \mathbf{L}^T\mathbf{L}$. This allows for unconstrained optimization of the objective function with respect to the decomposed matrix $\mathbf{L_k}$; the matrix $\mathbf{L_k}$ can then be used to generate an appropriate metric $\mathbf{M_k}$.

The gradient of the LM$^3$L optimization function can be shown as:

$$\frac{\partial J}{\partial \mathbf{L}_k} = 2\mathbf{L}_k \left[ w_k^p \sum_{i,j} y_{ij} h'[z] \mathbf{C}_{ij}^k + \lambda \sum_{l=1, l \neq k}^{K} \sum_{i,j} \left( 1 - \frac{d_{\mathbf{M}_k}^2(x_i^k, x_j^k)}{d_{\mathbf{M}_l}^2(x_i^l, x_j^l)} \right) \mathbf{C}_{ij}^k \right], \quad (3.27)$$

where $\mathbf{C}_{ij}^k = \left(x_i^k - x_j^k\right)\left(x_i^k - x_j^k\right)^T$ is the outer product of the differences and $h'[z]$ defines the derivative of the hinge loss function. The algorithm operates as a two-step process (alternating optimization) between the optimization of the decomposed metrics $\mathbf{L}_k$ and the weighting between the views $w_k$. First, the *weights are fixed*, $w = [w_{1,} w_{2,} ... w_k]$, and the metrics $\mathbf{M}_k$ are updated. The iterative update to the $\mathbf{L}_k$ estimate is generated via gradient for each view:

$$\mathbf{L}_k^{(t+1)} = \mathbf{L}_k^{(t)} - \beta \frac{\partial J}{\partial \mathbf{L}_k}. \quad (3.28)$$

Second, the metrics $\mathbf{M}_k$ *are fixed with the updated values* and *the individual weights*, $w = [w_{1,} w_{2,} ... w_k]$, are estimated. To estimate the update for the weights, a Lagrange function is constructed in equation 3.29:

$$\begin{aligned} La(w, \eta) = {} & \textstyle\sum_{k=1}^{K} w_k^p I_k \\ & + \lambda \textstyle\sum_{k,l=1,k<l}^{K} \sum_{i,j} \left(d_{\mathbf{M}_k}(x_i^k, x_j^k) - d_{\mathbf{M}_l}(x_i^l, x_j^l)\right)^2 - \eta \left(\sum_{k=1}^{K} w_k - 1\right), \end{aligned} \quad (3.29)$$

The Lagrange equation is optimized and each weight is estimated using equation 3.30:

$$w_k = \frac{(1/I_k)^{1/(p-1)}}{\sum_{k=1}^{K} (1/I_k)^{1/(p-1)}}. \quad (3.30)$$

These two steps are then repeated for each iteration until $\left|J^{(t)} - J^{(t-1)}\right| < \varepsilon$, that is, some minimum is reached. The full derivation of this algorithm is outlined by Hu et al. [2014], and the algorithm for optimization for LM$^3$L is given as their Algorithm 1; presented here is just the high-level outline.[17]

[17] fit.astro.vsa.utilities.ml.metriclearning.l3ml

# Chapter 4

# System Design and Performance of an Automated Classifier

This chapter will focus on the construction and application of a supervised pattern classification algorithm for the identification of variable stars. Given the reduction of a survey of stars into a standard feature space, the problem of using prior patterns to identify new observed patterns can be reduced to time–tested classification methodologies and algorithms. Such supervised methods, so called because the user trains the algorithms prior to application using patterns with known classes or labels, provides a means to probabilistically determine the estimated class type of new observations. These methods have two large advantages over hand classification procedures: the rate at which new data is processed is dependent only on the computational processing power available, and the performance of a supervised classification algorithm is quantifiable and consistent. Thus the algorithm produces rapid, efficient, and consistent results[Johnston, 2018].

This section will be structured as follows. First, the data and feature space to be implemented for training will be reviewed. Second, we will discuss the class labels to be used and the meaning behind them. Third, a set of classifiers (multi-layer perceptron, random forest, k-nearest neighbor, and support vector machine) will be trained and tested on the extracted feature space. Fourth, performance statistics will be generated for each classifier and a comparing and contrasting of the methods will be discussed with a champion classification method being selected. Fifth, the champion classification method will be applied to the new observations to be classified. Sixth, an anomaly detection algorithm will be generated using the so called one-class support vector machine and will be applied to the new observations. Lastly, based on the anomaly detection algorithm, and the supervised training algorithm a set of populations per class type will be generated. The result will be a highly reliable set of new populations per class type derived from the LINEAR survey.[1].

## 4.1 Related Work

The idea of constructing a supervised classification algorithm for stellar classification is not unique to this paper (see Dubath et al. 2011 for a review), nor is the construction of a classifier for time variable stars. Methods pursued include the construction of a detector to determine variability (two-class classifier Barclay et al. 2011), the design of random forests for the detection of photometric redshifts in spectra Carliles et al. [2010], the detection of transient events Djorgovski et al.

---

[1]Lightly Edited from original paper: Johnston, K. B., & Oluseyi, H. M. (2017). Generation of a supervised classification algorithm for time-series variable stars with an application to the LINEAR dataset. *New Astronomy, 52*, 35–47

[2012], and the development of machine-assisted discovery of astronomical parameter relationships Graham et al. [2013a]. Debosscher [2009] explored several classification techniques for the supervised classification of variable stars, quantitatively comparing the performed in terms of computational speed and performance which they took to mean accuracy. Likewise, other efforts have focused on comparing speed and robustness of various methods (e.g. Blomme et al. 2011, Pichara et al. 2012, Pichara and Protopapas 2013). These methods span both different classifiers and different spectral regimes, including IR surveys (Angeloni et al. 2014 and Masci et al. 2014), RF surveys [Rebbapragada et al., 2012], and optical [Richards et al., 2012].

## 4.2 Data

The procedure outlined in this paper will follow the standard philosophy for the generation of a supervised pattern classification algorithm as professed in Duda et al. [2012] and Hastie et al. [2004], i.e. exploratory data analysis, training and testing of supervised classifier, comparison of classifiers in terms of performance, application of classifier. Our training data is derived from a set of three well known variable star surveys: the ASAS survey [Pojmanski et al., 2005], the Hipparcos survey [Perryman et al., 1997], and the OGLE dataset [Udalski et al., 2002]. Data used for this study must meet a number of criteria:

1. Each star shall have differential photometric data in the *ugriz* system
2. Each star shall have variability in the optical channel (band) that exceeds some fixed threshold with respect to the error in amplitude measurement

3. Each star shall have a consistent class label, should multiple surveys address the same star

### 4.2.1 Sample Representation

These requirements reduce the total training set down to 2,054 datasets with 32 unique class labels. The features extracted are based on Fourier frequency domain coefficients [Deb and Singh, 2009], statistics associated with the time domain space, and differential photometric metrics; for more information see Richards et al. [2012] for a table of all 68 features with descriptions. The 32 unique class labels can be further generalized into four main groups: eruptive, multi-star, pulsating, and "other" [Debosscher, 2009], the breakdown of characterizations for the star classes follows the following classifications:

- *Pulsating*
  - *Giants*: Mira, Semireg RV, Pop. II Cepheid, Multi. Mode Cepheid
  - *RR Lyrae*: FO, FM, and DM
  - *"Others"* : Delta Scuti, Lambda Bootis, Beta Cephei, Slowly Pulsating B, Gamma Doradus, SX Phe, Pulsating Be
- *Erupting*: Wolf-Rayet, Chemically Peculiar, Per. Var. SG, Herbig AE/BE, S Doradus, RCB and Classical T-Tauri
- *Multi-Star*: Ellipsoidal, Beta Persei, Beta Lyrae, W Ursae Maj.
- *Other*: Weak-Line T-Tauri, SARG B, SARG A, LSP, RS Cvn

The *a priori* distribution of stellar classes is given in Table 4.1 for the broad classes and in Table 4.2 for the unique classes.

Table 4.1: Broad Classification of Variable Types in the Training and Testing Dataset

| Type | Count | % Dist |
|---|---|---|
| Multi-Star | 514 | 0.25 |
| Other | 135 | 0.07 |
| Pulsating | 1179 | 0.57 |
| Erupting | 226 | 0.11 |

Table 4.2: Unique Classification of Variable Types in the Training and Testing Dataset

| Class Type | % Dist | Class Type | % Dist |
|---|---|---|---|
| a. Mira | 8.0% | m. Slowly Puls. B | 1.5% |
| b1. Semireg PV | 4.9% | n. Gamma Doradus | 1.4% |
| b2. SARG A | 0.7% | o. Pulsating Be | 2.4% |
| b3. SARG B | 1.4% | p. Per. Var. SG | 2.7% |
| b4. LSP | 2.6% | q. Chem. Peculiar | 3.7% |
| c. RV Tauri | 1.2% | r. Wolf-Rayet | 2.0% |
| d. Classical Cepheid | 9.9% | r1. RCB | 0.6% |
| e. Pop. II Cepheid | 1.3% | s1. Class. T Tauri | 0.6% |
| f. Multi. Mode Cepheid | 4.8% | s2. Weak-line T Tauri | 1.0% |
| g. RR Lyrae FM | 7.2% | s3. RS CVn | 0.8% |
| h. RR Lyrae FO | 1.9% | t. Herbig AE/BE | 1.1% |
| i. RR Lyrae DM | 2.9% | u. S Doradus | 0.3% |
| j. Delta Scuti | 6.5% | v. Ellipsoidal | 0.6% |
| j1. SX Phe | 0.3% | w. Beta Persei | 8.7% |
| k. Lambda Bootis | 0.6% | x. Beta Lyrae | 9.8% |
| l. Beta Cephei | 2.7% | y. W Ursae Maj. | 5.9% |

It has been shown [Rifkin and Klautau, 2004] that how the classification of a multi-class problem is handled can affect the performance of the classifier; i.e. if the classifier is constructed to process all 32 unique classes as the same time, or if 32 different classifiers (detectors) are trained individually and the results are combined after application, or if a staged approach is best where a classifier is trained on the four broad classes first, then a secondary classifier is trained on the unique class labels in each broad class [Debosscher, 2009]. The *a priori* distribution of classes, the number of features to use, and the number of samples in the training set are key factors in determining which classification procedure to use. This dependence is often best generalized as the "curse of dimensionality" [Bellman, 1961], a set of problems that arise in machine learning that are tied to attempting to quantify a signature pattern for a given class, when the combination of a low number of training samples and high feature dimensionality results in a sparsity of data. Increasing sparsity results in a number of performance problems with the classifier, most of which amount to decrease generality (over-trained classifier) and decreased performance (low precision or high false alarm rate). Various procedures have been developed to address the curse of dimensionality, most often some form of dimensionality reduction technique is implemented or a general reframing of the classification problem is performed. For this effort, a reframing of the classification problem will be performed to address these issues.

### 4.2.2 Feature Space

Prior to the generation of the supervised classification algorithm, an analysis of the training dataset is performed. This exploratory data analysis [EDA, Tukey, 1977] is used here to understand the class separability prior to training, and to help the developer gain some insight into what should be expected in terms of performance of the final product. For example if during the course of the EDA it is found that the classes are linearly separable in the given dimensions using the training data, then we would expect a high performing classifier to be possible. Likewise, initial EDA can be useful in understanding the distribution of the classes in the given feature space answering questions like: are the class distributions multi-dimensional Gaussian? Do the class distributions have erratic shapes? Are they multi-modal? Not all classifiers are good for all situations, and often an initial qualitative EDA can help narrow down the window of which classifiers should be investigated and provide additional intuition to the analyst.

#### 4.2.2.1 Exploratory Data Analysis

Principle Component Analysis (PCA) is one of many methods [Duda et al., 2012], and often the one most cited, when EDA of multi-dimensional data is being performed. Via orthogonal transformation, PCA rotates the feature space into a new representation where the feature dimensions are organized such that the first dimension (the principle component) has the largest possible variance, given the feature space. This version of PCA is the most simple and straight-forward; there are numerous variants, all of which attempt a similar maximization process (e.g., of variance, of correlation, of between group variance) but may also employ an additional transformation (e.g., manifold mapping, using the "kernel trick" , etc.). Using the broad categories defined for the variable star populations, PCA is performed in R using the FactoMineR package [Lê et al., 2008], and the first two components are plotted (see Figure 4.1).

The PCA transformation is not enough to separate out the classes, however the graphical representation of the data does provide some additional insight about the feature space and the distribution of classes. The eruptive and multi-star populations appear to have a single mode in the dimensions presented in Figure 4.1, while the pulsating and the "other" categories appear to be much more irregular in shape. Further analysis addressing just the pulsating class shows that the dis-

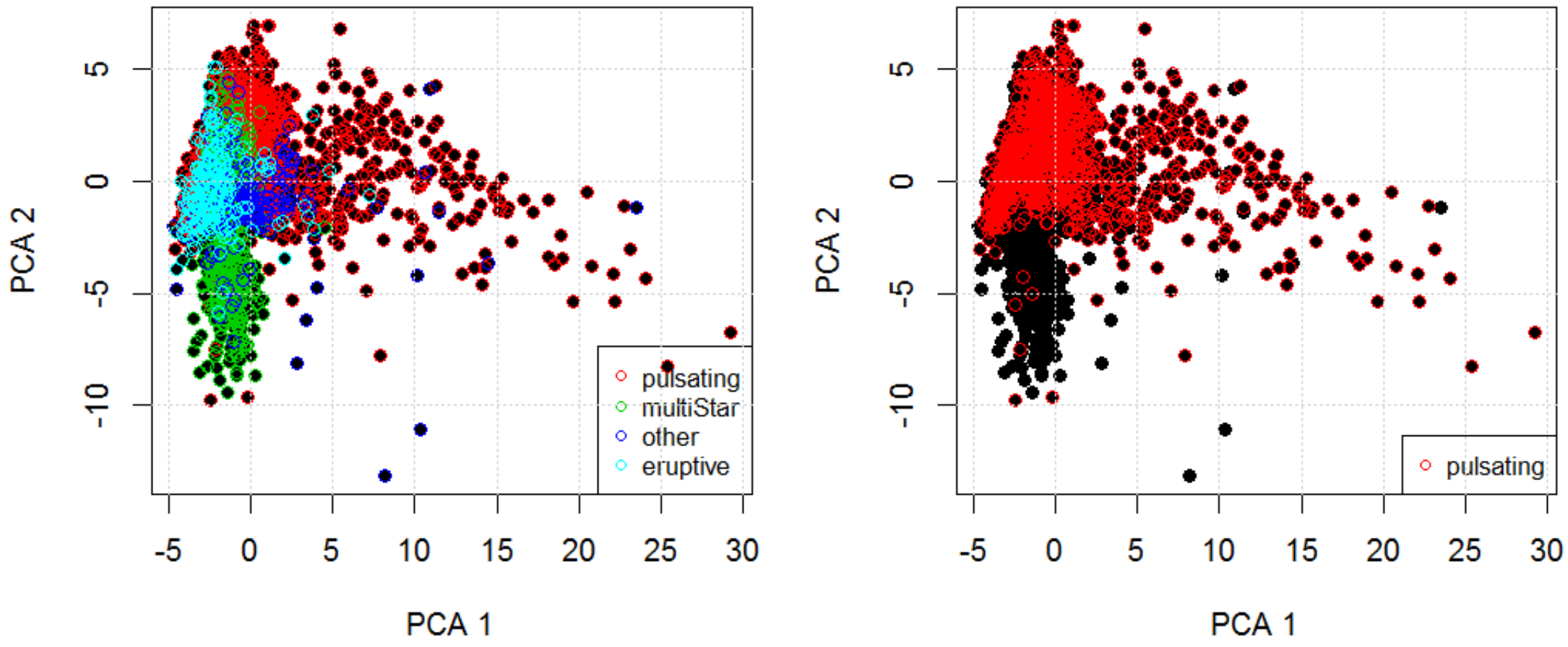

Figure 4.1: Left: PCA applied to the ASAS+Hipp+OGLE dataset, with the broad class labels identified and the first two principle components plotted. Right: Only the stars classified as pulsating are highlighted

tribution of stars with this label is spread across the whole of the feature space (Figure 4.1).

### 4.2.3 Effectiveness of Feature Space

In this representation of the feature space there is a significant overlap across all classes. Even if other methods of dimensionality reduction were implemented, for example Supervised-PCA [Bair et al., 2006], linear separation of classes without dimensional transformation is not possible. Application of SPCA results in the Figure 4.2, and is also provided in movie form as digitally accessible media[2].

---

[2] https://github.com/kjohnston82/LINEARSupervisedClassification/AohSPCA.mp4

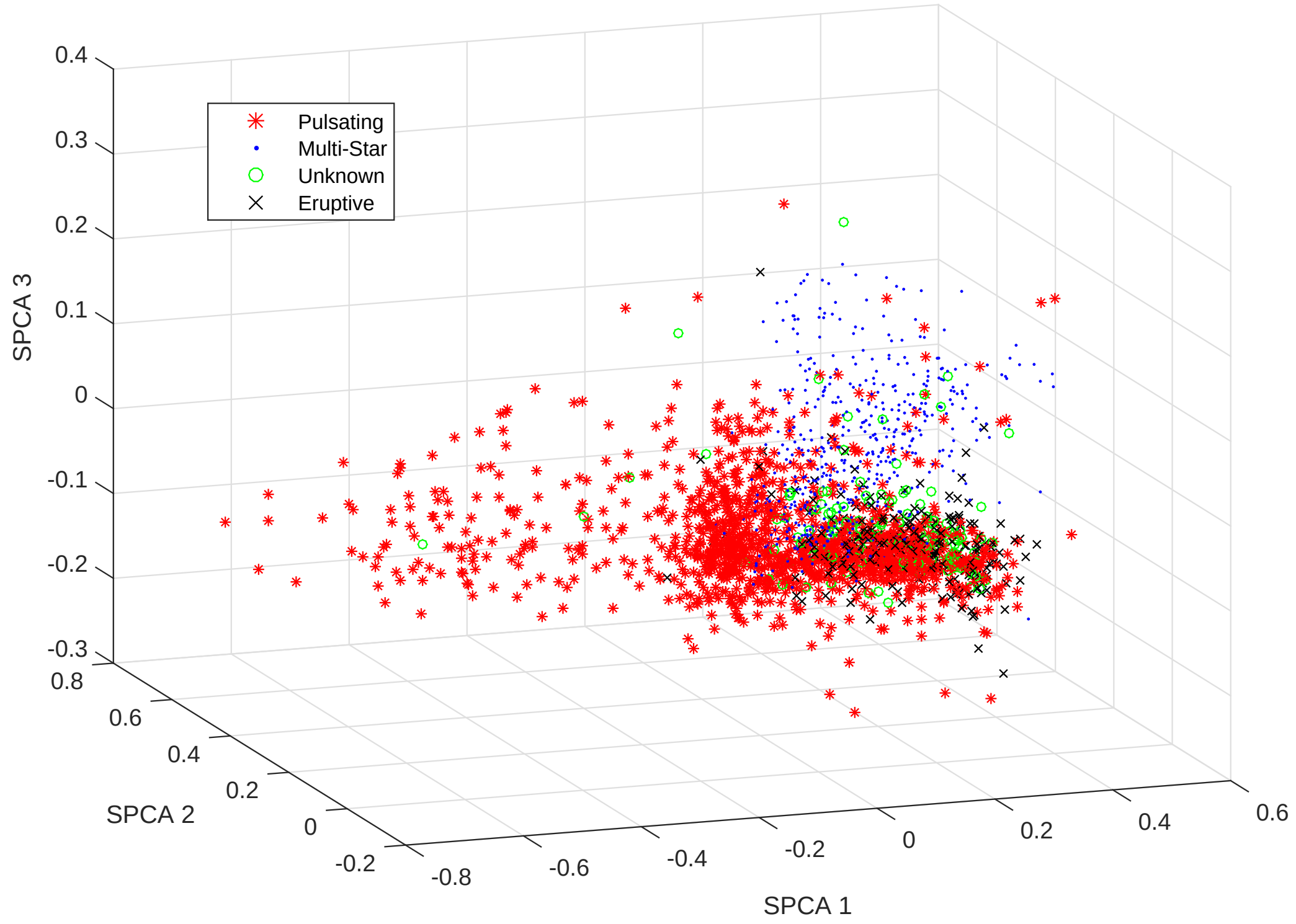

Figure 4.2: SPCA applied to the ASAS+Hipp+OGLE dataset

This non-Gaussian, non-linear separable class space requires further transformation to improve separation of classes or a classifier which performs said mapping into a space where the classes have improved separability. Four classifiers are briefly discussed which address these needs.

## 4.3 Supervised Classification Development

All algorithms are implemented in the R language, version 3.1.2 (2014-10-31) “Pumpkin–Helmet”, and operations are run on x86_64-w64-mingw32/x64 (64-bit) platform. Four classifiers are initially investigated: k-Nearest Neighbor (k-NN), support vector machine (SVM), random forest (RF) and multi-layer perceptron (MLP).

The k-Nearest Neighbor algorithm implemented is based on the k-NN algorithm outlined by Duda et al. [2012] and Altman [1992], with allowance for distance measurements using both $L_1$ (taxi cab) and $L_2$ (Euclidean distance) (see equation 4.1).

$$\|x\|_p = (|x_1|^p + |x_2|^p + ... + |x_n|^n)^{1/p}, p = 1, 2, ..., \infty. \tag{4.1}$$

The testing set is implemented to determine both optimal distance method to be used and $k$ value, i.e. number of nearest neighbors to count.

A number of SVM packages exist [Karatzoglou et al., 2005], the *e1071* package [Dimitriadou et al., 2008] is used in this study and was first implementation of SVM in R. It has been shown to perform well and contains a number of additional SVM utilities beyond the algorithm trainer that make it an ideal baseline SVM algorithm for performance testing. SVM decisions lines are hyperplanes, linear cuts that split the feature space into two sections; the optimal hyperplane is the one that has the larger distance to the nearest training data (i.e., maximum margin). Various implementations of the original algorithm exist, including the Kernel SVM [Boser et al., 1992] used here for this study with the Gaussian Kernel. KSVM uses the so called "Kernel Trick" to project the original feature space into a higher dimension, resulting in hyperplane decision lines that are non-linear, a beneficial functionality should one find that the classes of interest are not linearly separable.

The multilayer perceptron supervised learning algorithm (MLP) falls into the family of neural network classifiers. The classifier can be simply described as layers (stages) of perceptron (nodes), where each perceptron performs a different transformation on the same dataset. These perceptrons often employ simple transformation (i.e., logit, sigmod, etc.), to go from the original input feature space, into a set of posterior probabilities. The construction of these layers and the transformations is beyond the scope of this article, and for more information on neural networks, backpropagation, error minimization, and design of the classifier the reader is invited to review such texts as Rhumelhart et al. [1986]. This study makes use of the R library *RSNNS*, for the construction and analysis of the MLP classifier used; see Bergmeir and Benítez [2012].

Lastly, random forests are the conglomeration of sets of classification and regression trees (CARTs). The CART algorithm, made popular by Breiman et al. [1984], generates decisions spaces by segmenting the feature space dimension by dimension. Given an initial training set, the original CART is trained such that each decision made maximally increases the purity of resulting two new populations. Each subsequent node following either similarly divides the received population into two new populations with improved class purity or is a terminal node, where no further splits are made and instead a class estimate is provided.

A detailed discussion of how the CART algorithm is trained, the various varieties of impurity that can be used in the decision making process, addressed in [Breiman et al., 1984] as well as other standard pattern classification text (Hastie et al., 2004, Duda et al., 2012). Random Forests are the conglomeration of these CART classification algorithms, trained on variation of the same training set [Breiman, 2001]. This ensemble classifier constructs a set of CART algorithms, each one trained on a reduction of the original training set, this variation results in each CART algorithm in the set being slightly different. Given a new observed pattern applied to the set of CART classifiers, a set of decisions is generated. The Random Forest classifier combines these estimated class labels to generate a unified class estimate. This study makes use of the *randomForest* package in R, see Liaw and Wiener [2002].

### 4.3.1 Training and Testing

The training of all four classifier types proceeds with roughly the same procedure; following the one-vs.-all methodology for multi-class classification, a class type of interest is identified as either broad or unique, the original training set is split equally into a training set and a testing set with the *a priori* population distributions approximately equal to the population distribution of the combined training set. Adjustable parameters for each classifier are identified: the RF number of trees, the Kernel-SVM kernel spread, the k-NN k value and p value, the MLP number of units in the hidden layers, and then the classifier is initially trained and tested against the testing population. Parameters are then adjusted and subsequent classifiers are trained, and misclassification error is found as a function of the parameter adjustments. Those parameters resulting in a trained classifier with a minimal amount of error are implemented.

For each classifier, two quantifications of performance are generated: a receiver operating characteristic (ROC) curve and a precision recall (PR) curve. Fawcett [2006] outlines both, and discusses the common uses of each. Both concepts plot two performance statistics for a given classification algorithm, given some changing threshold value, which will for this study be a critical probability that the posterior probability of the class of interest (the target stellar variable) is compared against. These curves can be generated when the classifier is cast as a "two-class" problem, where one of the classes is the target (class of interest) while the other is not.

For any two-class classifier the metrics highlighted here can be generated and are a function of the decision space selected by the analyst. Frequently the acceptance threshold, i.e. the hypothesized class must have a posterior probability greater than some $\lambda$, is selected based on the errors of the classifier. Many generic classification algorithms are designed such that the false positive (fp) rate and 1-true positive (tp) rate are both minimized. Often this practice is ideal; however the problem faced in the instances addressed in this article require additional considerations. We note two points:

1. When addressing the unique class types, there are a number of stellar variable populations which are relatively much smaller than others. This so-called class imbalance has been shown [Fawcett, 2006] to cause problems with performance analysis if not handled correctly. Some classification algorithms adjust for this imbalance, but often additional considerations must be made, specifically when reporting performance metrics.

2. Minimization of both errors, or minimum-error-rate classification, is often based on the zero-one loss function. In this case, it is assumed that the cost of a false positive (said it was, when it really was not) is the same as a false negative (said it was not, when it really was). If the goal of this study is to produce a classifier that is able to classify new stars from very large surveys, some of which are millions of stars big, the cost of returning a large number of false alarms is much higher than the cost of missing some stars in some classes. Especially when class separation is small, if the application of the classifier results in significant false alarms the inundation of an analyst with bad decisions will likely result in a general distrust of the classifier algorithm.

The ROC curve expresses the adjustment of the errors as a function of the decision criterion. Likewise, the PR curve expresses the adjustment of precision (the percentage of true positives out of all decisions made) and recall (true positive rate) as a function of the decision criterion. By sliding along the ROC or PR curve, we can change the performance of the classifier. Note that increasing the true positive rate causes an increase in the false positive rate as well (and vice-versa). Often a common practice is to fix [Scharf, 1991] one of the metrics, false positive rate, of all classifiers used.

Similar to the ROC curve, the PR curve demonstrates how performance varies between precision and recall for a given value of the threshold. It is apparent that the PR and ROC curves are related [Davis and Goadrich, 2006], both have a true-positive rate as an axis (TP Rate and Recall are equivalent), both are functions of the threshold used in the determination of estimated class for a new patter (discrimination), both are based on the confusion matrix and the associated performance metrics. Thus fixing the false alarm rate, not only fixes the true positive rate, but also the precision of the classifier.

If the interest was a general comparison of classifiers, instead of selecting a specific performance level,Fawcett [2006] suggests that the computation of Area-Under-the-Curve quantifies either the PR and ROC curve into a single "performance" estimate that represents the classifier as a whole. The ROC-AUC of a classifier is equivalent to the probability that the classifier will rank a randomly chosen positive instance higher than a randomly chosen negative instance. The PR-AUC of a classifier is roughly the mean precision of the classifier. Both ROC and PR curves should be considered when evaluating a classifier [Davis and Goadrich, 2006], especially when class imbalances exist. For this study, the best performing classifier will be the one that maximizes both the ROC-AUC and the PR-AUC. Likewise, when the final performance of the classifier is proposed, false positive rate and precision will be reported and used to make assumptions about the decisions made by the classification algorithm.

### 4.3.2 Performance Analysis — Supervised Classification

Based on the foundation of performance analysis methods, ROC and PR curves, and AUC discussed, the study analyzes classification algorithms applied to both the broad and unique (individual) class labels.

#### 4.3.2.1 Broad Classes — Random Forest

Initially an attempt was made to adjust both the number of trees (*ntree*) and the number of variables randomly sampled as candidates at each split (*mtry*). Based on Breiman et al. [1984] recommendation, *mtry* was set to $\sqrt{M}$, where $M$ is the number of features. The parameters *ntree* was set to 100, based on the work performed by Debosscher [2009]. Classifiers were then generated based on the training sample, and the testing set was used to generate the ROC and PR AUC for each one-vs-all classifier. The associated curves are given in Appendix A, the resulting AUC estimates are in Table 4.3:

Table 4.3: ROC/PR AUC Estimates based on training and testing for the Random Forest Classifier.

| AUC | Pulsating | Eruptive | Multi-Star | Other |
|---|---|---|---|---|
| ROC | 0.971 | 0.959 | 0.992 | 0.961 |
| PR | 0.979 | 0.788 | 0.986 | 0.800 |

#### 4.3.2.2 Broad Classes — Kernel SVM

Instead of using the “hard” class estimates, common with SVM usage, the “soft” estimates, i.e. posterior probabilities, are used. This allows for the thresholding necessary to construct the PR and ROC curves. Kernel spreads of 0.001, 0.01, and 0.1 were tested (set as the variable gamma in R), the associated PR and ROC curve are given in Appendix A. It was found that 0.1 was optimal for the feature space (using Gaussian Kernels). The associated curves are given in Appendix A, the resulting AUC estimates are in Table 4.4:

Table 4.4: ROC/PR AUC Estimates based on training and testing for the Kernel SVM Classifier.

| AUC | Pulsating | Eruptive | Multi-Star | Other |
|---|---|---|---|---|
| ROC | 0.938 | 0.903 | 0.979 | 0.954 |
| PR | 0.952 | 0.617 | 0.968 | 0.694 |

#### 4.3.2.3 Broad Classes — k-NN

It was found that for the training set used, that increasing performance was gained with increasing values of k-nearest neighbors. Gains in performance were limiting after $k = 4$, and a value of $k = 10$ was selected to train with. The value of the polynomial defined in the generation of distance (via $L^p$-norm) was varied between 1 and 3, with decreasing performance found for $p > 3$. The associated PR and ROC curves were generated for values of $p < 4$. The associated curves are given in Appendix A, the resulting AUC estimates for $p < 3$ are in Table 4.5:

Table 4.5: ROC/PR AUC Estimates based on training and testing for the k-NN Classifier.

| | AUC | Pulsating | Eruptive | Multi-Star | Other |
|---|---|---|---|---|---|
| p-1 | ROC | 0.919 | 0.847 | 0.980 | 0.928 |
| | PR | 0.931 | 0.480 | 0.959 | 0.597 |
| p-2 | ROC | 0.901 | 0.802 | 0.967 | 0.877 |
| | PR | 0.918 | 0.368 | 0.931 | 0.519 |

#### 4.3.2.4 Broad Classes — MLP

There are two variables associated with MLP algorithm training: the number of units in the hidden layers (size) and the number of parameters for the learning function to use (*learnParam*). It was found that for the dataset: The *learnParam* value had little effect on the performance of the classifier, and it was taken to be 0.1 for implementation here. The variable size did have an effect, an initial study of values between 4 and 18 demonstrated that the best performance occurred between 4 and 8. PR and ROC curves for these values were generated and are in Appendix A, the resulting AUC estimates for values 4, 6 and 8 are in Table 4.6:

Table 4.6: ROC/PR AUC Estimates based on training and testing for the MLP Classifier.

| *learnParam* | AUC | Pulsating | Eruptive | Multi-Star | Other |
|---|---|---|---|---|---|
| 4 | ROC | 0.928 | 0.694 | 0.914 | 0.585 |
| | PR | 0.916 | 0.120 | 0.869 | 0.183 |
| 6 | ROC | 0.933 | 0.751 | 0.888 | 0.473 |
| | PR | 0.909 | 0.139 | 0.797 | 0.123 |
| 8 | ROC | 0.920 | 0.706 | 0.914 | 0.529 |
| | PR | 0.903 | 0.175 | 0.854 | 0.159 |

#### 4.3.2.5 Unique Classes

Analysis of the broad classes provided insight into the potential of a staged classifier. The performance of the broad classification algorithms does not suggest that the supervised variable star classification problem would be benefited by a staged design. The RF classifier performed best across all broad classes and against all other classifiers, but still had significant error; had the broad classes perfectly separated, further analysis into the staged design would have been warranted. Instead, two-class classifier designed based on the unique classes are explored.

Similar to the broad classification methodology, the training sample is separated into a training set and a testing data set for each unique class type for training in a two-class classifier. Again, the testing data is used to minimize the misclassification error and find optimal parameters for each of the classifiers. Each classifier is then optimal for the particular class of interest. With the change of design, a change of performance analysis is also necessary. With nearly ten times the number of classes, a comparison of ROC and PR curves per classifier type and per class type requires a methodology that allows the information plotted on a single plot (direct comparison). Keeping with the discussion outlined by Davis and Goadrich [2006], we plot ROC vs. PR for each classifier (Figure 4.3 as an example).

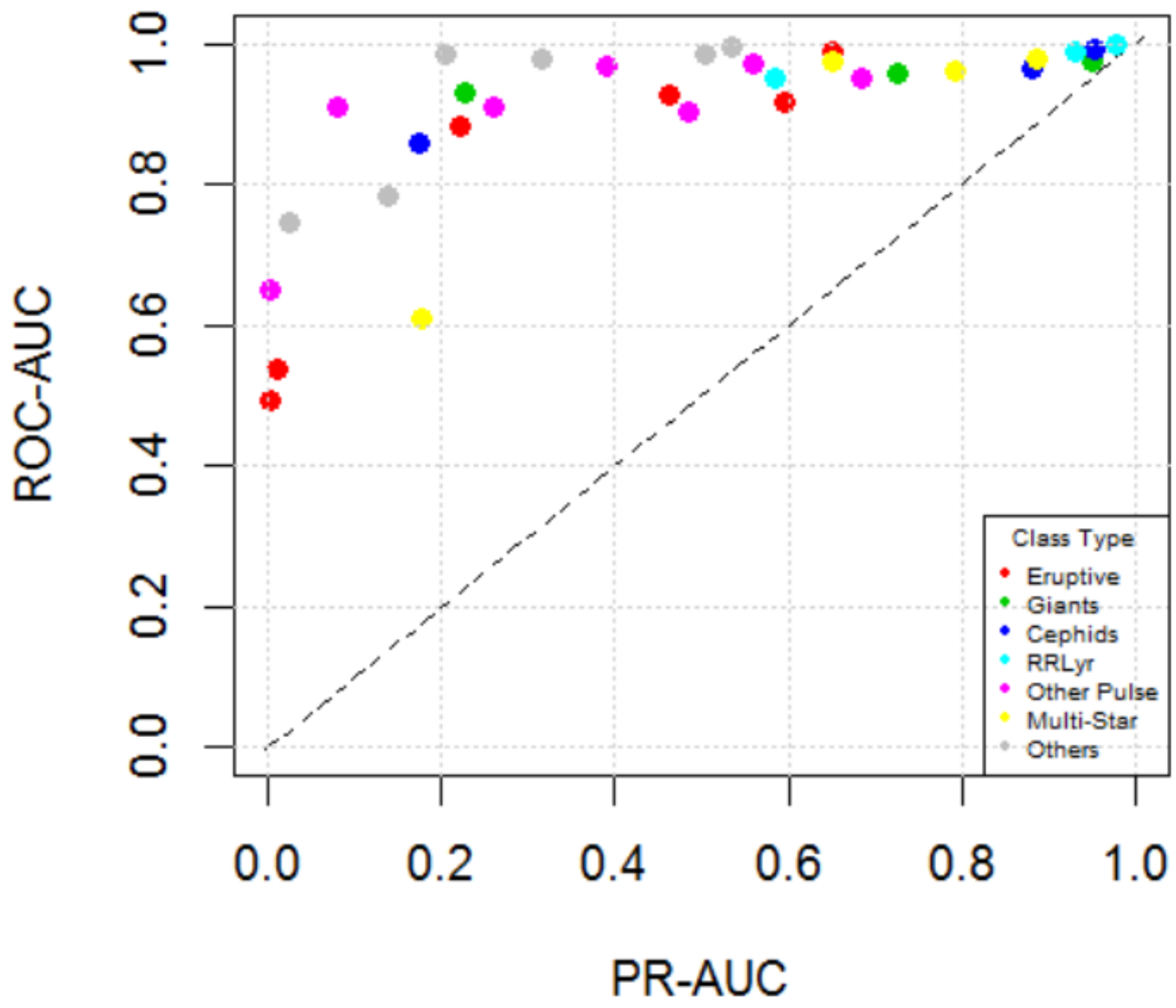

Figure 4.3: ROC vs. PR AUC Plot for the Multi-Layer Perceptron Classifier, generic class types (Eruptive, Giants, Cepheids, RRlyr, Other Pulsing, Multi-star, and other) are colored. The line y = x is plotted for reference (dashed line)

The set of these performance analysis graphs are in Appendix A. A comparison of the general performance of each classifier can be derived from the generation of AUC for each of the performance curves. Here, the quantification of general performance for a classifier is given as either mean precision across all class types or via non-parametric analysis of the AUC. The non-parametric analysis used is compiled as follows: for each class of stars, the average performance across classifiers is found, if for a classifier the performance is greater than the mean, an assignment of +1 is given, else -1. Over all classes, the summation of assignments is taken and given in Table 4.7.

Table 4.7: Performance Analysis of Individual Classifiers

| | ROC-AUC | | PR-AUC | |
|---|---|---|---|---|
| | Mean | Non-Para. | Mean | Non-Para. |
| KNN-Poly-1 | 0.884 | 2 | 0.530 | 8 |
| SVM | 0.905 | -4 | 0.407 | -26 |
| MLP | 0.894 | 2 | 0.470 | -8 |
| RF | 0.948 | 22 | 0.595 | 14 |

It is apparent that the RF classifier out-performs the other three classification algorithms, using both the mean of precision as well as a non-parametric comparison of the AUC statistics. The plot comparing ROC-AUC and PR-AUC for the Random Forest classifier is presented in Figure 4.4.

Based on Figure 4.4, it is observed that star populations of similar class types do not necessarily cluster together. Additionally it is apparent that the original size of the population in the training set, while having some effect on the ROC-AUC, has a major effect on the resulting PR-AUC. Figure 4.4 b demonstrates that for those classes with an initial population of 55 (empirically guessed value), the precision is expected to be greater than 70%. Surprisingly though for classes with an initial population of 55 or less, the limits of precision are less predictable and in fact appear to be random with respect to class of interest training size. Thus,

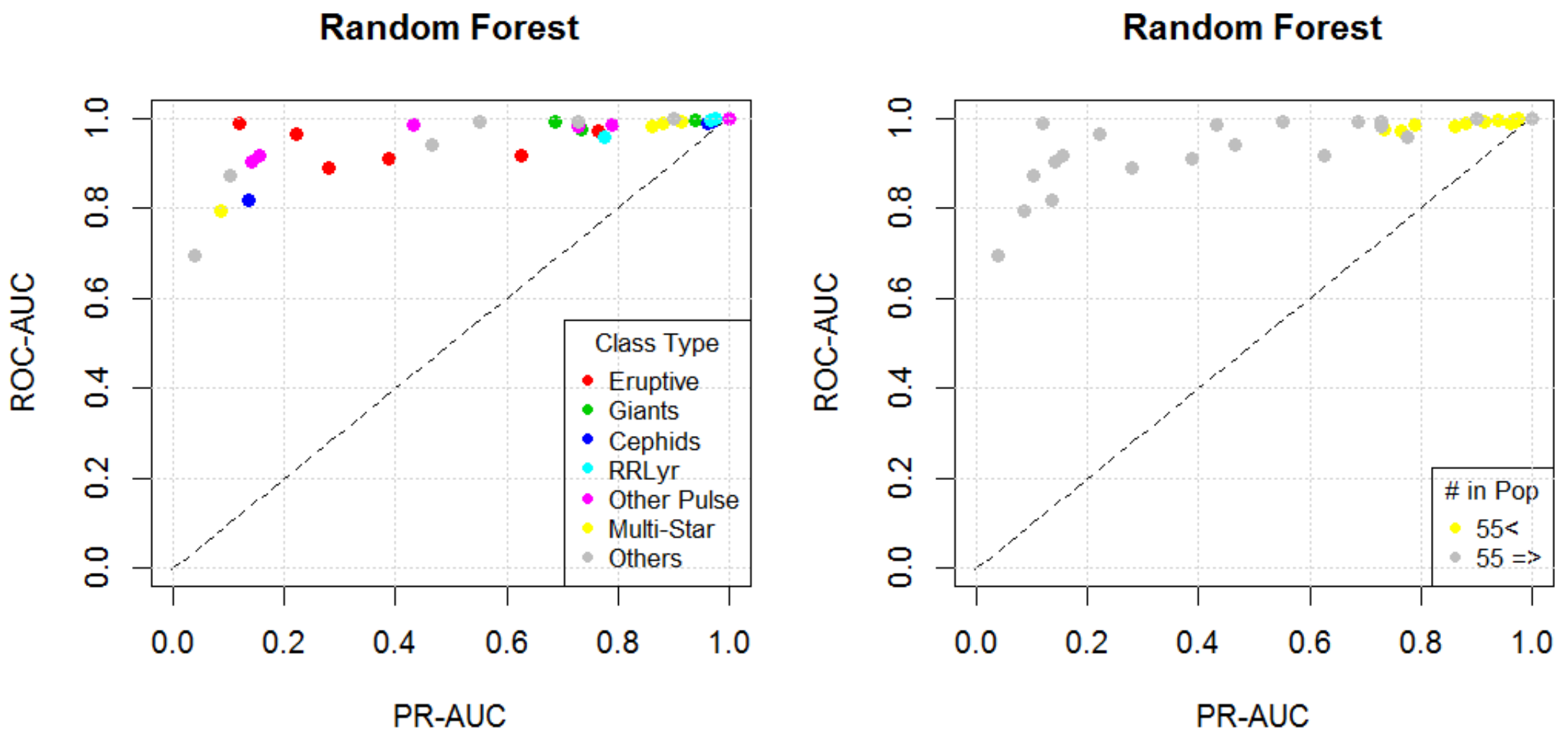

Figure 4.4: ROC vs. PR AUC Plot for the Random Forest Classifier, generic class types (Eruptive, Giants, Cepheids, RRlyr, Other Pulsing, Multi-star, and other) are colored. The line y = x is plotted for reference (dashed line). Left: shows the break down per generic class type, Right: shows the difference between populations with more then 55 members in the initial training dataset

without further training data or feature space improvements, the performance statistics graphed in Figure 4.4 are the statistics that will be used as part of the application of the classifier to the LINEAR dataset, performance statistics for the other classifiers are given in Appendix A.

### 4.3.3 Performance Analysis — Anomaly Detection

In addition to the pattern classification algorithm outlined, the procedure outlined here includes the construction of a One-Class Support Vector Machine (OC-SVN) for use as an anomaly detector. The pattern classification algorithms presented and compared as part of this analysis partition the entire decision space. For the random forest, kNN, MLP, and SVM two-class classifier algorithms, there is no consideration for deviations of patterns beyond the training set observed, i.e. absolute distance from population centers. All of the algorithms investigated consider relative distances, i.e. is the new pattern P closer to the class center of B or A? Thus, despite that an anomalous pattern is observed by a new survey, the classifier will attempt to estimate a label for the observed star based on the labels it knows.

In addition to the pattern classification algorithm outlined, the procedure outlined here includes the construction of a One-Class Support Vector Machine (OC-SVN) for use as an anomaly detector. The pattern classification algorithms presented and compared as part of this analysis partition the entire decision space. For the random forest, kNN, MLP, and SVM two-class classifier algorithms, there is no consideration for deviations of patterns beyond the training set observed, i.e. absolute distance from population centers. All of the algorithms investigated consider relative distances, i.e. is the new pattern P closer to the class center of B or A?

Thus, despite that an anomalous pattern is observed by a new survey, the classifier will attempt to estimate a label for the observed star based on the labels it knows. To address this concern, a one-class support vector machine is implemented as an anomaly detection algorithm. Lee and Scott [2007] describe the design and construction of such an algorithm. Similar to the Kernel- SVM discussed prior, the original dimensionality is expanded using the Kernel trick (Gaussian Kernels) allowing complex regions to be more accurately modeled. For the OC-SVM, the training data labels are adjusted such that all entered data is of class type one (+1). A single input pattern at the origin point is artificially set as class type two ( 1). The result is the lassoing or dynamic encompassing of known data patterns.

The lasso boundary represents the division between known (previously ob-served) regions of feature space and unknown (not-previously observed) regions. New patterns observed with feature vectors occurring in this unknown region are considered anomalies or patterns without support, and the estimated labels returned from the supervised classification algorithms should be questioned, despite the associated posterior probability of the label estimate [Schölkopf et al., 2001]. The construction of the OC-SVM to be applied as part of this analysis starts with the generation of two datasets (training and testing) from the ASAS + Hipp + OGLE training data. The initial training set is provided to the OC-SVM [Lee and Scott, 2007] algorithm which generates the decision space (lasso). This decision space is tested against the training data set; and the fraction of points declared to be anomalous is plotted against the spread of the Kernel used in the OC-SVM (Figure 4.5).

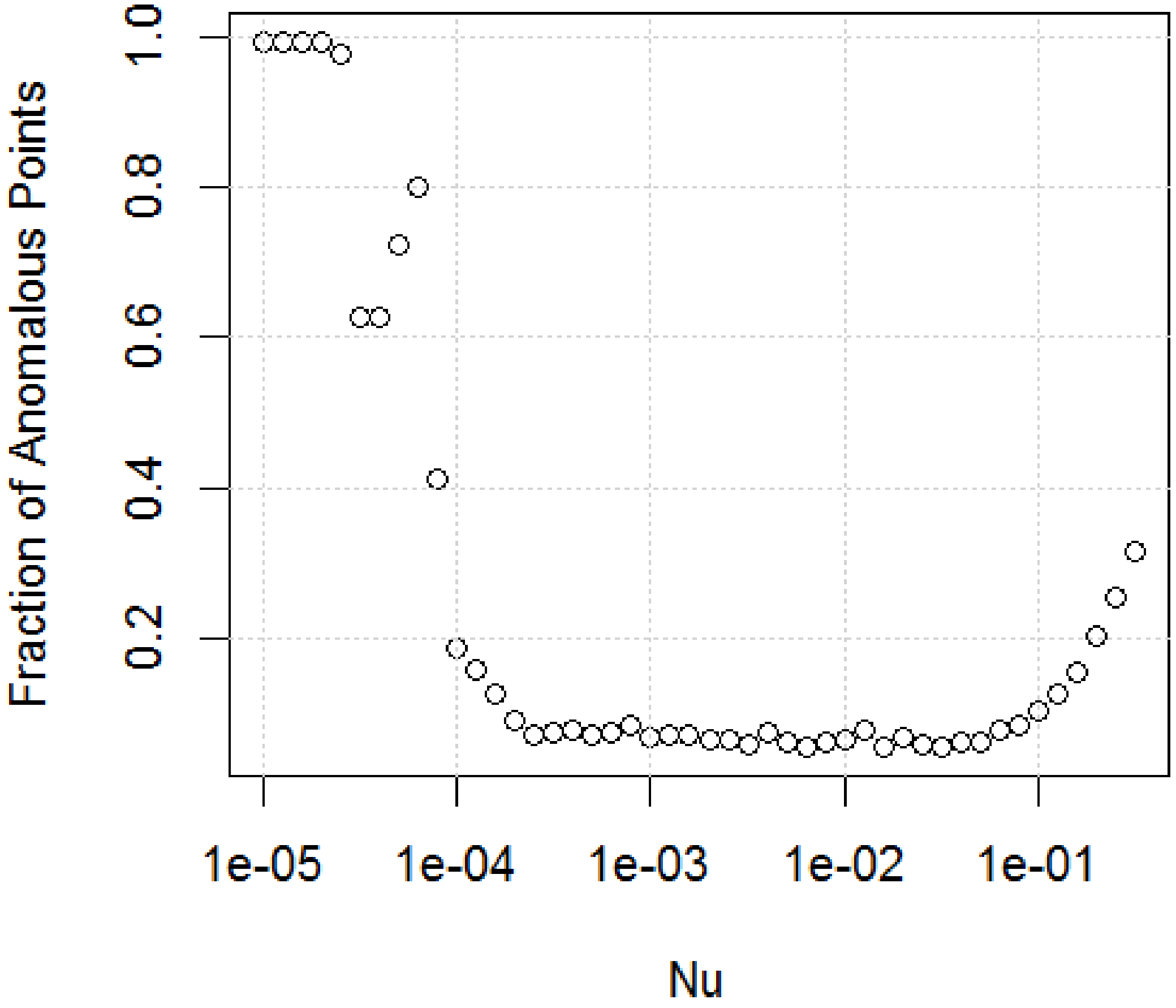

Figure 4.5: Fraction of Anomalous Points Found in the Training Dataset as a Function of the Gaussian Kernel Spread Used in the Kernel-SVM

Because of the hyper-dimensionality, the OC-SVM algorithm is unable to perfectly encapsulate the training data; however a minimization can be found and estimated. The first two principle components of the training data feature space are plotted for visual inspection (Figure 4.6), highlighting those points that were called “anomalous” based on a $nu$ value (kernel spread) of 0.001. Less than 5% of the points are referred to as anomalies (˜falsely).

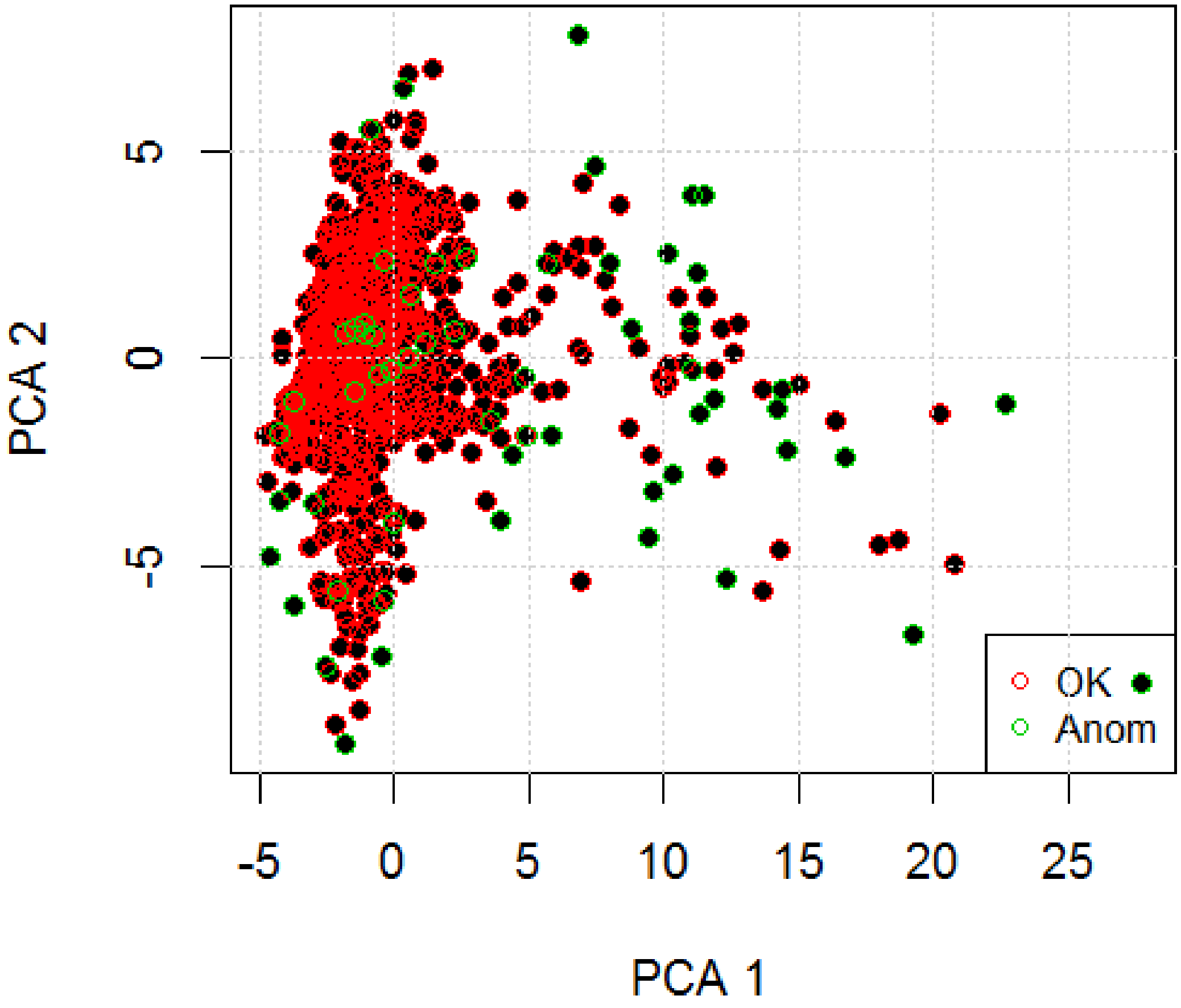

Figure 4.6: Plot of OC-SVM Results Applied to Training Data Only

Further testing is performed on the anomaly space, using the second dataset generated. As both datasets originate from the same parent population, the OC-SVM algorithm parameter ($nu$) is tuned to a value that maximally accepts the testing points (Figure 4.7).

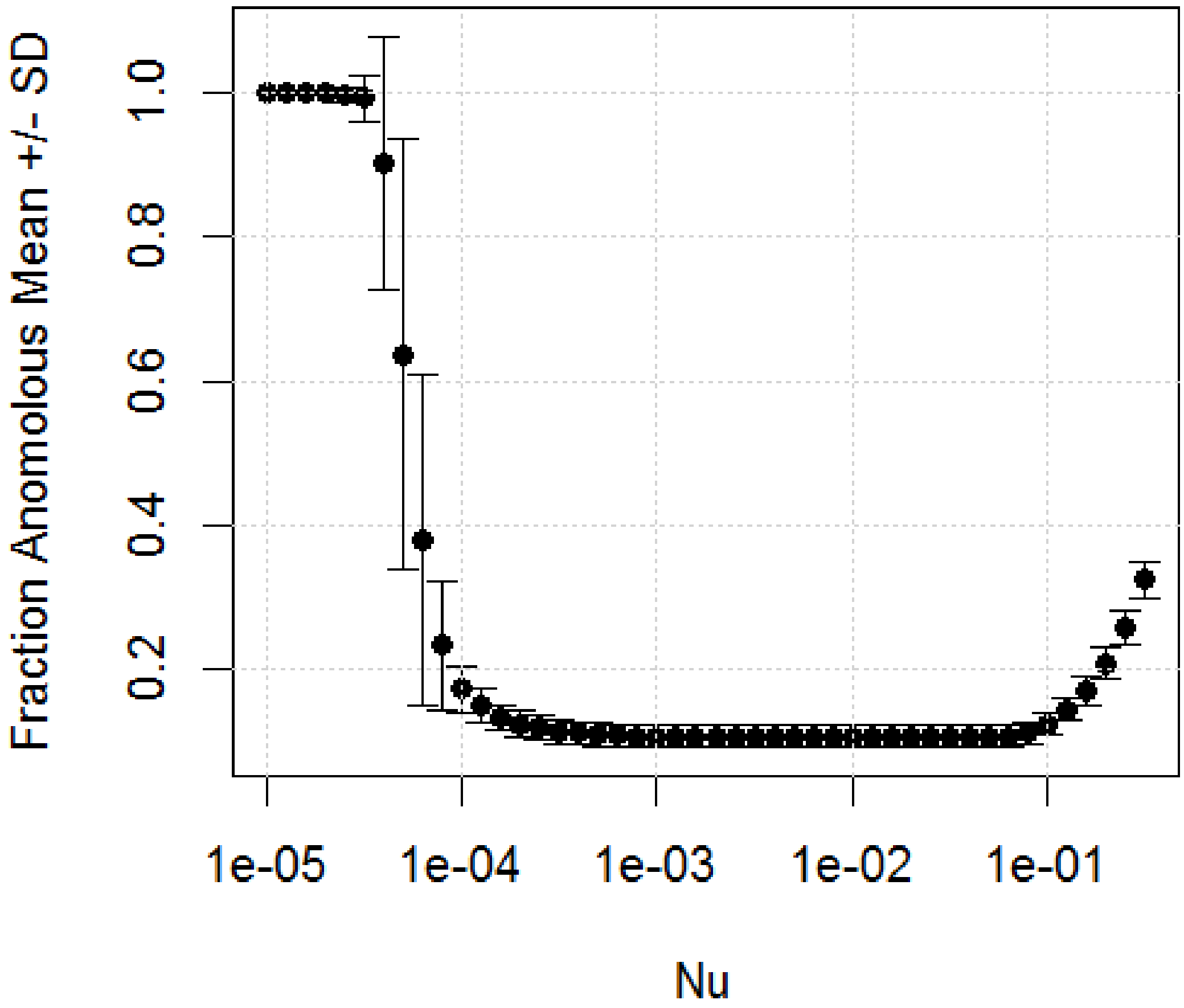

Figure 4.7: OC-SVM testing of the Testing data

The minimum fraction was found at a *nu* of 0.03. The OC-SVM was applied to the LINEAR dataset with the optimal kernel spread. All 192,744 datasets were processed, with 58,312 (False, or “anomalous” ) and 134,432 (True, or “expected” ) decisions made, i.e. 30% of the LINEAR dataset is considered anomalous based on the joint ASAS+HIPP+OGLE training dataset feature space.

## 4.4 Application of Supervised Classifier to LINEAR Dataset

### 4.4.1 Analysis and Results

For application to the LINEAR dataset, a RF classifier is constructed based on the training set discussed prior. The classifiers are designed using the one-vs.-all methodology, i.e. each stellar class has its own detector (i.e. overlap in estimated class labels is possible), therefore 32 individual two-class classifiers (detectors) are generated. The individual classification method (one-vs.-all) allows for each given star to have multiple estimated labels (e.g. multiple detectors returning a positive result for the same observation). The one-vs.-all methodology also allows the training step of the classification to be more sensitive to stars who might have been under-represented in the training sample, improving the performance of the detector overall.

Based on the testing performance results (ROC and PR curves) presented for the individual classifiers, the critical statistic used for the RF decision process was tuned such that a 0.5% false positive rate is expected when applied to the LINEAR dataset. In addition to the RF classifier, an OC-SVM anomaly detection algorithm was trained and used to determine if samples from the LINEAR dataset are anomalous with respect to the joint ASAS+OGLE+HIPP dataset. Applying the RF classifier(s) and the OC-SVM algorithm to the LINEAR dataset the following was found using a threshold setting corresponding to a false alarm rate of 0.5% (see ROC curve analysis). Given an initial set of LINEAR data (192,744 samples), the following table was constructed based on the results of the application of the isolated one-vs.-all RF classifiers only, see Table 4.8:

Table 4.8: Initial results from the application of the RF classifier(s)

| Class Type | Est. Pop | Class Type | Est. Pop |
|---|---|---|---|
| a. Mira | 3256 | m. Slowly Puls. B | 2 |
| b1. Semireg PV | 7 | n. Gamma Doradus | 2268 |
| b2. SARG A | 4291 | o. Pulsating Be | 14746 |
| b3. SARG B | 30 | p. Per. Var. SG | 284 |
| b4. LSP | 10 | q. Chem. Peculiar | 10 |
| c. RV Tauri | 5642 | r. Wolf-Rayet | 3970 |
| d. Classical Cepheid | 31 | r1. RCB | 1253 |
| e. Pop. II Cepheid | 326 | s1. Class. T Tauri | 17505 |
| f. Multi. Mode Cepheid | 556 | s2. Weak-line T Tauri | 4945 |
| g. RR Lyrae FM | 13470 | s3. RS CVn | 40512 |
| h. RR Lyrae FO | 1276 | t. Herbig AE/BE | 1358 |
| i. RR Lyrae DM | 9800 | u. S Doradus | 2185 |
| j. Delta Scuti | 493 | v. Ellipsoidal | 132 |
| j1. SX Phe | 9118 | w. Beta Persei | 481 |
| k. Lambda Bootis | 69 | x. Beta Lyrae | 2 |
| l. Beta Cephei | 2378 | y. W Ursae Maj. | 1365 |

103,628 stars were not classified (˜54%) and of those 11,619 were considered "Anomalous" . 57,848 stars were classified only once (˜30%) and of those 23,397 were considered "Anomalous" . 31,268 stars were classified with multiple labels (˜16%) and of those 23296 were considered "Anomalous" . The set of stars that were both classified once and did not have anomalous patterns (34,451) are broken down by class type in Table 4.9.

Table 4.9: Initial results from the application of the RF classifier(s) and the OC-SVM anomaly detection algorithm, classes that are major returned classes (> 1% of the total return set) are in bold

| Class Type | Est. Pop | % Total | Class Type | Est. Pop | % Total |
|---|---|---|---|---|---|
| a. Mira | 15 | 0.04% | m. Slowly Puls. B | 2 | 0.002% |
| b1. Semireg PV | 1 | 0.002% | **n. Gamma Doradus** | **2268** | **3.8%** |
| **b2. SARG A** | **1362** | **4.0%** | o. Pulsating Be | 14746 | 0.61% |
| b3. SARG B | 0 | 0% | p. Per. Var. SG | 284 | 0.26% |
| b4. LSP | 1 | 0.002% | q. Chem. Peculiar | 10 | 0% |
| **c. RV Tauri** | **538** | **1.6%** | **r. Wolf-Rayet** | **3970** | **6.2%** |
| d. Classical Cepheid | 2 | 0.006% | r1. RCB | 1253 | 0.01% |
| e. Pop. II Cepheid | 50 | 0.15% | **s1. Class. T Tauri** | **17505** | **5.4%** |
| f. Multi. Mode Cepheid | 286 | 0.83% | **s2. Weak-line T Tauri** | **4945** | **3.3%** |
| **g. RR Lyrae FM** | **2794** | **8.1%** | **s3. RS CVn** | **40512** | **46.6%** |
| **h. RR Lyrae FO** | **710** | **2.1%** | t. Herbig AE/BE | 1358 | 0.33% |
| **i. RR Lyrae DM** | **2350** | **6.8%** | **u. S Doradus** | **2185** | **1.7%** |
| j. Delta Scuti | 8 | 0.02% | v. Ellipsoidal | 132 | 0.08% |
| **j1. SX Phe** | **1624** | **4.7%** | w. Beta Persei | 481 | 0.42% |
| k. Lambda Bootis | 1 | 0.002% | x. Beta Lyrae | 2 | 0.006% |
| l. Beta Cephei | 25 | 0.07% | **y. W Ursae Maj.** | **1365** | **3.1%** |

The listing of individual discovered populations are provided digitally[3]. Two classes were not detected confidently out of the LINEAR dataset: SARG B and Chemically Peculiar. This does not mean that these stars are not contained in the LINEAR dataset. Similarly, those stars that were not classified are not necessarily in a "new" class of stars. There are a number of possibilities why these stars were not found in the survey including:

[3]https://github.com/kjohnston82/LINEARSupervisedClassification

1. Poor separation between the class of interest (for a given detector) and other stars. Poor separation could result in either the posterior probability not being high enough to detect the star, or more likely the star being classified as two different types at the same time.

2. Poor initial quantification of the signature class pattern in the training set feature space. If the training sample representing a given class type spanned only a segment of the signature class pattern region, the potential for an under-sampled or poorly bounding feature space exists. Furthermore, application of the anomaly detection algorithm, or any of the pattern classification algorithms, would result in decision-lines lassoing the under-sampled feature space, cutting through the “true class pattern region” . New observations of that class type, if they occurred outside of the original under-sampled space, would likely be flagged by the anomaly detection algorithm or as a different class.

Thus, those stars positively classified by the set of detectors used represent the set of LINEAR observations that have patterns that are consistent with those observed in the training set. As part of the testing process we estimate both a false alarm rate (FAR) of 0.5% across all classes and a precision rate from the PR curve. Then, each one-vs.-all detector will have a different precision rate, since the FAR is fixed. The precision rate estimates based on testing are given in Table 4.10. An adjusted estimate of “true” returned population sizes can be estimated by considering the precision rate, i.e., if 15 Mira stars were detected, and the Mira detector had a precision of ˜94% percent, then potentially 1 of those detections is a false positive.

Table 4.10: Precision Rate Estimates Per Class Type (in fractions), Bolded Classes are those with Precision < 80%

| Class Type | Precision | Est. Pop | Class Type | Precision | Est. Pop |
|---|---|---|---|---|---|
| a. Mira | 0.94 | 14 | m. Slowly Puls. B | 0.91 | 0 |
| b1. Semireg PV | 0.97 | 0 | n. Gamma Doradus | 0.88 | 1159 |
| **b2. SARG A** | **0.76** | 1035 | o. Pulsating Be | 0.91 | 192 |
| b3. SARG B | 0.94 | 0 | p. Per. Var. SG | 0.94 | 85 |
| b4. LSP | 0.91 | 0 | q. Chem. Peculiar | 0.94 | 0 |
| c. RV Tauri | 0.86 | 462 | r. Wolf-Rayet | 0.91 | 1939 |
| d. Classical Cepheid | 0.94 | 1 | **r1. RCB** | **0.73** | 2 |
| e. Pop. II Cepheid | 0.87 | 43 | **s1. Class. T Tauri** | **0.75** | 1383 |
| f. Multi. Mode Cepheid | 0.87 | 248 | **s2. Weak-line T Tauri** | **0.75** | 843 |
| g. RR Lyrae FM | 0.91 | 2542 | **s3. RS CVn** | **0.74** | 11850 |
| h. RR Lyrae FO | 0.88 | 624 | t. Herbig AE/BE | 0.86 | 96 |
| **i. RR Lyrae DM** | **0.73** | 1715 | **u. S Doradus** | **0.67** | 387 |
| j. $\delta$ Scuti | 0.95 | 7 | **v. Ellipsoidal** | **0.78** | 21 |
| **j1. SX Phe** | **0.73** | 1185 | w. Beta Persei | 0.97 | 138 |
| k. Lambda Bootis | 0.91 | 0 | x. Beta Lyrae | 0.97 | 1 |
| l. Beta Cephei | 0.91 | 22 | y. W Ursae Maj. | 0.91 | 986 |

## 4.5 Conclusions

This paper has demonstrated the construction and application of a supervised classification algorithm on variable star data. Such an algorithm will process observed stellar features and produce quantitative estimates of stellar class label. Using a hand-process (verified) dataset derived from the ASAS, OGLE, and Hipparcos survey, an initial training and testing set was derived. The trained one-vs.-all algorithms were optimized using the testing data via minimization of the misclassification rate. From application of the trained algorithm to the testing data, performance estimates can be quantified for each one-vs.-all algorithm. The Random Forest supervised classification algorithm was found to be superior for the feature space and class space operated in. Similarly, a one-class support vector machine was trained in a similar manner, and designed as an anomaly detector.

With the classifier and anomaly detection algorithm constructed, both were applied to a set of 192,744 LINEAR data points. Of the original samples, setting the threshold of the RF classifier using a false alarm rate of 0.5%, 34,451 unique stars were classified only once in the one-vs.-all scheme and were not identified by the anomaly detection algorithm. The total population is partitioned into the individual stellar variable classes; each subset of LINEAR ID corresponding to the matched patterns is stored in a separate file and accessible to the reader. While less than 18% of the LINEAR data was classified, the class labels estimated have a high level of probability of being the true class based on the performance statistics generated for the classifier and the threshold applied to the classification process.

### 4.5.1 Future Research

Further improvement in both the initial training dataset is necessary, if the requirements of the supervised classification algorithm are to be met (100% classification of new data). Larger training data, with more representation (support) is needed to improve the class space representation used by the classifier and reduce the size of the "anomalous" decision region. Specifically, additional example of the under-sampled variable stars, enough to perform k-fold cross-validation would yield improved performance and increased generality of the classifier. An improved feature space could also benefit the process, if new features were found to provide additional linear separation for certain classes, such as those presented in [Johnston and Peter, 2017].

However, additional dimensionality without reduction of superfluous features is warned against as it may only worsen the performance issues of the classifier. Instead, investigation into the points found to be anomalous in under-sampled classes, and determination if they are indeed of the class reported by the classifier designed here would be of benefit, as these points would serve to not only bolster the number of training points used in the algorithm, but they would also increase the size (and support) of the individual class spaces. Implementation of these concepts, with a mindfulness of the changing performance of the supervised classification algorithm, could result in performance improvements across the class space.

# Chapter 5

# Novel Feature Space Implementation

A methodology for the reduction of stellar variable observations (time-domain data) into a novel feature space representation is introduced. The proposed methodology, referred to as Slotted Symbolic Markov Modeling (SSMM), has a number of advantages over other classification approaches for stellar variables. SSMM can be applied to both folded and unfolded data. Also, it does not need time-warping for alignment of the waveforms. Given the reduction of a survey of stars into this new feature space, the problem of using prior patterns to identify new observed patterns can be addressed via classification algorithms. These methods have two large advantages over manual-classification procedures: the rate at which new data is processed is dependent only on the computational processing power available and the performance of a supervised classification algorithm is quantifiable and consistent[Johnston and Peter, 2018].

The remainder of this paper is structured as follows. First, the data, prior efforts, and challenges uniquely associated to classification of stars via stellar variability is reviewed. Second, the novel methodology, SSMM, is outlined including the feature space and signal conditioning methods used to extract the unique time-domain signatures. Third, a set of classifiers (random forest/bagged decisions tree, k-nearest neighbor, and Parzen window classifier) is trained and tested on the extracted feature space using both a standardized stellar variability dataset and the LINEAR dataset. Fourth, performance statistics are generated for each classifier and a comparing and contrasting of the methods is discussed. Lastly, an anomaly detection algorithm is generated using the so called one-class Parzen Window Classifier and the LINEAR dataset. The result will be the demonstration of the SSMM methodology as being a competitive feature space reduction technique, for usage in supervised classification algorithms[1].

## 5.1 Related Work

Many prior studies on time-domain variable star classification [Debosscher, 2009, Barclay et al., 2011, Blomme et al., 2011, Dubath et al., 2011, Pichara et al., 2012, Pichara and Protopapas, 2013, Graham et al., 2013a, Angeloni et al., 2014, Masci et al., 2014]. rely on periodicity domain feature space reductions. Debosscher [2009] and Templeton [2004] review a number of feature spaces and a number of efforts to reduce the time domain data, most of which implement Fourier techniques, primarily the Lomb–Scargle (L-S) Method [Lomb, 1976, Scargle, 1982], to estimate the primary periodicity [Eyer and Blake, 2005, Deb and Singh, 2009,

[1]Lightly Edited from original paper: Johnston, K. B., & Peter, A. M. (2017). Variable star signature classification using slotted symbolic Markov modeling. *New Astronomy, 50*, 1–11.

Richards et al., 2012, Park and Cho, 2013, Ngeow et al., 2013]. Lomb–Scargle is favored because of the flexibility it provides with respect to observed datasets; when sample rates are irregular and drop outs are common in the data being observed. Long et al. [2014] advance L-S even further, introducing multi-band (multidimensional) generalized L-S, allowing the algorithm to take advantage of information across filters, in cases where multi-channel time-domain data is available. There have also been efforts to estimate frequency using techniques other than L-S such as the Correntropy Kernelized Periodogram, [Huijse et al., 2011] or MUlti SIgnal Classificator [Tagliaferri et al., 2003].

The assumption of the light curve being periodic, or even that the functionality of the signal being represented in the limited Fourier space that Lomb–Scargle uses, has been shown [Barclay et al., 2011, Palaversa et al., 2013] to result in biases and other challenges when used for signature identification purposes. Supervised classification algorithms implementing these frequency estimation algorithms do so to generate an estimate of primary frequency used to fold all observations resulting in a plot of magnitude vs. phase, something Deb and Singh [2009] refer to as "reconstruction" . After some interpolation to place the magnitude vs. phase plots on similar regularly sampled scales, the new folded time series can be directly compared (1-to-1) with known folded time series. Comparisons can be performed via distance metric [Tagliaferri et al., 2003], correlation [Protopapas et al., 2006], further feature space reduction [Debosscher, 2009] or more novel methods [Huijse et al., 2012].

It should be noted that the family of stars with the label "stellar variable" is a large and diverse population: eclipsing binaries, irregularly pulsating variables, nova (stars in outburst), multi-model variables, and many others are frequently processed using the described methods despite the underlying stellar variability functionality not naturally lending itself to Fourier decomposition and the associated assumptions that accompany the said decomposition. Indeed this is why Szatmary et al. [1994], Barclay et al. [2011], Palaversa et al. [2013] and others suggest using other decomposition methods such as discrete wavelet transformations, which have been shown to be powerful in the effort to decompose a time series into the time-frequency (phase) space for analysis [Torrence and Compo, 1998, Bolós and Benítez, 2014, Rioul and Vetterli, 1991]. It is noted that the digital signal processing possibilities beyond Fourier domain analysis time series comparison and wavelet transformation are too numerous to outline here; however the near complete review by Fulcher et al. [2013] is highly recommended.

## 5.2 Slotted Symbolic Markov Modeling

The discussion of the Slotted Symbolic Markov Modeling (SSMM) algorithm encompasses the analysis, reduction and classification of data. The *a priori* distribution of class labels are roughly evenly distributed for both studies, therefore the approach uses a multi-class classifier. Should the class labels with additional data become unbalanced, other approaches are possible [Rifkin and Klautau, 2004]. Data specific challenges, associated with astronomical time series observations, have been identified as needing to be addressed as part of the algorithm design.

### 5.2.1 Algorithm Design

Stellar variable time series data can roughly be described as passively observed time series snippets, extracted from what is a contiguous signal (star shine) over multiple nights or sets of observations. The time series signatures have the potential to change over time, and new observations allow for the increased opportunity for an unstable signature over the long term. Astronomical time series data is also frequently irregular, i.e., there is often no associated fixed $\Delta t$ over the whole of the data that is consistent with the observation. Even when there is a consistent observation rate, this rate is often broken up because of observational constraints. The stellar variable moniker covers a wide variety of variable types: stationary (consistently repeating identical patterns), non-stationary (patterns that increase/decrease in frequency over time), non-regular variances (variances that change over the course of time, shape changes), as well as both Fourier and non-Fourier sequences/patterns. Pure time-domain signals do not lend themselves to signature identification and pattern matching, as their domain is infinite in terms of potential discrete data (dimensionality). Not only must a feature space representation be found, but the dimensionality should not increase with increasing data.

Based on these outlined data/domain specific challenges (continuous time series, irregular sampling, and varied signature representations) this paper will attempt to develop a feature space extraction methodology that will construct an analysis of stellar variables and characterize the shape of the periodic stellar variable signature. A number of methods have been demonstrated that fit this profile [Grabocka et al., 2012, Fu, 2011, Fulcher et al., 2013], however many of these methods focus on identifying a specific time series shape sequence in a long(er) continuous time series, and not necessarily on the differentiation between time series sequences. To address these domain specific challenges, the following methodology outline is implemented:

1. To address the irregular sampling rate, a slotting methodology is used [Re-

hfeld et al., 2011]: Gaussian kernel window slotting with overlap. The slotting methodology is used to generate estimates of amplitudes at regularized points, with the result being a up-sampled conditioned waveform. This has been shown to be useful in the modeling and reconstruction of variability dynamics[Rehfeld and Kurths, 2014, Kovačević et al., 2015, Huijse et al., 2011], and is similar to the methodologies used to perform Piecewise Aggregate Approximation [Keogh et al., 2001].

2. To reduce the conditioned time series into a usable feature space, the amplitudes of the conditioned time series will be mapped to a discrete state space based on a standardized alphabet. The result is the state space representation of the time domain signal, and is similar to the methodologies used to perform Symbolic Aggregate Approximation [Lin et al., 2007].

3. The state space transitions are then modeled as a first order Markov Chain Ge and Smyth [2000], and the state transition probability matrix (Markov Matrix) is generated, a procedure unique to this study. It will be shown that a mapping of the transitions from observation to observation will provide an accurate and flexible characterization of the stellar variability signature.

The Markov Matrix is vectorized into a vector, and is the signature pattern (feature vector) used in the classification of time-domain signals for this study. It should be noted, if the underlying signature was too sparsely sampled the feature space transform would not capture the shapes or features of interest, as it would be true with any time-domain transform (i.e., Fourier methods will not capture frequency content over the Nyquist). Many of the transforms that are commonly used combat the problems of irregular samples or low sample density, make addi-tional assumptions such as the underlying waveform shape (e.g. Box–fitting Least Squares Siverd et al., 2012) or the frequency of occurrence (based on physical parameters). Many of these assumptions are oriented towards specific target detection; as such, the false alarm and missed detection rates are tied specifically to those assumptions. A collection and comparison of these frequency sampling methods and the associated assumptions can be found in Graham et al. [2013b].

### 5.2.2 Slotting (Irregular Sampling)

Each waveform is conditioned using the slotting resampling methodology for irregularly sampled waveforms outlined in Rehfeld et al. [2011]. The slotting procedure acts as follows:

1. A set of evenly spaced (in time) windows, with size $w$ are generated, the windows can be overlapping or adjacent
2. Within each window (slot) are observed samples, the difference in time between the observation point and the center of the window is computed
3. a Gaussian kernel with width $s$ weights the contribution of the observation, a estimate of the amplitude for the new point is generated from the weighted amplitudes of the contributing observations.

For this implementation, an overlapping slot (75% overlap) was used, meaning that the window width is larger then the distance between slot centers. This methodology is effectively Kernel Smoothing with Slotting [Li and Racine, 2007], with a slot width of $w$. This $w$ is optimized, via cross-validation of the data; anecdotally however, median sample rate of the waveforms is often a best estimate; as such the rate would capture at least one point in each slot when applied to continuous observation data. Initial testing (ANOVA) demonstrated that the misclassification rate varied little with changes in $w$ about the median.

Let the set of $\{y(t_n)\}_{n=1}^{N}$ samples, where $t_1 < t_2 < t_3 < ... < t_N$ and there are N samples, be the initial time series dataset. The observed time series data is standardized (subtract the mean, divide by the standard deviation), and then the slotting procedure is applied. If $x[i] \leftarrow y(t_i)_{i=1}^{N}$, then the algorithm to generate the slotted time domain data is given in Algorithm 1.

**Algorithm 1** Gaussian Kernel Slotting

```
procedure GaussianKernelSlotting(x[i], t[i], w, λ)

    x_prime[i] ← (x[i] − mean(x[i]))/std(x[i])          ▷ Standardize Amplitudes
    t[i] ← t[i] − min(t[i])                              ▷ Start at Time Origin
    slotCenters ← 0 : w/4 : max(t[i]) + w                ▷ Make Slot Locations
    timeSeriesSets = []                                  ▷ Initialize Time Series Sets
    slotSet = []                                         ▷ Make an Empty Slot Set

    while i < length(slotCenters) do                     ▷ Compute Slots
        idx ← all t in interval [slotCenters − w, slotCenters + w]
        inSlotX ← x[idx]
        inSlotT ← t[idx]

        if inSlot is empty then                          ▷ There is a Gap
            if slotSet is empty then                     ▷ Move to Where Data is
                currentPt ← find next t > slotCenters + w
                i ← find last slotCenters < t[currentPt]
            else                                         ▷ Store the Slotted Estimates
                add slotSet to structure timeSeriesSets
                slotSet ← []
            end if
        else
            weights ← exp(−((inSlot − slotCenters)^2 ∗ λ))
            meanAmp ← sum(weights ∗ inSlotX)/sum(weights)
            add meanAmp to the current slotSet
        end if
        i++
    end while
end procedure
```

The time series, with irregular sampling and large gaps is conditioned by the Gaussian slotting method. Gaps in the waveform are defined as regions where a slot contains no observations. Continuous observations (segments) are the set of observations between the gaps. This results in a irregularly sampled waveform potentially resulting in a set of regularly sampled waveforms for various lengths depending on the original survey parameters. This conditioning is also similar to the Piecewise Aggregation Approximation [Lin et al., 2003, Keogh et al., 2001]. Instead of down-sampling the time domain datasets as PAA does however, the data is up-sampled using the slotting methodology. This is necessary because of the sparsity of the time domain sampling of astronomical data.

### 5.2.3 State Space Representation

If it is assumed that the conditioned standardized waveform segments have an amplitude distribution that approximates a Gaussian distribution (which they won't, but that is irrelevant to the effort), then using a methodology similar to Symbolic Aggregate Approximation [Lin et al., 2007, 2012] methodologies, an alphabet (state space) is defined based on our assumptions as an alphabet extending between $\pm 2\sigma$ and will encompass 95% of the amplitudes observed. This need not always be the case, but the advantage of the standardization of the waveform is that, with some degree of confidence the information from the waveform is contained roughly between $\pm 2\sigma$. Figure 5.1 demonstrates a eight state translation; the alphabet will be significantly more resolved then this for astronomical waveforms.

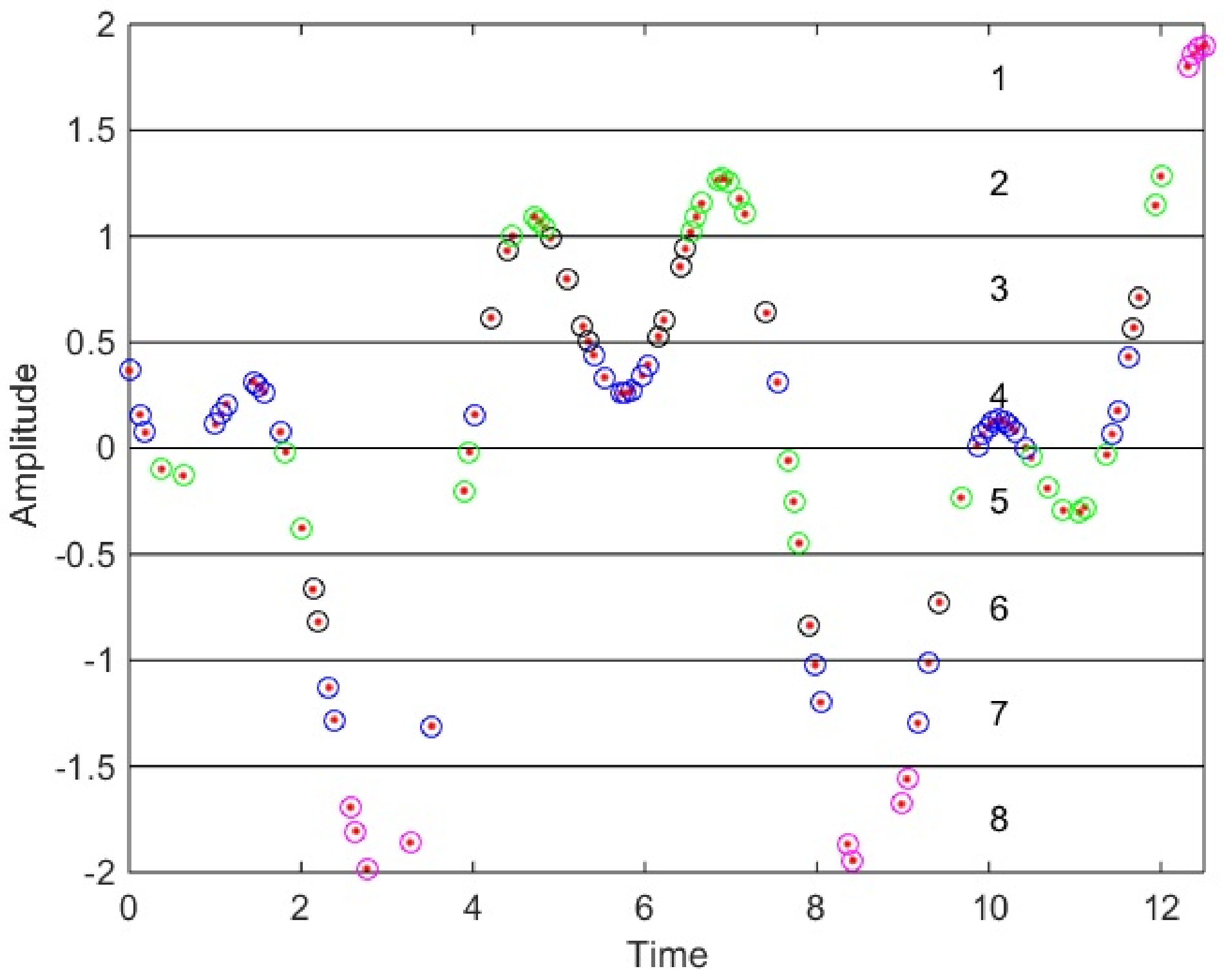

Figure 5.1: Example State Space Representation

The resolution of the alphabet granularity is to be determined via cross-validation to determine an optimal resolution for a given survey. The set of state transitions, the transformation of the conditioned signal, is used to populate a transition probability matrix or first order Markov Matrix.

### 5.2.4 Transition Probability Matrix (Markov Matrix)

The transition state frequencies are estimated for signal measured between empty slots, transitions are not evaluated between day-night periods, or between slews (changes in observation directions during a night) and only evaluated for continuous observations. Each continuous set of conditioned waveforms (with Slotting and State Approximation applied) is used to populate the empty matrix $P$, with dimensions equal to $r \times r$, where $r$ is the number of states, is built. The matrix is populated using the following rules:

- $N_{ij}$ is the number of observation pairs $x[t]$ and $x[t+1]$ with $x[t]$ is state $s_i$ and $x[t+1]$ in state $r_j$
- $N_i$ is the number of observation pairs $x[t]$ and $x[t+1]$ with $x[t]$ in state $s_i$ and $x[t+1]$ in any one of the states $j = 1, ..., r$

The now populated matrix $P$ is a transition frequency matrix, with each row $i$ representing a frequency distribution (histogram) of transitions out of the state $s_i$. The transition probability matrix is approximated by converting the elements of P by approximating the transition probabilities using $P_{ij} = {}^{N_{ij}}/_{N_i}$. The resulting matrix is often described as a first order Markov Matrix [Ross, 2013]. State changes are based on only the observation-to-observation amplitude changes; the matrix is a representation of the linearly interpolated sequence [Ge and Smyth, 2000]. Furthermore, the matrix is vectorized similar to image analysis methods into a feature space vector, with dimensions depend on the resolution and bounds of the states. The algorithm to process the time-domain conditioned data is given in Algorithm 2.

**Algorithm 2** Markov Matrix Generation

```
procedure MarkovMatrixGeneration(timeSeriesSets, s)
    markovMatrix = []
    for i := 1 to length of timeSeriesSets do
        markovMatrixPrime ← []
        currentSlotSet ← markovMatrixPrime[j]
        for k := 1 to length of currentSlotSet do
            idxIn ← find state containing currentSlotSet[k − 1]
            idxOut ← find state containing currentSlotSet[k]
            markovMatrixPrime[idxIn, idxOut] + +
        end for
        markovMatrix ← markovMatrix + markovMatrixPrime
    end for
    N_i = sum along row of markovMatrix
    for j := 1 to length of s do
        if N_i ≠ 0 then
            markovMatrix[:, j] ← N_ij / N_i        ▷ Estimate Markov Matrix
        end if
    end for
end procedure
```

The resulting Markov Matrix is vectorized into a feature vector given by:

$$\mathbf{P}_i = \begin{bmatrix} p_{11} & p_{12} & \cdots & p_{1r} \\ p_{21} & p_{22} & \cdots & \cdots \\ \vdots & \vdots & \ddots & \vdots \\ p_{r1} & p_{21} & \cdots & p_{rr} \end{bmatrix} \Rightarrow vec(\mathbf{P}_i) = \begin{bmatrix} p_{11} & p_{12} & \dots & p_{21} & \dots & p_{rr} \end{bmatrix}, \quad (5.1)$$

where $\mathbf{P}_i$ is the Markov Chain of the $i^{th}$ input training set, and $x_i$ is the $i^{th}$ input vectorized training pattern. When using the Markov matrix representation, the resolution of the state set needs to be small to avoid loss of information resulting from over generalization. However, if the state resolution is too small the sparsity of the transition matrix will result in a shape signature that is too dependent on noise and the " individualness" of specific waveform to be of any use. Thus additional processing is necessary for further analysis; even a small set of states (12 x 12) will result in a feature vector with high dimensionality (144 dimensions). While a window and overlap size is assumed for the slotting to address the irregular sampling of the time series data, there are two adjustable features associated with this analysis: the kernel width associated with the slotting and the state space (alphabet) resolution. It is apparent that a range of resolutions and kernel width need to be tested to determine best performance given a generic supervised classifier.

### 5.2.5 Feature Space Reduction (ECVA)

For these purposes a rapid initial classification algorithm, General Quadratic Discriminate Analysis [Duda et al., 2012], was implemented to estimate the misclassification rate (wrong decisions/total decisions). To reduce the large, sparse, feature vector resulting from the unpacking of the Markov Matrix we applied a supervised dimensionality reduction technique commonly referred to as canonical variate analysis (ECVA) [Nørgaard et al., 2006]. The methodology for ECVA has roots in principle component analysis (PCA). PCA is a procedure performed on large multidimensional datasets with the intent of rotating what is a set of possibly correlated dimensions into a set of linearly uncorrelated variables [Scholz, 2006].

The transformation results in a dataset, where the first principle component (dimension) has the largest possible variance. PCA is an unsupervised methodology, *a priori* labels for the data being processed are not taken into consideration, and while a reduction in feature dimensionality is obtained and a maximization in the the variance will occur, the operation may not maximize the linear separability of the class space.

In contrast to PCA, Canonical Variate Analysis does take class labels into considerations. The variation between groups is maximized resulting in a transformation that benefits the goal of separating classes. Given a set of data $\mathbf{x}$ with: $g$ different classes, $n_i$ observations of each class, and $r \times r$ dimensions in each observation; following Johnson et al. [1992], the within-group and between-group covariance matrix is defined as:

$$\mathbf{S}_{within} = \frac{1}{n-g}\sum_{i=1}^{g}\sum_{j=1}^{n_i}(\mathbf{x}_{ij} - \bar{\mathbf{x}}_{ij})(\mathbf{x}_{ij} - \bar{\mathbf{x}}_i)' \tag{5.2}$$

$$\mathbf{S}_{between} = \frac{1}{g-1}\sum_{i=1}^{g} n_i(\mathbf{x}_i - \bar{\mathbf{x}})(\mathbf{x}_i - \bar{\mathbf{x}})'; \tag{5.3}$$

where $n = \sum_{i=1}^{g} n_i$, $\bar{\mathbf{x}}_i = \frac{1}{n_i}\sum_{j=1}^{n_i}\mathbf{x}_{ij}$, and $\bar{\mathbf{x}} = \frac{1}{n}\sum_{j=1}^{n_i} n_i\mathbf{x}_i$. CVA attempts to maximize the function:

$$J(\mathbf{w}) = \frac{\mathbf{w}'\mathbf{S}_{between}\mathbf{w}}{\mathbf{w}'\mathbf{S}_{within}\mathbf{w}}. \tag{5.4}$$

Which is solvable so long as $\mathbf{S}_{within}$ is non-singular, which need not be the case, especially when analyzing multicollinear data. When the case arises that the dimensions of the observed patterns are multicollinear additional considerations need to be made. Nørgaard et al. [2006] outlines a methodology for handling these cases in CVA. Partial least squares analysis, PLS2 [Wold, 1939], is used to solve the above linear equation, resulting in an estimate of $\mathbf{w}$, and given that, an estimate of the canonical variates (the reduced dimension set). The application of ECVA to our vectorized Markov Matrices results in a reduced feature space of dimension $g - 1$.

## 5.3 Implementation of Methodology

### 5.3.1 Datasets

Two datasets are addressed here, the first is the STARLIGHT dataset from the UCR time series database, the second is published data from the LINEAR survey. The UCR time series dataset is used to base line the time-domain dataset feature extraction methodology proposed, it is compared to the results published on the UCR website. The UCR time series data contains only time domain data that has already been folded and put into magnitude phase space, no photometric data from either SDSS or 2MASS, nor star identifications for these data, could be recovered, and only three class types are provided which are not defined besides by number. The second dataset, the LINEAR survey, provides an example of a modern large scale astronomical survey, contains time-domain data that has not been folded or otherwise manipulated, is already associated with SDSS and 2MASS photometric values, and has five identified stellar variable types. For each dataset, the state space resolution and the kernel widths for the slotting methods will be optimized using 5-fold cross-validation. The performances of three classifiers on only the time-domain dataset for the UCR data, and on the mixture of time-domain data and color data for the LINEAR survey, are estimated using 5-fold cross-validation and testing. The performances of the classifiers will be compared. Finally an anomaly detection algorithm will be trained and tested, for the LINEAR dataset.

### 5.3.2 Pattern Classification Algorithm

The training set is used for 5-Fold cross-validation, and a set of three classification algorithms are tested [Hastie et al., 2009, Duda et al., 2012]: k-Nearest Neighbor (k-NN), Parzen Window Classifier (PWC) and Random Forest (RF). Cross-validation is used to determine optimal classification parameters (e.g., kernel width) for each of the classification algorithms. The first three algorithms implemented were designed by the authors in MATLAB, based on Duda et al. [2012] and Hastie et al. [2009] (k-NN and PWC) algorithm outlines. Code is accessible via github[2].

#### 5.3.2.1 k-NN

The k-nearest neighbor algorithm is a non-parametric classification method; it uses a voting scheme based on an initial training set to determine the estimated label[Altman, 1992]. For a given new observation, the $L_2$ Euclidean distance is found between the new observation and all points in the training set. The distances are sorted, and the $k$ closest training sample labels are used to determine the new observed sample estimated label (majority rule). Cross-validation is used to find an optimal $k$ value, where $k$ is any integer greater than zero.

[2]https://github.com/kjohnston82/SSMM

#### 5.3.2.2 PWC

Parzen windows classification is a technique for non-parametric density estimation, which is also used for classification [Parzen, 1962, Duda et al., 2012]. Using a given kernel function, the technique approximates a given training set distribution via a linear combination of kernels centered on the observed points. As the PWC algorithm (much like a k-NN) does not require a training phase, as the data points are used explicitly to infer a decision space. Rather than choosing the $k$ nearest neighbors of a test point and labeling the test point with the weighted majority of its neighbor's votes, one can consider all points in the voting scheme and assign their weight by means of the kernel function. With Gaussian kernels, the weight decreases exponentially with the square of the distance, so far away points are practically irrelevant. Cross-validation is necessary however, to determine an optimal value of $h$, the "width" of the radial basis function (or whatever kernel is being used).

#### 5.3.2.3 Random Forest Classifier

To generate the random forest classifier, the *TreeBagger* algorithm in MATLAB is implemented. The algorithm generates $n$ decision trees on the provided training sample. The $n$ decision trees operate on any new observed pattern, and the decision made by each tree are conglomerated together (majority rule) to generate a combined estimated label. To generate Breiman's random forest algorithm [Breiman et al., 1984], the value *NVarToSample* is provided a value (other than all) and a random set of variables is used to generate the decision trees; see the MATLAB *TreeBagger* documentation for more information.

### 5.3.3 Comparison to Standard Set (UCR)

The UCR time domain datasets are used to basis classification methodologies [Keogh et al., 2011]. The UCR time domain datasets [Protopapas et al., 2006], are derived from a set of Cepheid, RR Lyrae, and Eclipsing Binary Stars. The time-domain datasets have been phased (folded) via the primary period and smoothed using the SUPER-SMOOTHER algorithm [Reimann, 1994] by the Protopapas study prior to being provided to the UCR database. The waveforms received from UCR are amplitude as a function of phase; the SUPER-SMOOTHER algorithm was also used [Protopapas et al., 2006] to produce regular samples (in the amplitude vs. phase space). The sub-groups of each of the three classes are combined together in the UCR data (i.e., RR (ab) + RR (c) = RR), similarly the data is taken from two different studies (OGLE and MACHO). A plot of the phased light curves is given in Figure 5.2.

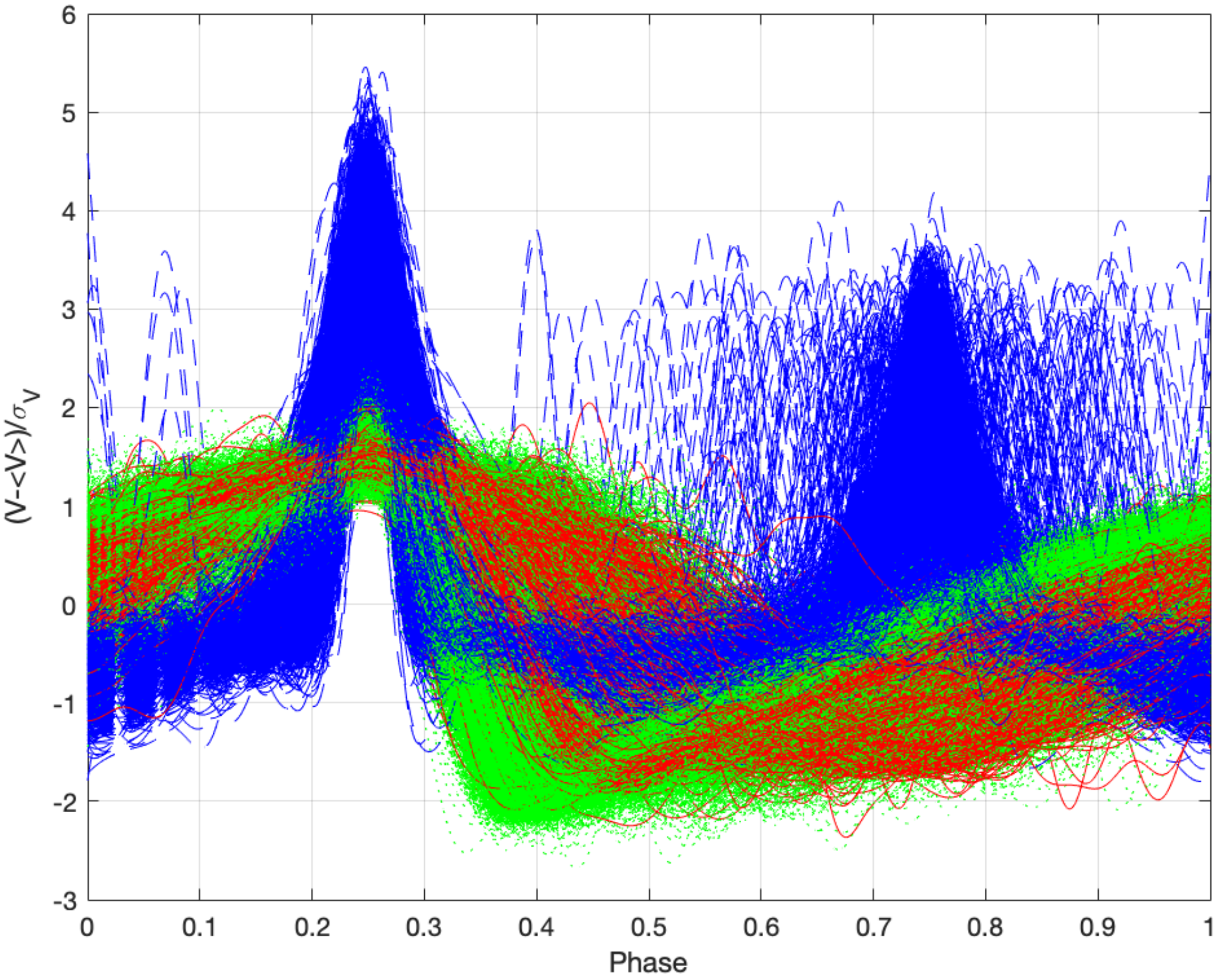

Figure 5.2: UCR Phased Light Curves. Classes are given by number only: 1 = Blue Line (Eclipsing Binaries), 2 = Green Small Dashed Line (Cepheid), 3 = Red Big Dashed Line (RR Lyr)

Class analysis is a secondary effort when applying the methodology outlined to the UCR dataset, the primary concern is a demonstration of performance of the supervised classification methodology with respect to the baseline performance reported by UCR implementing a simple waveform nearest neighbor algorithm.

#### 5.3.3.1 Analysis

The folded waveforms are treated identical to the unfolded waveforms in terms of the processing presented. Values of phase were generated to accommodate the slotting technique, thereby allowing the functionally developed to be used for both amplitude vs. time (LINEAR) as well as amplitude vs. phase (UCR). The slotting, State Space Representation, Markov Matrix and ECVA flow is implemented exactly the same. As there are only three classes in the dataset, the ECVA algorithm results in a dimensionality of only two $(g - 1)$. No accompanying color data with the time-domain data is available, so only the time-domain data will be focused on for this analysis. The resulting ECVA plot is presented in Figure 5.3.

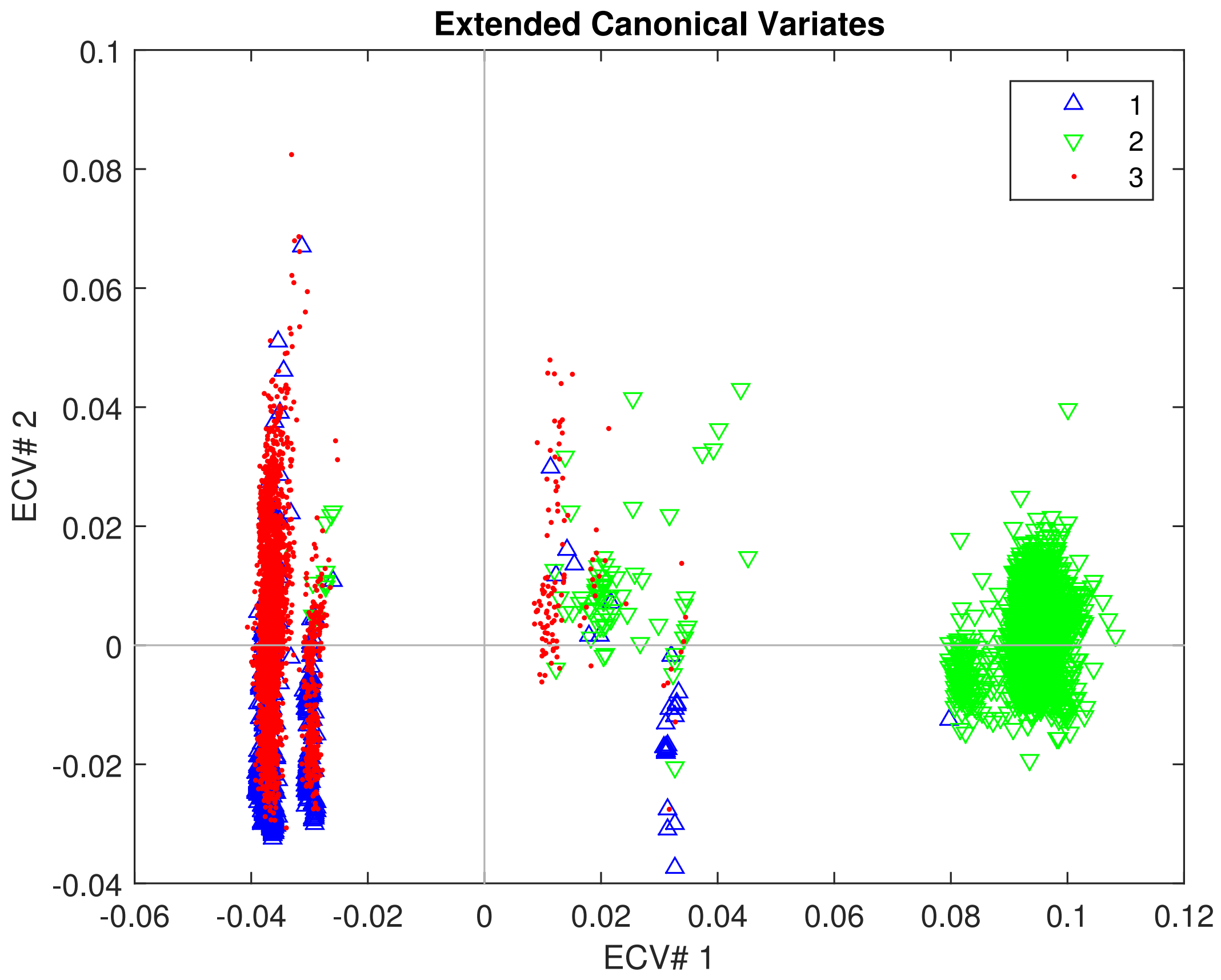

Figure 5.3: ECVA reduced feature space using the UCR Star Light Curve Data

Each classifier is then trained only on the ECVA reduced time-domain feature space. The resulting optimization analysis, based on the 5-fold cross-validation is presented in Figures B.1a, B.1b and B.1c. Depending on the methodology used, cross-validation estimates a misclassification error of $< 10\%$. The UCR website reports the following error estimates for this dataset, note that all methods reported use direct distance to generate a feature space (direct comparison of curves): 1-NN Euclidean Distance (15.1%), 1-NN Best Warping Window DTW (9.5%) and 1-NN DTW, no warping window (9.3%). For a more detailed comparison, the confusion matrix for each of the optimized classifiers is presented in Appendix B, Tables B.1a, B.1b, and B.1c.

#### 5.3.3.2 Discussion

The SSMM methodology presented does no worse than the 1-NN presented by Keogh et al. [2011] and appears to provide some increase in performance. The procedure described operates on folded data as well as unfolded data and does not need time-warping for alignment of the waveform, demonstrating the flexibility of the method. The procedure not only separated out the classes outlined, but in addition found additional clusters of similarity in the dataset. Whether these clusters correspond to the sub-groupings reported by the original generating source (RR (ab) and RR (c), etc.) is not known, as object identification is not provided by the UCR dataset.

### 5.3.4 Application to New Set (LINEAR)

For the analysis of the proposed algorithm design, the LINEAR dataset is parsed into training, cross-validation and test sets on time series data from the LINEAR survey that has been verified, and for which accurate photometric values are available [Sesar et al., 2011, Palaversa et al., 2013]. From the starting sample of 7,194

LINEAR variables, a clean sample of 6,146 time series datasets and their associated photometric values were used for classification. Stellar class type is limited further to the top five most populous classes: RR Lyr (ab), RR Lyr (c), Delta Scuti / SX Phe, Contact Binaries and Algol-Like Stars with 2 Minima; resulting in a set of 6,086 observations. The distribution of stellar classes is presented in Table 5.1.

Table 5.1: Distribution of LINEAR Data across Classes

| Type | Count | Percentage |
|---|---|---|
| Algol | 287 | 5.6 |
| Contact Binary | 1805 | 35.6 |
| Delta Scuti | 68 | 1.3 |
| RRab | 2189 | 43.0 |
| RRc | 737 | 14.5 |

#### 5.3.4.1 Non-Variable Artificial Data

In support of the supervised classification algorithm, artificial datasets have been generated and introduced into the training/testing set. These artificial datasets are a representation of stars with-out variability. This introduction of artificial data is done for the same reasons the training of the anomaly detection algorithm is performed:

- The LINEAR dataset implemented only represents five of the most populous variable star types [Richards et al., 2012], thus the class space defined by the classes is incomplete.

- Even if the class space was complete, studies such as Debosscher [2009], Dubath et al. [2011] have all shown that many stellar variable populations are under-sampled.

- Similarly, many of the studies focus on stellar variables only, and do not include non-variable stars. While filters are often applied to separate variable and non-variable stars—Chi-Squared specifically, Sesar et al. 2013)— these are not necessarily perfect methods for removing non-variable populations, and could result in an increase in false alarms.

This artificial time series is generated with a Gaussian Random amplitude distribution. In addition to the time-domain information randomly generated, photometric information is also generated. The photometric measurements used to classify the stars are used to generate empirical distributions (histograms) of each of the feature vectors. These histograms are turned into cumulative distribution functions (CDFs). The artificially generated photometric patterns are generated via sampling from these generated empirical distribution functions. Sampling is performed via the Inverse Transform method [Law and Kelton, 1991] . These artificial datasets are treated identical in processing to the other observed waveforms.

#### 5.3.4.2 Time Domain and Color Feature Space

In addition to the time domain data, color data is obtainable for the LINEAR dataset, resulting from the efforts of large photometric surveys such as SDSS and 2MASS. These additional features are merged with the reduced time domain feature space, resulting in an overall feature space. For this study, the optical SDSS filters ($ugriz$) and the IR filters ($JK$) are used to generate the color features:

$u - g$, $g - i$, $i - K$, and $J - K$. The color magnitudes are corrected for the ISM extinction using $E(B - V)$ from the SFD maps and the extinction curve shape from Berry et al. [2012]. In addition to these color features, bulk time domain statistics are also generated: $logP$ is the log of the primary period derived from the Fourier domain space, $magMed$ is the median LINEAR magnitude, $ampl$, $skew$, and $kurt$ are the amplitude, skewness, and kurtosis for the observed light curve distribution. These additional features will be included for the analysis of the LINEAR dataset. See electronic supplement—Combined LINEAR Features, Extra-Figure-CombinedLINEARFeatures.fig—for a plot matrix of the combined feature space.

#### 5.3.4.3 Analysis

It is assumed that the state space resolution that minimizes the misclassification rate using QDA, will likewise minimize the misclassification rate using any of the other classification algorithms. The slot width was taken to be 0.015 and a kernel spread of 0.01 was used. Using the optimal amplitude state space resolution (0.03), a three dimensional plot (the first three ECVA parameters) is constructed; see the electronic supplement for the associated movie (ECVA Feature LINEAR Movie, ExtendedCanonicalVariates.mp4). Figure 5.4 is a plot of the first two extended canonical variates:

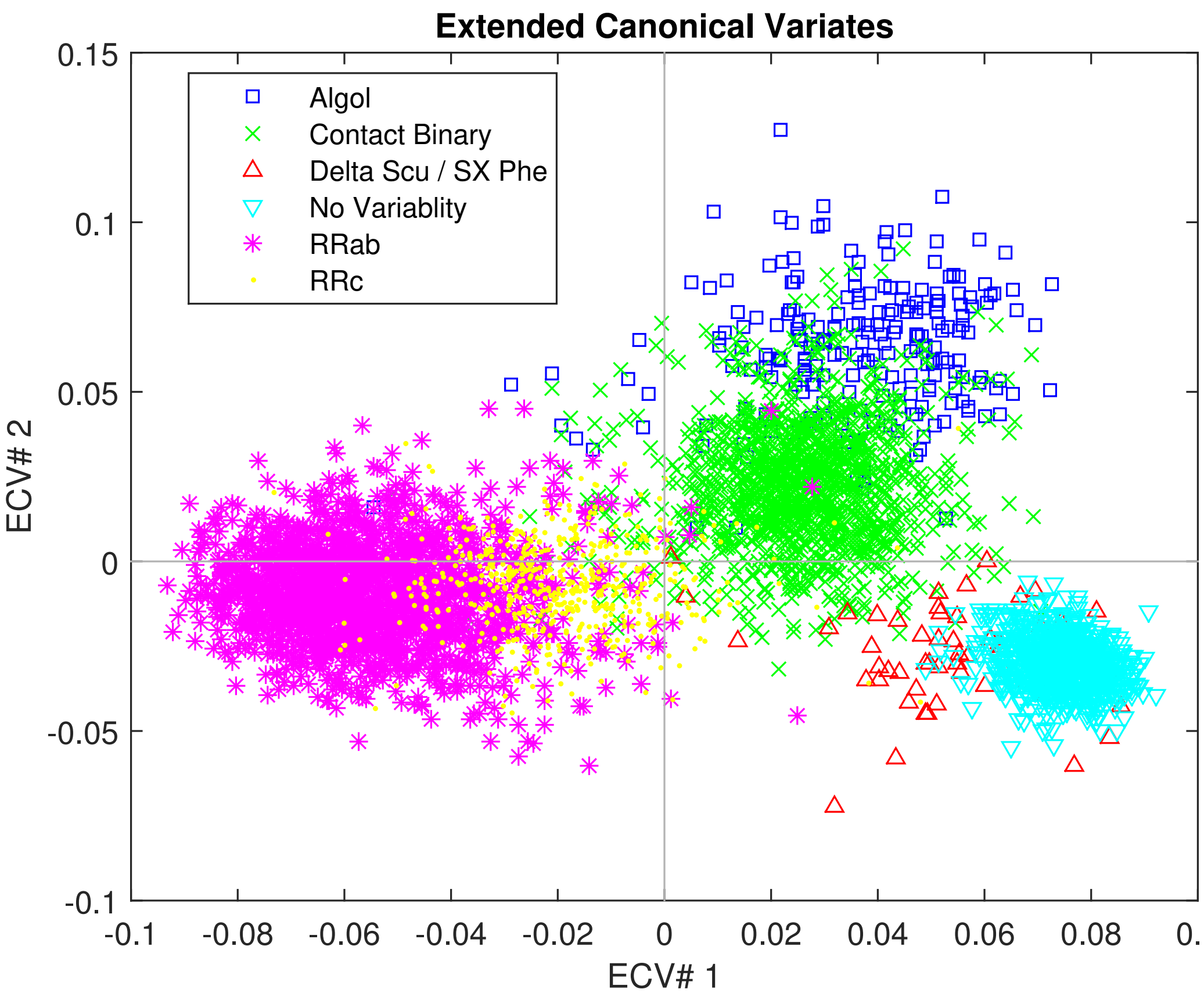

Figure 5.4: First two Extended Canonical Variates for the Time-Domain Feature Space

Based on the merged feature space, the optimal parameters for the k-NN, PWC and Random Forest Classifier are generated. The cross-validation optimization figures for each are presented in Figures B.2a, B.2b and B.2c respectively. Testing was performed on a pre-partitioned set, separate from the training and cross-validation populations.

The transformation applied to the training and cross-validation data were also applied to the testing data. After optimal parameters have been found for both the resolution of the Markov Model and the classification algorithms, the testing set is used to estimate the confusion matrix. Confusion matrices are generated—and given in in Appendix B— True Labels are shown on the left column and Estimated Label are shown on the top row (Tables B.2a, B.2b and B.2c). Further analysis was performed comparing the classification capability of a supervised classifier with only the SSMM features (post-ECVA analysis), with only the traditional feature spaces and a classifier with all three feature spaces (photometric data, frequency and time statistics, and SSMM ) to show the SSMM relative performance (Table : 5.2).

Table 5.2: Misclassification Rates of Feature Spaces from Testing Data

| | 1-NN | PWC | RF |
|---|---|---|---|
| All Features | 0.01 | 0.01 | 0.01 |
| Color and Frequency | 0.01 | 0.02 | 0.01 |
| SSMM Only | 0.03 | 0.04 | 0.03 |

Comparable performance is obtained just using the SSMM feature space compared to the color and frequency space, and for PWC, a small increase is obtained when the features are combined.

#### 5.3.4.4 Anomaly Detection

In addition to the pattern classification algorithm outlined, the procedure outlined includes the construction of an anomaly detector. The pattern classification algorithm presented as part of this analysis, partition the entire decision space based on the known class type provided in the LINEAR dataset. For many supervised classifier algorithms, and indeed those presented here, there is no consideration for deviations of patterns beyond the training set observed, i.e. absolute distance from population centers. All of the algorithms investigated consider relative distances, i.e. is the new pattern $P$ closer to the class center of B or A? Thus, despite that an anomalous pattern is observed by a new survey, the classifier will attempt to estimate a label for the observed star based on the labels it knows. To address this concern, a one-class anomaly detection algorithm is implemented.

Anomaly Detection and Novelty Detection methods are descriptions of similar processes with the same intent, i.e., the detection of new observations outside of the class space established by training. These methods have been proposed for stellar variable implementations prior to this analysis [Protopapas et al., 2006]. Tax [2001] and Tax and Muller [2003] outline the implementation of a number of classifiers for One-Class (OC) classification, i.e., novel or anomaly detection. Here, the PWC algorithm (described earlier) is transformed into an OC anomaly detection algorithm. The result is the "lassoing" or dynamic encompassing of known data patterns.

The lasso boundary represents the division between known (previously observed) regions of feature space and unknown (not-previously observed) regions. New patterns observed with feature vectors occurring in this unknown region are considered anomalies or patterns without support, and the estimated labels returned from the supervised classification algorithms should be questioned, despite the associated posterior probability of the label estimate. This paper implements the DD Toolbox designed by Tax and implements the PR toolbox [Duin et al., 2007]. The resulting error curve generated from the cross-validation of the PWC-OC algorithm resembles a threshold model (probit), the point which minimizes the error and minimizes the kernel width is found (Figure 5.5).

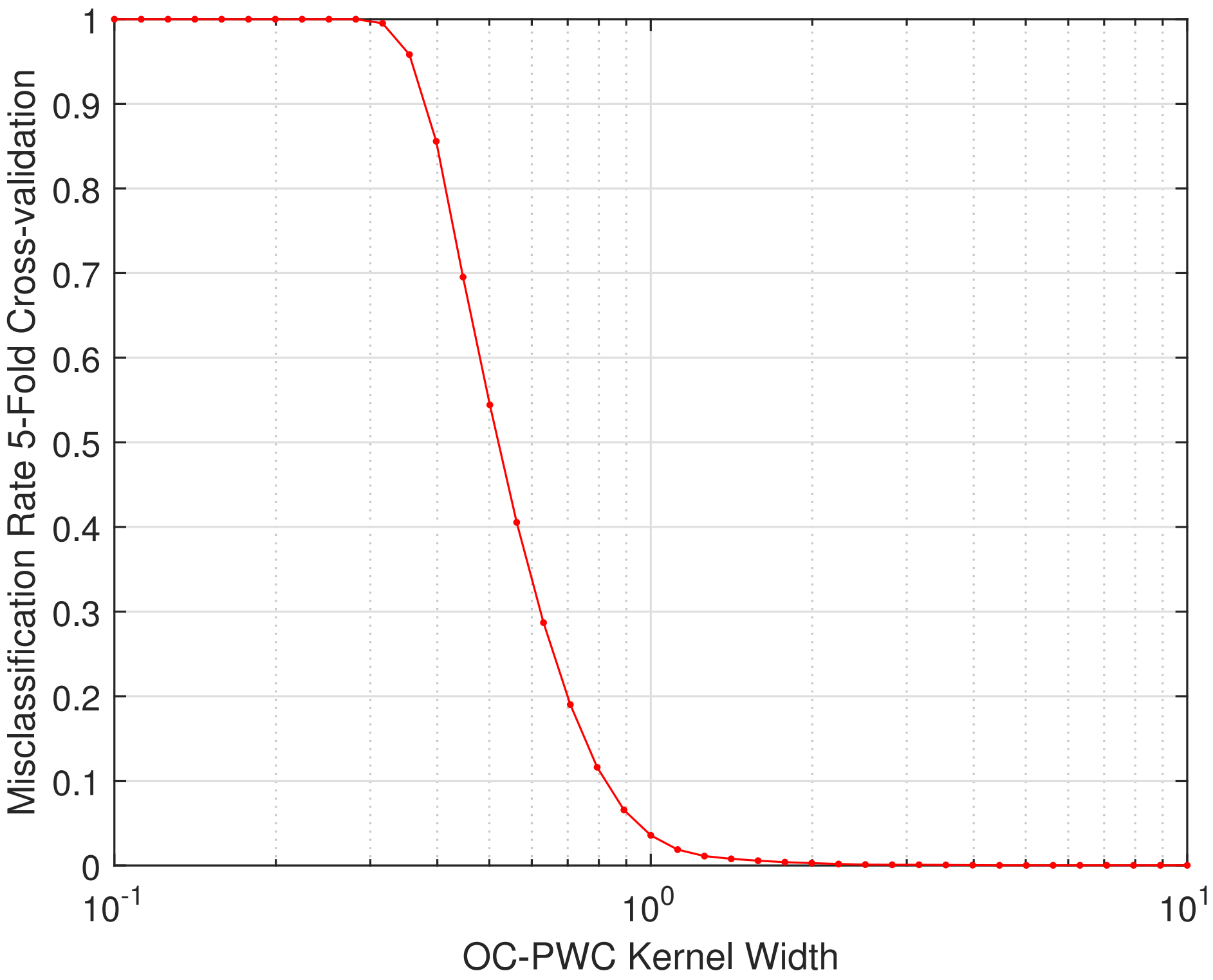

Figure 5.5: OC-PWC Kernel Width Optimization for LINEAR Data

This point (minimization of error and kernel width) is the optimal kernel width (2.5). Estimated misclassification rate of the detector is determined via evaluation of the testing set and found to be 0.067%.

#### 5.3.4.5 Discussion

Given only time series data (no photometric data, frequency data, or time domain statistics), for the classes and the LINEAR observations made (resolution of amplitude and frequency rate of observations) a $\sim 2\%$ misclassification rate with the various, more general, classifiers. Kernel width of the slots used to account for irregular sampling and state space resolution are major factors in performance, as mentioned we have assumed a slotting with that is a function of the survey (median of the continuous sample rate). This has been found to work optimally for the LINEAR dataset, however cross-validation could be performed with other dataset to determine if addition optimization is possible.

With the addition of photometric data, the misclassification rate is reduced by another $\sim 1\%$, and results in a nearly separable class space, depending on the methodology used to determine the estimated class. An anomaly detection algorithm is trained and tested on the time series data and photometric data. An expected misclassification rate, with both the anomaly detection algorithm and supervised classification algorithm, of $\sim 0.07\%$ is found.

## 5.4 Conclusions

The Slotted Symbolic Markov Modeling (SSMM) methodology developed has been able to generate a feature space which separates variable stars by class (supervised classification). This methodology has the benefit of being able to accommodate irregular sampling rates, dropouts and some degree of time-domain variance. It also provides a fairly simple methodology for feature space generation, necessary for classification. One of the major advantages of the methodology used is that a signature pattern (the transition state model) is generated and updated with new observations. The transition frequency matrix for each star is accumulated, given new observations, and the probability transition matrix is re-estimated. The methodology's ability to perform is based on the input data sampling rate, photometric error and most importantly the uniqueness of the time-domain patterns expressed by variable stars of interest.

The analysis presented has demonstrated the SSMM methodology performance is comparable to the UCR baseline performance analysis, if not slightly better. In addition, the translation of the feature space has demonstrated that the original suggestion of three classes might not be correct; a number of additional clusters are revealed as are some potential misclassifications in the training set. The performance of four separate classifiers trained on the UCR dataset is examined. It has been shown that the methodology presented is comparable to direct distance methods (UCR base line). It is also shown that the methodology presented is more flexible. The LINEAR dataset provides more opportunity to demonstrate the proposed methodology. The larger class space, unevenly sampled data with dropouts and color data all provide additional challenges to be addressed. After optimization, the misclassification rate is roughly $\sim 1\%$, depending on the classifier implemented. An anomaly detection algorithm is trained and tested on the time series data and color data as well, the combined algorithm has an expected misclassification rate of $\sim 0.07\%$. The effort represents the construction of a supervised classification algorithm.

### 5.4.1 Future Research

Further research is outlined in three main focus topics: dataset improvement, methodology improvement, simulation/performance analysis. The limited dataset and class space used for this study is known. Future efforts will include a more complete class space, as well as more data to support under-represented class types. Specifically datasets such as the Catalina Real Time Transient Survey [Drake et al., 2009], will provide greater depth and completeness as a prelude to the data sets that will be available from the Panoramic Survey Telescope & Rapid Response

System and the Large Synoptic Survey Telescope (LSST).

In addition to improving the underlying training data used, the methodology outline will also be researched to determine if more optimal methods are available. Exploring the effects of variable size state space for the translation could potentially yield performance improvements, as could a comparison of slotting methods (e.g. box slots vs. Gaussian slots vs. other kernels or weighting schemes). Likewise, implementations beyond supervised classification (e.g., unsupervised classification) were not explored as part of this analysis. How the feature space outlined in this analysis would lend itself to clustering or expectation-maximization algorithms is yet to be determined.

In future work, how sampling rates and photometric errors affect the ability to represent the underlying time-domain functionality using synthetic time-domain signals will be explored. Simulation of the expected time domain signals will allow for an estimation of performance of other spectral methods (DWT/DFT for irregular sampling), which will intern allow for and understanding of the benefits and drawbacks of each methodology, relative to both class type and observational conditions. This type of analysis would require the modeling and development of synthetic stellar variable functions to produce reasonable (and varied) time domain signature.

# Chapter 6

# Detector for O'Connell Type EBs

With the rise of large-scale surveys, such as Kepler, the Transiting Exoplanet Survey Satellite (TESS), the Kilodegree Extremely Little Telescope (KELT), the Square Kilometre Array, the Large Synoptic Survey Telescope (LSST), and Pan-STARRS, a fundamental working knowledge of statistical data analysis and data management to reasonably process astronomical data is necessary. The ability to mine these data sets for new and interesting astronomical information opens a number of scientific windows that were once closed by poor sampling, in terms of both number of stars (targets) and depth of observations (number of samples).

This section focuses on the development of a novel, modular time-domain signature extraction methodology and its supporting supervised pattern detection algorithm for variable star detection. The design could apply to any number of variable star types that exhibit consistent periodicity (cyclostationary) in their flux; examples include most Cepheid-type stars (RR Lyr, SX Phe, Gamma Dor, etc...) as well as other eclipsing binary types. Nonperiodic variables would require a different feature space [Johnston and Peter, 2017], but the underlying detection scheme could still be relevant. Herein we present the design's utility, by its targeting of eclipsing binaries that demonstrate a feature known as the O'Connell effect.

We have selected O'Connell effect eclipsing binaries (OEEBs) to demonstrate initially our detector design. We highlight OEEBs here because they compose a subclass of a specific type of variable star (eclipsing binaries). Subclass detection provides an extra layer of complexity for our detector to try to handle. We demonstrate our detector design on Kepler eclipsing binary data from the Villanova catalog, allowing us to train and test against different subclasses in the same parent variable class type. We train our detector design on Kepler eclipsing binary data and apply the detector to a different survey—the Lincoln Near-Earth Asteroid Research asteroid survey [LINEAR, Stokes et al., 2000]—to demonstrate the algorithm's ability to discriminate and detect our targeted subclass given not just the parent class but other classes as well.

Classifying variable stars relies on proper selection of feature spaces of interest and a classification framework that can support the linear separation of those features. Selected features should quantify the telltale signature of the variability—the structure and information content. Prior studies to develop both features and classifiers include expert selected feature efforts [Debosscher, 2009, Sesar et al., 2011, Richards et al., 2012, Graham et al., 2013a, Armstrong et al., 2016, Mahabal et al., 2017, Hinners et al., 2018], automated feature selection efforts [McWhirter et al., 2017, Naul et al., 2018], and unsupervised methods for feature extraction [Valenzuela and Pichara, 2018, Modak et al., 2018]. The astroinformatics community-standard features include quantification of statistics associated with the time-domain photometric data, Fourier decomposition of the data, and color information in both the optical and IR domains [Nun et al., 2015, Miller et al., 2015]. The number of individual features commonly used is upward of 60 and growing [Richards et al., 2011] as the number of variable star types increases, and as a result of further refinement of classification definitions [Samus' et al., 2017]. We seek here to develop a novel feature space that captures the signature of interest for the targeted variable star type.

The detection framework here maps time-domain stellar variable observations to an alternate distribution field (DF) representation [Sevilla-Lara and Learned-Miller, 2012] and then develops a metric learning approach to identify OEEBs. Based on the matrix-valued DF feature, we adopt a metric learning framework to directly learn a distance metric [Bellet et al., 2015] on the space of DFs. We can then utilize the learned metric as a measure of similarity to detect new OEEBs based on their closeness to other OEEBs. We present our metric learning approach as a competitive push–pull optimization, where DFs corresponding to known OEEBs influence the learned metric to measure them as being nearer in the DF space. Simultaneously, DFs corresponding to non-OEEBs are pushed away and result in large measured distances under the learned metric.

This section is structured as follows. First, we review the targeted stellar variable type, discussing the type signatures expected. Second, we review the data used in our training, testing, and discovery process as part of our demonstration of design. Next, we outline the novel proposed pipeline for OEEB detection; this review includes the feature space used, the designed detector/classifier, and the associated implementation of an anomaly detection algorithm [Chandola et al., 2009]. Then, we apply the algorithm, trained on the expertly selected/labeled Villanova Eclipsing Binary catalog OEEB targets, to the rest of the catalog with the purpose of identifying new OEEB stars. We present the results of the discovery process using a mix of clustering and derived statistics. We apply the Villanova Eclipsing Binary catalog trained classifier, without additional training, to the LINEAR data set. We provide results of this cross-application, i.e., the set of discovered OEEBs. For comparison, we detail two competing approaches. We develop training and testing strategies for our metric learning framework, and finally, we conclude with a summary of our findings and directions for future research.

## 6.1 Eclipsing Binaries with OConnell Effect

The O'Connell effect [O'Connell, 1951] is defined for eclipsing binaries as an asymmetry in the maxima of the phased light curve (see Figure 6.2). This maxima asymmetry is unexpected, as it suggests an orientation dependency in the brightness of the system.

Similarly, the consistency of the asymmetric over many orbits is also surprising, as it suggests that the maxima asymmetry has a dependence on the rotation of the binary system. While the cause of the O'Connell effect is not fully understood, researchers have offered a number of explanations. Additional data and modeling are necessary for further investigation [McCartney, 1999].

### 6.1.1 Signatures and Theories

Several theories propose to explain the effect, including starspots, gas stream impact, and circumstellar matter [McCartney, 1999]. The work by [Wilsey and Beaky, 2009] outlines each of these theories and demonstrates how the observed effects are generated by the underlying physics.

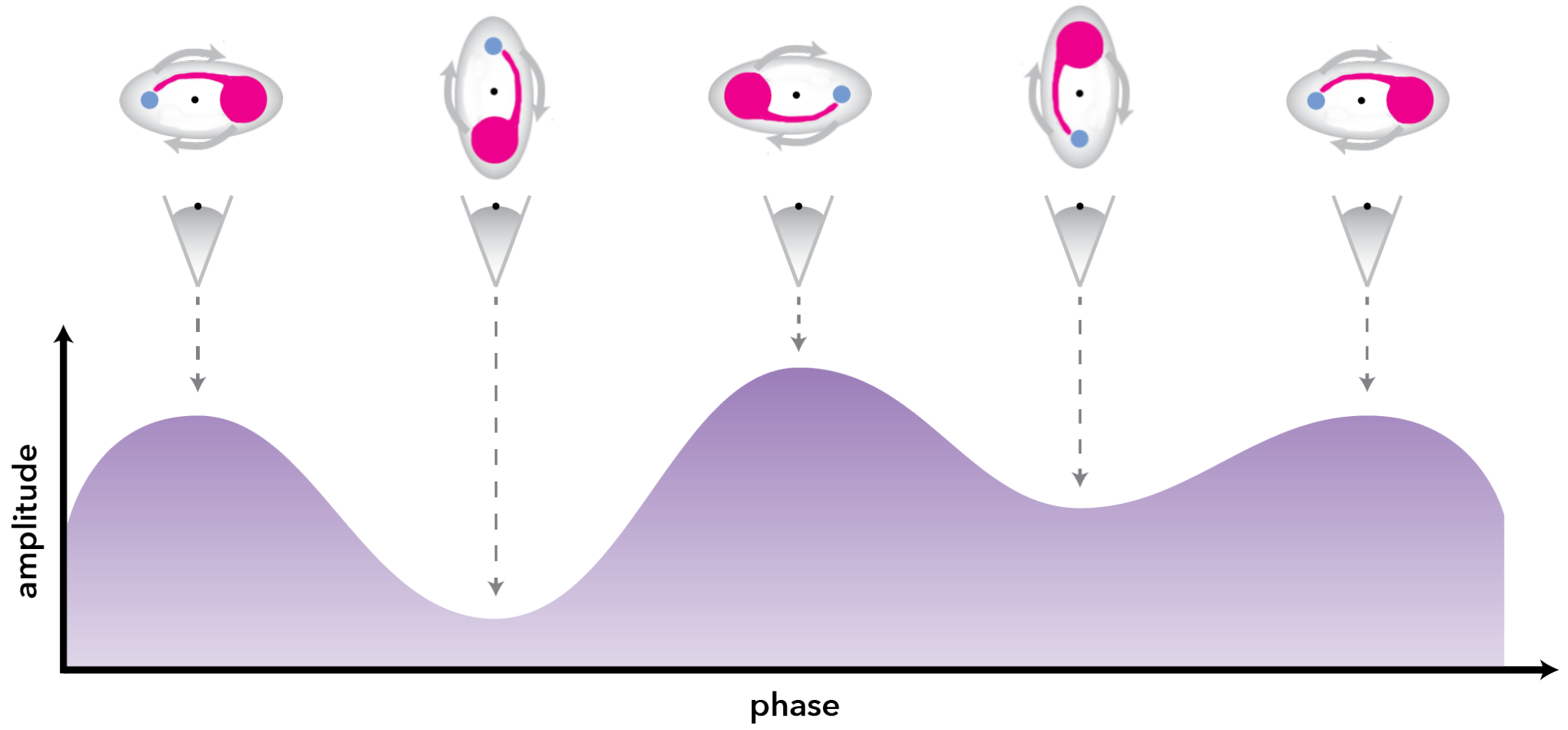

Figure 6.1: Conceptual overview of a $\beta$ Lyrae—EB type—eclipsing binary (Semi-Detached Binary) that has the O'Connell Effect, an asymmetry of the maxima. The figure is a representation of the binary orientation with respect to a viewer and the resulting observed light curve.

- Starspots result from chromospheric activity, causing a consistent decrease in brightness of the star when viewed as a point source. While magnetic surface activity will cause both flares (brightening) and spots (darkening), flares tend to be transient, whereas spots tend to have longer-term effects on the observed binary flux. Thus, between the two, starspots are the favored hypothesis for causing long-term consistent asymmetry; often binary simulations (such as the Wilson–Devinney code) can be used to model O'Connell effect binaries via including an often large starspot [Zboril and Djurasevic, 2006].

- Gas stream impact results from matter transferring between stars (smaller to larger) through the L1 point and onto a specific position on the larger star, resulting in a consistent brightening on the leading/trailing side of the secondary/primary.

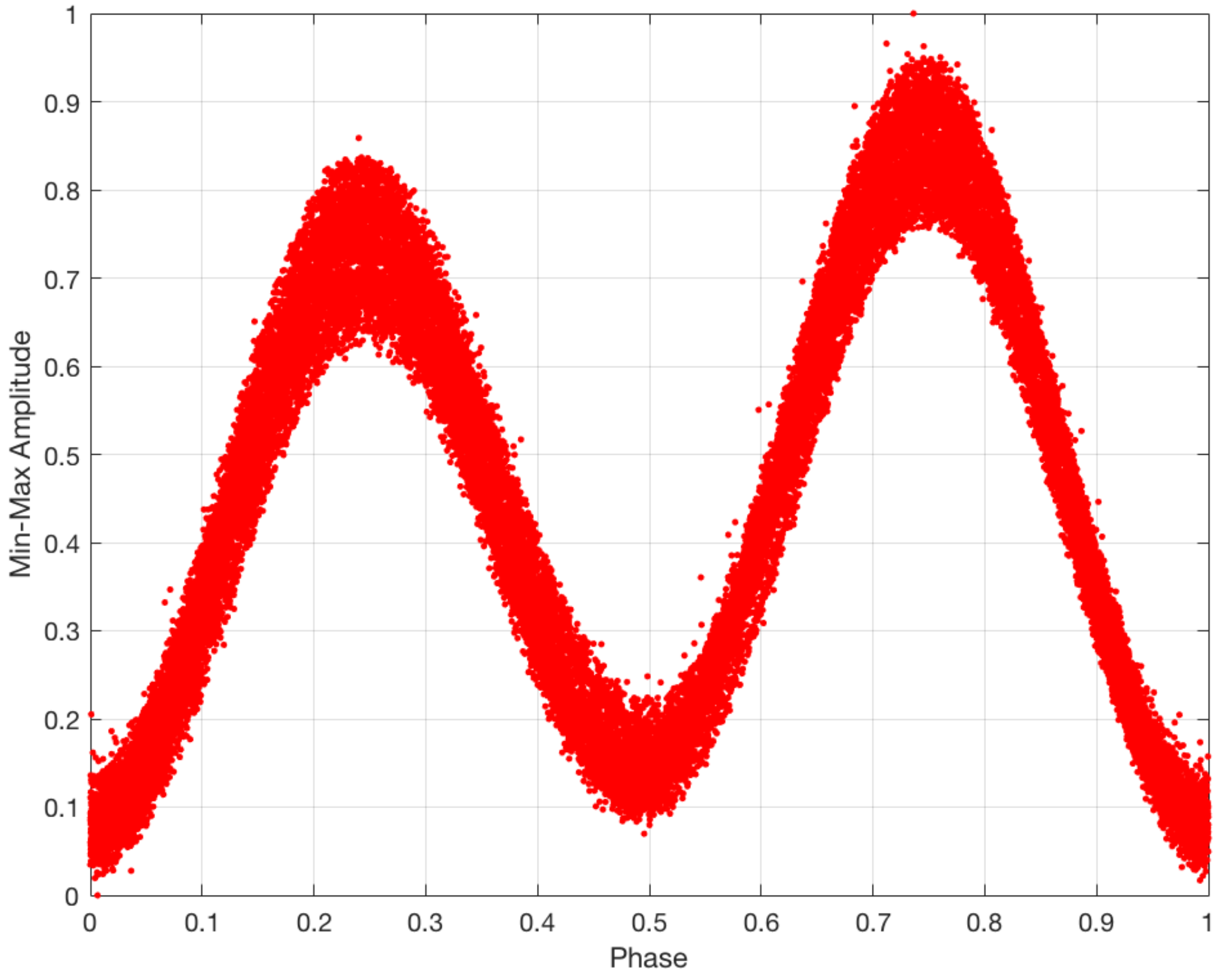

Figure 6.2: An example phased light curve of an eclipsing binary with the O'Connell effect (KIC: 10861842). The light curve has been phased such that the global minimum (cooler in front of hotter) is at lag 0 and the secondary minimum (hotter in front of cooler) is at approximately lag 0.5. The side-by-side binary orientations are at approximately 0.25 and 0.75. Note that the maxima, corresponding to the side-by-side orientations, have different values.

- The circumstellar matter theory proposes to describe the increase in brightness via free-falling matter being swept up, resulting in energy loss and heating, again causing an increase in amplitude. Alternatively, circumstellar matter in orbit could result in attenuation, i.e., the difference in maximum magnitude of the phased light curve results from dimming and not brightening.

In the study McCartney [1999], the authors limited the sample to only six star systems: GSC 03751-00178, V573 Lyrae, V1038 Herculis, ZZ Pegasus, V1901 Cygni, and UV Monocerotis. Researchers have used standard eclipsing binary simulations [Wilson and Devinney, 1971] to demonstrate the proposed explanations for each light curve instance and estimate the parameters associated with the physics of the system. [Wilsey and Beaky, 2009] noted other cases of the O'Connell effect in binaries, which have since been described physically; in some cases, the effect varied over time, whereas in other cases, the effect was consistent over years of observation and over many orbits. The effect has been found in both overcontact, semidetached, and near-contact systems.

While one of the key visual differentiators of the O'Connell effect is $\Delta m_{\max}$, this alone could not be used as a means for detection, as the targets trained on or applied to are not guaranteed to be (a) eclipsing binaries and (b) periodic. One of the goals we are attempting to highlight is the transformation of expert qualitative target selection into quantitative machine learning methods.

### 6.1.2 Characterization of OEEB

We develop a detection methodology for a specific target of interest—OEEB—defined as an eclipsing binary where the light curve (LC) maxima are consistently at different amplitudes over the span of observation. Beyond differences in maxima, and a number of published examples, little is defined as a requirement for identifying the O'Connell effect [Wilsey and Beaky, 2009, Knote, 2015].

McCartney [1999] provide some basic indicators/measurements of interest in relation to OEEB binaries: the O'Connell effect ratio (OER), the difference in maximum amplitudes ($\Delta m$), the difference in minimum amplitudes, and the light curve asymmetry (LCA). The metrics are based on the smoothed phased light curves. The OER is calculated as Equation 6.1:

$$\mathrm{OER} = \frac{\sum_{i=1}^{n/2} (I_i - I_1)}{\sum_{i=n/2+1}^{n} (I_i - I_1)}, \tag{6.1}$$

where the min-max amplitude measurements for each star are grouped into phase bins ($n = 500$), where the mean amplitude in each bin is $I_i$. An $OER > 1$ corresponds to the front half of the light curve having more total flux; note that for the procedure we present here, $I_1 = 0$. The difference in max amplitude is calculated as Equation 6.2:

$$\Delta m = \max_{t<0.5} \left(f(t)_N\right) - \max_{t\geq 0.5} \left(f(t)_N\right), \tag{6.2}$$

where we have estimated the maximum in each half of the phased light curve. The LCA is calculated as Equation 6.3:

$$\mathrm{LCA} = \sqrt{\sum_{i=1}^{n/2} \frac{\left(I_i - I_{(n+1-i)}\right)^2}{I_i^2}}. \tag{6.3}$$

As opposed to the measurement of OER, LCA measures the deviance from symmetry of the two peaks. Defining descriptive metrics or functional relationships (i.e., bounds of distribution) requires a larger sample than is presently available. An increased number of identified targets of interest is required to provide the sample size needed for a complete statistical description of the O'Connell effect. The quantification of these functional statistics allows for the improved understanding of not just the standard definition of the targeted variable star but also the population distribution as a whole. These estimates allow for empirical statements to be made regarding the differences in light curve shapes among the variable star types investigated. The determination of an empirically observed distribution, however, requires a significant sample to generate meaningful descriptive statistics for the various metrics.

In this effort, we highlight the functional shape of the phased light curve as our defining feature of OEEB stars. The prior metrics identified are selected or reduced measures of this functional shape. We propose here that, as opposed to training a detector on the preceding indicators, we use the functional shape of the phased light curve by way of the distribution field to construct our automated system.

## 6.2 Variable Star Data

As a demonstration of design, we apply the proposed algorithm to a set of predefined, expertly labeled eclipsing binary light curves. We focus on two surveys of interest: first, the Kepler Villanova Eclipsing Binary catalog, from which we derive our initial training data as well as our initial discovery (unlabeled) data, and second, the Lincoln Near-Earth Asteroid Research, which we treat as unlabeled data.

### 6.2.1 Kepler Villanova Eclipsing Catalog

Leveraging the Kepler pipeline already in place, and using the data from the Villanova Eclipsing Binary catalog [Kirk et al., 2016], this study focuses on a set of predetermined eclipsing binaries identified from the Kepler catalog. From this catalog, we developed an initial, expertly derived, labeled data set of proposed targets of interest identified as OEEB. Likewise, we generated a set of targets identified as “not interesting” based on our expert definitions, i.e., intuitive inference.

Using the Eclipsing Binary catalog [Kirk et al., 2016], we identified a set of 30 targets of interest and 121 targets of noninterest via expert analysis—by-eye selection based on researchers’ interests. Specific target identification is listed in a supplementary digital file at the project repository.[1] We use this set of 151 light curves for training and testing.

---

[1] ./supplement/KeplerTraining.xlsx

#### 6.2.1.1 Light Curve/Feature Space

Prior to feature space processing, the raw observed photometric time domain data are conditioned and processed. Operations include long-term trend removal, artifact removal, initial light curve phasing, and initial eclipsing binary identification; we performed these actions prior to the effort demonstrated here, by the Eclipsing Binary catalog (our work uses all 2875 long-cadence light curves available as of the date of publication). The functional shape of the phased light curve is selected as the feature to be used in the machine learning process, i.e., detection of targets of interest. While the data have been conditioned already by the Kepler pipeline, added steps are taken to allow for similarity estimation between phased curves. Friedman's SUPERSMOOTHER algorithm [Friedman, 1984, VanderPlas and Ivezić, 2015] is used to generate a smooth 1-D functional curve from the phased light curve data. The smoothed curves are normalized via the min-max scaling equation 6.4:

$$f(\phi)_N = \frac{f(\phi) - \min(f(\phi))}{\max\left(f(\phi)\right) - \min(f(\phi))}, \tag{6.4}$$

where $f(\phi)$ is the smoothed phased light curve, $f$ is the amplitude from the database source, $\phi$ is the phase where $\phi \in [0, 1]$, and $f(\phi)_N$ is the min-max scaled amplitude. (Note that we will use the terms $f(\phi)_N$ and min-max amplitude interchangeably throughout this article.) We use the minimum of the smoothed phased light curve as a registration marker, and both the smoothed and unsmoothed light curves are aligned such that lag/phase zero corresponds to minimum amplitude (eclipse minima; see [McCartney, 1999]).

#### 6.2.1.2 Training/Testing Data

The labeled training data are provided as part of the supplementary digital project repository. We include the SOI and NON-SOI Kepler identifiers here (KIC).

Table 6.1: Collection of KIC of Interest (30 Total)

| | | | | | | | |
|---|---|---|---|---|---|---|---|
| 10123627 | 11924311 | 5123176 | 8696327 | 11410485 | 7696778 | 7516345 | 9654476 |
| 10815379 | 2449084 | 5282464 | 8822555 | 7259917 | 6223646 | 4350454 | 9777987 |
| 10861842 | 2858322 | 5283839 | 9164694 | 7433513 | 9717924 | 5820209 | 7584739 |
| 11127048 | 4241946 | 5357682 | 9290838 | 8394040 | 7199183 | | |

Table 6.2: Collection of KIC Not of Interest (121 Total)

| | | | | | | | |
|---|---|---|---|---|---|---|---|
| 10123627 | 11924311 | 5123176 | 8696327 | 11410485 | 10007533 | 10544976 | 11404698 |
| 10024144 | 10711646 | 11442348 | 12470530 | 3954227 | 10084115 | 10736223 | 11444780 |
| 10095469 | 10794878 | 11652545 | 2570289 | 4037163 | 10216186 | 10802917 | 12004834 |
| 10253421 | 10880490 | 12108333 | 3127873 | 4168013 | 10257903 | 10920314 | 12109845 |
| 10275747 | 11076176 | 12157987 | 3344427 | 4544587 | 10383620 | 11230837 | 12216706 |
| 10485137 | 11395117 | 12218858 | 3730067 | 4651526 | 9007918 | 8196180 | 7367833 |
| 9151972 | 8248812 | 7376500 | 6191574 | 4672934 | 9179806 | 8285349 | 7506446 |
| 9205993 | 8294484 | 7518816 | 6283224 | 4999357 | 9366988 | 8298344 | 7671594 |
| 9394601 | 8314879 | 7707742 | 6387887 | 5307780 | 9532219 | 8481574 | 7879399 |
| 9639491 | 8608490 | 7950964 | 6431545 | 5535061 | 9700154 | 8690104 | 8074045 |
| 9713664 | 8758161 | 8087799 | 6633929 | 5606644 | 9715925 | 8804824 | 8097553 |
| 9784230 | 8846978 | 8155368 | 7284688 | 5785551 | 9837083 | 8949316 | 8166095 |
| 9935311 | 8957887 | 8182360 | 7339345 | 5956776 | 9953894 | 3339563 | 4474193 |
| 12400729 | 3832382 | 12553806 | 4036687 | 2996347 | 4077442 | 3557421 | 4554004 |
| 6024572 | 4660997 | 6213131 | 4937217 | 6370361 | 5296877 | 6390205 | 5374999 |
| 6467389 | 5560831 | 7119757 | 5685072 | 7335517 | 5881838 | | |

### 6.2.2 Data Untrained

The 2,000+ eclipsing binaries left in the Kepler Eclipsing Binary catalog are left as unlabeled targets. We use our described detector to "discover" targets of interest, i.e., OEEB. The full set of Kepler data is accessible via the Villanova Eclipsing Binary website (http://keplerebs.villanova.edu/).

For analyzing the proposed algorithm design, the LINEAR data set is also leveraged as an unknown "unlabeled" data set ripe for OEEB discovery [Sesar et al., 2011, Palaversa et al., 2013]. From the starting sample of 7,194 LINEAR variables, we used a clean sample of 6,146 time series data sets for detection. Stellar class type is limited further to the top five most populous classes—RR Lyr (ab), RR Lyr (c), Delta Scuti / SX Phe, Contact Binaries, and Algol-Like Stars with two minima—resulting in a set of 5,086 observations.

Unlike the Kepler Eclipsing Binary catalog, the LINEAR data set contains targets other than (but does include) eclipsing binaries; the data set we used [Johnston and Peter, 2017] includes Algols (287), Contact Binaries (1805), Delta Scuti (68), and RR Lyr (ab-2189, c-737). The light curves are much more poorly sampled; this uncertainty in the functional shape results from lower SNR (ground survey) and poor sampling. The distribution of stellar classes is presented in Table 5.1. The full data sets used at the time of this publication from the Kepler and LINEAR surveys are available from the associated public repository.[2]

## 6.3 PMML Classification Algorithm

Relying on previous designs in astroinformatics to develop a supervised detection algorithm [Johnston and Oluseyi, 2017], we propose a design that tailors the requirements specifically toward detecting OEEB-type variable stars.

---

[2]github.com/kjohnston82/OCDetector

### 6.3.1 Prior Research

Many prior studies on time-domain variable star classification [Debosscher, 2009, Barclay et al., 2011, Blomme et al., 2011, Dubath et al., 2011, Pichara et al., 2012, Pichara and Protopapas, 2013, Graham et al., 2013a, Angeloni et al., 2014, Masci et al., 2014] rely on periodicity domain feature space reductions. Debosscher [2009] and Templeton [2004] review a number of feature spaces and a number of efforts to reduce the time-domain data, most of which implement Fourier techniques, primarily the Lomb–Scargle (L-S) method [Lomb, 1976, Scargle, 1982], to estimate the primary periodicity [Eyer and Blake, 2005, Deb and Singh, 2009, Richards et al., 2012, Park and Cho, 2013, Ngeow et al., 2013].

The studies on classification of time-domain variable stars often further reduce the folded time-domain data into features that provide maximal-linear separability of classes. These efforts include expert selected feature efforts [Debosscher, 2009, Sesar et al., 2011, Richards et al., 2012, Graham et al., 2013a, Armstrong et al., 2016, Mahabal et al., 2017, Hinners et al., 2018], automated feature selection efforts [McWhirter et al., 2017, Naul et al., 2018], and unsupervised methods for feature extraction [Valenzuela and Pichara, 2018, Modak et al., 2018]. The astroinformatics community-standard features include quantification of statistics associated with the time-domain photometric data, Fourier decomposition of the data, and color information in both the optical and IR domains [Nun et al., 2015, Miller et al., 2015]. The number of individual features commonly used is upward of 60 and growing [Richards et al., 2011] as the number of variable star types increases and as a result of further refinement of classification definitions [Samus' et al., 2017]. Curiously, aside from efforts to construct a classification algorithm from the time-domain data directly [McWhirter et al., 2017], few efforts in astroinformat-ics have looked at features beyond those described here—mostly Fourier domain transformations or time domain statistics. Considering the depth of possibility for time-domain transformations [Fu, 2011, Grabocka et al., 2012, Cassisi et al., 2012, Fulcher et al., 2013], it is surprising that the community has focused on just a few transforms. Similarly, the astroinformatics community has focused on only a few classifiers, limited mostly to standard classifiers and, specifically, decision tree algorithms, such as random forest–type classifiers.

Here we propose an implementation that simplifies the traditional design: limiting ourselves to a one versus all approach [Johnston and Oluseyi, 2017] targeting a variable type of interest; limiting ourselves to a singular feature space—the distribution field of the phased light curve—based on Helfer et al. [2015] as a representation of the functional shape; and introducing a classification/detection scheme that is based on similarity with transparent results [Bellet et al., 2015] that can be further extended, allowing for the inclusion of an anomaly detection algorithm.

### 6.3.2 Distribution Field

As stated, this analysis focuses on detecting OEEB systems based on their light curve shape. The OEEB signature has a cyclostationary signal, a functional shape that repeats with a consistent frequency. The signature can be isolated using a process of period finding, folding, and phasing [Graham et al., 2013b]; the Villanova catalog provides the estimated "best period." The proposed feature space transformation will focus on the quantification or representation of this phased functional shape. This particular implementation design makes the most intuitive sense, as visual inspection of the phased light curve is the way experts identify these unique sources.

As discussed, prior research on time-domain data identification has varied between generating machine-learned features [Gagniuc, 2017], implementing generic features [Masci et al., 2014, Palaversa et al., 2013, Richards et al., 2012, Debosscher, 2009], and looking at shape- or functional-based features [Haber et al., 2015, Johnston and Peter, 2017, Park and Cho, 2013]. This analysis will leverage the distribution field transform to generate a feature space that can be operated on; a distribution field (DF) is defined as [Helfer et al., 2015, Sevilla-Lara and Learned-Miller, 2012] Equation 6.5:

$$\mathrm{DF}_{ij} = \frac{\sum \left[y_j < f\left(x_i \leq \phi \leq x_{i+1}\right)_N < y_{j-1}\right]}{\sum \left[y_j < f\left(\phi\right)_N < y_{j-1}\right]}, \tag{6.5}$$

where [ ] is the Iverson bracket [Iverson, 1962], given as

$$[P] = \begin{cases} 1 & P = \text{true} \\ 0 & \text{otherwise,} \end{cases} \tag{6.6}$$

and $y_j$ and $x_i$ are the corresponding normalized amplitude and phase bins, respectively, where $x_i = 0, 1/n_x, 2/n_x, \ldots, 1$, $y_i = 0, 1/n_y, 2/n_y, \ldots, 1$, $n_x$ is the number of time bins, and $n_y$ is the number of amplitude bins. The result is a right stochastic matrix, i.e., the rows sum to 1. Bin number, $n_x$ and $n_y$, is optimized by cross-validation as part of the classification training process. Smoothed phased data —generated from SUPERSMOOTHER—are provided to the DF algorithm.

We found this implementation to produce a more consistent classification process. We found that the min-max scaling normalization, when outliers are present, can produce final patterns that focus more on the outlier than the general functionality of the light curve. Likewise, we found that using the unsmoothed data in the DF algorithm resulted in a classification that was too dependent on the scatter of the phased light curve. Although at first glance, that would not appear to be an issue, this implementation resulted in light curve resolution having a large impact on the classification performance—in fact, a higher impact than the shape itself. An example of this transformation is given in Figure 6.3.

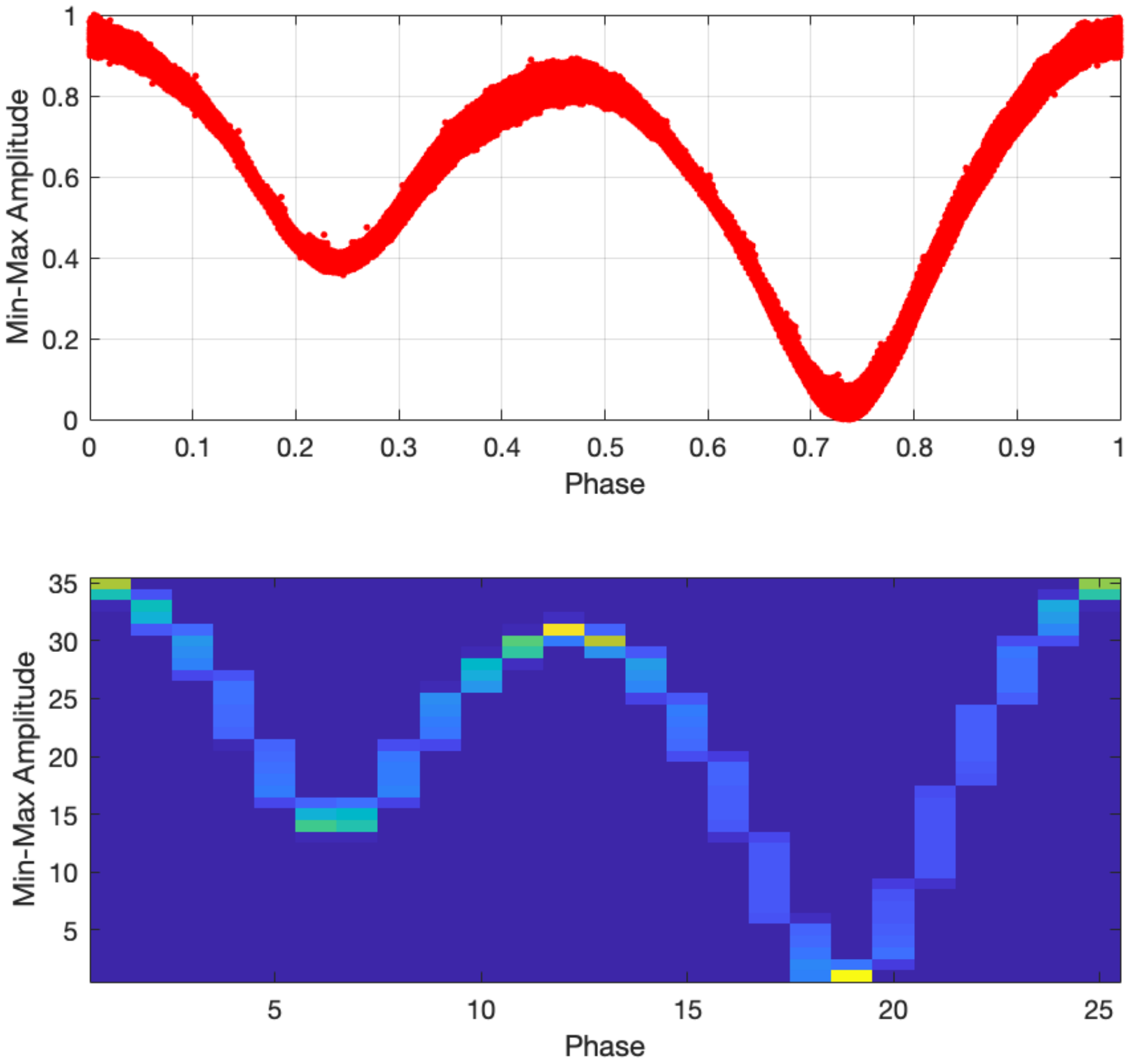

Figure 6.3: An example phased light curve (top) and the transformed distribution field (bottom) of an Eclipsing Binary with the O'Connell effect (KIC: 7516345).

Though the DF exhibits properties that a detection algorithm can use to identify specific variable stars of interest, it alone is not sufficient for our ultimate goal of automated detection. Rather than vectorizing the DF matrix and treating it as a feature vector for standard classification techniques, we treat the DF as the matrix-valued feature that it is [Helfer et al., 2015]. This allows for the retention of row and column dependence information that would normally be lost in the vectorization process [Ding and Dennis Cook, 2018].

### 6.3.3 Metric Learning

At its core, the proposed detector is based on the definition of similarity and, more formally, a definition of distance. Consider the example triplet "$x$ is more similar to $y$ than to $z$," i.e., the distance between $x$ and $y$ in the feature space of interest is smaller than the distance between $x$ and $z$. The field of metric learning focuses on defining this distance in a given feature space to optimize a given goal, most commonly the reduction of error rate associated with the classification process. Given the selected feature space of DF matrices, the distance between two matrices $X$ and $Y$ [Bellet et al., 2015, Helfer et al., 2015] is defined as Equation 6.7:

$$d(X,Y) = \|X - Y\|_M^2 = tr\left\{(X - Y)^T M\,(X - Y)\right\}. \tag{6.7}$$

$M$ is the metric that we will be attempting to optimize, where $M \succeq 0$ (positive semi-definite). The PMML procedure outlined in [Helfer et al., 2015] is similar to the metric learning methodology LMNN [Weinberger et al., 2009], save for its implementation on matrix-variate data as opposed to vector-variate data. We summarize it here. The developed objective function is given in Equation 6.8:

$$E = \frac{1-\lambda}{N_c - 1} \sum_{i,j} \left\| \mathrm{DF}_c^i - \mathrm{DF}_c^j \right\|_M^2 - \frac{\lambda}{N - N_c} \sum_{i,k} \left\| \mathrm{DF}_c^i - \mathrm{DF}_c^k \right\|_M^2 + \frac{\gamma}{2} \left\| M \right\|_F^2 , \quad (6.8)$$

where $N_c$ is the number of training data in class $c$; $\lambda$ and $\gamma$ are variables to control the importance of push versus pull and regularization, respectively; and the triplet $\left\{\mathrm{DF}_c^i, \mathrm{DF}_c^j, \mathrm{DF}_c^k\right\}$ defines the relationship between similar and dissimilar observations, i.e., $\mathrm{DF}_c^i$ is similar to $\mathrm{DF}_c^j$ and dissimilar to $\mathrm{DF}_c^k$, as per the definitions outlined in Bellet et al. [2015]. Clearly there are three basic components: a pull term, which is small when the distance between similar observations is small; a push term, which is small when the distance between dissimilar observations is larger; and a regularization term, which is small when the Frobenius norm ($\left\| M \right\|_F^2 = \sqrt{Tr(MM^H)}$) of $M$ is small. Thus the algorithm attempts to bring similar distribution fields closer together, while pushing dissimilar ones farther apart, while attempting to minimize the complexity of the metric $M$. The regularizer on the metric $M$ guards against overfitting and consequently enhances the algorithm's ability to generalize, i.e., allow for operations across data sets.This regularization strategy is similar to popular regression techniques like lasso and ridge [Hastie et al., 2009].

Additional parameters $\lambda$ and $\gamma$ weight the importance of the push–pull terms and metric regularizer, respectively. These free parameters are typically tuned via standard cross-validation techniques on the training data. The objective function represented by Equation 6.8 is quadratic in the unknown metric $M$; hence it is possible to obtain the following closed-form solution to the minimization of the

Equation 6.8 objective function as:

$$M = \frac{\lambda}{\gamma\left(N - N_c\right)} \sum_{i,k} \left(\mathrm{DF}_c^i - \mathrm{DF}_c^k\right)\left(\mathrm{DF}_c^i - \mathrm{DF}_c^k\right)^T - \frac{1-\lambda}{\gamma\left(N_c - 1\right)} \sum_{i,k} \left(\mathrm{DF}_c^i - \mathrm{DF}_c^j\right)\left(\mathrm{DF}_c^i - \mathrm{DF}_c^j\right)^T . \quad (6.9)$$

Equation 6.9 does not guarantee that $M$ is positive semi-definite (PSD). To ensure this property, we can apply the following straightforward projection step after calculating $M$ to ensure the requirement of $M \succeq 0$:

1. perform eigen decomposition: $M = U^T \Lambda U$;
2. generate $\Lambda_+ = \max(0, \Lambda)$, i.e., select positive eigenvalues;
3. reconstruct the metric $M$: $M = U^T \Lambda_+ U$.

This projected metric is used in the classification algorithm. The metric learned from this push–pull methodology is used in conjunction with a standard k-nearest neighbor (k-NN) classifier.

### 6.3.4 k-NN Classifier

The traditional k-NN algorithm is a nonparametric classification method; it uses a voting scheme based on an initial training set to determine the estimated label [Altman, 1992]. For a given new observation, the $L_2$ Euclidean distance is found between the new observation and all points in the training set. The distances are sorted, and the $k$ closest training sample labels are used to determine the new observed sample estimated label (majority rule). Cross-validation is used to find an optimal $k$ value, where $k$ is any integer greater than zero.

The k-NN algorithm estimates a classification label based on the closest samples provided in training. For our implementation, the distance between a new pattern $\mathrm{DF}^i$ and each pattern in the training set is found using the optimized metric as opposed to the identity metric that would been used in the $L_2$ Euclidean distance case. The new pattern is classified depending on the majority of the closest $k$ class labels. The distance between patterns is in Equation 6.7, using the learned metric $M$.

## 6.4 Results of the Classification

The new OEEB systems discovered by the method of automated detection proposed here can be used to further investigate their frequency of occurrence, provide constraints on existing light curve models, and provide parameters to look for these systems in future large-scale variability surveys like LSST.

### 6.4.1 Kepler Trained Data

The algorithm implements fivefold cross-validation [Duda et al., 2012]; the algorithmic details associated with the cross-validation process are beyond the scope of this article, but in short, (1) the algorithm splits each class in the labeled data in half, with one half used in training and the other in testing; (2) the training data are further subdivided into five partitions; and (3) as the algorithm is trained, these partitions are used to generate a training set using four of the five partitions and a cross-validation set using the fifth.

The minimization of misclassification rate is used to optimize floating parameters in the design, such as the number of $x$-bins, the number of $y$-bins, and k-values. Some parameters are more sensitive than others; often this insensitivity is related to the loss function or the feature space, or the data themselves. For example, the $\gamma$ and $\lambda$ values weakly affected the optimization, while the bin sizes and k-values had a stronger effect.

The cross-validation process was then reduced to optimizing the $n_x$, $n_y$, and $k$ values; $n_x$ and $n_y$ values were tested over values 20–40 in steps of 5. The set of optimized parameters is given as $\gamma$ =1.0, $\lambda = 0.75$, $n_x = 25$, $n_y = 35$, and $k = 3$. Given the optimization of these floating variables in all three algorithms, the testing data are then applied to the optimal designs.

### 6.4.2 Kepler Untrained Data

The algorithm is applied to the Villanova Eclipsing Binary catalog entries that were not identified as either "Of Interest" or "Not of Interest," i.e., unlabeled for the purposes of our targeted goal. The trained and tested data sets are combined into a single training set for application; the primary method (push–pull metric classification) is used to optimize a metric based on the optimal parameters found during cross-validation and to apply the system to the entire Villanova Eclipsing Binary data (2875 curves).

#### 6.4.2.1 Design Considerations

On the basis of the results demonstrated in [Johnston and Oluseyi, 2017], the algorithm additionally conditions the detection process based on a maximal distance allowed between a new unlabeled point and the training data set in the feature space of interest.

This anomaly detection algorithm is based on the optimized metric; a maximum distance between data points is based on the training data set, and we use a fraction (0.75) of that maximum distance as a limit to determine “known” versus “unknown.” The value of the fraction was initially determined via trial and error, based on our experiences with the data set and the goal of minimizing false alarms (which were visually apparent). This further restricts the algorithm to classifying those targets that exist only in “known space.” The k-NN algorithm generates a distance dependent on the optimized metric; by restricting the distances allowed, we can leverage the algorithm to generate the equivalent of an anomaly detection algorithm.

The resulting paired algorithm (detector + distance limit) will produce estimates of “interesting” versus “not interesting,” given new—unlabeled—data. Our algorithm currently will not produce confidence estimates associated with the label. Confidence associated with detection can be a touchy subject, both for the scientists developing the tools and for the scientists using them. Here we have focused on implementing a k-NN algorithm with optimized metric (i.e., metric learning); posterior probabilities of classification can be estimated based on k-NN output [Duda et al., 2012] and can be found as $(k_c/(n * \text{volume}))$; linking these posterior probability estimates to “how confident am I that this is what I think this is” is not often the best choice of description.

Confidence in our detections will be a function of the original PPML classification algorithm performance, the training set used and the confidence in the labeling process, and the anomaly detection algorithm we implemented. Even $(k_c/(n * \text{volume}))$ would not be a completely accurate description in our scenario.

Some researchers [Dalitz, 2009] have worked on linking "confidence" in k-NN classifiers with distance between the points. Our introduction of an anomaly detection algorithm into the design thus allows a developer/user the ability to limit the false alarm rate by introducing a maximum acceptable distance thus allowing for some control of confidence in our classification results; see [Johnston and Oluseyi, 2017] for more information.

#### 6.4.2.2 Results

Once we remove the discovered targets that were also in the initial training data, the result is a conservative selection of 124 potential targets of interest listed in a supplementary digital file at the project repository.[3] We here present an initial exploratory data analysis performed on the phased light curve data. At a high level, the mean and standard deviation of the discovered curves are presented in Figure 6.4.

A more in-depth analysis as to the meaning of the distribution functional shapes is left for future study. Such an effort would include additional observations (spectroscopic and photometric additions would be helpful) as well as analysis using binary simulator code such as Wilson–Devinney [Prša and Zwitter, 2005]. It is noted that in general, there are some morphological consistencies across the discovered targets:

1. In the majority of the discovered OEEB systems, the first maximum following the primary eclipse is greater than the second maximum following the secondary eclipse.

[3] ./supplement/AnalysisOfClusters.xlsx

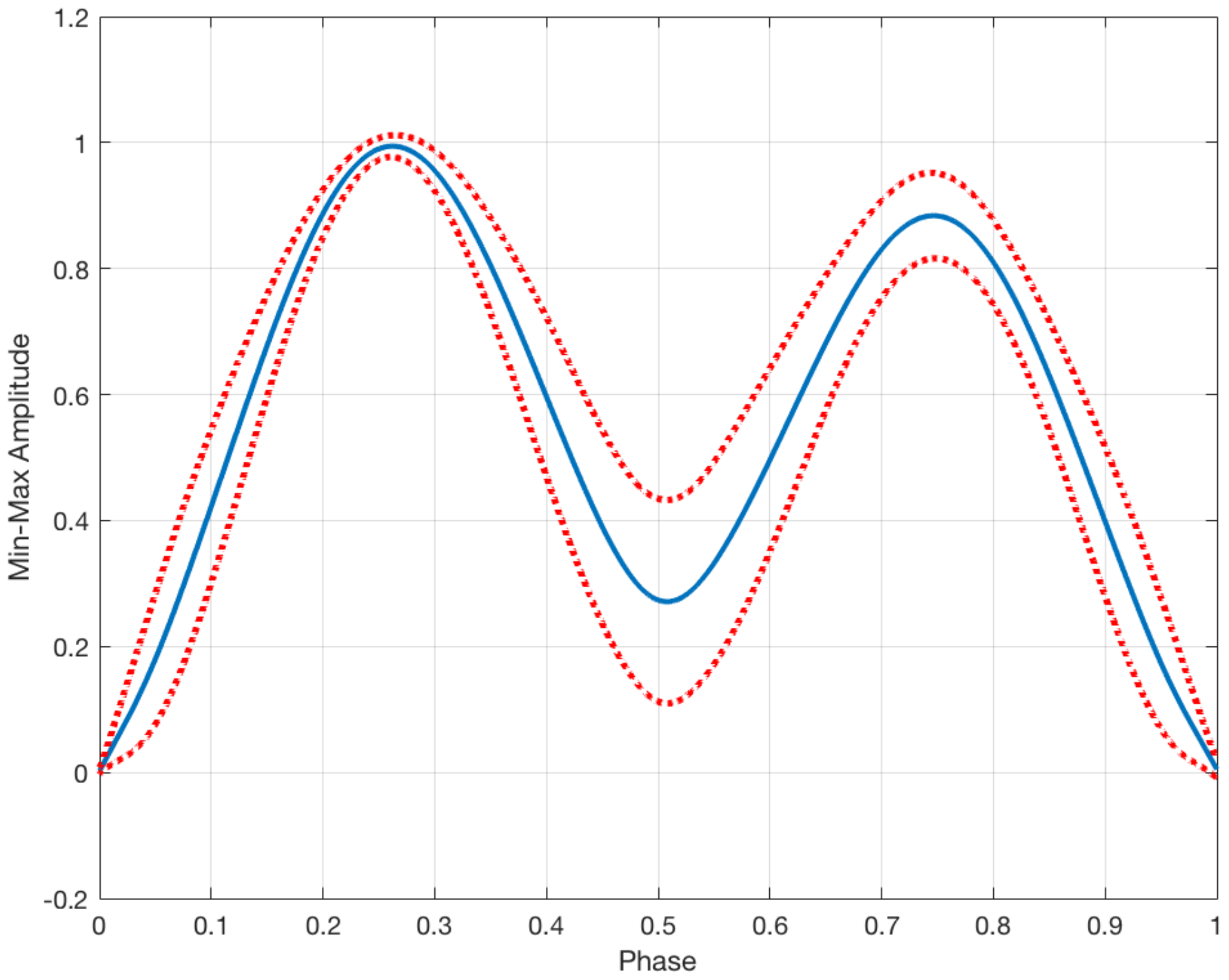

Figure 6.4: The mean (solid) and a $1-\sigma$ standard deviation (dashed) of the distribution of O'Connell effect Eclipsing Binary phased light curves discovered via the proposed detector out of the Kepler Eclipsing Binary catalog.

2. The light curve relative functional shape from the primary eclipse (minima) to primary maxima is fairly consistent across all discovered systems.

3. The difference in relative amplitude between the two maxima does not appear to be consistent, nor is the difference in relative amplitude between the minima.

We perform additional exploratory data analysis on the discovered group via subgrouping partitioning with unsupervised clustering. The k-means clustering algorithm with matrix-variate distances presented as part of the comparative methodologies is applied to the discovered data set (their DF feature space). This clustering is presented to provide more detail on the discovered group morphological shapes. The associated 1-D curve generated by the SUPERSMOOTHER algorithm is presented with respect to their respective clusters (clusters 1–8) in Figure 6.5.

The clusters generated were initialized with random starts, thus additional iterations can potentially result in different groupings. The calculated metric values and the clusters numbers for each star are presented in the supplementary digital file. A plot of the measured metrics as well as estimated values of period and temperature (as reported by the Villanova Kepler Eclipsing Binary database), are given with respect to the cluster assigned by k-means.[4] Following figure 4.6 in [McCartney, 1999], plot of OER versus Δm is isolated and presented in Figure 6.6.

#### 6.4.2.3 Subgroup Analysis

The linear relationship between OER and $\Delta m$ reported in [McCartney, 1999] is apparent in the discovered Kepler data as well. The data set here extends from OER $\sim (0.7, 1.8)$ and $\Delta m \sim (-0.3, 0.4)$, not including the one sample from cluster 3 that is extreme. This is comparable to the reported range in [McCartney, 1999] of OER $\sim (0.8, 1.2)$ and $\Delta m \sim (-0.1, 0.05)$—a similar OER range, but our Kepler data span a much larger $\Delta m$ domain, likely resulting from our additional application of min-max amplitude scaling. The gap in $\Delta m$ between $-0.08$ and $0.02$ is caused by the bias in our training sample and algorithm goal: we only include O'Connell effect binaries with a user-discernible $\Delta m$.

---

[4] ./Figures/ReducedFeaturesKeplerAll_Temp.png

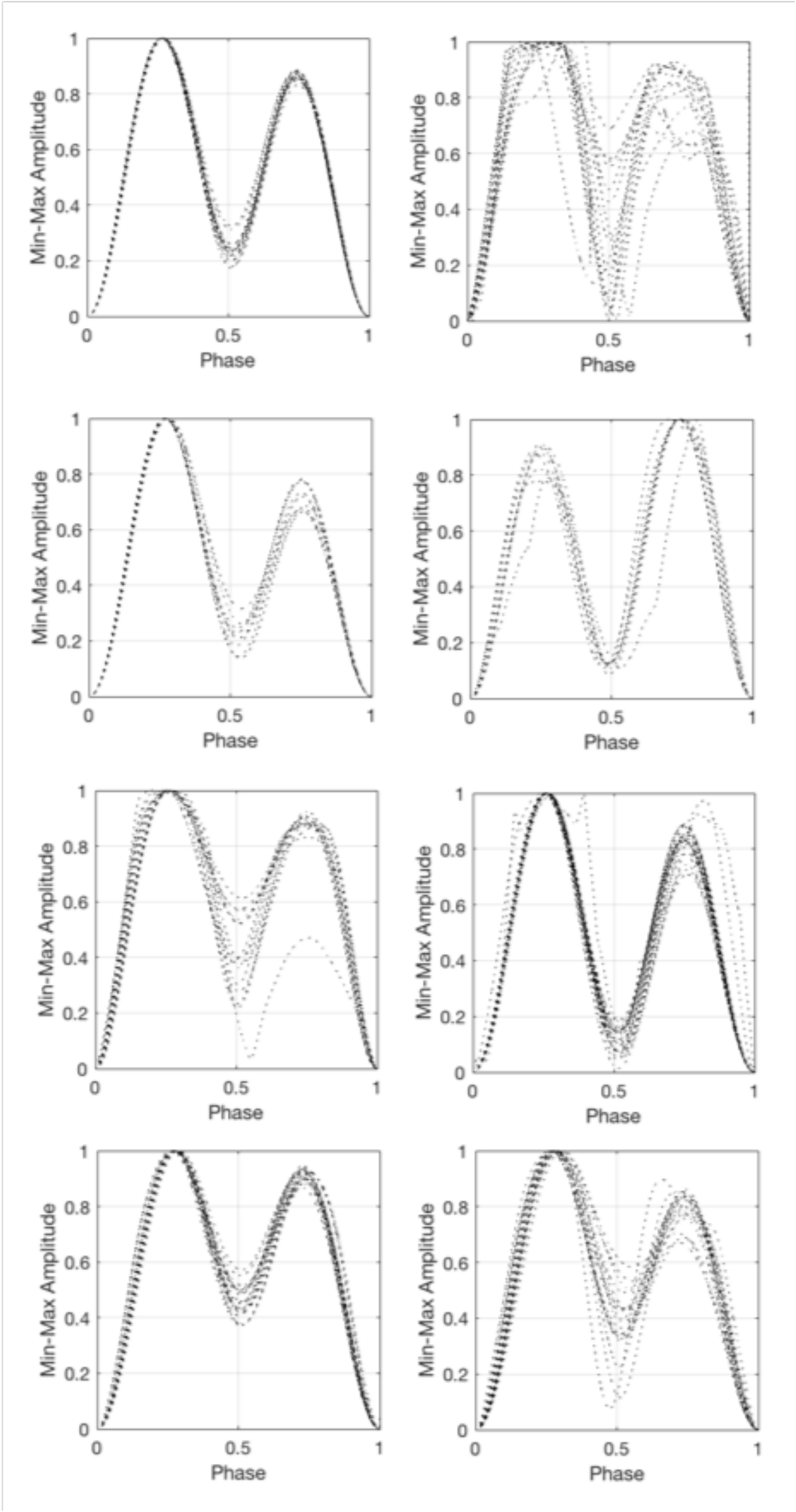

Figure 6.5: The phased light curves of the discovered OEEB data from Kepler, clustered via k-mean applied to the DF feature space. Cluster number used is based on trial and error, and the unsupervised classification has been implemented here only to highlight morphological similarities. The left four plots represent clusters 1–4 (top to bottom), and the right four plots represent clusters 5–8 (top to bottom).

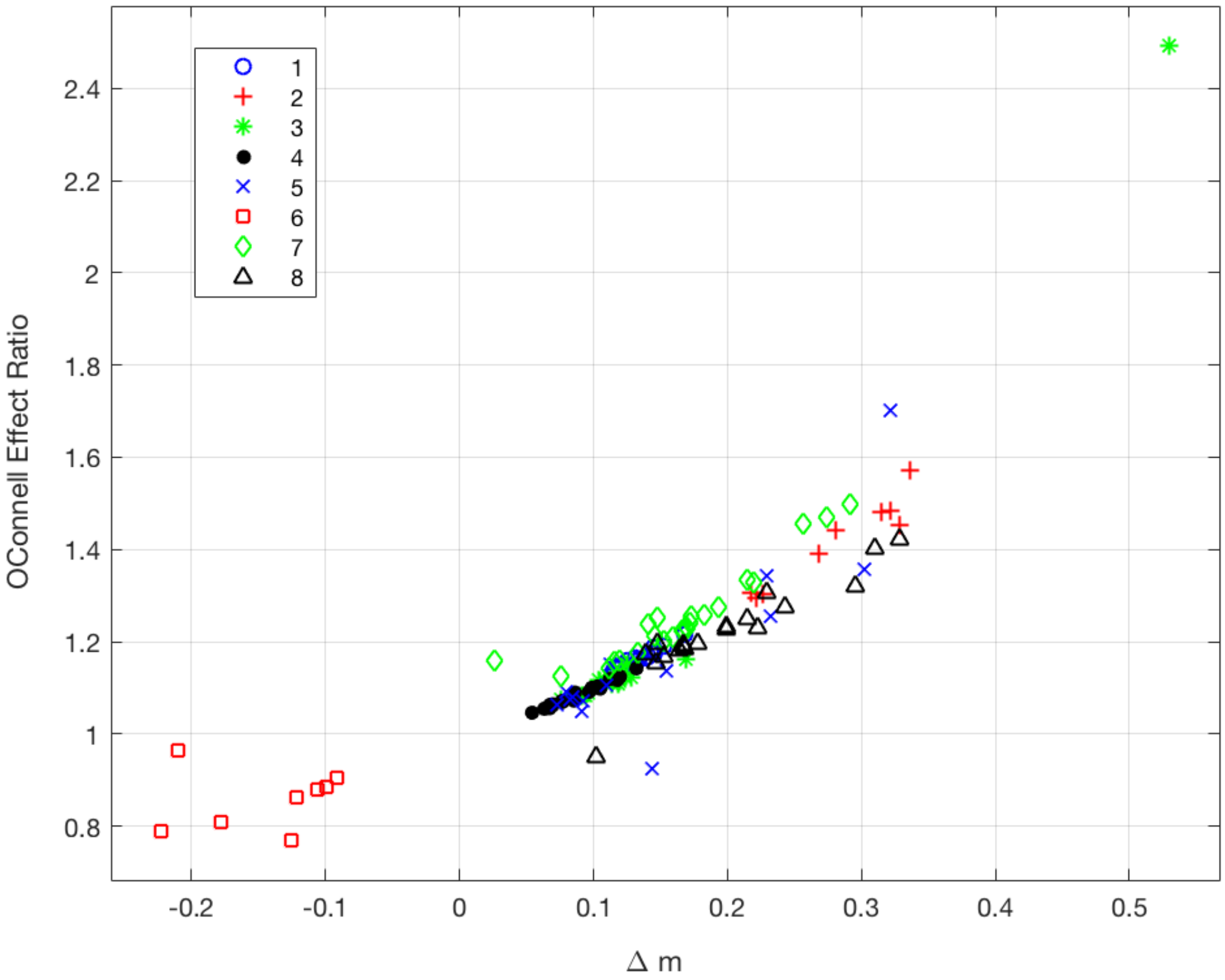

Figure 6.6: OER versus $\Delta$m for discovered Kepler O'Connell effect Eclipsing Binaries. This relationship between OER and $\Delta$m was also demonstrated in [McCartney, 1999].

Table 6.3: Metric Measurements from the Discovered O'Connell Effect Eclipsing Binaries from the Kepler Data Set

| Cluster | $\Delta m$ | $\sigma_{\Delta m}/\Delta m$ | OER | $\sigma_{\mathrm{OER}}/\mathrm{OER}$ | LCA | $\sigma_{\mathrm{LCA}}/\mathrm{LCA}$ | # |
|---|---|---|---|---|---|---|---|
| 1 | 0.13 | 0.11 | 1.16 | 0.02 | 7.62 | 0.25 | 17 |
| 2 | 0.28 | 0.17 | 1.41 | 0.07 | 8.92 | 0.16 | 9 |
| 3 | 0.14 | 0.78 | 1.20 | 0.30 | 7.13 | 0.25 | 15 |
| 4 | 0.09 | 0.24 | 1.08 | 0.02 | 6.95 | 0.23 | 22 |
| 5 | 0.15 | 0.55 | 1.17 | 0.16 | 8.54 | 0.58 | 15 |
| 6 | −0.14 | −0.36 | 0.86 | 0.08 | 8.36 | 0.19 | 8 |
| 7 | 0.17 | 0.36 | 1.25 | 0.08 | 9.41 | 0.82 | 24 |
| 8 | 0.20 | 0.31 | 1.22 | 0.08 | 8.03 | 0.36 | 19 |

*Note.* Metrics are based on [McCartney, 1999] proposed values of interest.

The clusters identified by the k-mean algorithm applied to the DF feature space roughly correspond to groupings in the OER/$\Delta$m feature space (clustering along the diagonal). The individual cluster statistics (mean and relative error) with respect to the metrics measured here are given in Table 6.3. All of the clusters have a positive mean $\Delta m$, save for cluster 6. The morphological consistency within a cluster is visually apparent in Figure 6.5 but also in the relative error of LCA, with clusters 5 and 7 being the least consistent. The next step will include applications to other surveys.

### 6.4.3 LINEAR Catalog

We further demonstrate the algorithm with an application to a separate independent survey. Machine learning methods have been applied to classifying variable stars observed by the LINEAR survey [Sesar et al., 2011], and while these methods have focused on leveraging Fourier domain coefficients and photometric measurements $\{u, g, r, i, z\}$ from SDSS, the data also include best estimates of period, as all of the variable stars trained on had cyclostationary signatures. It is then trivial to extract the phased light curve for each star and apply our Kepler trained detector to the data to generate "discovered" targets of interest.

Table 6.4: Discovered OEEBs from LINEAR

| | | | | | | | |
|---|---|---|---|---|---|---|---|
| 13824707 | 19752221 | 257977 | 458198 | 7087932 | 4306725 | 23202141 | 15522736 |
| 1490274 | 21895776 | 2941388 | 4919865 | 8085095 | 4320508 | 23205293 | 17074729 |
| 1541626 | 22588921 | 346722 | 4958189 | 8629192 | 6946624 | | |

The discovered targets are aligned, and the smoothed light curves are presented in Figure 6.7. Note that the LINEAR IDs are presented in Table 6.4 and as a supplementary digital file at the project repository.[5]

Application of our Kepler trained detector to LINEAR data results in 24 "discovered" OEEBs. These include four targets with a negative O'Connell effect. Similar to the Kepler discovered data set, we plot OER/$\Delta m$ features using lower-resolution phased binnings ($n = 20$) and see that the distribution and relationship from [McCartney, 1999] hold here as well (see Figure 6.8).

## 6.5 Discussion on the Methodology

### 6.5.1 Comparative Studies

The pairing of DF feature space and push–pull matrix metric learning represents a novel design; thus it is difficult to draw conclusions about performance of the design, as there are no similar studies that have trained on this particular data set, targeted this particular variable type, used this feature space, or used this classifier. As we discussed earlier, classifiers that implement matrix-variate features

[5] ./supplement/LINEARDiscovered.xlsx

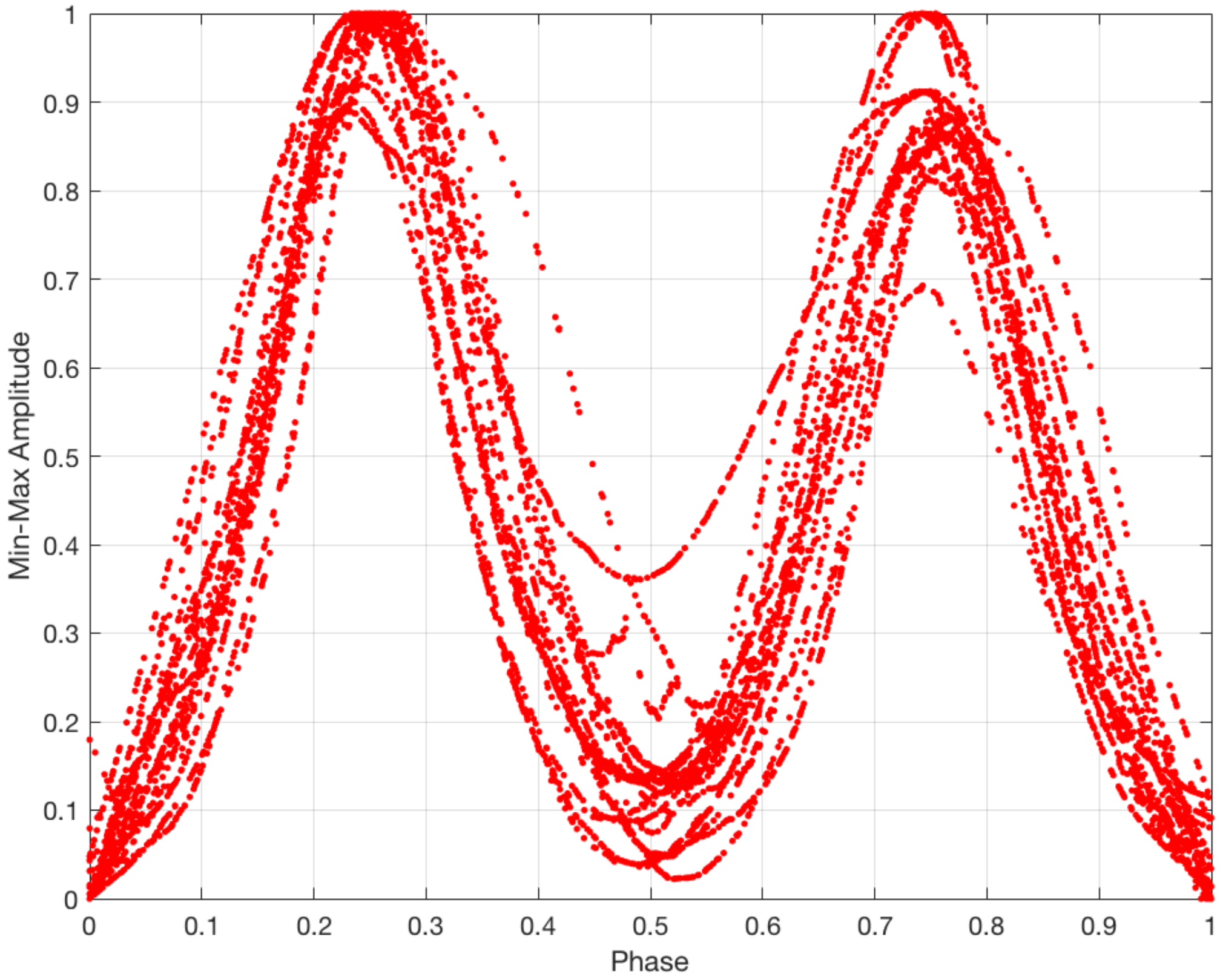

Figure 6.7: Distribution of phased–smoothed light curves from the set of discovered LINEAR targets that demonstrate the OEEB signature. LINEAR targets were discovered using the Kepler trained detector.

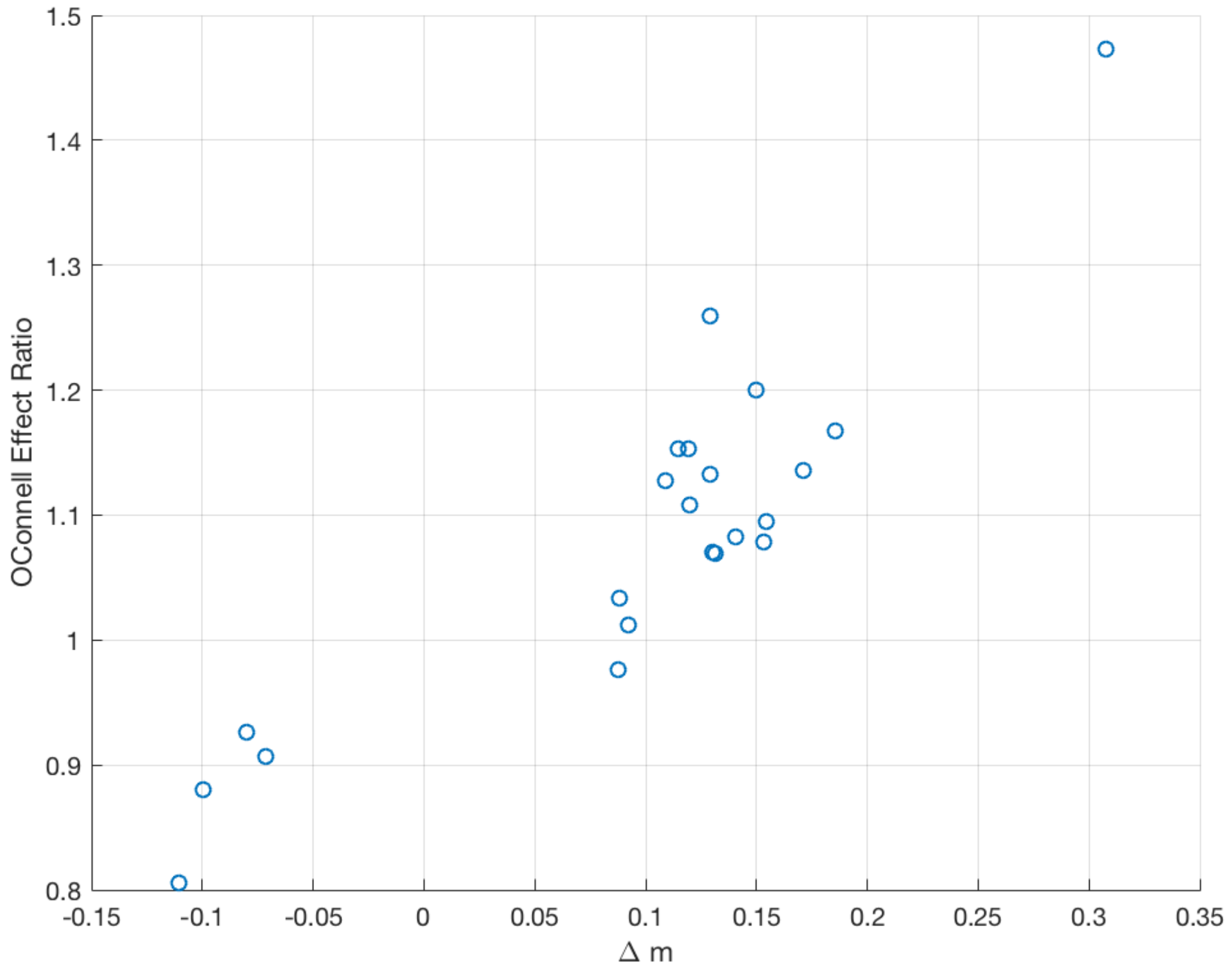

Figure 6.8: OER versus $\Delta$m for the discovered OEEB in the LINEAR data set. This relationship between OER and $\Delta$m was also demonstrated in [McCartney, 1999] and is similar to the distribution found in Figure 6.6.

directly are few and far between and almost always not off the shelf. We have developed here two hybrid designs—off-the-shelf classifiers mixed with feature space transform—to provide context and comparison.

These two additional classification methodologies implement more traditional and well-understood features and classifiers: k-NN using $L^2$ distance applied to the phased light curves (Method A) and k-means representation with quadratic discriminant analysis (QDA) (Method B). Method A is similar to the UCR [Chen et al., 2015b] time series data baseline algorithm, reported as part of the database. Provided here is a direct k-NN classification algorithm applied directly to the smoothed, aligned, regularly sampled phased light curve. This regular sampling is generated via interpolation of the smoothed data set and is required because of the nature of the nearest neighbor algorithm requiring one-to-one distance. Standard procedures can then be followed [Hastie et al., 2009]. Method B borrows from Park et al. [2003], transforming the matrix-variate data into vector-variate data via estimation of distances between our training set and a smaller set of exemplar means DFs that were generated via unsupervised learning. Distances were found using the Frobenius norm of the difference between the two matrices.

Whereas Method A uses neither the DF feature representation nor the metric learning methodology, Method B uses DF feature space but not the metric learning methodology. This presents a problem, however, as most standard out-of-the-box classification methods require a vector input. Indeed, many methodologies, even when faced with a matrix input, choose to vectorize the matrix. An alternative to this implementation is a secondary transformation into a lower-dimensional feature space. Following the work of [Park et al., 2003], we implement a matrix distance k-means algorithm (e.g., k-means with a Frobenius norm) to generate estimates of clusters in the DF space. Observations are transformed by finding the Euclidean distance between each training point and each of the k-mean matrices discovered. The resulting set of k-distances is treated as the input pattern, allowing the use of the standard QDA algorithm [Duda et al., 2012]. The performances of both the proposed methodology and the two comparative methodologies are presented in Table 6.5. The algorithms are available as open source code, along with our novel implementation, at the project repository.

Table 6.5: Comparison of Performance Estimates across the Proposed Classifiers (Based on Testing Data)

| | PPML | Method A | Method B |
|---|---|---|---|
| Error rate | 12.5% | 15.6% | 12.7% |

We present the performance of the main novel feature space/classification pairing as well as the two additional implementations that rely on more standard methods. Here we have evaluated performance based on misclassification rate, i.e., 1-accuracy given by Fawcett [2006] as $1 - \text{correct}/\text{total}$. The method we propose has a marginally better misclassification rate (Table 6.5) and has the added benefit of (1) not requiring unsupervised clustering, which can be inconsistent, and (2) providing nearest neighbor estimates allowing for demonstration of direct comparison. These performance estimate values are dependent on the initial selected training and testing data. They have been averaged and optimized via cross-validation; however, with so little initial training data and with the selection process for which training and testing data are randomized, performance estimates may vary. Of course, increases in training data will result in increased confidence in performance results.

We have not included computational times as part of this analysis, as they tend to be dependent on the system operated on. We can anecdotally discuss that, on the system implemented as part of this research (MacBook Pro, 2.5 GHz Intel i7, 8 GB RAM), the training optimization of our proposed feature extraction and PPML classification total took less than 5–10 min to run—variation depending on whatever else was running in the background. Use of the classifiers on unlabeled data resulted in a classification in fractions of seconds per star. However, we should note that this algorithm will speed up if it is implemented on a parallel processing system, as much of the time taken in the training process resulted from linear algebra operations that can be parallelized.

### 6.5.2 Strength of the Tools

The DF representation maps deterministic, functional stellar variable observations to a stochastic matrix, with the rows summing to unity. The inherently probabilistic nature of DFs provides a robust way to mitigate interclass variability and handle irregular sampling rates associated with stellar observations. Because the DF feature is indifferent to sampling density so long as all points along the functional shape are represented, the trained detection algorithm we generate and demonstrate in this article can be trained on Kepler data but directly applied to the LINEAR data, as shown in section 6.4.3.

The algorithm, including comparison methodologies, designed feature space transformations, classifiers, utilities, and so on, is publicly available at the project repository;[6] all code was developed in MATLAB and was run on MATLAB 9.3.0 (R2017b). The operations included here can be executed either via calling indi-vidual functions or using the script provided (ImplementDetector.m). Likewise, a Java version of all of the individual computational functions has been generated [see JVarStar, Johnston et al., 2019] and is included in the project repository.[7]

---

[6]https://GitHub.com/kjohnston82/OCDetector

### 6.5.3 Perspectives

This design is modular enough to be applied as is to other types of stars and star systems that are cyclostationary in nature. With a change in feature space, specifically one that is tailored to the target signatures of interest and based on prior experience, this design can be replicated for other targets that do not demonstrate a cyclostationary signal (i.e., impulsive, nonstationary, etc.) and even to targets of interest that are not time variable in nature but have a consistent observable signature (e.g., spectrum, photometry, image point-spread function, etc.). One of the advantages of attempting to identify the O'Connell effect Eclipsing Binary is that one only needs the phased light curve—and thus the dominant period allowing a phasing of the light curve—to perform the feature extraction and thus the classification. The DF process here allows for a direct transformation into a singular feature space that focuses on functional shape.

For other variable stars, a multiview approach might be necessary; either descriptions of the light curve signal across multiple transformations (e.g., Wavelet and DF), or across representations (e.g. polarimetry and photometry) or across frequency regimes (e.g. optical and radio) would be required in the process of properly defining the variable star type. The solution to this multiview problem is neither straightforward nor well understood [Akaho, 2006]. Multiple options have been explored to resolve this problem: combination of classifiers, canonical correlation analysis, posterior-probability blending, and multimetric classification. The computational needs of the algorithm have only been roughly studied, and a more thorough review is necessary in the context of the algorithm proposed and the needs of the astronomy community. The k-NN algorithm dependence on pairwise difference, while one of its strong suits is also one of the more computationally demanding parts of the algorithm. Functionality such as $k-d$ **trees** as well as other feature space partitioning methods have been shown to reduce the computational requirements.

[7] https://GitHub.com/kjohnston82/VariableStarAnalysis

## 6.6 Conclusion

The method we have outlined here has demonstrated the ability to detect targets of interest given a training set consisting of expertly labeled light curve training data. The procedure presents two new functionalities: the distribution field, a shape-based feature space, and the push–pull matrix metric learning algorithm, a metric learning algorithm derived from LMNN that allows for matrix-variate similarity comparisons. A comparison to less novel, more standard methods was demonstrated on a Kepler eclipsing binary sub-dataset that was labelled by an expert in the field of O'Connell effect binary star systems. The performance of the three methods is presented, the methodology proposed (DF + Push-Pull Metric Learning) is comparable to or outperforms the other methods. As a demonstration, the design is applied to Kepler eclipsing binary data and LINEAR data. Furthermore, the increase in the number of systems, and the presentation of the data, allows us to make additional observations about the distribution of curves and trends within the population. Future work will involve the analysis of these statistical distributions, as well as an inference as to their physical meaning.

The new OEEB systems we discovered by the method of automated detection proposed here can be used to further investigate their frequency of occurrence, provide constraints on existing light curve models, and provide parameters to look for these systems in future large-scale variability surveys like LSST. Although the effort here targets OEEB as a demonstration, it need not be limited to those particular targets. We could use the DF feature space along with the push–pull metric learning classifier to construct a detector for any variable stars with periodic variability. Furthermore, any variable star (e.g., supernova, RR Lyr, Cepheids, eclipsing binaries) can be targeted using this classification scheme, given the appropriate feature space transformation allowing for quantitative evaluation of similarity. This design could be directly applicable to exo-planet discovery; either via light curve detection (e.g., to detect eclipsing exo-planets) or via machine learning applied to other means (e.g., spectral analysis).

# Chapter 7

# Multi-View Classification of Variable Stars Using Metric Learning

The classification of variable stars relies on a proper selection of feature spaces of interest and a classification framework that can support the linear separation of those features. Features should be selected that quantify the tell-tale signature of the variability — the structure and information content. Prior studies generated feature spaces such as: Slotted Symbolic Markov Model [SSMM, Johnston and Peter, 2017], Fourier transform [Deb and Singh, 2009], wavelet transformation, Distribution Field [DF, Johnston et al., forthcoming], and so on; they attempt to completely differentiate or linearly-separate various type of variable stars classes. These efforts include: expert selected feature efforts [Debosscher, 2009, Sesar et al., 2011, Richards et al., 2012, Graham et al., 2013a, Armstrong et al., 2016, Mahabal et al., 2017, Hinners et al., 2018], automated feature selection efforts [McWhirter et al., 2017, Naul et al., 2018], and unsupervised methods for feature extraction [Valenzuela and Pichara, 2018, Modak et al., 2018].

The astroinformatics-community standard features include quantification of statistics associated with the time domain photometric data, Fourier decomposition of the data, and color information in both the optical and infrared domain [Nun et al., 2015, Miller et al., 2015]. The number of individual features commonly used is upwards of 60+ and growing [Richards et al., 2011] as the number of variable star types increases, and as a result of further refinement of classification definitions [Samus' et al., 2017].

Similarly, these efforts are limited in either size or scope based on: the survey goals from which the data being trained on was originally derived [Angeloni et al., 2014], the developer/scientists research interests [Pérez-Ortiz et al., 2017, McCauliff et al., 2015], or a subset of the top five to ten most frequent class-types Kim and Bailer-Jones [2016], Pashchenko et al. [2018], Naul et al. [2018]. In our research, no efforts were found in the literature that address all variables identified by Samus' et al. [2017]—most address some subset. For a informative breakdown of different types of variable stars, see Eyer and Blake [2005]. As surveys become more complete and more dense in observations, the complexity of the problem is likely to grow [Bass and Borne, 2016].

Here in lies the complication of expertly selected feature sets; their original function is keyed to the original selection of variable stars of interest. Additionally the features selected are often co-linear, resulting in little to no new information or separability despite the increase in dimensionality and additional increase in computational power needed to manage the data [D'Isanto et al., 2016]. Growing the feature dimensionality, via either additional feature space transformations or addition of information resulting from multi-messenger astronomy, results in increasing the sparsity of the training data representation of class feature distribution. This requires increasingly more complex classifier designs to both support the dimensionality as well as the potential non-linear class-space separation.

Curiously, aside from efforts to construct a classification algorithm from the time domain data directly [McWhirter et al., 2017], few efforts in astroinformatics have looked at other features beyond those described above—mostly Fourier domain transformations or time domain statistics. Considering the depth of possibility for time domain transformations [Fu, 2011, Grabocka et al., 2012, Cassisi et al., 2012, Fulcher et al., 2013], it is surprising that the community has focused on just a few transforms. Similarly, the astroinformatics-community has focused on just a few classifiers as well, limited to mostly standard classifiers, and specifically decision tree algorithms such as random forest type classifiers.

Prior studies have initially addressed the potential of using metric learning as a means for classification of variable stars [Johnston et al., forthcoming]. Metric learning has a number of benefits that are advantageous to the astronomer:

- Metric learning uses nearest neighbors (k-NN) classification to generate the decision space [Hastie et al., 2009, Duda et al., 2012], k-NN provides instant clarity into the reasoning behind the classifiers decision (based on similarity, "$x_i$ is closer to $x_j$ than $x_k$" ).

- Metric learning leverages side information (the supervised labels of the training data) to improve the metric, i.e. a transformation of the distance between points that favors a specific goal: similar closer together, different further apart, simplicity of the metric, feature dimensionality reduction, etc.. This side data is based on observed prior analyzed data, thus decisions have a grounding in expert identification as opposed to black-box machine learning [Bellet et al., 2015]. Dimensionality reduction in particular can be helpful for handling feature spaces that are naturally sparse.

- k-NN can be supported by other algorithm structures such as data partitioning methods to allow for a rapid response time in assigning labels to new observations, despite relying upon a high number of training data [Faloutsos et al., 1994].

- The development of an anomaly detection functionality Chandola et al. [2009], which has been shown to be necessary to generate meaningful classifications [see: Johnston and Peter, 2017, Johnston and Oluseyi, 2017], is easily constructed from the k-NN metric learning framework.

## 7.1 Procedure

The following procedure presented requires only processed (artifact removed) time domain data, i.e. light curves. The features extracted and used for classification are based on only the time domain data. This paper outlines a number of novel developments in the area of time-domain variable star classification that are of major benefit to the developer/researcher. First, we demonstrate both the Slotted Symbolic Markov Model Johnston and Peter [2017] and the Distribution Field Johnston et al. [forthcoming] transforms as viable feature spaces to use for classification of variable stars on their own; SSMM requires no phasing of the time domain data but still provides a feature that is shape based, DF allows for the consideration of the whole phased waveform without additional picking and choosing of metrics from the waveform [i.e., see Richards et al., 2012].

Second, we demonstrate leveraging metric learning as a viable means of classification of variable stars that has dramatic benefits to the user. Metric learning decisions have an implicit traceability: the ability to follow from the classifier's decision, to the weights associated with each individual feature used as part of the classification, to the nearest-neighbors used in making the decision provide a clear idea of why the classifier made the decision. This direct comparison of newly observed with prior observations, and the justification via historical comparison, make this method ideal for astronomical—and indeed scientific—applications.

Lastly, this paper will introduce Multi-view learning as a methodology that can provide a major benefit to the astronomical community. Astronomy often deals with multiple transformations (e.g., Fourier Domain, Wavelet Domain, statistical...etc) and multiple domains of data types (visual, radio frequency, high energy, particle, etc.). The ability to handle, and just as importantly co-train an optimization algorithm on, multiple domain data will be necessary as the multitude of data grows. The project software is provided publicly at the associated GitHub repository [1].

This design will be generic enough that it can be transferred from project to project, survey to survey, and class space to class space, with a minimal change in features while still being able to maximize performance of the classifier with respect top targeted project goals. In this paper will be organized as follows: (1) summarize current stellar variable classification efforts, features currently in use, and machine learning methodologies exercised (2) review the features used (statis-tics, color, DF and SSMM) (3) review the classification methodologies used (metric learning, LM$^3$L, and LM$^3$L-MV) (4) demonstrate our optimization of feature extraction algorithm for our datasets, leveraging "simple" classification methods (k-NN) and cross-validation processes (5) demonstrate our optimization of classification parameters for LM$^3$L and LM$^3$L-MV via cross-validation and (6) report on the performance of the feature/classifier pairing. Our proposal is an implementation of both the feature extraction and classifier for the purposes of multi-class identification, that can handle raw observed data.

[1] https://github.com/kjohnston82/VariableStarAnalysis

## 7.2 Theory and Design

We present an initial set of time domain feature extraction methods; the design demonstrated is modular in nature, allowing for a user to append or substitute feature spaces that an expert has found to be of utility in the identification of variable stars. Although our initial goal is variable star identification, given a separate set of features this method could be applied to other astroinformatics problems (i.e., image classification for galaxies, spectral identification for stars or comets, etc.). While we demonstrate the classifier has a multi-class classification design, which is common in the astroinformatics references we have provided, the design here can easily be transformed into a one-vs-all design [Johnston and Oluseyi, 2017] for the purposes of generating a detector or classifier designed specifically to a user's needs [Johnston et al., forthcoming].

### 7.2.1 Signal Conditioning

Required are features that can respond to the various signal structures that are unique to the classes of interest, i.e. phased light shape, frequency distribution, phase distribution, etc.). Our implementation starts with raw data (such as astronomical light curves) as primary input, which are then mapped into a specific feature space. To support these transformations, a set of signal conditioning methods are implemented for the two new feature space presented below. These techniques are based on the methods presented in Johnston and Peter [2017] and are fairly common in the industry. The data that is leveraged— with respect to classification of the waveform—is on the order of hundreds of observations over multiple cycles. While the data is not cleaned as part of the upfront process, the features that are implemented are robust enough to not be affected by intermittent noise. The raw waveform is left relatively unaffected, however smoothing does occur on the phased waveform to generate a new feature vector, i.e. a phased smoothed waveform.

The phased waveform is generated via folding the raw waveform about a period found to best represent the cyclical process [Graham et al., 2013b]. The SUPER-SMOOTHER algorithm [Friedman, 1984] is used to smooth the phased data into a functional representation. Additionally in some cases, the originating survey/mission will perform some of these signal conditioning processes as part of their analysis pipeline (e.g., Kepler). This includes outlier removal, period finding, and long term trend removal. Most major surveys include a processing pipeline, our modular analysis methods provide a degree of flexibility that allow the implementer to take advantage of these pre-applied processes. Specifically of use, while our feature extraction SSMM does not require a phased waveform, the DF feature does, thus period finding methods are of importance.

Most of the period finding algorithms are methods of spectral transformation with an associated peak/max/min finding algorithm and include such methods as: discrete Fourier transform, wavelets decomposition, least squares approximations, string length, auto-correlation, conditional entropy and auto-regressive methods. Graham et al. [2013b] review these transformation methods (with respect to period finding), and find that the optimal period finding algorithm is different for different types of variable stars. The Lomb–Scargle method was selected as the main method for generating a primary period for this implementation. For more information, our implementation of the Lomb–Scargle algorithm is provided as part of the Variable Star package[2].

### 7.2.2 Feature Extraction

For our investigation we have selected feature spaces that quantify the functional shape of repeated signal—cyclostationary signal—but are dynamic enough to handle impulsive type signals (e.g., supernova) as well. This particular implementation design makes the most intuitive sense, visual inspection of the light curve is how experts identify these sources. Prior research on time domain data identification has varied between generating machine learned features [Bos et al., 2002, Broersen, 2009, Blomme et al., 2011, Bolós and Benítez, 2014, Gagniuc, 2017], implementing generic features [e.g. Fourier domain features; Debosscher, 2009, Richards et al., 2012, Palaversa et al., 2013, Masci et al., 2014], and looking at shape or functional based features [e.g. DF, SSMM; Park and Cho, 2013, Haber et al., 2015].

[2] fit.astro.vsa.analysis.feature.LombNormalizedPeriodogram

We implement two novel time domain feature space transforms: SSMM and DF. It is not suggested that these features are going to be the best in all cases, nor are they the only choice as is apparent from Fulcher et al. [2013]. Any feature space, so long as it provides separability, would be usable here. One need only think of how to transform the observable (time domain, color, spectra, etc.) into something that is a consistent signature for stars in given class-type (i.e., variable star type).

#### 7.2.2.1 Slotted Symbolic Markov Models (SSMM)

Slotted Symbolic Markov Models (SSMM) is useful in the differentiation between variable star types [Johnston and Peter, 2017]. The time domain slotting described in Rehfeld et al. [2011] is used to regularize the sampling of the photometric observations. The resulting regularized sampled waveform is transformed into a state space [Lin et al., 2007, Bass and Borne, 2016]; thus the result of the conditioning is the stochastic process $\{y_n,\, n = 1, 2, ...\}$. The stochastic process is then used to populate the empty matrix $\mathbf{P}$ [Ge and Smyth, 2000], the elements of $\mathbf{P}$ are populated as the transition state probabilities (equation 7.1).

$$\mathbf{P}\{y_{n+1} = j |\, y_n = i,\, y_{n-1} = i_{n-1}, ...,\, y_1 = i_1,\, y_0 = i_0\} = P_{ij} \tag{7.1}$$

The populated matrix $\mathbf{P}$ is the SSMM feature; and is often described as a first order Markov Matrix.

#### 7.2.2.2 Distribution Field (DF)

A distribution field (DF) is an array of probability distributions, where probability at each element is defined as [Helfer et al., 2015] equation 7.2.

$$\mathrm{DF}_{ij} = \frac{\sum \left[ y_j < x' \left( x_i \leq p \leq x_{i+1} \right) < y_{j-1} \right]}{\sum \left[ y_j < x' \left( p \right) < y_{j-1} \right]} \tag{7.2}$$

Note, [ ] is the Iverson Bracket [Iverson, 1962], $y_j$ and $x_i$ are the corresponding normalized amplitude and phased time bins, respectively. The result is a 2-D histogram that is a right stochastic matrix, i.e. the rows sum to one. Bin numbers, are optimized by cross-validation as part of the classification training process. Separately, SSMM itself is an effective feature for discriminating variable star types as shown by Johnston and Peter [2017]. Similarly, DF has been shown to be a valuable feature for discriminating time domain signatures, see Helfer et al. [2015] and Johnston et al. [forthcoming].

### 7.2.3 Classification and Metric Learning

The classification methodology known as metric learning has its roots in the understanding of how and why observations are considered similar. The very idea of similarity is based around the numerical measurement of distance, and the computation of a distance is generated via application of a distance function. Bellet et al. [2015] define the metric distance as equation 7.3

$$d(x, x') = \sqrt{\left( x - x' \right)^T \mathbf{M} \left( x - x' \right)}; \tag{7.3}$$

where $X \subseteq \mathbb{R}^d$ and the metric is required to be $\mathbf{M} \in \mathbb{S}^d_+$. $\mathbb{S}^d_+$ is the cone of symmetric positive semi-definite (PSD) $d \times d$ real valued matrices. Metric learning seeks to optimize this distance, via manipulation of the metric $\mathbf{M}$, based on available side data. How the optimization occurs, what is focused on and what is considered important, i.e. the construction of the objective function, is the underlying difference between the various metric learning algorithms.

The side information is defined as the set of labeled data $\{x_i, y_i\}_{i=1}^n$. Furthermore the triplet is defined as $(x_i, x_j, x_k)$ where $x_i$ and $x_j$ have the same label but $x_i$ and $x_k$ do not. It is expected then, based on the definition of similarity and distance, that $d(x_i, x_j) < d(x_i, x_k)$, i.e., that the distances between similar labels is smaller than the distances between dissimilar ones. Methods such as LMNN [Weinberger et al., 2009] use this inequality to defined an objective function that optimizes the metric to bring similar things closer together, while pushing dissimilar things further apart.

Given the metric learning optimization process, the result is a tailored distance metric and associated distance function (equation 6.7). This distance function is then used in a standard k-NN classification algorithm. The k-NN algorithm estimates a classification label based on the closest samples provided in training [Altman, 1992]. If $x_n$ is a set of training data $n$ big, then we find the distance between a new pattern $x_i$ and each pattern in the training set. The new pattern is classified depending on the majority of class labels in the closest $k$ data points.

## 7.3 Challenges Addressed

In the application of the LM$^3$L algorithm to our data we found a number of challenges not specified by the original paper that required attention. Some of these challenges were a direct result of our views (vectorization of matrix-variate data) and some of these challenges were resulting from practical application (hinge loss functionality and step-size optimization).

### 7.3.1 Hinge Loss Functionality

While the original LM$^3$L paper does not specify details with respect to the implementation of the hinge loss functionality used, the numerical implementation of both the maxima and the derivative of the maxima are of critical importance. For the implementation here, the hinge-loss functionality is approximated using Generalized Logistic Regression [Zhang and Oles, 2001, Rennie and Srebro, 2005]. Should a different approximation of hinge loss be requested, care should be given to the implementation, as definitions from various public sources are not consistent. For purposes here, the Generalized Logistic Regression is used to approximate the hinge loss ($h[x] \approx g_+(z, \phi)$) and is defined as equation 7.4:

$$g_+(z,\phi) = \begin{cases} 0.0 & z \leq -10 \\ z & z \geq 10 \\ \frac{1}{\phi}\log\left(1 + \exp\left(z\phi\right)\right) & -10 < z < 10 \end{cases} \tag{7.4}$$

the derivative of the Generalized Logistic Regression is then given as Equation 7.5:

$$\frac{\partial g_+(z,\phi)}{\partial z} = \begin{cases} 0.0 & z \leq -10 \\ 1 & z \geq 10 \\ \frac{\exp(\phi z)}{1+\exp(\phi z)} & -10 < z < 10 \end{cases} \tag{7.5}$$

For practical reasons (underflow/overflow) the algorithm is presented as a piece-wise function, in particular this is necessary because of the exponential in the functionality. In addition, the public literature is not consistent on the definition of the hinge-loss functionality approximation, specifically the relationship between the notations: $[z]_+$, $h[z]$, $\max(z, 0)$, and $g_+(z, \phi)$; usually the inconsistency is with respect to the input i.e. $z$, $-z$, or $1 - z$. We have explicitly stated our implementation here to eliminate any confusion.

### 7.3.2 Step-Size Optimization

While LM$^3$L provides an approximate "good" step size to use, in practice it was found that a singular number was not necessarily useful. While the exact reasons of why a constant step size was not beneficial were not investigated; the following challenges were identified:

1. The possibility of convergence was very sensitive to the step size.
2. Small step sizes that did result in a consistent optimization, resulted in a very slow convergence.
3. While an attempt could be made to find an optimal step size with respect to all views, it seems unlikely this would occur given the disparate nature of the views we have selected (distribution field, photometric color, time domain statistics, etc.).

4. For the metric learning methods used here (in both the standard and the proposed algorithms) the objective function magnitude scales with the number of training data sets, view dimensions and the number of views, as is apparent from the individual component of LM$^3$L: $\sum_{i,j,} h\left[\tau_k - y_{ij}\left(\mu_k - d^2_{\mathbf{M}_k}(x_i^k, x_j^k)\right)\right]$. With increasing number of training data, the objective function will increase and the gradient component ($w_k^p \sum_{i,j} y_{ij} h'[z]\mathbf{C}_{ij}^k$) will similarly be effected. This means that computational overflows could occur just by increasing the number of training data used.

In lieu of a singular estimate, we propose a dynamic estimate of the step-size per iteration per view. A review of step-size and gradient descent optimization methods [Ruder, 2016] suggest a number of out-of-the-box solutions to the question of speed (specifically methods such as Mini-Batch gradient descent).

The question of dynamic step size requires more development, in particular while methods exists, these are almost entirely focused on vector variate optimization. Barzilai and Borwein [1988] outline a method for dynamic step size estimation that has its' basis in secant root finding, the method described is extended here to allow for matrix variate cases. The gradient descent update for our metric learning algorithm is given as Equation 7.6.

$$\mathbf{L}^{(t+1)} = \mathbf{L}^{(t)} - \beta \frac{\partial J}{\partial \mathbf{L}} \tag{7.6}$$

In the spirit of Barziliai and Borwein, here in known as the BB-step method,

the descent algorithm is reformulated as Equation 7.7:

$$\lambda_k = \arg\min_{\lambda} \|\Delta\mathbf{L} - \lambda\Delta g(\mathbf{L})\|_F^2 \tag{7.7}$$

where $\lambda_k$ is a dynamic step size to be estimated per iteration and per view, $\triangle g(\mathbf{L}) = \nabla f\left(\mathbf{L}^{(t)}\right) - \nabla f\left(\mathbf{L}^{(t-1)}\right)$ and $\Delta\mathbf{L} = \mathbf{L}^{(t)} - \mathbf{L}^{(t-1)}$. The Forbinus norm can be defined as $\|A\|_F^2 = Tr(A \cdot A^H)$, the BB-step method can be found as Equation 7.8:

$$\frac{\partial}{\partial\lambda} Tr\left[(\Delta\mathbf{L} - \lambda\Delta g(\mathbf{L}))\,(\Delta\mathbf{L} - \lambda\Delta g(\mathbf{L}))^H\right] = 0 \tag{7.8}$$

Based on the Matrix Cookbook [Petersen et al., 2008], Equation 7.8 can be transformed into Equation 7.9.

$$Tr\left[-\Delta g(L)\,[\Delta\mathbf{L} - \lambda\Delta g(\mathbf{L})]^H - [\Delta\mathbf{L} - \lambda\Delta g(\mathbf{L})]\,\Delta g(L)^H\right] = 0 \tag{7.9}$$

With some algebra, Equation 7.9 can be turned into a solution for our approximation of optimal step size, given here as Equation 7.10.

$$\hat{\lambda} = \frac{1}{2} \cdot \frac{Tr\left[\Delta g(\mathbf{L}) \cdot \Delta\mathbf{L}^H + \Delta\mathbf{L} \cdot \Delta g(\mathbf{L})^H\right]}{Tr\left[\Delta g(\mathbf{L}) \cdot \Delta g(\mathbf{L})^H\right]} \tag{7.10}$$

It is elementary to show that our methodology can be extended for $\triangle g(\mathbf{L}_k) = \nabla f\left(\mathbf{L}_k^{(t)}\right) - \nabla f\left(\mathbf{L}_k^{(t-1)}\right)$ and $\Delta\mathbf{L}_k = \mathbf{L}_k^{(t)} - \mathbf{L}_k^{(t-1)}$; likewise we can estimate $\hat{\lambda}_k$ per view, so long as the estimates of both gradient and objective function are monitored at each iteration. While this addresses our observations, it should be noted that the fourth challenged outlined (scaling with increasing features and training data) was only partially addressed. Specifically, the above methodology does not address an initial guess of $\lambda_k$; in multiple cases it was found that this initial value was set to high, causing our optimization to diverge. Providing an initial metric in the form of $\sigma\mathbb{I}$ where $0 < \sigma < 1$ , was found to improve the chances of success, where the $\sigma$ was used to offset a $J$ value (from the objective function) that was too high (overflow problems). Care should be taken to set both the initial $\lambda_k$ and $\sigma$ to avoid problems.

### 7.3.3 Vectorization and ECVA

The features focused on as part of our implementation include both vector variate and matrix variate views. The matrix variate views requires transformation from their matrix domain to a vectorized domain for implementation in the LM$^3$L framework. The matrix-variate to vector-variate transformation implemented here is outlined in Johnston and Peter [2017]. The matrix is transformed $\text{vec}(X_i^k) = x_i^k$ to a vector domain. A dimensionality reduction process is implemented as some of the matrices are large enough to result in large sparse vectors (i.e., $20 \times 20$ DF matrix $= 400$ element vector). To reduce the large sparse feature vector resulting from the unpacking of matrix, we applied a supervised dimensionality reduction technique commonly referred to as extended canonical variate analysis (ECVA) [Nørgaard et al., 2006].

The methodology for ECVA has roots in principle component analysis (PCA). PCA is a procedure performed on large multidimensional datasets with the intent of rotating what is a set of possibly correlated dimensions into a set of linearly uncorrelated variables [Scholz, 2006]. The transformation results in a dataset, where the first principle component (dimension) has the largest possible variance. PCA is an unsupervised methodology, i.e. known labels for the data being processed is not taken into consideration, thus a reduction in feature dimensionality will occur, and while it maximizes the variance, it might not maximize the linear separability of the class space. In contrast to PCA, Canonical Variate Analysis does take class labels into considerations. The variation between groups is maximized resulting in a transformation that benefits the goal of separating classes. Given a set of data $\mathbf{x}$ with: $g$ different classes, $n_i$ observations of each class; following Johnson et al. [1992], the within-group and between-group covariance matrix is defined as Equations 7.11 and 7.12 respectfully.

$$\mathbf{S}_{within} = \frac{1}{n-g} \sum_{i=1}^{g} \sum_{j=1}^{n_i} (x_{ij} - \bar{x}_{ij})(x_{ij} - \bar{x}_i)' \tag{7.11}$$

$$\mathbf{S}_{between} = \frac{1}{g-1} \sum_{i=1}^{g} n_i (x_i - \bar{x})(x_i - \bar{x})' \tag{7.12}$$

Where $n = \sum_{i=1}^{g} n_i$, $\bar{x}_i = \frac{1}{n_i} \sum_{j=1}^{n_i} x_{ij}$, and $\bar{x} = \frac{1}{n} \sum_{j=1}^{n_i} n_i x_i$. CVA attempts to maximize the Equation 7.13.

$$J(\mathbf{w}) = \frac{\mathbf{w}' \mathbf{S}_{between} \mathbf{w}}{\mathbf{w}' \mathbf{S}_{within} \mathbf{w}} \tag{7.13}$$

The equation is solvable so long as $\mathbf{S}_{within}$ is non-singular, which need not be the case, especially when analyzing multicollinear data. When the case arises that the dimensions of the observed patterns are multicollinear, additional considerations need to be made. Nørgaard et al. [2006] outlines a methodology (ECVA) for handling these cases in CVA. Partial least squares analysis, PLS2 [Wold, 1939], is used to solve the above linear equation, resulting in an estimate of $\mathbf{w}$, and given that, an estimate of the canonical variates (the reduced dimension set). The application of ECVA to our vectorized matrices results in a reduced feature space of dimension $g - 1$, this reduced dimensional feature space, per view, is then used in the $\text{LM}^3\text{L}$ classifier.

#### 7.3.3.1 Multi-View Learning

We address the following classification problem: given a set of expertly labeled side data containing $C$ different classes (e.g., variable star types), where measurements can be made on the classes in question to extract a set of feature spaces for each observation, how do we define a distance metric that optimizes the misclassification rate? Specifically, how can this be done within the context of variable star classification based on the observation of photometric time-domain data? We have identified a number of features that may provide utility in discriminating between various types of stellar variables. We review how to combine this information together and generate a decision space; or rather, how to define the distance $d_{ij} = (x_i - x_j)'\mathbf{M}(x_i - x_j)$, when $x_i$ contains two matrices (SSMM or DF in our case). Specifically we attempt to construct a distance metric based on multiple attributes of different dimensions (e.g. $\mathbb{R}^{m\times n}$ and $\mathbb{R}^{m\times 1}$ ).

To respond to this challenge we investigate the utility of multi-view learning. For our purposes here we specify each individual measurement as the feature, and the individual transformations or representations of the underlying measurement as the feature space. Views, are the generic independent collections of these features or feature space. Thus, if provided the color of a variable star in *ugriz*, the individual measurements of $u - g$ or $r - i$ shall be referred to here as the features but the collective set of colors is the feature space. Our methodology here allows us to defined sets of collections of these feature and/or feature spaces as independent views, for example: all of *ugriz* measurements, the vectorized DF measurement, the concatenation of time-domain statistics and colors, the reduced (selected) sam-pling of Fourier spectra, could all be individual views. The expert defined these views *a priori*.

Xu et al. [2013], Kan et al. [2016] review multi-view learning and outline some basic definitions. Multi-view learning treats the individual views separately, but also provides some functionality for joint learning where the importance of each view is dependent on the others. As an alternative to multi-view learning, the multiple views could be transformed into a single view, usually via concatenation. The cost–benefit analysis of concatenated single-view vs. multi-view learning are discussed in Xu et al. [2013] and are beyond the scope of this paper.

Classifier fusion [Kittler et al., 1998, Tax, 2001, Tax and Muller, 2003] could be considered as an alternative to multi-view learning, with each view independently learned, and resulting in an independent classification algorithm. The result of the set of these classifiers are combined together (mixing of posterior probability) to result in a singular estimate of classification/label; this is similar to the operation of a Random Forest classifier, i.e. results from multiple individual trees combined together to form a joint estimate. We differentiate between the single-view learning with concatenation, multi-view learning, and classifier fusion designs based on when the join of the views is considered in the optimization process: before, during, or after.

Multi-view learning can be divided into three topics: 1) co-training, 2) multiple-kernel learning, and 3) subspace learning. Each method attempts to consider all views during the training process. Multiple-kernel learning algorithms attempt to exploit kernels that naturally correspond to different views and combine kernels either linearly or non-linearly to improve learning performance [Gönen and Alpaydın, 2011, Kan et al., 2016].

Sub-space learning uses canonical correlation analysis (CCA), or a similar method, to generate an optimal latent representation of two views which can be trained on directly. The CCA method can be iterated multiple times based on the number of views, this process will frequently result in a dimensionality that is lower then the original space [Hotelling, 1936, Akaho, 2006, Zhu et al., 2012, Kan et al., 2016].

This work will focus on a method of co-training, specifically metric co-training. Large Margin Multi-Metric Learning [Hu et al., 2014, 2018] is an example of metric co-training; the designed objective function minimizes the optimization of the individual view, as well as the difference between view distances, simultaneously. The full derivation of this algorithm is outlined in Hu et al. [2014], and the algorithm for optimization for $\text{LM}^3\text{L}$ is given as their Algorithm 1. This algorithm is implemented Java and is available as part of the software distribution.

Our implementation also includes additional considerations not discussed in the original reference. These considerations were found to be necessary based on challenges discovered when we applied the $\text{LM}^3\text{L}$ algorithm to our data. The challenges and our responses are discussed in Appendix A.

In addition to the implementation of $\text{LM}^3\text{L}$, we have developed a matrix variate version as well (section 7.3.4). This matrix variate classifier is novel with respect to multi-view learning methods and is one of two metric learning methods that we know of, the other being Push-Pull Metric Learning [Helfer et al., 2015].

### 7.3.4 Large Margin Multi-Metric Learning with Matrix Variates

The literature on metric learning methods is fairly extensive (see Bellet et al. [2015] for a review), however all of the methods presented so far focus on the original definition that is based in $X \subseteq \mathbb{R}^{d \times 1}$ , i.e. vector-variate learning. While the handling of matrix-variate data has been addressed here, the method require a transformation—vec($x$) and then ECVA—which ignores the problem of directly operating on matrix-variate data. The literature on matrix-variate classification and operations is fairly sparse. The idea of a metric learning supervised classification methodology based on matrix-variate data is novel.

Most of the matrix-variate research has some roots in the work by Hotelling [1936] and Dawid [1981]. There are some key modern references to be noted as well: Ding and Cook [2014] and Ding and Dennis Cook [2018] address matrix-variate PCA and matrix variate regression (matrix predictor and response), Dutilleul [1999] and Zhou et al. [2014] address the mathematics of the matrix normal distribution, and Safayani and Shalmani [2011] address matrix-variate CCA.

Developing a matrix-variate metric learning algorithm requires a formal definition of distance for matrix-variate observations, i.e. where $X \subseteq \mathbb{R}^{p \times q}$. Glanz and Carvalho [2013] define the matrix normal distribution as $X_i \sim MN(\mu, \mathbf{\Sigma}_s, \mathbf{\Sigma}_c)$, where $X_i$ and $\mu$ are $p \times q$ matrices, $\mathbf{\Sigma}_s$ is a $p \times p$ matrix defining the row covariance, and $\mathbf{\Sigma}_c$ is a $q \times q$ matrix defining the column covariance. Equivalently the relationship between the matrix normal distribution and the vector normal distribution is given as equation 7.14,

$$\mathrm{vec}\left(X_i\right) \sim N\left(\mathrm{vec}\left(\mu\right), \mathbf{\Sigma}_c \otimes \mathbf{\Sigma}_s\right). \tag{7.14}$$

The matrix-variate normal distribution is defined as equation 7.15 [Gupta and Nagar, 2000]

$$\begin{aligned} P\left(X_i; \mu, \mathbf{\Sigma}_s, \mathbf{\Sigma}_c\right) = (2\pi)^{-\frac{pq}{2}} \left|\left(\mathbf{\Sigma}_c \otimes \mathbf{\Sigma}_s\right)^{-1}\right|^{\frac{1}{2}} \\ \exp\left\{-\frac{1}{2}\mathrm{vec}\left(X_i - \mu\right)\left(\mathbf{\Sigma}_\mathrm{c} \otimes \mathbf{\Sigma}_\mathrm{s}\right)^{-1}\mathrm{vec}\left(X_i - \mu\right)\right\}. \end{aligned} \tag{7.15}$$

This distribution holds for the features that we are using as part of this study, at least within the individual classes. The Mahalanobis distance between our observations is then defined for the Matrix-Variate case as equations 7.16 to 7.18:

$$d_{\mathbf{\Sigma}}(X, X') \doteq \mathrm{vec}\left(X - X'\right)\left(\mathbf{\Sigma}_\mathrm{c} \otimes \mathbf{\Sigma}_\mathrm{s}\right)^{-1}\mathrm{vec}\left(X - X'\right), \tag{7.16}$$

$$= \mathrm{vec}\left(X - X'\right)^T \mathrm{vec}\left(\mathbf{\Sigma}_s^{-1}\left(X - X'\right)\mathbf{\Sigma}_c^{-1}\right), \tag{7.17}$$

$$= \mathrm{tr}\left[\mathbf{\Sigma}_c^{-1}\left(X - X'\right)^T \mathbf{\Sigma}_s^{-1}\left(X - X'\right)\right]. \tag{7.18}$$

This last iteration of the distance between matrices is used in our development of a metric learning methodology. Similar to the development of $\mathrm{LM}^3\mathrm{L}$ and the outline of Torresani and Lee [2007], we develop a metric learning algorithm for matrix-variate data. First the Mahalanobis distance for the matrix-variate multi-view case is recast as equation 7.19

$$d_{\mathbf{U}_k,\mathbf{V}_k}(X_i^k, X_j^k) = \mathrm{tr}\left[\mathbf{U}_k \left(X_i^k - X_j^k\right)^T \mathbf{V}_k \left(X_i^k - X_j^k\right)\right]; \qquad (7.19)$$

where $\mathbf{U}_k$ and $\mathbf{V}_k$ represent the inverse covariance of the column and row respectively. The individual view objective function is constructed similar to the LMNN [Weinberger et al., 2009] methodology; we define a push (equation 7.20) and pull (equation 7.21) as:

$$push_k = \gamma \sum_{j \rightsquigarrow i,l} \eta_{ij}^k \left(1 - y_{il}\right) \cdot h\left[d_{\mathbf{U}_k,\mathbf{V}_k}(X_i^k, X_j^k) - d_{\mathbf{U}_k,\mathbf{V}_k}(X_i^k, X_l^k) + 1\right], \qquad (7.20)$$

$$pull_k = \sum_{i,j} \eta_{ij}^k \cdot d_{\mathbf{U}_k,\mathbf{V}_k}(X_i^k, X_j^k); \qquad (7.21)$$

where $y_{il} = 1$ if and only if $y_i = y_l$ and $y_{il} = 0$ otherwise; and $\eta_{ij}^k = 1$ if and only if $x_i$ and $x_j$ are targeted neighbors of similar label $y_i = y_j$. For a more in-depth discussion of target neighbor, see Torresani and Lee [2007].

Furthermore, we include regularization terms [Schultz and Joachims, 2004] with respect to $\mathbf{U}_k$ and $\mathbf{V}_k$ as part of the objective function design; these are defined as $\lambda \left\|\mathbf{U}_k\right\|_F^2$ and $\lambda \left\|\mathbf{V}_k\right\|_F^2$, respectively. The inclusion of regularization terms in our objective function help promote sparsity in the learned metrics. Favoring sparsity can be beneficial when the dimensionality of the feature spaces is high, and can help lead to a more generic and stable solution.

The sub-view objective function is then equation 7.22:

$$\min_{\mathbf{U}_k,\mathbf{V}_k} I_k = \sum_{i,j} \eta_{ij}^k \cdot d_{\mathbf{U}_k,\mathbf{V_k}}(X_i^k, X_j^k) + \\ \gamma \sum_{j \rightsquigarrow i,l} \eta_{ij}^k (1 - y_{il}) \cdot h \left[ d_{\mathbf{U}_k,\mathbf{V}_k}(X_i^k, X_j^k) - d_{\mathbf{U}_k,\mathbf{V}_k}(X_i^k, X_l^k) + 1 \right] + \\ \lambda \left\| \mathbf{U}_k \right\|_F^2 + \lambda \left\| \mathbf{V}_k \right\|_F^2 ; \quad (7.22)$$

where $\lambda > 0$ and controls the importance of the regularization. From LM$^3$L the objective function is equation 7.23:

$$\min_{\mathbf{U}_k,\mathbf{V}_k} J_k = w_k I_k + \\ \mu \sum_{q=1,q\neq k}^{K} \sum_{i,j} \left( d_{\mathbf{U}_k,\mathbf{V}_k}(X_i^k, X_j^k) - d_{\mathbf{U}_l,\mathbf{V}_l}(X_i^q, X_j^q) \right)^2 ; \quad (7.23)$$

where $\sum_{k=1}^{K} w_k = 1$ and the first term is the contribution of the individual $k^{th}$ view, while the second term is designed to minimize the distance difference between attributes.

This objective design is solved using a gradient descent solver operation. To enforce the requirements of $\mathbf{U}_k \succ 0$ and $\mathbf{V}_k \succ 0$, the metrics are decomposed—$\mathbf{U}_k = \mathbf{\Gamma}_k^T \mathbf{\Gamma}_k$ and $\mathbf{V}_k = \mathbf{N}_k^T \mathbf{N}_k$. The gradient of the objective function with respect to the decomposed matrices $\mathbf{\Gamma}_k$ and $\mathbf{N}_k$ is estimated. The unconstrained optimum is found using the gradient of the decomposed matrices; the $\mathbf{U}_k$ and $\mathbf{V}_k$ matrices are then reconstituted at the end of the optimization process. We reformulate the matrix variate distance as equation 7.24:

$$d_{\mathbf{\Gamma}_k,\mathbf{N}_k}(\Delta_{ij}^k) = \mathrm{tr} \left[ \mathbf{\Gamma}_k^T \mathbf{\Gamma}_k \left( \Delta_{ij}^k \right)^T \mathbf{N}_k^T \mathbf{N}_k \left( \Delta_{ij}^k \right) \right] ; \quad (7.24)$$

for ease we make the following additional definitions: $d_{\mathbf{U}_k,\mathbf{V}_k}(X_i^k, X_j^k) = d_{ij}^k$, $X_i^k - X_j^k = \Delta_{ij}^k$, $\mathbf{A}_{ij}^k = \left(\Delta_{ij}^k\right)^T \mathbf{N}_k^T \mathbf{N}_k \left(\Delta_{ij}^k\right)$ , and $\mathbf{B}_{ij}^k = \Delta_{ij}^k \mathbf{\Gamma}_k^T \mathbf{\Gamma}_k \left(\Delta_{ij}^k\right)^T$. Note that $\mathbf{A}_{ij}^k = \left(\mathbf{A}_{ij}^k\right)^T$ and $\mathbf{B}_{ij}^k = \left(\mathbf{B}_{ij}^k\right)^T$. Additionally we identify the gradients as equations 7.25 and 7.26:

$$2\mathbf{\Gamma}_k \mathbf{A}_{ij}^k = \frac{\partial d_{ij}^k}{\partial \mathbf{\Gamma}_k} \tag{7.25}$$

$$2\mathbf{N}_k \mathbf{B}_{ij}^k = \frac{\partial d_{ij}^k}{\partial \mathbf{N}_k}, \tag{7.26}$$

as being pertinent for derivation. We give the gradient of the individual view objective $I_k$ as equations 7.27 and 7.28:

$$\frac{\partial I_k}{\partial \mathbf{\Gamma}_k} = 2\mathbf{\Gamma}_k \left( (1-\gamma) \sum_{i,j} \eta_{ij}^k \cdot \mathbf{A}_{ij}^k + \gamma \sum_{j \rightsquigarrow i,l} \eta_{ij}^k \left(1 - y_{il}\right) \cdot h'[z] \cdot \left[\mathbf{A}_{ij}^k - \mathbf{A}_{il}^k\right] + \lambda \mathbf{I} \right) \tag{7.27}$$

$$\frac{\partial I_k}{\partial \mathbf{N_k}} = 2\mathbf{N}_k \left( (1-\gamma) \sum_{i,j} \eta_{ij}^k \cdot \mathbf{B}_{ij}^k + \gamma \sum_{j \rightsquigarrow i,l} \eta_{ij}^k \left(1 - y_{il}\right) \cdot h'[z] \cdot \left[\mathbf{B}_{ij}^k - \mathbf{B}_{il}^k\right] + \lambda \mathbf{I} \right), \tag{7.28}$$

and the gradient of the joint objective as equations 7.29 and 7.30:

$$\frac{\partial J_k}{\partial \mathbf{\Gamma}_k} = w_k^p \frac{\partial I_k}{\partial \mathbf{\Gamma}_k} + 4\mu \mathbf{\Gamma}_k \sum_{q=1, q \neq k}^{K} \sum_{i.j} \left(d_{ij}^k - d_{ij}^q\right) \mathbf{A}_{ij}^k \tag{7.29}$$

$$\frac{\partial J_k}{\partial \mathbf{N_k}} = w_k^p \frac{\partial I_k}{\partial \mathbf{N}_k} + 4\mu \mathbf{N}_\mathrm{k} \sum_{q=1, q \neq k}^{K} \sum_{i.j} \left(d_{ij}^k - d_{ij}^q\right) \mathbf{B}_{ij}^k. \tag{7.30}$$

To estimate the update for the weights, we solve for the Lagrange function given equation 7.31:

$$La(w,\eta) = \sum_{k=1}^{K} w_k^p I_k + \\ \lambda \sum_{k,l=1,k<l}^{K} \sum_{i,j} \left(d_{ij}^k - d_{ij}^l\right)^2 - \eta \left(\sum_{k=1}^{K} w_k - 1\right); \quad (7.31)$$

we estimate the weights as equation 7.32:

$$w_k = \frac{(1/I_k)^{1/(p-1)}}{\sum_{k=1}^{K} (1/I_k)^{1/(p-1)}}. \quad (7.32)$$

The implementation of distance in the multi-view case, i.e. implementation of distance used in the k-NN algorithm, is given as equation 7.33:

$$d(X_i, X_j) = \sum_{k=1}^{K} w_k \mathrm{tr}\left[\mathbf{U}_k \left(X_i^k - X_j^k\right)^T \mathbf{V}_k \left(X_i^k - X_j^k\right)\right] \quad (7.33)$$

We note the following about the algorithm:

1. Similar to $\mathrm{LM}^3\mathrm{L}$ we optimize in two stages at each iteration: freezing the weights and optimizing $\mathbf{\Gamma}_k$ and $\mathbf{N}_k$ with respect to the primary objective function, then freezing the estimates of $\mathbf{\Gamma}_k$ and $\mathbf{N}_k$ and optimizing $w_k$ given the Lagrangian

2. The generation of the gradient for the objective is $\nabla J_k\left(X_i^k; \mathbf{U}_k, \mathbf{V}_k\right) = \left[\frac{\partial J_k}{\partial \Gamma_k}, \frac{\partial J_k}{\partial \mathbf{N}_k}\right]$; simultaneous estimate of the gradient is possible—there no need for flip-flopping the order of operation unlike the estimate of the sample covariance matrices themselves as shown in Glanz and Carvalho [2013].

3. The step sizes for each iteration are estimated using our BB method generated, step sizes for $\mathbf{U}_k$ and $\mathbf{V}_k$ are found independently from each other and from each view, i.e. the equations 7.34 and 7.35

$$\hat{\beta}_k = \frac{1}{2} \cdot \frac{Tr\left[\Delta g(\mathbf{\Gamma}_k) \cdot \Delta\mathbf{\Gamma}_k^H + \Delta\mathbf{\Gamma}_k \cdot \Delta g(\mathbf{\Gamma}_k)^H\right]}{Tr\left[\Delta g(\mathbf{\Gamma}_k) \cdot \Delta g(\mathbf{\Gamma}_k)^H\right]} \tag{7.34}$$

$$\hat{\kappa}_k = \frac{1}{2} \cdot \frac{Tr\left[\Delta g(\mathbf{N}_k) \cdot \Delta\mathbf{N}_k^H + \Delta\mathbf{N}_k \cdot \Delta g(\mathbf{N}_k)^H\right]}{Tr\left[\Delta g(\mathbf{N}_k) \cdot \Delta g(\mathbf{N}_k)^H\right]} \tag{7.35}$$

The algorithm recombines the decomposed matrices to produce the results $\mathbf{U}_k = \mathbf{\Gamma}_k^T\mathbf{\Gamma}_k$ and $\mathbf{V}_k = \mathbf{N}_k^T\mathbf{N}_k$ per view. The methodology is proposed in Algorithm 3.

---

**Algorithm 3** LM$^3$L-MV Algorithm Flow

---

**Require:** $\rho \geq 1$
**Ensure:** $X_k$

**while** $\left|J^{(t)} - J^{(t-1)}\right| < \epsilon$ **do**
  **for** $k = 1, ..., K$ **do**
    Solve $\nabla J_k\left(X_i^k; U_k, V_k\right) = \left[\frac{\partial J_k}{\partial \Gamma_k}, \frac{\partial J_k}{\partial \mathbf{N}_k}\right]$
    $\hat{\beta}_k^t = \frac{1}{2} \cdot \frac{Tr\left[\Delta g(\mathbf{\Gamma}_k) \cdot \Delta\mathbf{\Gamma}_k^H + \Delta\mathbf{\Gamma}_k \cdot \Delta g(\mathbf{\Gamma}_k)^H\right]}{Tr\left[\Delta g(\mathbf{\Gamma}_k) \cdot \Delta g(\mathbf{\Gamma}_k)^H\right]}$
    $\hat{\kappa}_k^t = \frac{1}{2} \cdot \frac{Tr\left[\Delta g(\mathbf{N}_k) \cdot \Delta\mathbf{N}_k^H + \Delta\mathbf{N}_k \cdot \Delta g(\mathbf{N}_k)^H\right]}{Tr\left[\Delta g(\mathbf{N}_k) \cdot \Delta g(\mathbf{N}_k)^H\right]}$
    $\mathbf{\Gamma}_k^{(t+1)} = \mathbf{\Gamma}_k^{(t)} - \beta_k^{(t)} \frac{\partial J_k}{\partial \mathbf{\Gamma}_k}$
    $\mathbf{N}^{(t+1)} = \mathbf{N}^{(t)} - \kappa_k^{(t)} \frac{\partial J_k}{\partial \mathbf{N}_k}$
  **end for**
  **for** $k = 1, ..., K$ **do**
    $w_k = \frac{(1/I_k)^{1/(p-1)}}{\sum_{k=1}^{K}(1/I_k)^{1/(p-1)}}$
    $J_k = w_k I_k$
    $+= \mu \sum_{q=1, q\neq k}^{K} \sum_{i.j} \left(d_{\mathbf{U}_k,\mathbf{V}_k}(X_i^k, X_j^k) - d_{\mathbf{U}_l,\mathbf{V}_l}(X_i^q, X_j^q)\right)^2$
    $J^{(t)} = J^{(t)} + J_k$
  **end for**
**end while**
  **return** $\mathbf{U}_k = \mathbf{\Gamma}_k^T\mathbf{\Gamma}_k$ and $\mathbf{V}_k = \mathbf{N}_k^T\mathbf{N}_k$

---

## 7.4 Implementation

We develop a supporting functional library in Java (java-jdk/11.0.1), and rely on a number of additional publicly available scientific and mathematical open source packages including the Apache foundation commons packages (e.g. Math Commons Foundation, 2018b and Commons Lang Foundation, 2018c) and the JSOFA package to support our designs. The overall functionality is supported at a high level by the following open source packages:

- Maven is used to manage dependencies, and produce executable functionality from the project Foundation [2018a]
- JUnit is used to support library unit test management [Team, 2018a]
- slf4j is used as a logging frame work [Team, 2017]
- MatFileRW is used for I/O handling [Team, 2018b]

We recommend reviewing the vsa-parent .pom file included as part of the software package for a more comprehensive review of the functional dependency. Versions are subject to upgrades as development proceeds beyond this publication. Execution of the code was performed on a number of platforms including a personal laptop (MacBook Pro, 2.5GHz Intel Core i7, macOS Mojave) and an institution high performance computer (Florida Institute of Technology, BlueShark HPC)[3]. The development of the library and functionality in Java allow for the functionality presented here to be applied regardless of platform. We are not reporting processing times as part of this analysis as the computational times varied depending on platform used. Our initial research included using the parallel computing functionality packaged with Java, in combination with the GPU functionality on the BlueShark computer. Further research is necessary to quantify optimal implementation with respect to convergence speed and memory usage.

[3]https://it.fit.edu/information-and-policies/computing/blueshark-supercomputer-hpc/

### 7.4.1 Optimization of Features

Similar to Johnston and Peter [2017], we use the University of California Riverside Time Series Classification Archive (UCR) and the Lincoln Near-Earth Asteroid Research (LINEAR) dataset to demonstrate the performance of our feature space classifier. The individual datasets are described as follows:

- **UCR**: We baseline the investigated classification methodologies [Keogh et al., 2011] using the UCR time domain datasets. The UCR time domain dataset STARLIGHT [Protopapas et al., 2006] is derived from a set of Cepheid, RR Lyra, and Eclipsing Binary Stars. This time-domain dataset is phased (folded) via the primary period and smoothed using the SUPER–SMOOTHER algorithm [Reimann, 1994] by the Protopapas study prior to being provided to the UCR database. Note that the sub-groups of each of the three classes are combined together in the UCR data (i.e., RR Lyr (ab) + RR Lyr (c) = RR).

- **LINEAR**: The original database LINEAR is subsampled; we select time series data that has been verified and for which accurate photometric values are available [Sesar et al., 2011, Palaversa et al., 2013]. This subsampled set is parsed into separate training and test sets. From the starting sample of 7,194 LINEAR variables, a clean sample of 6,146 time series datasets and their associated photometric values is used. Stellar class-type is limited further to the top five most populous classes: RR Lyr (ab), RR Lyr (c), $\delta$ Scuti / SX Phe, Contact Binaries and Algol-Like Stars with 2 Minima, resulting in a set of 6,086 observations.

Training data subsets are generated as follows: UCR already defines a training and test set, the LINEAR data is split into a training and test set using a predefined algorithm (random assignment, of nearly equal representation of classes in training and test). We used a method of 5-fold cross-validation both datasets; the partitions in 5-fold algorithm are populated via random assignment. For more details on the datasets themselves, a baseline for performance, and additional references, see Johnston and Peter [2017].

### 7.4.2 Feature Optimization

The time domain transformation we selected requires parameter optimization (resolution, kernel size, etc.); each survey can potentially have a slightly different optimal set of transformation parameters with respect to the underlying survey parameters (e.g. sample rate, survey frequency, number of breaks over all observations, etc.). While we could include the parameters optimization in the cross-validation process for the classifier, this will be highly computationally challenging, specifically for classifier that require iterations, as we would be handling an increasing number of permutations with each iteration, over an unknown number of iteration. To address this problem, the feature space is cross-validated on the training dataset, and k-NN classification is used (assuming a fixed temporary $k$ value allows little to no tuning) to estimate the misclassification error with the proposed feature space parameters. The optimized features are used as givens for the cross-validation process in optimizing the intended classifier. Likely some loss of performance will occur, but considering how the final classifier design is based on k-NN as well, it is expected to be minor.

Because of the multi-dimensional nature of our feature space, we propose the following method for feature optimization—per class we generate a mean representation of the feature (given the fraction of data being trained on), all data are then transformed (training and cross-validation data) via Park et al. [2003] into a distance representation, i.e. the difference of the observed feature and each of the means is generated. Note that for the matrix feature spaces, the Frobenius Norm is used. Alternatively we could have generated means based on unsupervised clustering (k-Means); while not used in this study, this functionality is provided as part of the code. We found that the performance using the unsupervised case was very sensitive to the initial number of $k$ used. For the LINEAR and UCR datasets, the results were found with respect to optimization of feature (DF and SSMM) parameters to be roughly the same. A k-NN algorithm is applied to the reduced feature space, 5-fold classification is then used to generate the estimate of error, and the misclassification results are presented a response map given feature parameters (Figures 7.1 and Figure 7.2):

We select the optimum values for each feature space, based on a minimization of both the LINEAR and UCR data. These values are estimated to be: DF Optimized (x, y) – $30 \times 25$, SSMM Optimized (res x scale) – $0.06 \times 35.0$

### 7.4.3 Large Margin Multi-Metric Learning

The implementation of $LM^3L$ is applied to the UCR and LINEAR datasets. Based on the number of views associated with each feature set, the underlying classifier

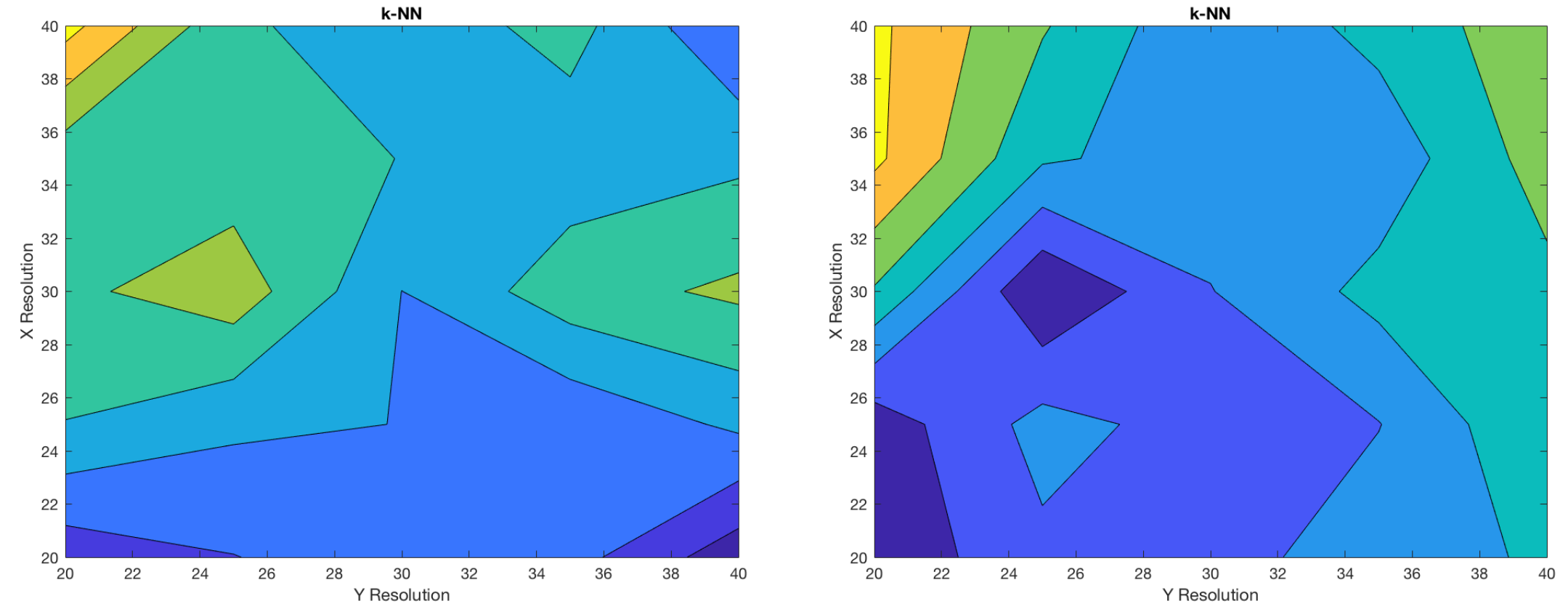

Figure 7.1: Parameter optimization of the Distribution Field feature space (Left: UCR Data, Right: LINEAR Data). Heat map colors represent misclassification error, (dark blue—lower, bright yellow—higher)

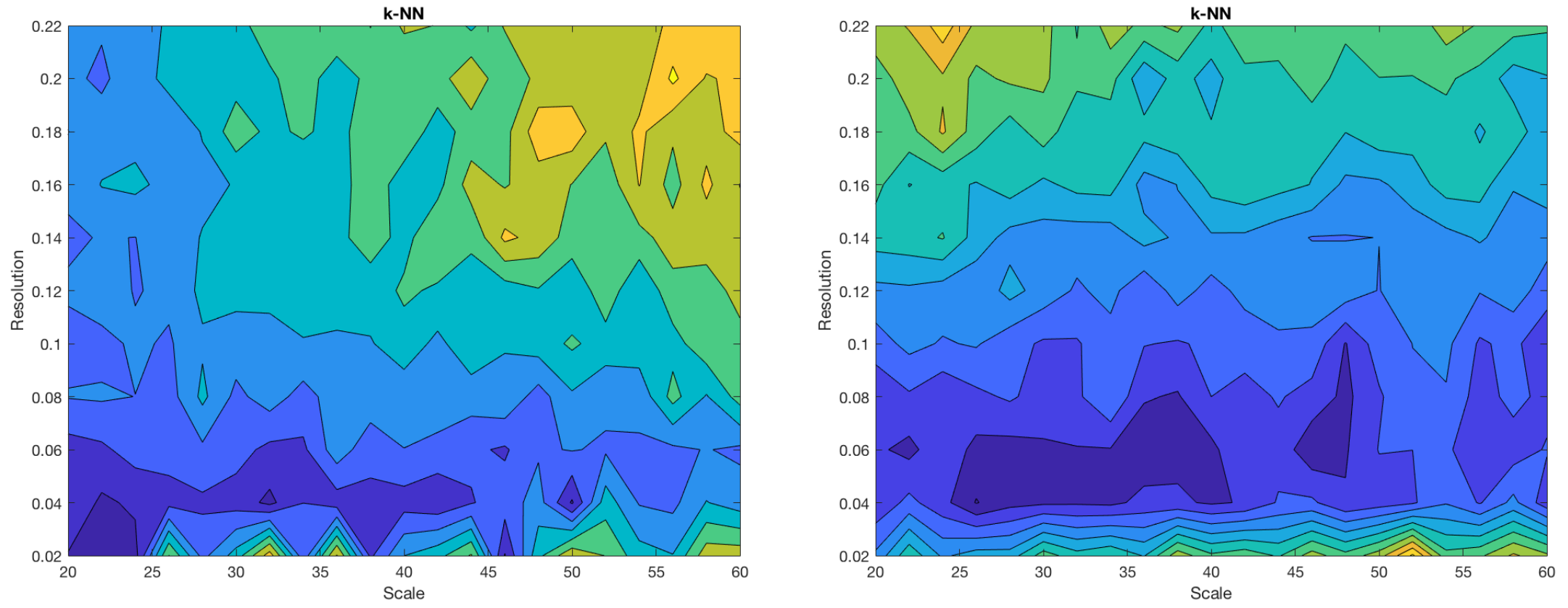

Figure 7.2: Parameter optimization of the SSMM feature space (Left: UCR Data, Right: LINEAR Data). Heat map colors represent misclassification error, (dark blue—lower, bright yellow—higher)

will be different (e.g. UCR does not contain color information and it also has only three classes compared to LINEAR's five). The features SSMM, DF, and Statistical Representations (mean, standard deviation, kurtosis, etc.) are computed for both datasets. Color and the time domain representations provided with the LINEAR data are also included as additional views.

To allow for the implementation of the vector-variate classifier, the dimensionality of the SSMM and DF features are reduced via vectorization of the matrix and then processing by the ECVA algorithm, resulting in a dimensionality that is $k-1$, where $k$ is the number of classes. We note that without this processing via ECVA, the times for the optimization became prohibitively long, this is similar to the implementation of IPMML given in Zhou et al. [2016]. SSMM and DF features are generated with respect to the LINEAR dataset—Park's transformation is not applied here—the feature space reduced via ECVA, the results are and given in Figure 7.3 (DF) and Figure 7.4 (SSMM).

Similarly, the SSMM and DF features are generated for the UCR dataset—Park's transformation is not applied here–and the feature space reduced via ECVA. The results are plotted and given in Figure 7.5 (DF–Left) and (SSMM–Right).

The dimensions given in the figures are reduced dimensions resulting from the ECVA transform and therefore they do not necessarily have meaningful descriptions (besides $x_1, x_2, x_3, ..., x_n$). These reduced feature spaces are used as input to the LM$^3$L algorithm.

The individual views are standardized (subtract by mean and divide by standard deviation). Cross-validation of LM$^3$L is used to optimize the three tunable parameters and the one parameter associated with the k-NN. The LM$^3$L authors recommend some basic parameter starting points; our analysis includes investigating the tunable values as well as an upper (+1) and lower (-1) level about each parameter, over a set of odd k-NN values [1,19]; the optimization only needs to occur for each set of tunable values, the misclassification given a k-Value can be evaluated separately, this experiment is outlined in Table 7.1.

Cross-validation is performed to both optimize for our application and investi-gate the sensitivity of the classifier to adjustment of these parameters. For a break down of the cross validation results, see the associated datasets and spreadsheet provided as part of the digital supplement.

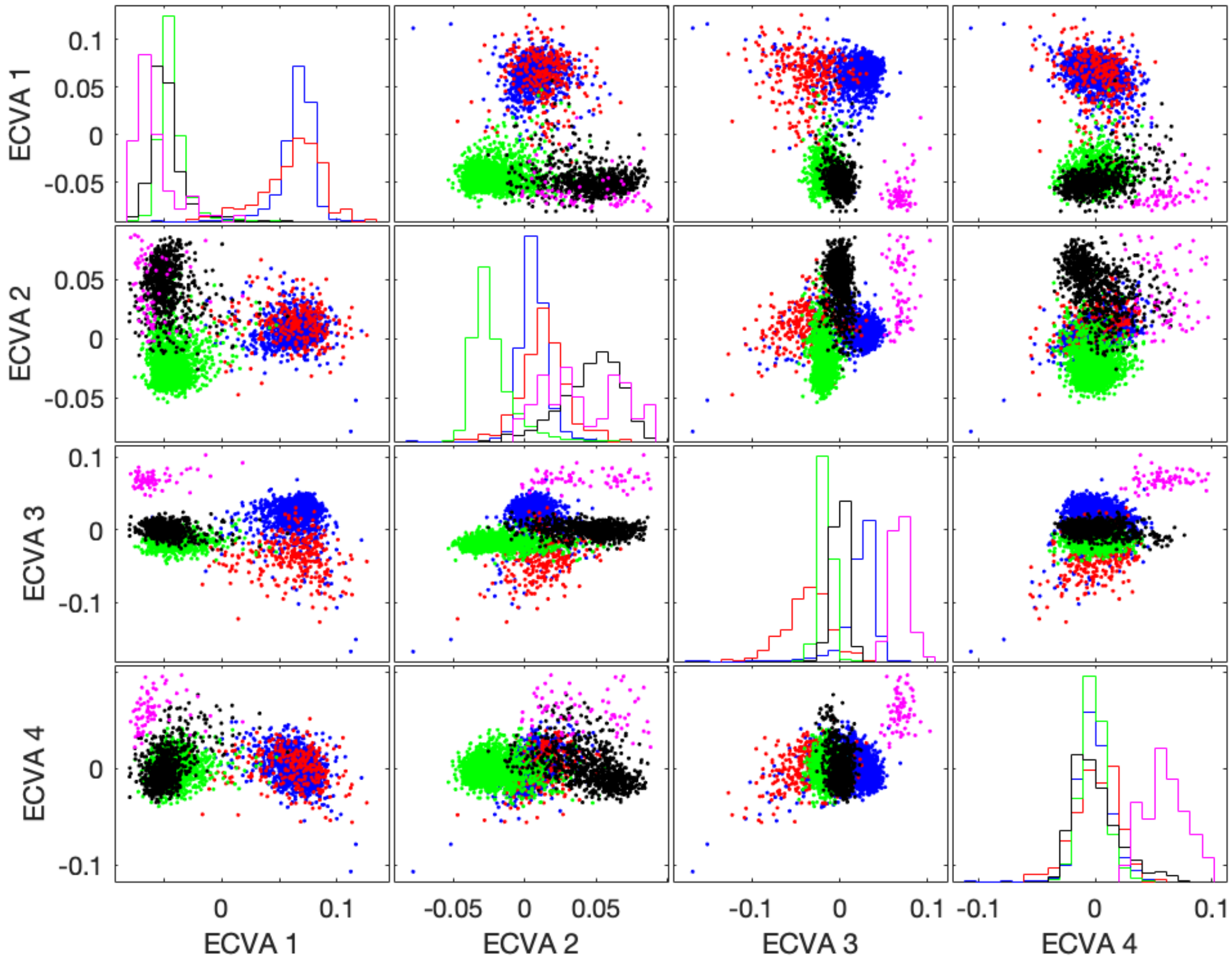

Figure 7.3: DF Feature space after ECVA reduction from LINEAR (Contact Binary/blue circle, Algol/ red +, RRab/green points, RRc in black squares, Delta Scu/SX Phe magneta diamonds) off-diagonal plots represent comparison between two different features, on-diagonal plots represent distribution of classes within a feature (one dimensional)

#### 7.4.3.1 Testing and Results (UCR & LINEAR)

Based on the cross-validation process, the following optimal parameters are found:

- *LINEAR*: k-NN(11), $\tau(1.0)$, $\mu(5.0)$, $\lambda(0.1)$

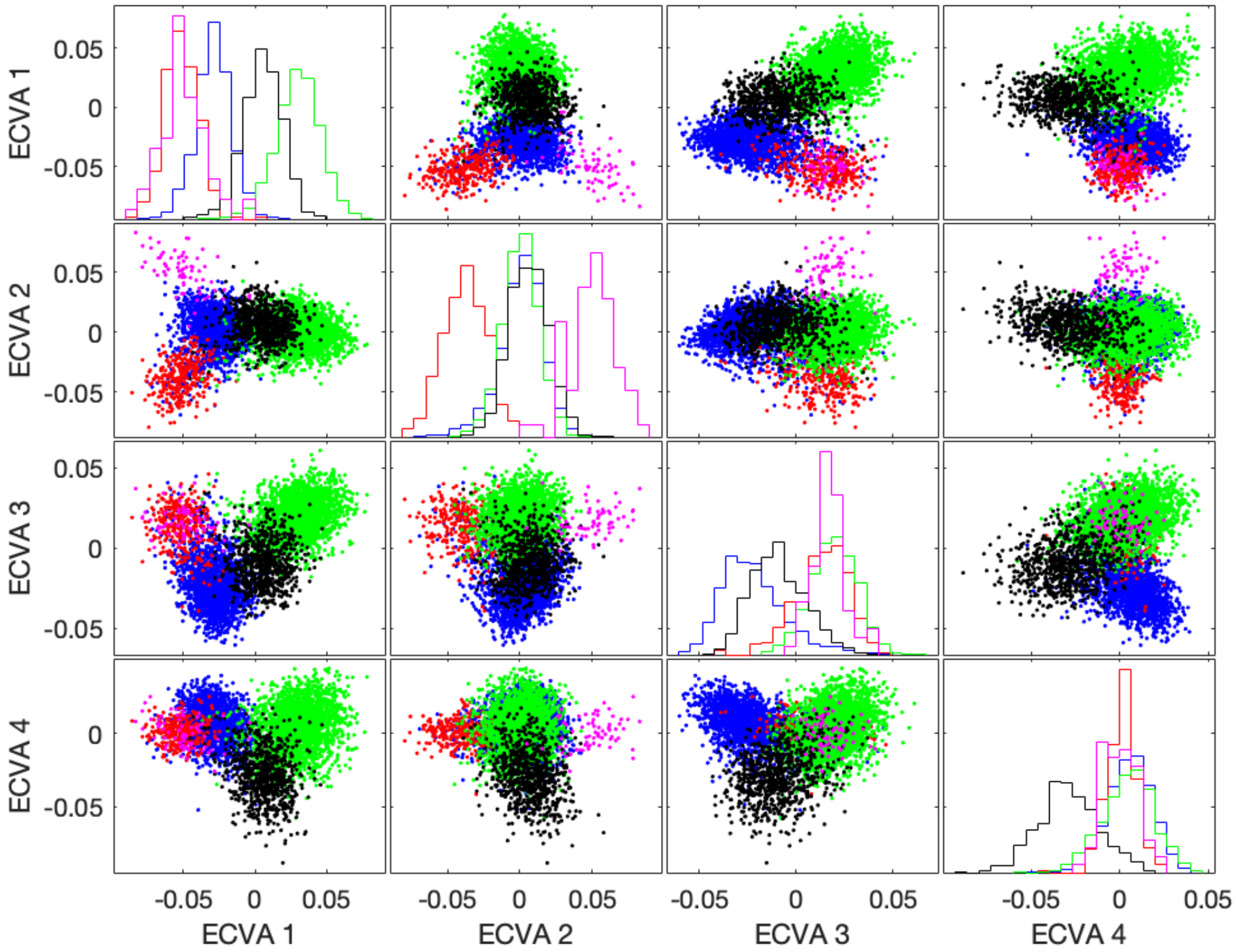

Figure 7.4: SSMM feature space after ECVA reduction LINEAR (Contact Binary/blue circle, Algol/ red +, RRab/green points, RRc in black squares, Delta Scu/SX Phe magneta diamonds) off-diagonal plots represent comparison between two different features, on-diagonal plots represent distribution of classes within a feature (one dimensional)

- *UCR*: k-NN(9), $\tau(1.0)$, $\mu(5.0)$, $\lambda(0.1)$

The classifier is then trained using the total set of training data along with the optimal parameters selected. As a reminder, the $\lambda$ parameter controls the importance of regularization, the $\mu$ parameter controls the importance of pairwise distance in the optimization process, and $\gamma$ controls the balance between push and pull.

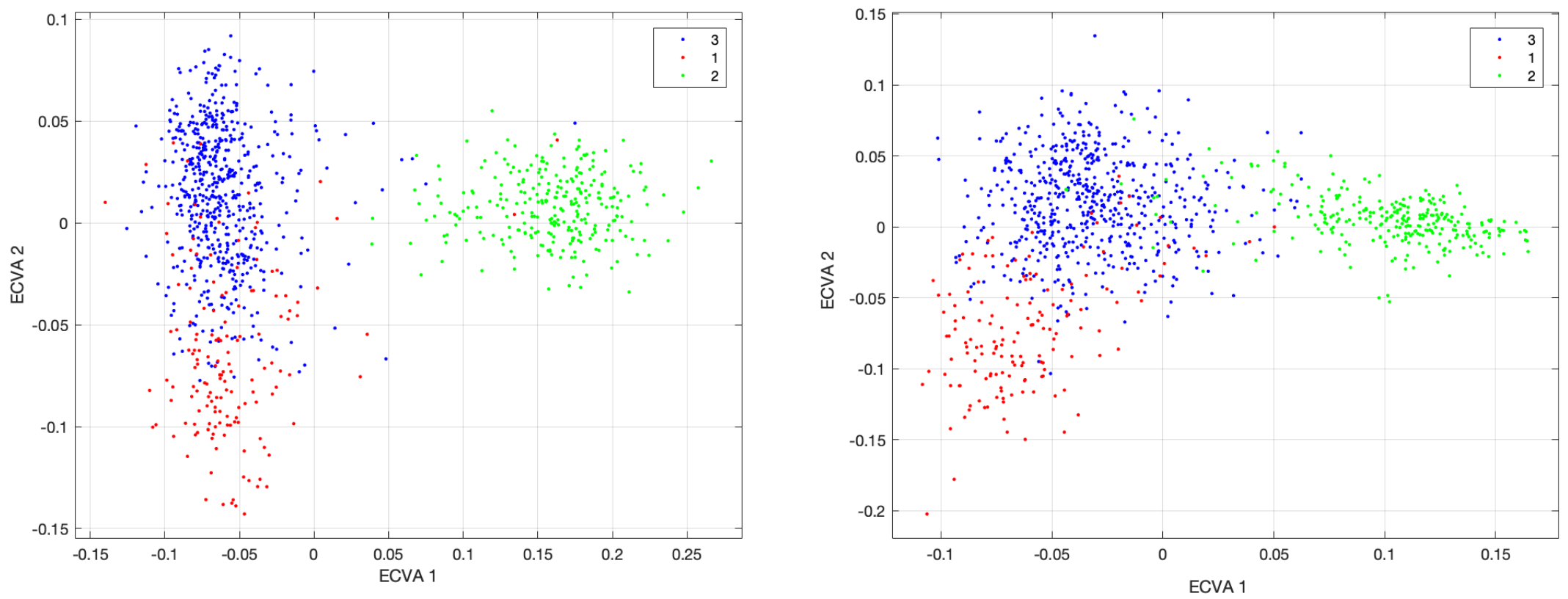

Figure 7.5: DF (Left) and SSMM (Right) feature space after ECVA reduction from UCR. Class names (1,2, and 3) are based on the classes provided by the originating source and the UCR database

Table 7.1: The cross-validation process for $LM^3L$ tunable values

| Variable | 1 | 2 | 3 | 4 | 5 | 6 | 7 |
|---|---|---|---|---|---|---|---|
| $\tau$ | 1 | 1.75 | 0.25 | 1 | 1 | 1 | 1 |
| $\mu$ | 5 | 5 | 5 | 8 | 2 | 5 | 5 |
| $\lambda$ | 0.5 | 0.5 | 0.5 | 0.5 | 0.5 | 1.0 | 0.1 |

The trained classifier is applied to the test data, the confusion matrices [Fawcett, 2006] resulting from the application are presented in Table 7.2 and Table 7.3:

Table 7.2: LINEAR confusion matrix via $LM^3L$ entries are counts (percent)

| Misclassification Rate | RR Lyr (ab) | $\delta$ Scu / SX Phe | Algol | RR Lyr (c) | Contact Binary | Missed |
|---|---|---|---|---|---|---|
| RR Lyr (ab) | 1081 (0.992) | 0 (0.000) | 0 (0.000) | 6 (0.006) | 1 (0.001) | 2 (0.002) |
| Delta Scu / SX Phe | 0 (0.000) | 23 (0.852) | 0 (0.000) | 2 (0.074) | 2 (0.074) | 0 (0.000) |
| Algol | 1 (0.007) | 0 (0.000) | 108 (0.788) | 0 (0.000) | 28 (0.204) | 0 (0.000) |
| RR Lyr (c) | 23 (0.062) | 0 (0.000) | 1 (0.003) | 343 (0.925) | 4 (0.011) | 0 (0.000) |
| Contact Binary | 3 (0.003) | 0 (0.000) | 29 (0.033) | 9 (0.010) | 832 (0.952) | 1 (0.001) |

### 7.4.4 Large Margin Multi-Metric Learning - MV

The implementation of $LM^3L$-MV is applied to the UCR and LINEAR datasets. The features SSMM, DF, and Statistical Representations (mean, standard deviation, kurtosis, etc.) are computed for both datasets. Similar to the $LM^3L$ proce-

Table 7.3: UCR confusion matrix via $LM^3L$ entries are counts (percent)

| Misclassification Rate | 2 | 3 | 1 | Missed |
|---|---|---|---|---|
| 2 | 2296 (0.996) | 9 (0.004) | 0 (0.000) | 0 (0.000) |
| 3 | 17 (0.004) | 4621 (0.972) | 116 (0.023) | 0 (0.000) |
| 1 | 8 (0.007) | 375 (0.319) | 794 (0.675) | 0 (0.000) |

dure, color and the time domain representations provided with the LINEAR data are also included as additional views. The implementation of the matrix-variate classifier, allows us to avoid the vectorization and feature reduction (ECVA) step. The individual views are standardized prior to optimization. Also similar to the LM$^3$L procedure, cross-validation of LM$^3$L-MV is used to optimize the three tunable parameters and the one parameter associated with the k-NN. The table of explored tunable parameters is given in Table 7.4:

Table 7.4: The cross-validation process for $LM^3L - MV$ tunable values

| Variable | 1 | 2 | 3 | 4 | 5 | 6 | 7 |
|---|---|---|---|---|---|---|---|
| $\lambda$ | 0.5 | 1.0 | 0.25 | 0.5 | 0.5 | 0.5 | 0.5 |
| $\mu$ | 0.5 | 0.5 | 0.5 | 1.0 | 0.25 | 0.5 | 0.5 |
| $\gamma$ | 0.5 | 0.5 | 0.5 | 0.5 | 0.5 | 1.0 | 0.25 |

For a break down of the results, see the associated datasets and spreadsheet provided as part of the digital supplement.

#### 7.4.4.1 Testing and Results (UCR & LINEAR)

Based on the cross-validation process, the following optimal parameters are found (and their cross-validation error estimates):

- *LINEAR*: k-NN(15), $\lambda(0.5)$, $\mu(0.5)$, $\gamma(0.5)$
- *UCR*: k-NN(19), $\lambda(0.5)$, $\mu(1.0)$, $\gamma(0.5)$

The classifier is then trained using the total set of training data along with the optimal parameters selected. The trained classifier is applied to the test data, the confusion matrices resulting from the application are presented in Table 7.5 and Table 7.6:

Table 7.5: LINEAR confusion matrix via $LM^3L-MV$ entries are counts (percent)

| Misclassification Rate | RR Lyr (ab) | Delta Scu / SX Phe | Algol | RR Lyr (c) | Contact Binary | Missed |
|---|---|---|---|---|---|---|
| RR Lyr (ab) | 1074 (0.985) | 0 (0.000) | 1 (0.001) | 15 (0.014) | 0 (0.000) | 0 (0.000) |
| Delta Scu / SX Phe | 1 (0.037) | 24 (0.889) | 0 (0.000) | 2 (0.074) | 0 (0.000) | 0 (0.000) |
| Algol | 3 (0.022) | 0 (0.000) | 104 (0.759) | 1 (0.007) | 29 (0.212) | 0 (0.000) |
| RR Lyr (c) | 23 (0.059) | 0 (0.000) | 1 (0.003) | 343 (0.930) | 4 (0.008) | 0 (0.000) |
| Contact Binary | 3 (0.003) | 0 (0.000) | 29 (0.035) | 9 (0.001) | 832 (0.958) | 1 (0.002) |

Table 7.6: UCR confusion matrix via $LM^3L-MV$ entries are counts (percent)

| Misclassification Rate | 2 | 3 | 1 | Missed |
|---|---|---|---|---|
| 2 | 2298 (0.997) | 6 (0.003) | 0 (0.000) | 1 (˜0.000) |
| 3 | 4 (0.001) | 4450 (0.936) | 300 (0.063) | 0 (0.000) |
| 1 | 3 (0.003) | 467 (0.397) | 707 (0.601) | 0 (0.000) |

### 7.4.5 Comparison

The matrix-variate and the vector-variate versions do not perform much different under the conditions provided given the data observed. However, as a reminder, the LM$^3$L implementation includes a feature reduction methodology (ECVA) that our LM$^3$L-MV does not. The ECVA front end was necessary as the dimensionality of the unreduced input vectors results in features and metrics which are prohibitively large (computationally). It is not entirely surprising that our two competitive methodologies perform similarly, with the LM$^3$L algorithm of having the benefit of being able to process the matrix-variate spaces ahead of time via ECVA and thus being able to process the SSMM and DF features spaces in a lower dimension ($c-1$ dimensions). For a quantitative comparison of our classifiers, we have computed precision and recall metrics for our presented classifiers [Fawcett, 2006, Sokolova and Lapalme, 2009]as well as an overall estimate of F1-score, the results of this analysis are presented in Table 7.7.

A direct one-to-one comparison to other pattern classification methods is difficult, as there are no other classifiers that we know of that are both multi-view and matrix-variate. We can however provide some context by looking at known alternatives that may partially address our particular situation. We have included results based on the implementation of a multi-view k-NN classifier (7.33) in both the matrix-variate and vector-variate domains, but with the metrics being the identity matrix (i.e. Euclidean and Forbinus distances respectively), as a baseline reference point. Similarly, we have included results based on the implementation of a single view k-NN classifier (7.33) applied to the vectorized and concatenated features (i.e. Euclidean distance).

For comparison we include classifiers generated from the main individual feature spaces; optimization was performed using Random Forest classification [Breiman et al., 1984]. In addition to these standard methods applied to the unreduced feature space/views, we have generated classifiers based on the dimensionally reduced feature space generated resulting from the ECVA algorithm applied to the DF and SSMM vectorized feature spaces. These reduced feature spaces/views are implemented using the Zhou et al. [2016] IPMML (i.e., multi-view algorithm), this implementation is the nearest similar implementation to both our $LM^3L$ and $LM^3L$-MV algorithm designs. Detailed computations associated with all analyses are included as part of the digital supplement.

It should be noted, that ECVA has its limitations; anecdotally on more then one occasion during the initial analysis, when the full dataset was provided to the algorithm, the memory of the machine was exceeded. Care was taken with the LINEAR dataset to develop a training dataset that was small enough that the out-of-memory error would not occur, but a large enough that each of the class-types was represented sufficiently. Similarly, the projection into lower dimensional space meant that the LM$^3$L implementation iterated at a much faster rate with the same amount of data, compared to the LM$^3$L-MV algorithm. The matrix multiplication operations associated with the matrix distance computation are more computationally expensive compared to the simpler vector metric distance computation, however many computational languages have been optimized for matrix multiplication (e.g., MATLAB, Mathmatica, CUDA, etc.). Again, the time the ECVA algorithm takes to operate upfront saves the LM$^3$L iterations time. In general, both algorithms perform well with respect to misclassification rate, but both also require concessions to handle the scale and scope of the feature spaces used. The cost of most of these concession can be mitigated with additional machine learning strategies, some of which we have begun to implement here—parallel computation for example.

Table 7.7: F1-Score metrics for the proposed classifiers with respect to LINEAR and UCR datasets

| F1-Score | UCR | LINEAR |
|---|---|---|
| $LM^3L$ | 0.904 | 0.918 |
| $LM^3L - MV$ | 0.860 | .916 |
| $IPMML$ | 0.900 | 0.916 |
| $k - NN$ Multi-view MV | 0.725 | 0.574 |
| $k - NN$ Multi-View | 0.691 | 0.506 |
| $k - NN$ Concatenated | 0.650 | 0.427 |
| $RFDF$ | 0.878 | 0.650 |
| $RFSSMM$ | 0.659 | 0.402 |
| $RFTimeStatistics$ | 0.678 | 0.787 |

## 7.5 Conclusions

The classification of variable stars relies on a proper selection of features of interest and a classification framework that can support the linear separation of those features. Features should be selected that quantify the signature of the variability, i.e. its structure and information content. Here, two features which have utility in providing discriminatory capabilities, the SSMM and DF feature spaces are studied. The feature extraction methodologies are applied to the LINEAR and UCR dataset, as well as a standard breakdown of time domain descriptive statistics, and in the case of the LINEAR dataset, a combination of *ugriz* colors. To support the set of high-dimensionality features, or views, multi-view metric learning is investigated as a viable design. Multi-view learning provides an avenue for integrating multiple transforms to generate a superior classifier. The structure of multi-view metric learning allows for a number of modern computational designs to be used to support increasing scale and scope (e.g., parallel computation); these considerations can be leveraged given the parameters of the experiment designed or the project in question.

Presented, in addition to an implementation of a standard multi-view metric learning algorithm (LM$^3$L) that works with a feature space that has been vectorized and reduced in dimension, is a multi-view metric learning algorithm designed to work with matrix-variate views. This new classifier design does not require transformation of the matrix-variate views ahead of time, and instead operates directly on the matrix data. The development of both algorithm designs (matrix-variate and vector-variate) with respect to the targeted experiment of interest (discrimination of time-domain variable stars) highlighted a number of challenges to be addressed prior to practical application. In overcoming these challenges, it was found that the novel classifier design ($LM^3L$-MV) performed on order of the staged (Vectorization + ECVA + $LM^3L$) classifier. Future research will include investigating overcoming high dimensionality matrix data (e.g. SSMM), improving the parallelization of the design presented, and implementing community standard workarounds for large dataset data (i.e., on-line learning, stochastic/batch gradient descent methods, k-d tree... etc.).

# Chapter 8

# Conclusions

We focused our efforts on the field of astroinformatics, specifically machine learning relating to time-domain features and variable star identification. We outlined a vertically integrated analysis: the collection of training data and the development of a time-domain feature space, a classification/detection optimized algorithm, and a performance analysis procedure that can properly represent the classifier performance. We assume that the survey will handle much of the signal conditioning and detection (front end logic). Most of what we are designing is the "back end" and includes both development and testing (verification/validation).

## 8.1 Variable Star Analysis

The development of the Variable Star Analysis (VarStar) library is a capstone for this research.[1] The VarStar library is a Java library and contains not only the novel analysis methods used within the research (SSMM and DF) and the developed machine learning code unique to this project (LM$^3$L-MV) but also the set of fundamental mathematics and supporting functions necessary for computation. The code can be split into three basic categories: bindings (data objects that act as containers for similar information), utilities (mathematics, machine learning, and other generic tools), and analysis (features and transforms used in our research as well as executable functions that were developed for the research). The design of the library functions flows from the math bundle, downward, with reliance on a multitude of third-party open source packages. The flow design is given in Figure 8.1.

[1] https://github.com/kjohnston82/VariableStarAnalysis

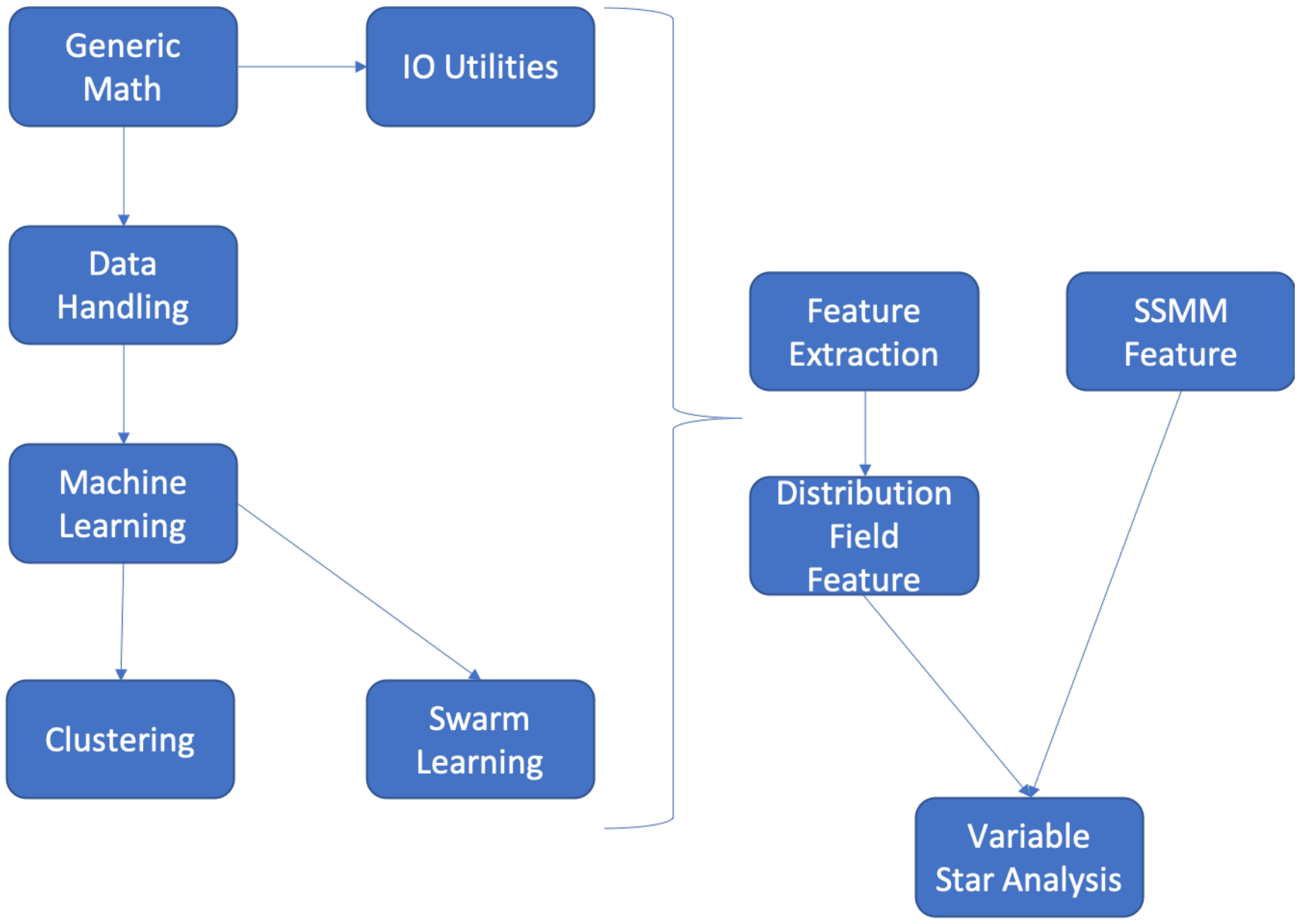

Figure 8.1: A rough outline of the Variable Star Analysis Library (JVarStar) bundle functional relationships. Notice that the generic math and utility bundles flow down to more specific functional designs such as the clustering algorithms.

We briefly outline the contents of the library functionality; for more detail, please see the code itself. The underlying library is actively developed in Java (java-jdk/11.0.1) and relies on a number of additional publicly available scientific and mathematic open source packages, including the Apache foundation commons packages (e.g., Math Commons [Foundation, 2018b] and Commons Lang [Foundation, 2018c]) and the JSOFA package [Harrison, 2016]. The Apache Math Commons package specifically provides the Java objects necessary for the handling of vector and matrix mathematics; linear algebra within the VSA library is dependent on this package and the **VectorOperations/MatrixOperations** classes that extend the Math Common's functionality for use.

The VSA math bundle contains low-level functionality, such as numeric tests (e.g., is even? is odd?) and numerical constants (e.g., $4\pi^2$) that are common in scientific applications. The math bundle also contains geometric functionality and algorithms useful in scientific applications, such as the oriented Graham scan [Graham, 1972] for generating 2-D convex hulls given a set of distributed points, algorithms for the random distribution generation on surfaces, and an implementation of the QuickHull3D algorithm [Barber et al., 1996] for generating convex hulls and intersection of convex hulls given a set of distributed points in ND space. Fundamental object and geometry mathematics is also included, and the bundle has classes for cone and plane shapes. Similarly, the math bundle contains linear and nonlinear solver methods, including a polynomial solver, a weighted multiple linear regression solver, and functionality that uses or ingests functionality inherent to the Apache Math Common's analysis bundle (see http://commons.apache.org/proper/commons-math/userguide/analysis.html for more information).

The I/O utilities bundle provides support for accessing and writing data into and out of multiple formats, including .mat file formats (MATLAB). The VSA design relies on the storage of data in a .mat format, which allows for easy analysis of results in a scripting language (MATLAB); the storage of data files in a .mat format is also efficient, as the MatFileRW functionality used for I/O handling [Team, 2018b] provides the ability to store structures and other fundamental data objects. More complex data I/O is also handled; this requires, however, that data objects be serializable.[2]

While there is a singular machine learning bundle, the clustering bundle, data handling bundle, and swarm optimization bundle also contain machine learning functionality for the user, but have been split apart for the sake of development. The machine learning bundle is a Java implementation of the MATLAB project MatLearn[3] and contains several standard machine learning algorithms, such as the classification and regression tree (CART), the logistic regression algorithm (LRC), the k-NN algorithm, linear and quadratic discriminate analysis (LDA/QDA), canonical variate analysis (CVA), Parzen window classsification (PWC), and a set of metric learning algorithms (LMNN, NCA, S&J, MMC, ITML, etc.). As discussed, this is also where we have developed the $LM^3L$-MV algorithm. The machine learning also contains fundamental functionalities, such as classes for distance measurement (e.g., MetricDistance), performance utilities for evaluation of classifiers such as confusion matrix functionality, and classes designed to support supervised classification (training, cross-validation, and testing).

The data handling bundle contains functionality for managing data relationships between pattern, label, and view. Classes here support the sorting and separation of data based on these three (pattern, label, and view) data dependencies. The clustering bundle contains classes that support k-means and EM-GMM clustering and includes some of the development work done to extend k-means into the matrix-variate space. In support of the development of the $LM^3L$-MV algorithm, some research was performed on the topic of particle swarm optimization as a means of improving the iterative optimization design to move away from the standard gradient descent algorithm. At the end of the day, this research was not pursued, but the functionality developed for the optimization remains in the Swarm Optimization bundle.

---

[2]https://docs.oracle.com/javase/tutorial/jndi/objects/serial.html

[3]https://github.com/kjohnston82/MatLearn

Beyond the fundamental building blocks that the utilities package represents is the analysis package, which contains both the unique feature spaces we have developed and more standard feature spaces, such as the Lomb–Scargle transform. These feature transformations are then tied together in the package Variable Star Analysis, and this bundle handles reading in of data, processing of the individual raw waveforms, application of the features to the data, data handling and workflow management of the patterns/label pairing, and training of the targeted supervised classification algorithm.

The Maven functionality that stitches all of the packages together handles also the dependency management and provides the ability to compile and generate executables. This allows for development and distribution of executable training algorithms and .jar functionality that can be transferred and batch run for training purposes. Furthermore, the Java library structure, and Maven project management, will allow others to interface with all or part of the VSA design based on user needs.

The project software is provided publicly at the associated GitHub repository.[4] The overall functionality is supported at a high level by the following open source packages: Maven is used to manage dependencies and produce executable functionality from the project Foundation [2018a], JUnit is used to support library unit test management [Team, 2018a], and slf4j is used as a logging framework [Team, 2017]. A more complete view of the dependencies and versions can be found as part of the VSA-parent .pom file included as part of the software package.

## 8.2 Future Development

As demonstrated, our research has produced new features of use, a new classifier, a review of design for supervised classification systems (including methods of performance analysis), and the application of these methods in the construction of a detector. Additionally, this has produced a body of code that has been made open to the public for further development and use. Avenues of future development include research of additional features, classifiers, and detector designs, expanding upon what we have produced so far. Additionally, we identify two specific efforts that are necessary to improving our designs: increasing the number of standard variable star data and developing a synthetic stellar variable waveform generator.

### 8.2.1 Standard Data Sets

With the development of new feature extraction methodologies and classification techniques, estimations of performance are necessary. However, we have found that there are no clear standard data sets against which proposed methods can be compared. We have used the UCR time series data set "Starlight" as a standard in our SSMM paper [Johnston and Oluseyi, 2017]. Proposed is the collection, labeling, and benchmarking of publicly available time-domain data from variable stars, allowing multiple services to train on similar data (and thus allowing for like comparisons). So far, we have limited ourselves to a few surveys. Our real interest is whether a feature space/classifier pairing performs well for each survey independently, regardless of survey parameters like limiting magnitude or sample rate.

[4]https://github.com/kjohnston82/VariableStarAnalysis

We have started the process of collecting public variable star survey data sets. The data sets must have the original light curve, where the filter and the relative times (from initial observation) must be known. The data must also be labeled; stars must have been identified as being of a particular type of variable. We intend to set up a database of labeled stellar variability data for the astronomical community to test against, similar to the UCR database for time-domain data. These data will be different from UCR. For example, the data set will include varying length data (different length light curves) per target/star, as would be the case with astronomical data. The data may have additional information as well—color, for example—and so would be multi-view. This is important, as these are challenges faced by any astronomer wanting to do time-domain machine learning.

The setup of a database of variable star training resources is a twofold effort. First, we have already begun to collect data to store. The process has been on us to collect, work with, and interview holders of historical labeled data sets. We intend to construct a publicly accessible database with a web-based front end to be hosted by Florida Tech (ICE) and ViziR. Second, one of the benefits of the UCR database is that baseline performance is provided to the users of the data. A standard methodology for training would be implemented for each data set, and the performance (misclassification rate) would be presented for each data set, allowing users to compare against a standard. The algorithm would implement a simple version of our multiclass classifier as a standard classifier and would be publicly available via open source code.

Based on our initial efforts on this front, we have identified a set of challenges. First, we have already run into issues acquiring publicly available training data. Finding variable star light data sets for which the original light curve areas are available (as opposed to reduced data sets) has been problematic. The locations and ownership of data are scattered (decentralized), and often the light curves for the original data are no longer available for certain machine learning studies. Second, collecting the data to host will be an effort, as will be hosting the data (making it publicly available). Considering the amount of data in any given survey, the storage and access logistics associated with such a database deserve some consideration.

### 8.2.2 Simulation

In lieu of real data, often it is beneficial to develop an empirical simulator to generate synthetic signatures that can be used to test the designed system. The development of a synthetic simulator has a number of benefits, training data are always available, and the development of a simulator often requires understanding of the defining qualities of a classification type (here variable type). We answer the question, what makes this particular variable type unique in observation? So, how do we test the performance of either the feature extraction methodologies or the supervised classification methodologies we are generating? Johnston and Oluseyi [2017] highlight standards in performance analysis methods for supervised classification (performance metrics). These performance estimates are dependent on the initial labels given for the training data. Training data labels are completely dependent on hand-labeling of survey data (supervised). We ask the question, how can we provide data to the supervised classification algorithm, where we know for certain the label? To address this challenge, we propose developing a synthetic stellar variable generation.

Developing such an algorithm would require understanding the basic definition of the stellar variable (what makes type A unique from other types). Proper synthesis requires understanding the distribution of features and codified descriptions of the variable star types. These require a methodology to formulate a distribution of the features. This would include both the random generation of scalar features and the development of a random generator to handle time-domain functions. The generated features would need to be correlated with one another as well (we are trying to interpolate between variable star examples). In addition to the generator, we also construct an algorithm for the removal of synthetic signals that approximate various survey conditions: time spent on target, day/night breaks, error in magnitude (noise model), and so on. We define two representations:

- the archetype: the fundamental representation of what makes type A of that type might be the "first" observed variable star of that type, a pinnacle example
- the generator: the set of feature distributions that describes the range of the variable star class type

We propose to construct such a setup to experiment with or test our supervised classification system. The injection of synthetic data would allow us to determine the extrema that will still produce a result in classification (minimum frequency, minimum amplitude changes, inconsistent cyclic pattern, etc.). Synthetic models can provide indications as to what features work when and what survey conditions are necessary for the classification algorithm to operate. To develop prior probability estimates and feature space likelihood distributions, we use labeled survey data from the identified (and available) surveys. We can also use stellar models to help guide our distribution estimates, especially when we are attempting to model the extrema of any given class.

We have attempted an initial effort; however, based on our initial research, we have identified a set of challenges. As discussed by Sterken and Jaschek [2005], there is no standard compendium of variable stars. The American Variable Star Association is in charge of monitoring, coordinating, and defining variable star types, but it does not manage a single "encyclopedia" of archetypes. Without a singular standard collection (and, in some cases, not even a standard definition) of some variable star types, the generation of archetypes for all variable star types is impractical (completeness issue). Similarly, interpolation across functional shapes is a leading-edge technology.

## 8.3 Results

We outline the following developments of this research:

1. System Design and Performance of an Automated Classifier

(a) *Publication.* Johnston, K. B., & Oluseyi, H. M. (2017). Generation of a supervised classification algorithm for time-series variable stars with an application to the LINEAR dataset. *New Astronomy, 52*, 35–47.

(b) *LSC [ascl:1807.033].* Supervised classification of time-series variable stars Johnston, Kyle B.

   i. LSC (LINEAR Supervised Classification) trains a number of classifiers, including random forest and K-nearest neighbor, to classify variable stars and compares the results to determine which classifier is most successful. Written in R, the package includes anomaly detection code for testing the application of the selected classifier to new data, thus enabling the creation of highly reliable data sets of classified variable stars[5].

(c) Results

   i. We have demonstrated the construction and application of a supervised classification algorithm on variable star data. Such an algorithm will process observed stellar features and produce quantitative estimates of stellar class labels. Using a hand-processed (verified) data set derived from the ASAS, OGLE, and Hipparcos surveys, an initial training and testing set was derived.

   ii. The trained one-vs.-all algorithms were optimized using the testing data via minimization of the misclassification rate. From application of the trained algorithm to the testing data, performance estimates can be quantified for each one-vs.-all algorithm. The Random Forest supervised classification algorithm was found to be superior for the feature space and class space operated in which we operated.

[5]https://github.com/kjohnston82/LINEARSupervisedClassification

iii. Similarly, a one-class support vector machine was trained in a similar manner and designed as an anomaly detector.

iv. The design was applied to a set of 192,744 LINEAR data points. Of the original samples, setting the threshold of the RF classifier using a false alarm rate of 0.5%, 34,451 unique stars were classified only once in the one-vs.-all scheme and were not identified by the anomaly detection algorithm.

v. The total population is partitioned into the individual stellar variable classes; each subset of LINEAR ID corresponding to the matched patterns is stored in a separate file and accessible to the reader.

2. Novel Feature Space Implementation

   (a) *Publication.* Johnston, K. B., & Peter, A. M. (2017). Variable star signature classification using slotted symbolic Markov modeling. *New Astronomy, 50*, 1–11.

   (b) *Poster.* Johnston, K. B., & Peter, A. M. (2016). *Variable star signature classification using slotted symbolic Markov modeling.* Presented at AAS 227, Kissimmee, FL.

   (c) *SSMM[ascl:1807.032].* Slotted symbolic Markov modeling for classifying variable star signatures Johnston, Kyle B.; Peter, Adrian, M.

i. SSMM (slotted symbolic Markov modeling) reduces time-domain stellar variable observations to classify stellar variables. The method can be applied to both folded and unfolded data and does not require time warping for waveform alignment. Written in MATLAB, the performance of the supervised classification code is quantifiable and consistent, and the rate at which new data are processed is dependent only on the computational processing power available[6].

(d) Results

i. The SSMM methodology developed has been able to generate a feature space that separates variable stars by class (supervised classification). This methodology has the benefit of being able to accommodate irregular sampling rates, dropouts, and some degree of time-domain variance. It also provides a fairly simple methodology for feature space generation, necessary for classification.

ii. One of the major advantages of the methodology used is that a signature pattern (the transition state model) is generated and updated with new observations.

iii. The performance of four separate classifiers trained on the UCR data set is examined. It has been shown that the methodology presented is comparable to direct distance methods (UCR baseline). It is also shown that the methodology presented is more flexible.

[6]https://github.com/kjohnston82/SSMM

iv. The LINEAR data set provides more opportunity to demonstrate the proposed methodology. The larger class space, unevenly sampled data with dropouts, and color data all provide additional challenges to be addressed. After optimization, the misclassification rate is $\sim$1%, depending on the classifier implemented. An anomaly detection algorithm is trained and tested on the time series data and color data, and the combined algorithm has an expected misclassification rate of $\sim$0.07%.

3. Detector for O'Connell-Type EBs (Using Metric Learning and DF Features)

   (a) *Publication.* Johnston, K. B., et al. (2019). A detection metric designed for O'Connell effect eclipsing binaries. *Computational Astrophysics and Cosmology.* Manuscript in review.

   (b) *Poster.* Johnston, K. B., et al. (2018). *Learning a novel detection metric for the detection of O'Connell effect eclipsing binaries.* Presented at AAS 231, National Harbor, MD.

   (c) *OCD.* O'Connell Effect Detector Using Push-Pull Learning Johnston, Kyle B.; Haber, Rana

      i. OCD (O'Connell effect detector using push-pull learning) detects eclipsing binaries that demonstrate the O'Connell effect. This time-domain signature extraction methodology uses a supporting supervised pattern detection algorithm. The methodology maps stellar variable observations (time-domain data) to a new representation known as distribution fields (DF), the properties of which enable efficient handling of issues such as irregular sampling and multiple

values per time instance. Using this representation, the code applies a metric learning technique directly on the DF space capable of specifically identifying the stars of interest; the metric is tuned on a set of labeled eclipsing binary data from the Kepler survey, targeting particular systems exhibiting the O'Connell effect. This code is useful for large-scale data volumes such as that expected from next-generation telescopes like LSST[7].

(d) Results

i. A modular design is developed that can be used to detect types of stars and star systems that are cyclostationary in nature. With a change in feature space, specifically one that is tailored to the target signatures of interest and based on prior experience, this design can be replicated for other targets that do not demonstrate a cyclostationary signal (e.g., impulsive, nonstationary) and even to targets of interest that are not time-variable in nature but have a consistent observable signature (e.g., spectrum, photometry, image point-spread function).

ii. The method outlined here has demonstrated the ability to detect targets of interest given a training set consisting of expertly labeled light curve training data.

iii. The procedure presents two new functionalities: the DF, a shape-based feature space, and the Push-Pull Matrix Metric Learning

---

[7]https://github.com/kjohnston82/OCDetector

algorithm, a metric learning algorithm derived from LMNN that allows for matrix-variate similarity comparisons.

iv. A comparison to less novel, more standard methods was demonstrated on a Kepler eclipsing binary subdata set that was labeled by an expert in the field of O'Connell effect binary star systems.

v. The design is applied to Kepler eclipsing binary data and LINEAR data. Furthermore, the increase in the number of systems and the presentation of the data allow us to make additional observations about the distribution of curves and trends within the population.

4. Multi-view Classification of Variable Stars Using Metric Learning

   (a) *Paper.* Johnston, K. B., et al. (2019). Variable star classification using multi-view metric learning. *Computational Astrophysics and Cosmology.* Manuscript in review.

   (b) *Poster.* Johnston, K. B., et al. (2018). *Variable star classification using multi-view metric learning.* Presented at ADASS Conference XXVIII, College Park, MD.

   (c) *Java Project VariableStarAnalysis.* Contains Java translations of code designed specifically for analysis and the supervised classification of variable stars[8]:

   (d) Results

[8]https://github.com/kjohnston82/VariableStarAnalysis

i. Two features that have utility in providing discriminatory capabilities, the SSMM and DF feature spaces, are studied. The feature extraction methodologies are applied to the LINEAR and UCR data sets, as well as to a standard breakdown of time-domain descriptive statistics and, in the case of the LINEAR data set, a combination of *ugriz* colors.

ii. To support the set of high-dimensionality features, or views, multi-view metric learning is investigated as a viable design. Multi-view learning provides an avenue for integrating multiple transforms to generate a superior classifier.

iii. The structure of multi-view metric learning allows for a number of modern computational designs to be used to support increasing scale and scope (e.g., parallel computation); these considerations can be leveraged given the parameters of the experiment designed or the project in question.

iv. A new classifier design that does not require transformation of the matrix-variate views ahead of time is presented. This classifier operates directly on the matrix data.

v. The development of both algorithm designs (matrix-variate and vector-variate) with respect to the targeted experiment of interest (discrimination of time-domain variable stars) highlighted a number of challenges to be addressed prior to practical application. In overcoming these challenges, it was found that the novel classifier design (LM$^3$L-MV) performed on order of the staged (vectorization + ECVA + LM$^3$L) classifier.

Our efforts in astroinformatics have yielded code, identified previously unlabeled stellar variables, proved out a new feature space and classifier, and established methodologies to be used in future variable star identification efforts.

# Appendix A

# Chapter 4: Broad Class Performance Results

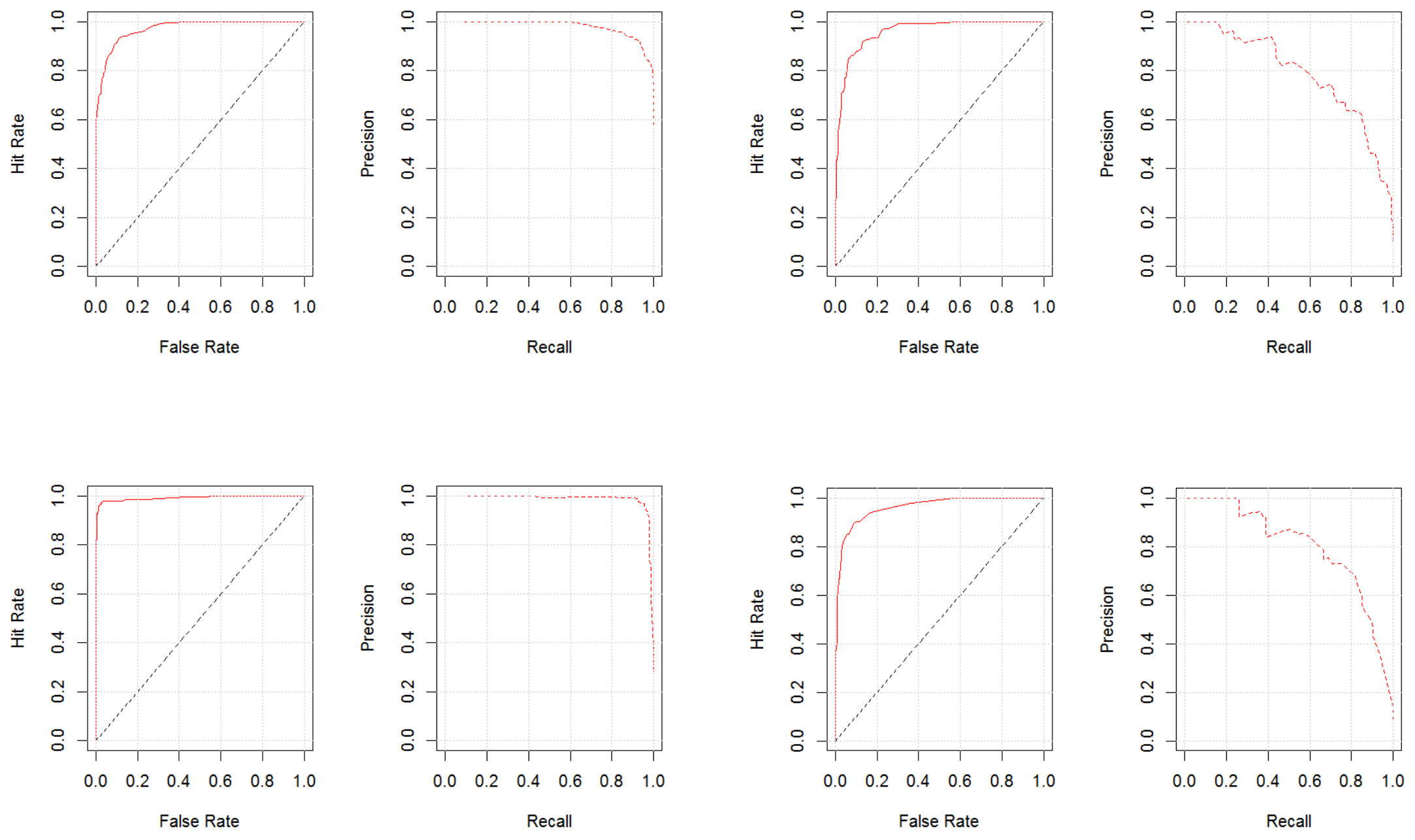

Figure A.1: Random Forest: mtry = 8, ntree = 100, (Top Left) Pulsating, (Top Right) Erupting, (Bottom Left) Multi-Star, (Bottom Right) Other

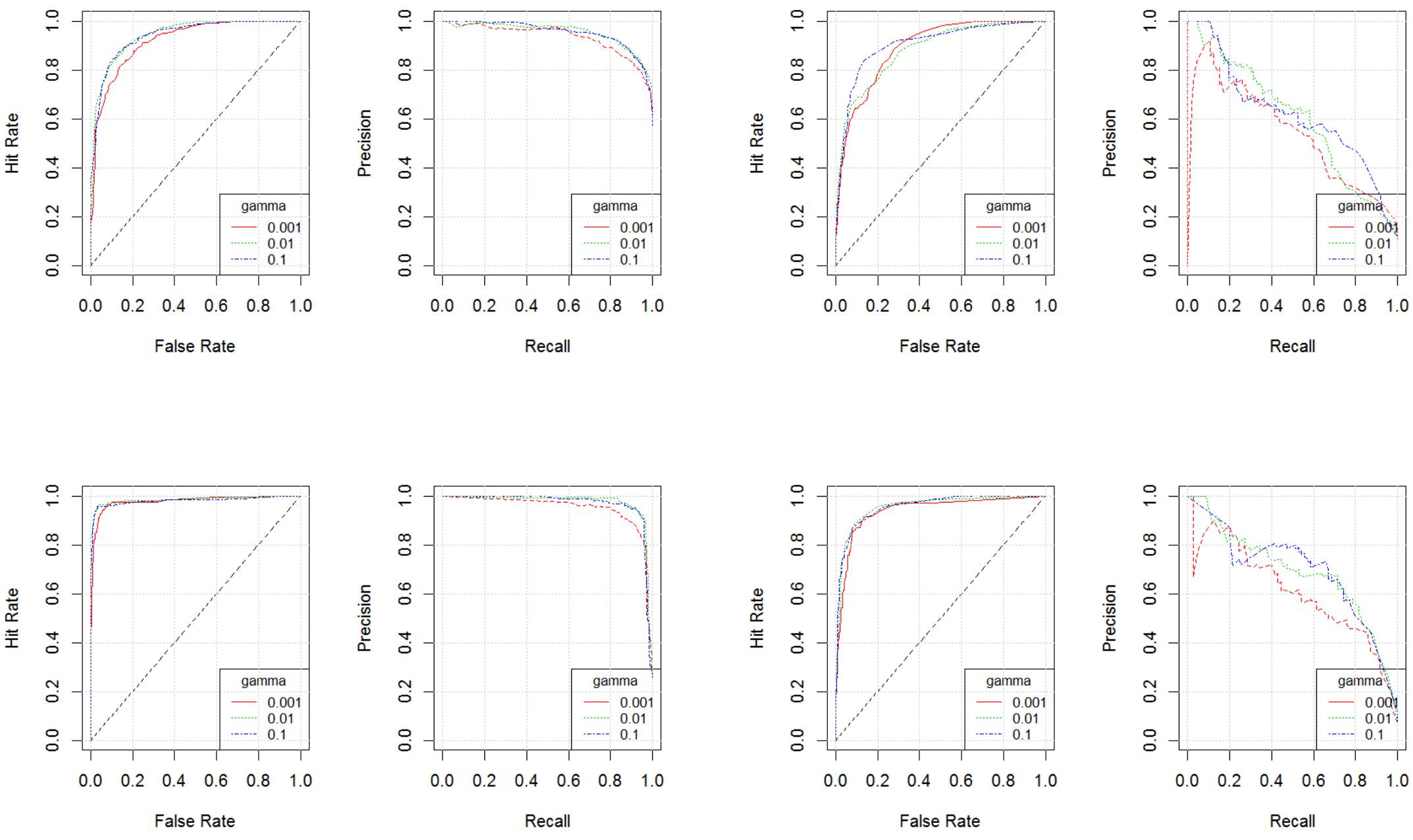

Figure A.2: SVM: (Top Left) Pulsating, (Top Right) Erupting, (Bottom Left) Multi-Star, (Bottom Right) Other

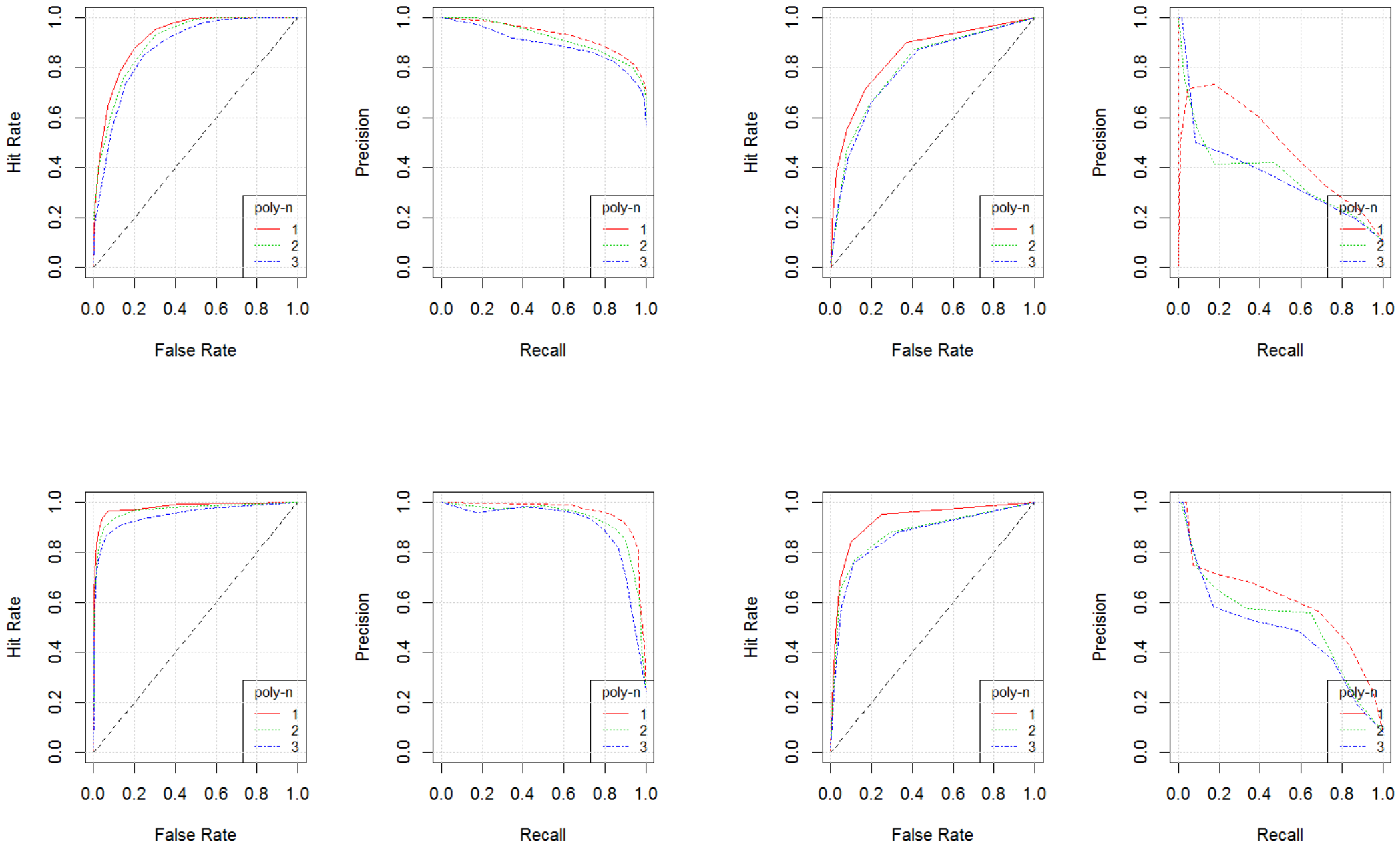

Figure A.3: kNN: (Top Left) Pulsating, (Top Right) Erupting, (Bottom Left) Multi-Star, (Bottom Right) Other

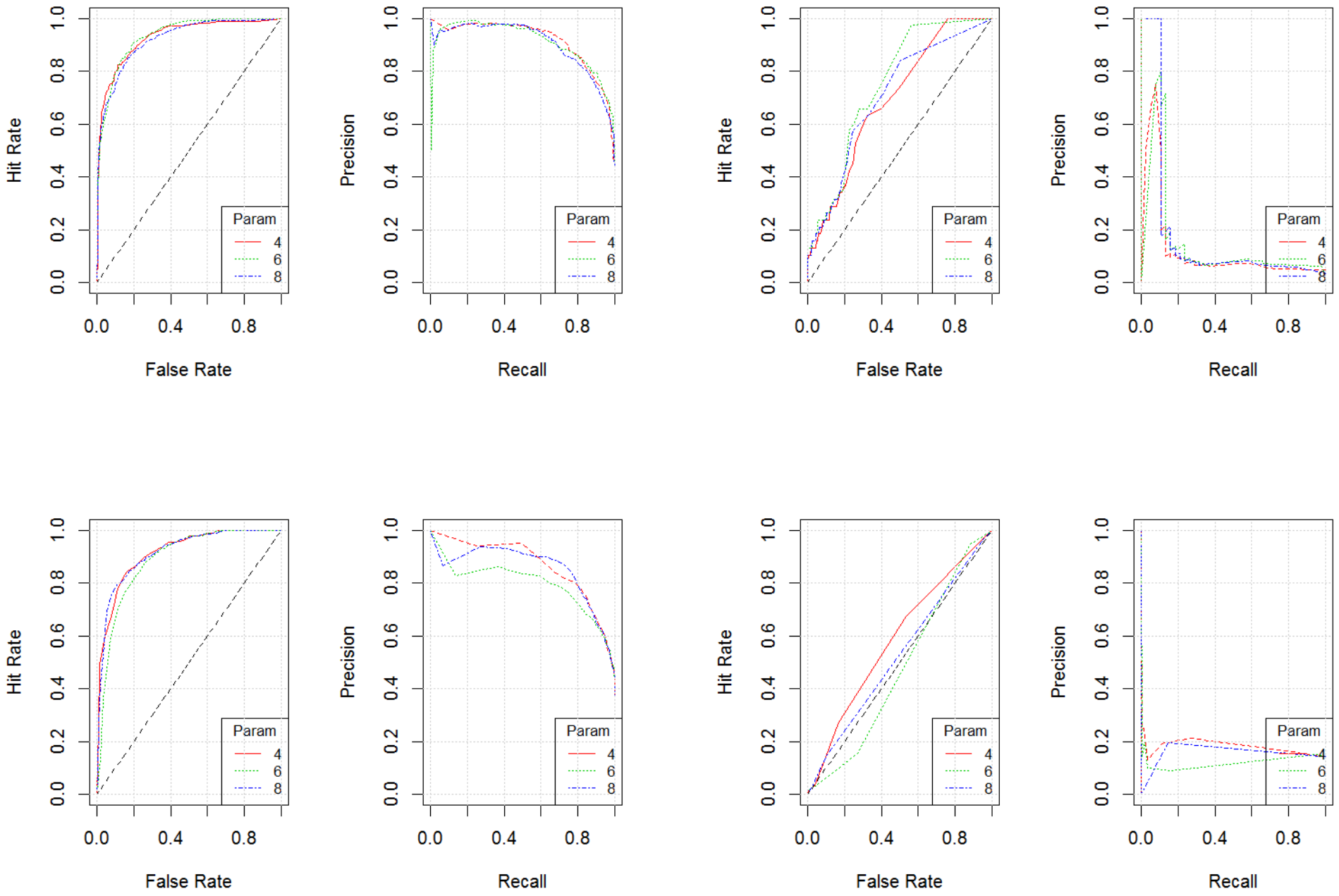

Figure A.4: MLP: (Top Left) Pulsating, (Top Right) Erupting, (Bottom Left) Multi-Star, (Bottom Right) Other

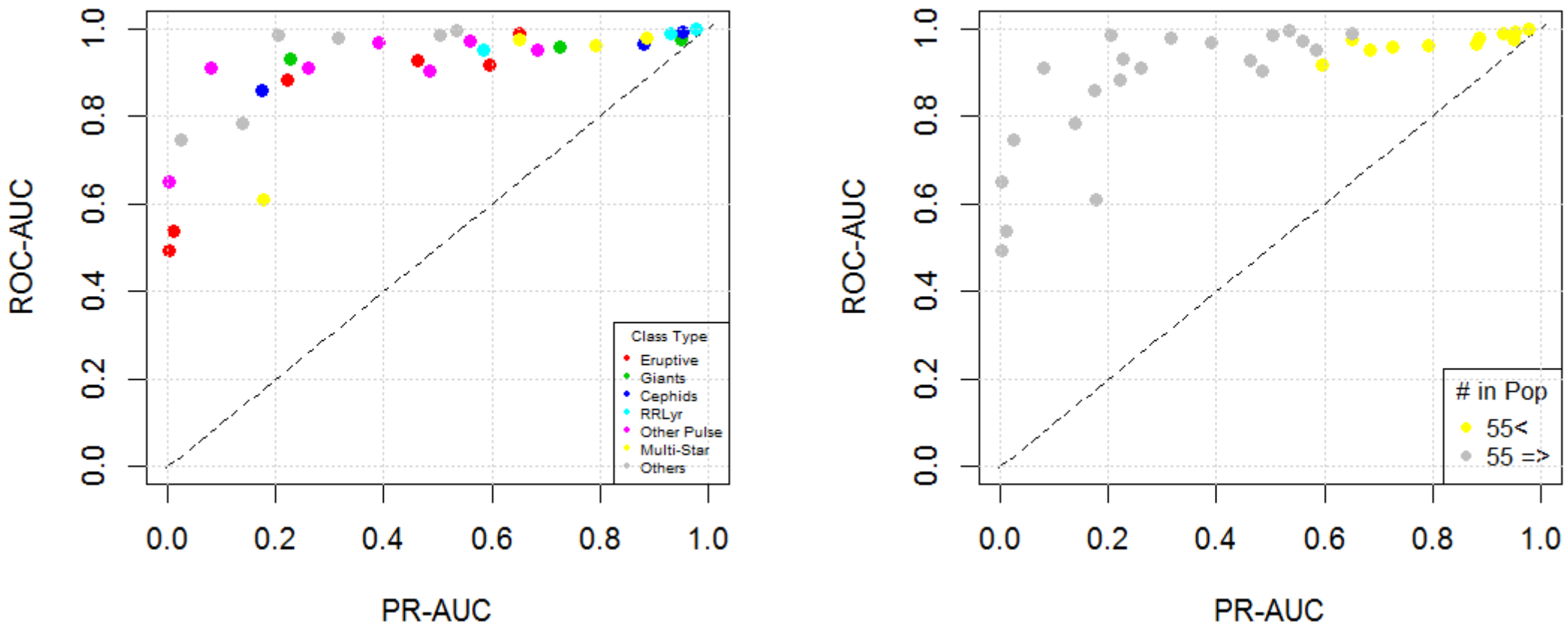

Figure A.5: MLP: Individual Classification, Performance Analysis

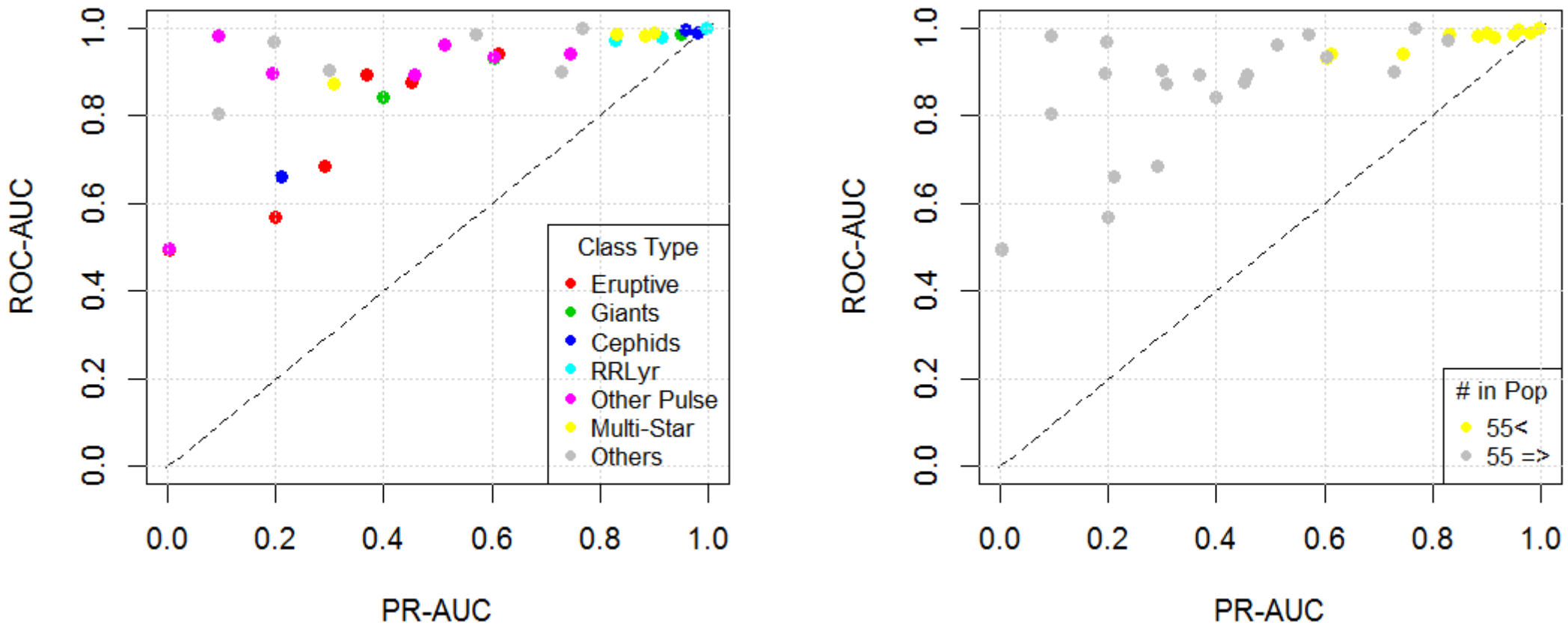

Figure A.6: kNN: Individual Classification, Performance Analysis

# Appendix B

# Chapter 5: Optimization Analysis Figures

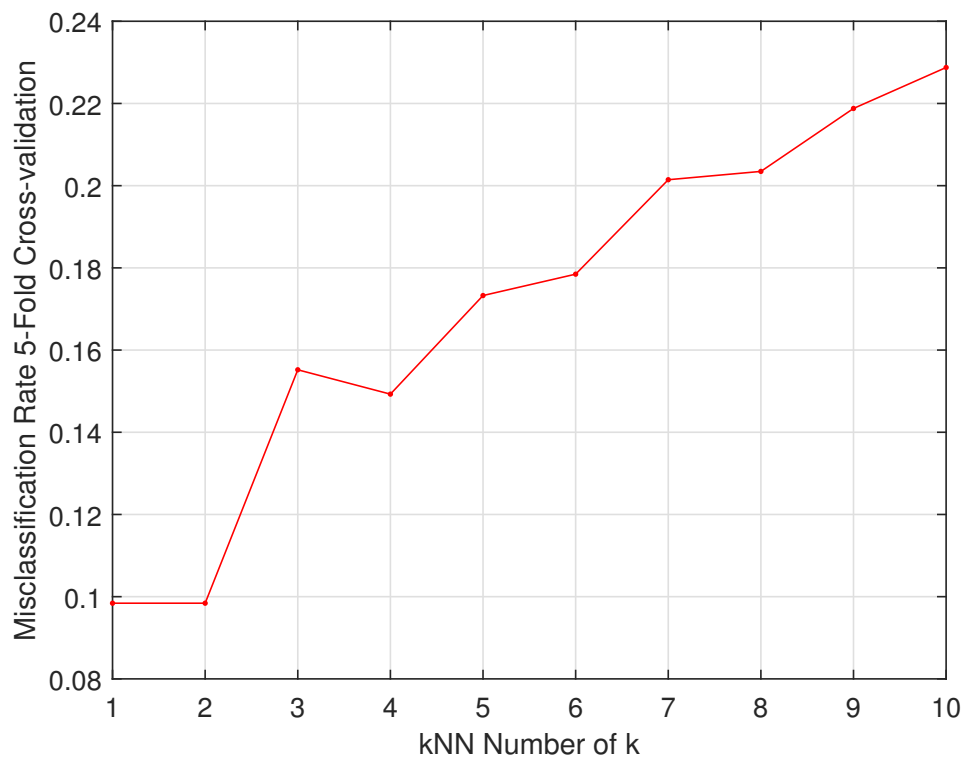

(a) Nearest Neighbor Classifiers

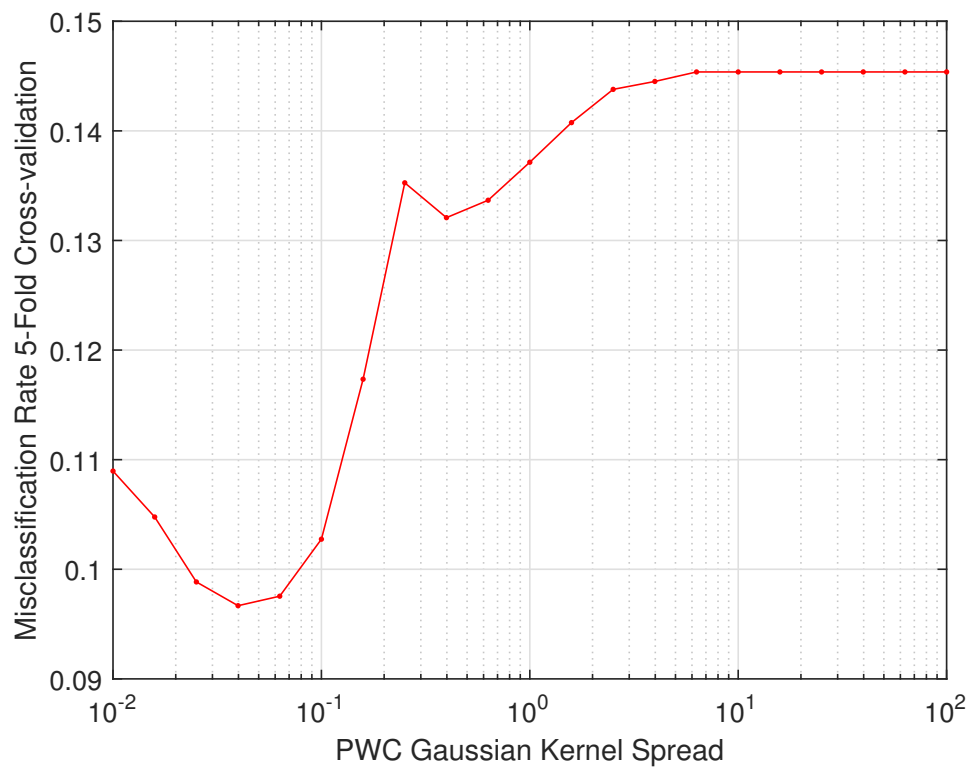

(b) Parzen Window Classifier

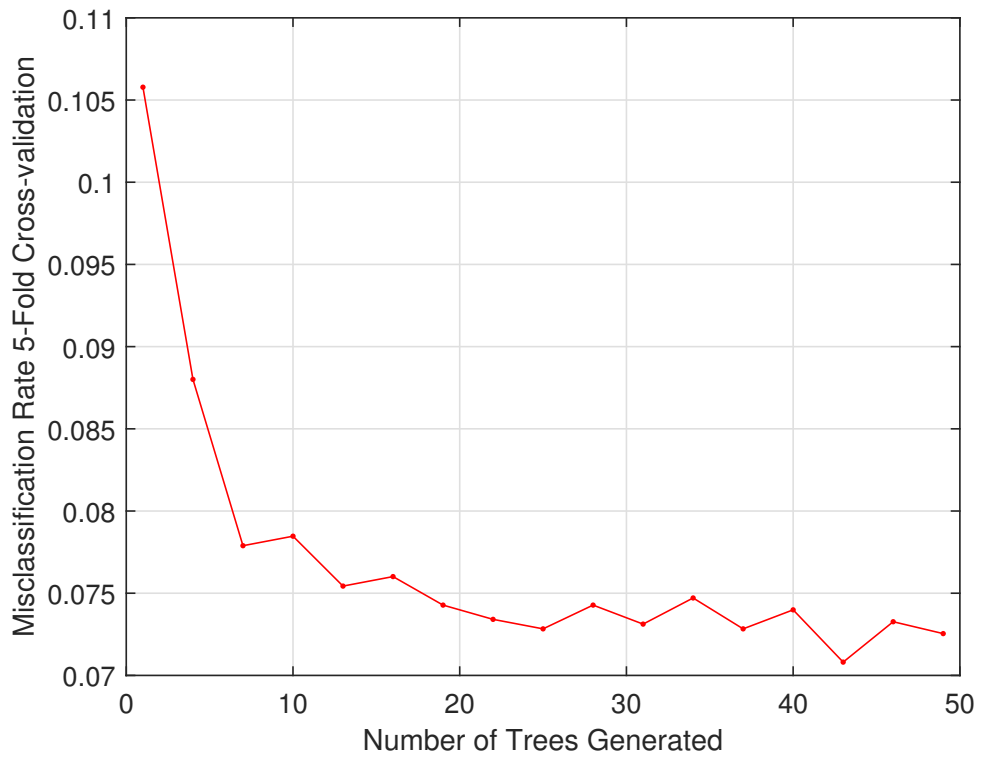

(c) Random Forest

Figure B.1: Classifier Optimization for UCR Data

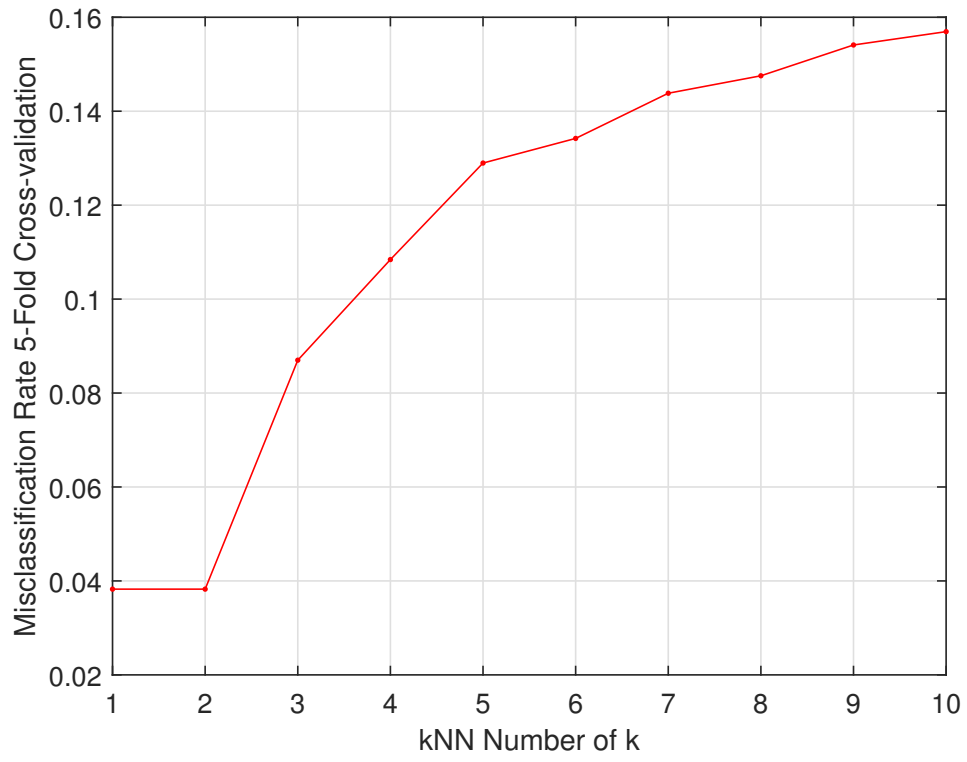

(a) Nearest Neighbor classifiers

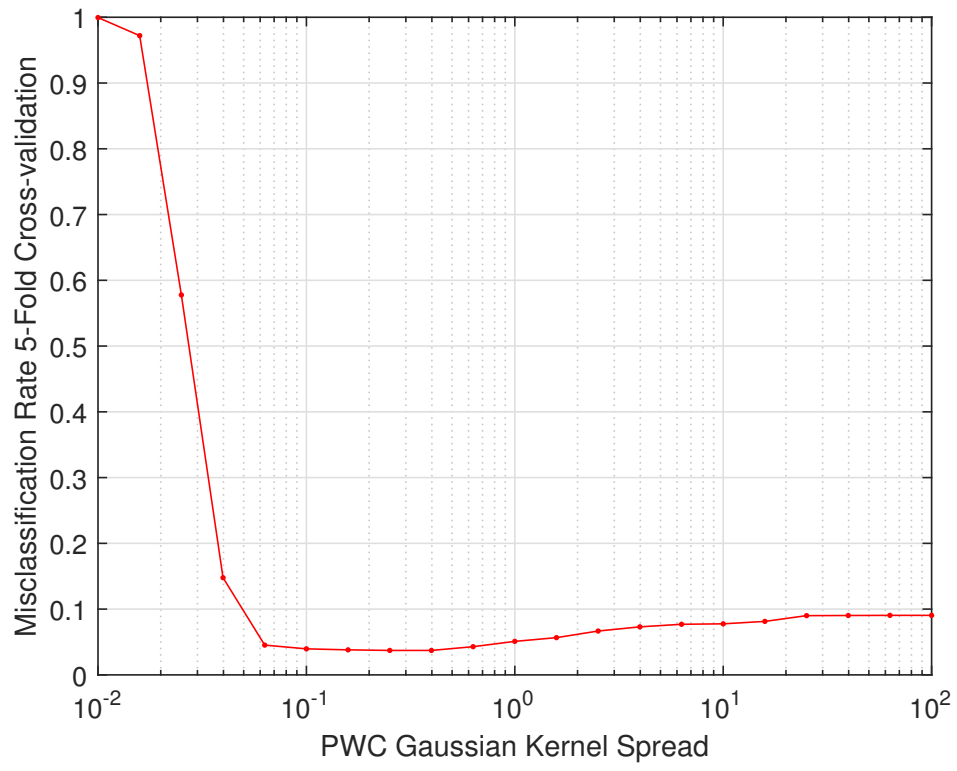

(b) Parzen window classifier

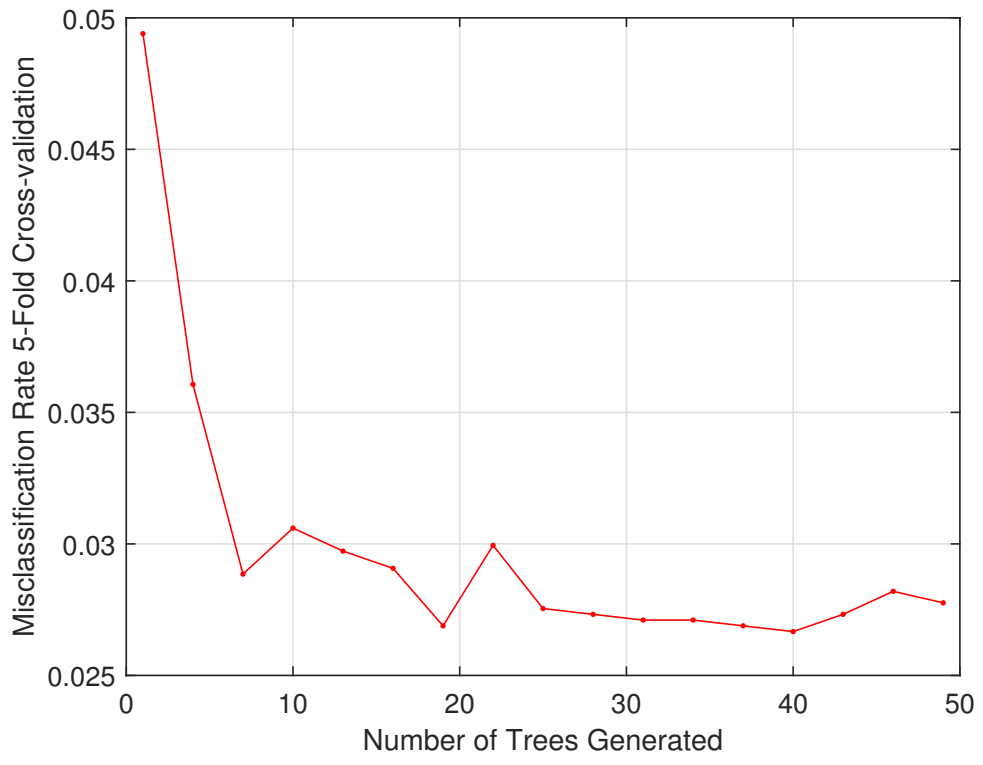

(c) Random Forest

Figure B.2: Classifier Optimization for LINEAR Data

## B.1 Chapter 5: Performance Analysis Tables

Table B.1: Confusion Matrix for Classifiers Based on UCR Starlight Data

(a) 1-NN

| True\Est | 1 | 2 | 3 |
|---|---|---|---|
| 1 | 0.86 | 0.003 | 0.13 |
| 2 | 0.0 | 0.99 | 0.008 |
| 3 | 0.031 | 0.002 | 0.97 |

(b) PWC

| True\Est | 1 | 2 | 3 |
|---|---|---|---|
| 1 | 0.82 | 0.003 | 0.18 |
| 2 | 0.00 | 0.97 | 0.035 |
| 3 | 0.16 | 0.004 | 0.84 |

(c) RF

| True\Est | 1 | 2 | 3 |
|---|---|---|---|
| 1 | 0.91 | 0.003 | 0.082 |
| 2 | 0.0 | 0.99 | 0.005 |
| 3 | 0.004 | 0.0007 | 0.99 |

Table B.2: Confusion Matrix for Classifiers Based on LINEAR Starlight Data

(a) 1-NN

| True\Est | Algol | Contact Binary | Delta Scuti | No Variation | RRab | RRc |
|---|---|---|---|---|---|---|
| Algol | 0.76 | 0.20 | 0.0 | 0.0 | 0.0 | 0.04 |
| Contact Binary | 0.03 | 0.95 | 0.005 | 0.005 | 0.01 | 0.0 |
| Delta Scuti | 0.0 | 0.0 | 0.88 | 0.12 | 0.0 | 0.0 |
| No Variation | 0.0 | 0.0 | 0.01 | 0.99 | 0.0 | 0.0 |
| RRab | 0.0 | 0.005 | 0.0 | 0.0 | 0.95 | 0.045 |
| RRc | 0.0 | 0.03 | 0.0 | 0.0 | 0.14 | 0.83 |

(b) PWC

| True\Est | Algol | Contact Binary | Delta Scuti | No Variation | RRab | RRc |
|---|---|---|---|---|---|---|
| Algol | 0.97 | 0.01 | 0.0 | 0.0 | 0.02 | 0.0 |
| Contact Binary | 0.0 | 0.99 | 0.0 | 0.0 | 0.0 | 0.01 |
| Delta Scuti | 0.0 | 0.0 | 0.94 | 0.06 | 0.0 | 0.0 |
| No Variation | 0.0 | 0.0 | 0.0 | 1.0 | 0.0 | 0.0 |
| RRab | 0.0 | 0.01 | 0.0 | 0.0 | 0.99 | 0.0 |
| RRc | 0.0 | 0.01 | 0.0 | 0.0 | 0.0 | 0.99 |

(c) RF

| True\Est | Algol | Contact Binary | Delta Scuti | No Variation | RRab | RRc |
|---|---|---|---|---|---|---|
| Algol | 0.93 | 0.07 | 0.0 | 0.0 | 0.0 | 0.04 |
| Contact Binary | 0.0 | 0.99 | 0.0 | 0.0 | 0.0 | 0.0 |
| Delta Scuti | 0.0 | 0.0 | 0.94 | 0.0 | 0.0 | 0.06 |
| No Variation | 0.0 | 0.02 | 0.0 | 0.98 | 0.0 | 0.0 |
| RRab | 0.0 | 0.0 | 0.0 | 0.0 | 1.0 | 0.05 |
| RRc | 0.0 | 0.0 | 0.0 | 0.0 | 0.0 | 1.0 |

# Appendix C

# Chapter 7: Additional Performance Comparison

Table C.1: LINEAR confusion matrix, $LM^3L$-MV - $LM^3L$

| Misclassification Rate | RR Lyr (ab) | Delta Scu / SX Phe | Algol | RR Lyr (c) | Contact Binary | Missed |
|---|---|---|---|---|---|---|
| RR Lyr (ab) | -7 | 0 | 1 | 9 | -1 | -2 |
| Delta Scu / SX Phe | 1 | 1 | 0 | 0 | -2 | 0 |
| Algol | 2 | 0 | -4 | 1 | 1 | 0 |
| RR Lyr (c) | -1 | 0 | 0 | 2 | -1 | 0 |
| Contact Binary | 0 | 0 | 2 | -8 | 5 | 1 |

Table C.2: UCR confusion matrix, LM$^3$L-MV - LM$^3$L

| Misclassification Rate | 2 | 3 | 1 | Missed |
|---|---|---|---|---|
| 2 | 2 | -3 | 0 | 1 |
| 3 | -13 | -171 | 184 | 0 |
| 1 | -5 | 92 | -87 | 0 |